\documentclass[11pt,a4paper]{article}
\usepackage{isabelle,isabellesym}

\usepackage{mdwlist}
\usepackage{amssymb}
\usepackage{amsmath}
\usepackage{bbding} 
\usepackage{pifont} 
\usepackage{stmaryrd} 
\usepackage{graphicx}
\usepackage{pdfsetup}

\isabellestyle{it}

\begin{document}

\newcommand{\idep}{\mathbb{I}^\mathcal{D}}
\newcommand{\odep}{\mathbb{O}^\mathcal{D}}
\newcommand{\instreams}{\mathbb{I}}
\newcommand{\outstreams}{\mathbb{O}}

\title{Formalisation and Analysis\\ of Component Dependencies}
\author{Maria Spichkova}
\maketitle

\begin{abstract}
This set of theories presents a formalisation in Isabelle/HOL+Isar of data dependencies between components. 
The approach allows to analyse system structure oriented towards efficient checking of system: 
it aims at elaborating for a concrete system, which parts of 
the system (or system model) are necessary to check a given property. 
\end{abstract}
\tableofcontents

\newpage
\section{Introduction}

The set of theories presented in this paper is an Isabelle/HOL+Isar \cite{npw,IsabelleManual} formalisation of data dependencies between components. This paper is organised as follows: first of all we give a  general introduction to our
approach for analyse system structure analysis oriented towards efficient checking of system: 
it aims at elaborating for a concrete system, which parts of 
the system (or system model) are necessary to check a given property. 
After that we present the Isabelle/HOL representation of these concepts and a small case study, 
where the dependency properties are verified formally using the Isabelle theorem prover  also applying its component Sledgehammer \cite{Sledgehammer,SledgehammerSpass}. 
 
In general,  we don't need  complete information about the system as  
to check its certain property.  An additional information about the system can slow  the whole process
down or even make it infeasible. In this theory we define constraints that allow to find/check
the minimal model (and the minimal extent of the system)
needed to verify a specific property.  
Our approach focuses on data
dependencies between system components. Dependencies' analysis results in a
decomposition that gives rise to a logical system architecture, which is the most appropriate for the
case of remote monitoring, testing and/or verification.

Let  $CSet$  be a set of components  
on a certain abstraction level $L$ of logical architecture 
(i.e. level of refinement/decom\-po\-sition, data type \emph{AbstrLevelsID} in our Isabelle formalisation).
We denote the sets of input and
output streams of a component $S$ by $\instreams(S)$ (function \emph{ IN ::  CSet $\Rightarrow$ chanID set} in Isabelle) and
$\outstreams(S)$ (function \emph{ OUT ::  CSet $\Rightarrow$ chanID set} in Isabelle).
The set of local variables of components is defined in Isabelle by VAR, and the function to map component identifiers to  the corresponding variables is defined by 
\emph{ VAR ::  CSet $\Rightarrow$ varID set}. 

Please note that concrete values for these functions cannot be specified in general, because they strongly depend on a concrete system. 
In this paper we present a small case study in the theories \emph{DataDependenciesConcreteValues.thy} (specification of the system architecture on several abstraction levels) 
and \emph{DataDependenciesCaseStudy.thy} (proofs of system architectures' properties).

Function \emph{ subcomp ::  CSet $\Rightarrow$ CSet set}  maps components to a (possibly empty) set of its subcomponents. 

We specify the components' dependencies by the  function 
\[
Sources^{L}: CSet^{L} \to  (CSet^{L})^*
\]
which returns for any component identifier $A$ the corresponding (possibly
empty) list of components (names) $B_1, \dots, B_{AN}$ that are the
sources for the input data streams of $A$ (direct or indirect):
\[
\begin{array}{l}
Sources^{L}(C) = \\
DSources^{L}(C)\ \cup\  \bigcup_{S \in  DSources^{L}(C)} \{  S_1 \mid S_1 \in  Sources^{L}(S) \} 
\end{array}
\]   
Direct data dependencies are defined by the function 
\[
DSources^{L}: CSet^{L} \to  (CSet^{L})^*
\]
\[ 
DSources^{L}(C) = 
\{  S \mid  \exists x \in \instreams(C) \wedge x \in \outstreams(S) 
\}  
\]
 %
For example, 
$C_1 \in DSources^{L}(C_2)$ means that at least one of the output channels of  $C_1$ 
is directly connected to some of input channels of   $C_2$.

$\idep(C, y)$ denotes the subset of $\instreams(C)$ that output channel $y$ depends upon, 
directly (specified in Isabelle by function \emph{ OUTfromCh::  chanID $\Rightarrow$ chanID set}
  or vial local variables (specified by function \emph{ OUTfromV::  chanID $\Rightarrow$ varID set}). 
For example, let the values of the output
channel $y$ of component $C$ depend only on the
value of the local variable $st$ that represents the current state of
$C$ and is updated depending to the input messages
the component receives via the channel $x$, then $\idep(C, y) = \{ x \}$.   
In Isabelle, $\idep(C, y)$  is specified by function \emph{ OUTfrom::  chanID $\Rightarrow$ varID set}.

Based on the definition above, we can decompose system's components to have for each component's output channel 
the minimal subcomponent  computing the corresponding results (we call them \emph{elementary components}). 
An elementary component either 
\begin{itemize}
\item
should have a single output channel (in this case this component can have no local variables), 
or 
\item
all it output channels are correlated, i.e. mutually depend on the same local variable(s).
\end{itemize}
If after these steps a single component is too complex, we can apply the
decomposition strategy presented in~\cite{spichkova2011decomp}. 
The result of the decomposition can be seen as a compositional refinement of the system \cite{broy_refinement2}. 
 
For any component $C$, the dual function $\odep$ returns the
corresponding set $\odep(C,x)$ of output channels depending on input
$x$. This is useful for tracing, e.g., 
if there are some changes in the specification, properties, constraints, etc.\ for $x$, 
we can trace which other channels can be affected by these changes.

\begin{figure}[ht!]
  \begin{center}
   \includegraphics[scale=0.15]{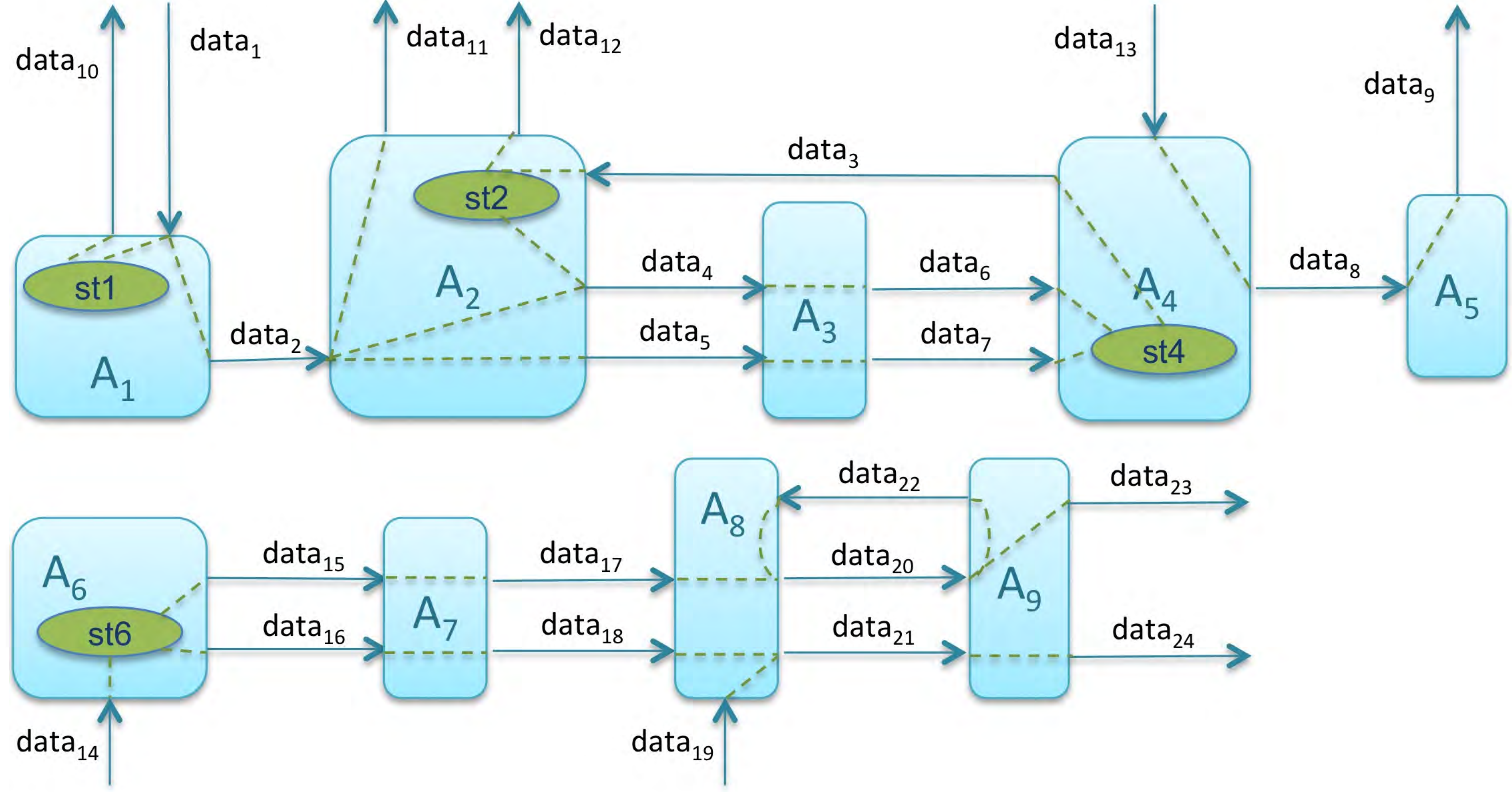}%
   \vspace{-3mm}
    \caption{System  $S$: Data dependencies and $\idep$ sets 
    }
    \label{fig:example_comm1}
  \end{center}
\end{figure}

If the input part of the component's interface is specified correctly in the
sense that the component does not have any ``unused'' input channels, the following relation
will hold: $ \forall x \in \instreams(C). ~ \odep(C, x) \neq
\emptyset $. 
We illustrate the presented ideas by a small case study:   
we show how system's components can be decomposed to optimise
the data dependencies within each single component, and after that we optimise architecture of the whole system. 
System
$S$ (cf.\ also Fig.~\ref{fig:example_comm1}) has 5 components, the set $CSet$ on the level $L_{0}$ is
defined by $\{A_1, \dots, A_9\}$. 
The sets $\idep$ of data dependencies between the components are defined in the theory \emph{DataDependenciesConcreteValues.thy}. 
We represent the dependencies graphically using dashed lines over the component box.  \\~

 ~\\

\begin{figure}[ht!]
  \begin{center}
   \includegraphics[scale=0.15]{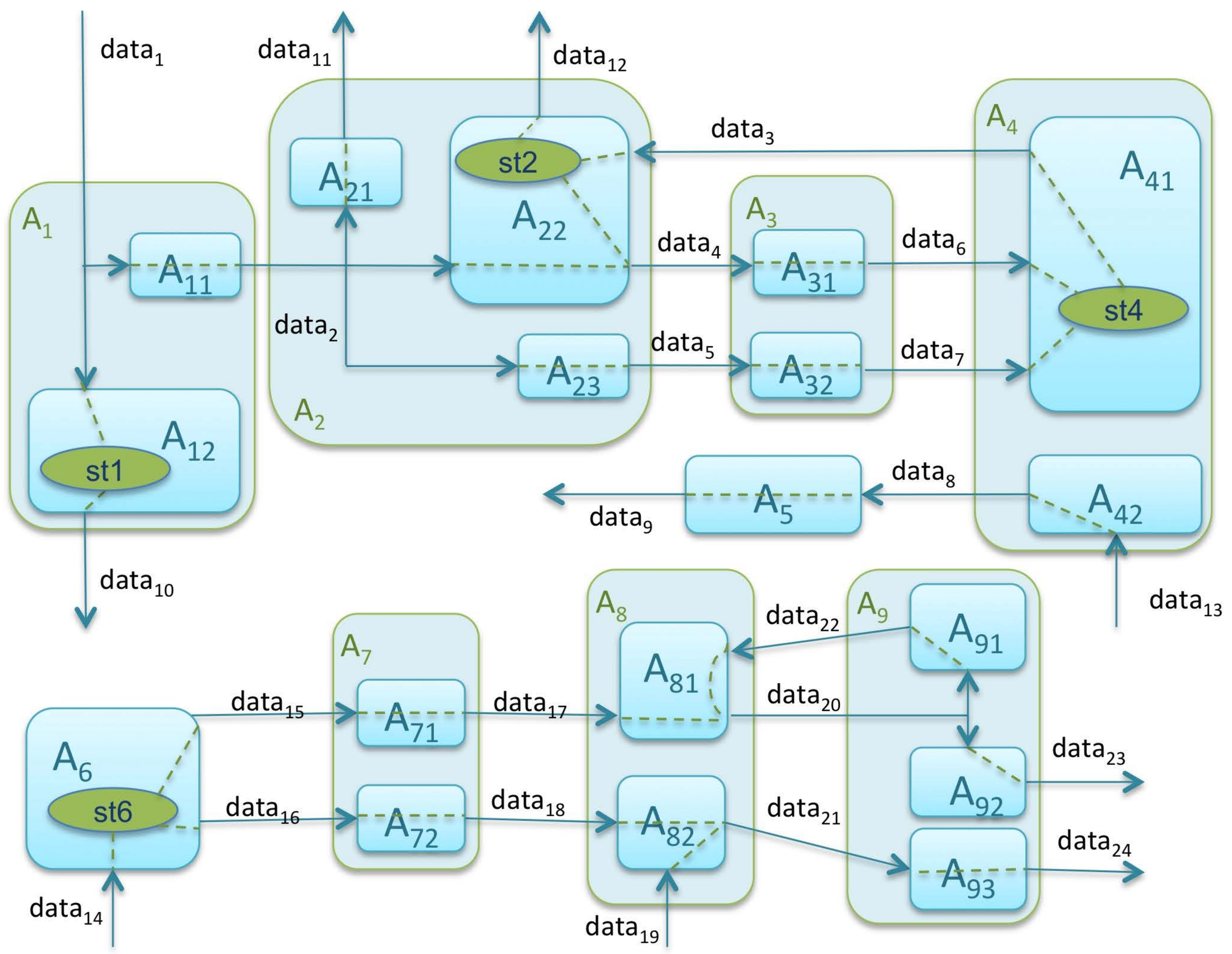}
      \vspace{-2mm}
    \caption{%
    Components' decomposition (level $L_{1}$)}
    \label{fig:example_comm2}
  \end{center}
\end{figure}

\begin{figure}[ht!]
  \begin{center}
   \includegraphics[scale=0.15]{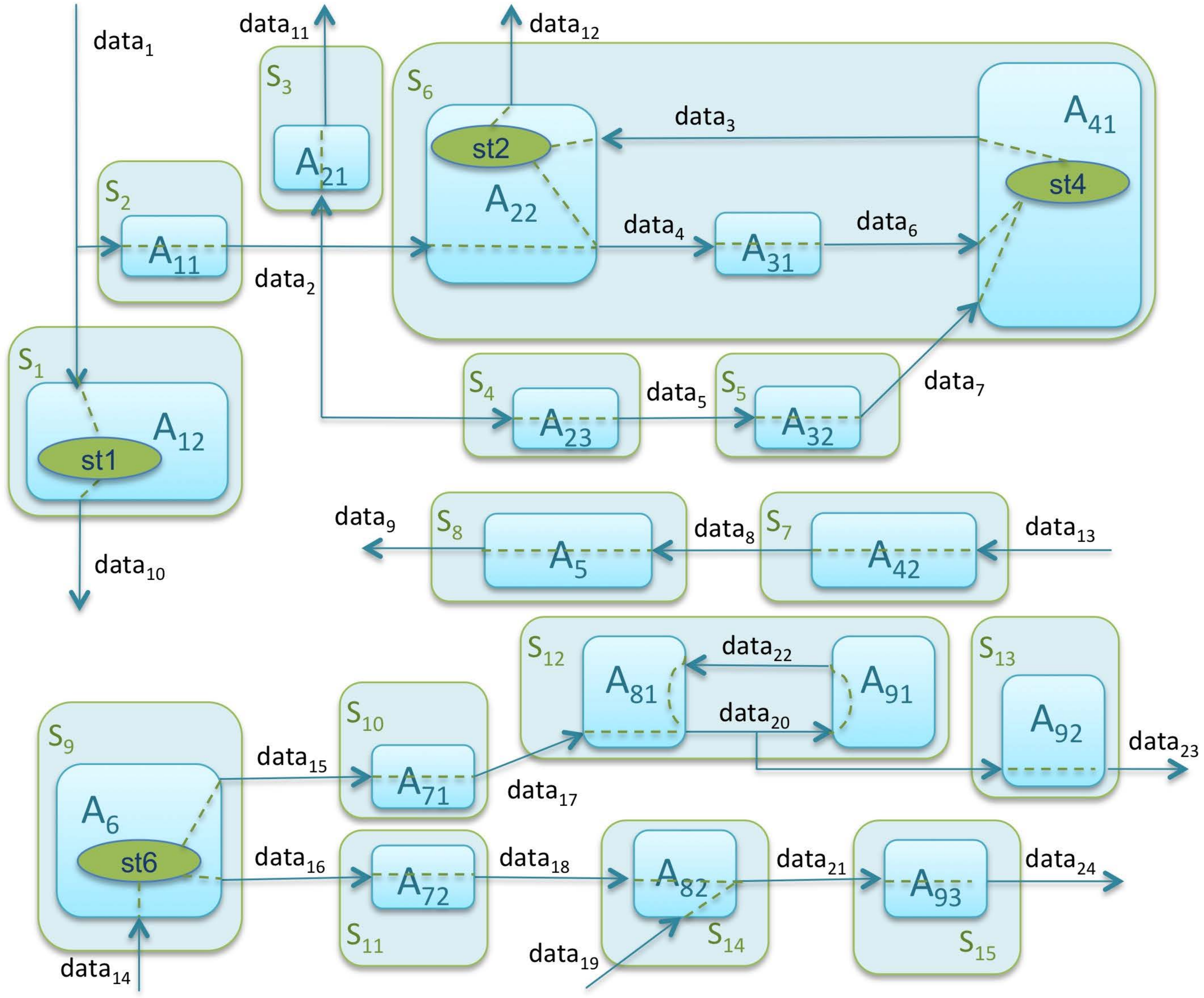}
    \caption{%
    Architecture of $S$ (level $L_{2}$)}
    \label{fig:L2a}
  \end{center}
\end{figure}

 \newpage 
 \noindent
 Now we can decompose the system's components according to the given  $\idep$ specification. This results into the next abstraction level $L_{1}$ of logical architecture (cf. Fig.~\ref{fig:example_comm2}), on which all components are elementary. Thus, we obtain a  (flat) architecture of system. 
 The main feature of this architecture is that  each output channel (within the system) 
belongs the minimal subcomponent of a system computing the corresponding results. 
 We represent this (flat) architecture  as a directed graph (components become vertices and channels become edges)  
and apply one of the existing distributed algorithms for the 
decomposition into its   strongly connected components,  e.g. FB~\cite{idetifyingSSCs}, OBF~\cite{OBF}, or the colouring algorithm~\cite{Orzan04ondistributed}.
Fig.~\ref{fig:L2a} presents the result of the architecture optimisation.

After optimisation of system's architecture, we can find the minimal part of the system
needed to check a specific property (cf. theory \emph{DataDependencies}).  
A property can be represented  by relations over data flows on the system's channels, 
and first of all we should check the property itself, whether it reflect a real relation within a system. 
Let for a relation $r$, $I_{r}$ $O_{r}$ be the sets of input and output channels of the system used in this relation.
For each channel from $O_{r}$ we recursively compute all the sets of the dependent components and corresponding 
input channels. 
Their union, restricted to the input channels of the system, 
 should be equal to $I_{r}$, otherwise we should check whether the property was specified correctly. 

Thus, from $O_{r}$ we obtain the set $outSetOfComponents$ of components having these channels as outputs, and
compute the union of corresponding sources' sets.
This union together with $outSetOfComponents$ give us  the minimal part of the system 
needed to check the property $r$: we formalise it in Isabelle  by the predicate $minSetOfComponents$.  

On the verification level this formalization is combinable with the Isabelle/HOL+Isar formalisation  of stream processing components \cite{FocusStreamsCaseStudies-AFP}, which aim is analysis of functional properties of systems and its components also using the idea of refinement-based verification \cite{spichkova2008refinement}.

For each channel and elementary component 
(i.e. for any component on the abstraction level $L_{1}$) we specify the following measures:
\begin{itemize}
\item
measure for costs of the data transfer/ upload to the cloud \emph{UplSize(f)}:\\
 size of  messages (data packages) within a data flow $f$ and  frequency they are produced. 
This measure can be defined on the level of logical modelling, 
where we already know the general type of the data and 
can also analyse the corresponding  component (or environment) model to estimate the frequency the data are produced;
\item
measure 
for requirement of using high-performance computing and cloud virtual machines, \emph{Perf(X)}: 
complexity of the computation within a component $X$, which can be estimated on the  level of logical modelling as well.  
\end{itemize}
On this basis, we build a system architecture, optimised for remote computation.  
The \emph{UplSize} measure should be analysed only for the channels that aren't local for the components on abstraction levels $L_{2}$ and $L_{3}$.

Using graphical representation, we denote the channels with \emph{UplSize} measure higher than a predefined value by thick red arrows (cf. also set \emph{UplSizeHighLoad} in
Isabelle theory \emph{DataDependenciesConcreteValues.thy}), and 
the components with  \emph{Perf} measure higher than a predefined value by   light green colour (cf. also set \emph{HighPerfSet} in
Isabelle theory \emph{DataDependenciesConcreteValues.thy}), 
where all other channel and components are marked  blue. 

\begin{figure}[ht!]
  \begin{center}
   \includegraphics[scale=0.15]{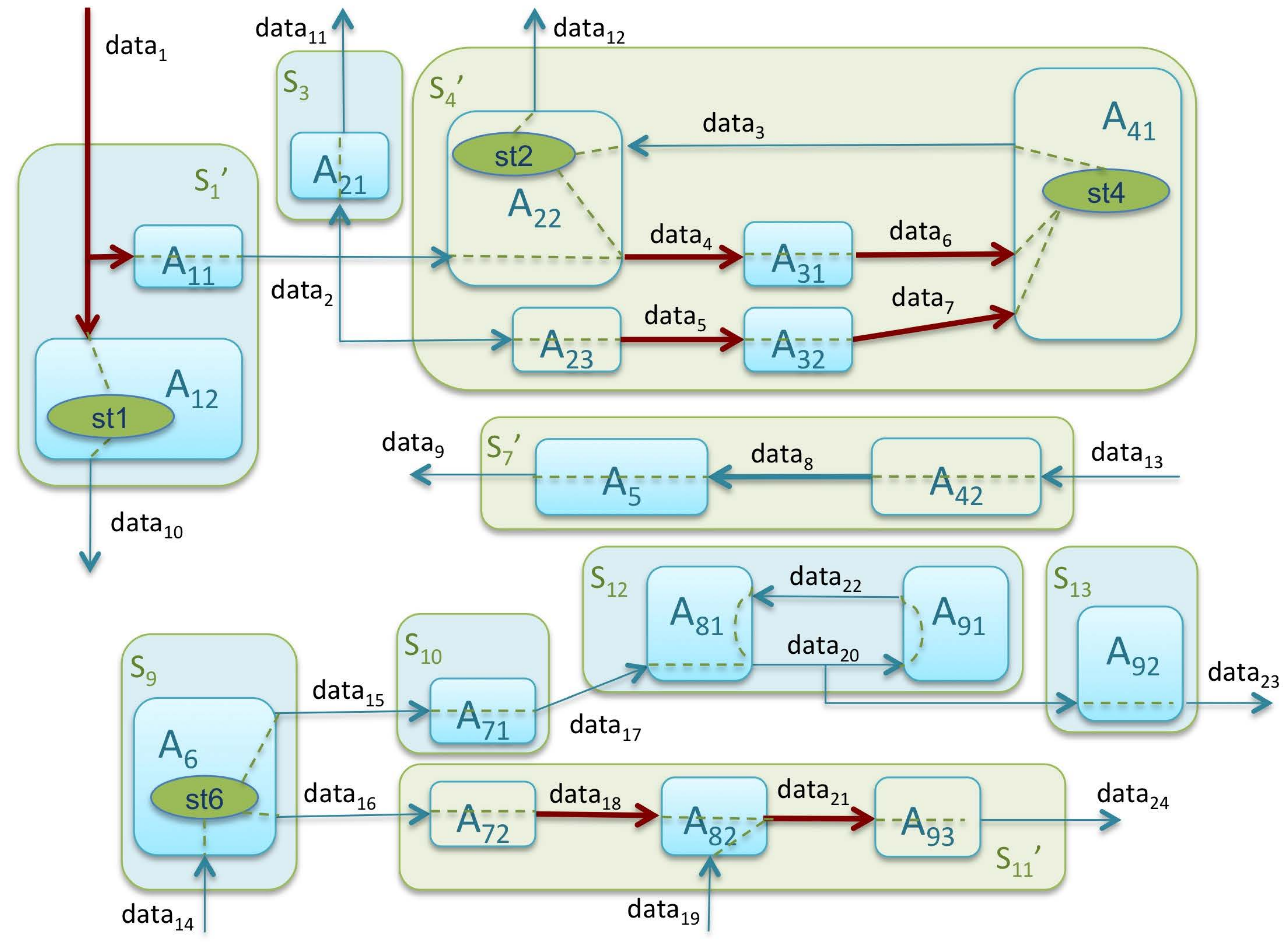}
    \caption{%
    Optimised architecture of $S$  (Level $L_{3}$)}
    \label{fig:remote}
  \end{center}
\end{figure}

Fig.~\ref{fig:remote} represents a system architecture, optimised for remote computation:
components from the abstraction level $L_{2}$ are composed together on the abstraction level $L_{3}$, if they are connected by at least one channel
 with \emph{UplSize} measure higher than a predefined value. 
The components $S_{4}'$ and $S_{7}'$ have  \emph{Perf} measure higher than a predefined value, i.e. using high-performance computing and cloud virtual machines is required.
\\
~
 
This approach can be used as a basis for the abstract modelling level within the development of cyber-physical systems, suggested in our previous work \cite{Spichkova_Campetelli2012,issec_cps2013}.

\parindent 0pt\parskip 0.5ex

\newpage
\begin{isabellebody}%
\def\isabellecontext{DataDependenciesConcreteValues}%
\isamarkupheader{Case Study: Definitions%
}
\isamarkuptrue%
\isadelimtheory
\endisadelimtheory
\isatagtheory
\isacommand{theory}\isamarkupfalse%
\ DataDependenciesConcreteValues\isanewline
\ \ \isakeyword{imports}\ Main\isanewline
\isakeyword{begin}%
\endisatagtheory
{\isafoldtheory}%
\isadelimtheory
\endisadelimtheory
\isanewline
\isanewline
\isacommand{datatype}\isamarkupfalse%
\ CSet\ {\isacharequal}\ sA{\isadigit{1}}{\isacharbar}\ sA{\isadigit{2}}{\isacharbar}\ sA{\isadigit{3}}{\isacharbar}\ sA{\isadigit{4}}{\isacharbar}\ sA{\isadigit{5}}{\isacharbar}\ sA{\isadigit{6}}{\isacharbar}\ sA{\isadigit{7}}{\isacharbar}\ sA{\isadigit{8}}{\isacharbar}\ sA{\isadigit{9}}{\isacharbar}\isanewline
\ \ \ \ \ \ \ \ \ \ \ \ \ \ \ sA{\isadigit{1}}{\isadigit{1}}{\isacharbar}\ sA{\isadigit{1}}{\isadigit{2}}{\isacharbar}\ sA{\isadigit{2}}{\isadigit{1}}{\isacharbar}\ sA{\isadigit{2}}{\isadigit{2}}{\isacharbar}\ sA{\isadigit{2}}{\isadigit{3}}{\isacharbar}\ \ sA{\isadigit{3}}{\isadigit{1}}{\isacharbar}\ sA{\isadigit{3}}{\isadigit{2}}{\isacharbar}\ sA{\isadigit{4}}{\isadigit{1}}{\isacharbar}\ sA{\isadigit{4}}{\isadigit{2}}{\isacharbar}\isanewline
\ \ \ \ \ \ \ \ \ \ \ \ \ \ \ sA{\isadigit{7}}{\isadigit{1}}{\isacharbar}\ sA{\isadigit{7}}{\isadigit{2}}{\isacharbar}\ sA{\isadigit{8}}{\isadigit{1}}{\isacharbar}\ sA{\isadigit{8}}{\isadigit{2}}{\isacharbar}\ sA{\isadigit{9}}{\isadigit{1}}{\isacharbar}\ sA{\isadigit{9}}{\isadigit{2}}{\isacharbar}\ sA{\isadigit{9}}{\isadigit{3}}{\isacharbar}\isanewline
\ \ \ \ \ \ \ \ \ \ \ \ \ \ \ sS{\isadigit{1}}{\isacharbar}\ sS{\isadigit{2}}{\isacharbar}\ sS{\isadigit{3}}{\isacharbar}\ sS{\isadigit{4}}{\isacharbar}\ sS{\isadigit{5}}{\isacharbar}\ sS{\isadigit{6}}{\isacharbar}\ sS{\isadigit{7}}{\isacharbar}\ sS{\isadigit{8}}{\isacharbar}\ sS{\isadigit{9}}{\isacharbar}\ sS{\isadigit{1}}{\isadigit{0}}{\isacharbar}\ sS{\isadigit{1}}{\isadigit{1}}{\isacharbar}\isanewline
\ \ \ \ \ \ \ \ \ \ \ \ \ \ \ sS{\isadigit{1}}{\isadigit{2}}\ {\isacharbar}sS{\isadigit{1}}{\isadigit{3}}{\isacharbar}\ sS{\isadigit{1}}{\isadigit{4}}{\isacharbar}\ sS{\isadigit{1}}{\isadigit{5}}{\isacharbar}\ sS{\isadigit{1}}opt\ {\isacharbar}\ sS{\isadigit{4}}opt\ {\isacharbar}\ sS{\isadigit{7}}opt\ {\isacharbar}\ sS{\isadigit{1}}{\isadigit{1}}opt\isanewline
\isanewline
\isacommand{datatype}\isamarkupfalse%
\ chanID\ {\isacharequal}\ data{\isadigit{1}}{\isacharbar}\ data{\isadigit{2}}{\isacharbar}\ data{\isadigit{3}}{\isacharbar}\ data{\isadigit{4}}{\isacharbar}\ data{\isadigit{5}}{\isacharbar}\ data{\isadigit{6}}{\isacharbar}\ data{\isadigit{7}}{\isacharbar}\ \isanewline
data{\isadigit{8}}{\isacharbar}\ data{\isadigit{9}}{\isacharbar}\ data{\isadigit{1}}{\isadigit{0}}{\isacharbar}\ data{\isadigit{1}}{\isadigit{1}}{\isacharbar}\ data{\isadigit{1}}{\isadigit{2}}{\isacharbar}\ data{\isadigit{1}}{\isadigit{3}}{\isacharbar}\ data{\isadigit{1}}{\isadigit{4}}{\isacharbar}\ data{\isadigit{1}}{\isadigit{5}}{\isacharbar}\ \isanewline
data{\isadigit{1}}{\isadigit{6}}{\isacharbar}\ data{\isadigit{1}}{\isadigit{7}}{\isacharbar}\ data{\isadigit{1}}{\isadigit{8}}{\isacharbar}\ data{\isadigit{1}}{\isadigit{9}}{\isacharbar}\ data{\isadigit{2}}{\isadigit{0}}{\isacharbar}\ data{\isadigit{2}}{\isadigit{1}}{\isacharbar}\ data{\isadigit{2}}{\isadigit{2}}{\isacharbar}\ data{\isadigit{2}}{\isadigit{3}}{\isacharbar}\ data{\isadigit{2}}{\isadigit{4}}\isanewline
\isanewline
\isacommand{datatype}\isamarkupfalse%
\ varID\ {\isacharequal}\ stA{\isadigit{1}}\ {\isacharbar}\ stA{\isadigit{2}}\ {\isacharbar}\ stA{\isadigit{4}}\ {\isacharbar}\ stA{\isadigit{6}}\isanewline
\isanewline
\isacommand{datatype}\isamarkupfalse%
\ AbstrLevelsID\ {\isacharequal}\ level{\isadigit{0}}\ {\isacharbar}\ level{\isadigit{1}}\ {\isacharbar}\ level{\isadigit{2}}\ {\isacharbar}\ level{\isadigit{3}}\isanewline
\isanewline
\isamarkupcmt{function IN maps component ID to the set of its input channels%
}
\isanewline
\isacommand{fun}\isamarkupfalse%
\ IN\ {\isacharcolon}{\isacharcolon}\ \ {\isachardoublequoteopen}CSet\ {\isasymRightarrow}\ chanID\ set{\isachardoublequoteclose}\isanewline
\isakeyword{where}\ \isanewline
\isanewline
\ \ \ {\isachardoublequoteopen}IN\ sA{\isadigit{1}}\ {\isacharequal}\ {\isacharbraceleft}\ data{\isadigit{1}}\ {\isacharbraceright}{\isachardoublequoteclose}\ \isanewline
{\isacharbar}\ {\isachardoublequoteopen}IN\ sA{\isadigit{2}}\ {\isacharequal}\ {\isacharbraceleft}\ data{\isadigit{2}}{\isacharcomma}\ data{\isadigit{3}}\ {\isacharbraceright}{\isachardoublequoteclose}\isanewline
{\isacharbar}\ {\isachardoublequoteopen}IN\ sA{\isadigit{3}}\ {\isacharequal}\ {\isacharbraceleft}\ data{\isadigit{4}}{\isacharcomma}\ data{\isadigit{5}}\ {\isacharbraceright}{\isachardoublequoteclose}\isanewline
{\isacharbar}\ {\isachardoublequoteopen}IN\ sA{\isadigit{4}}\ {\isacharequal}\ {\isacharbraceleft}\ data{\isadigit{6}}{\isacharcomma}\ data{\isadigit{7}}{\isacharcomma}\ data{\isadigit{1}}{\isadigit{3}}\ {\isacharbraceright}{\isachardoublequoteclose}\isanewline
{\isacharbar}\ {\isachardoublequoteopen}IN\ sA{\isadigit{5}}\ {\isacharequal}\ {\isacharbraceleft}\ data{\isadigit{8}}\ {\isacharbraceright}{\isachardoublequoteclose}\isanewline
{\isacharbar}\ {\isachardoublequoteopen}IN\ sA{\isadigit{6}}\ {\isacharequal}\ {\isacharbraceleft}\ data{\isadigit{1}}{\isadigit{4}}\ {\isacharbraceright}{\isachardoublequoteclose}\isanewline
{\isacharbar}\ {\isachardoublequoteopen}IN\ sA{\isadigit{7}}\ {\isacharequal}\ {\isacharbraceleft}\ data{\isadigit{1}}{\isadigit{5}}{\isacharcomma}\ data{\isadigit{1}}{\isadigit{6}}\ {\isacharbraceright}{\isachardoublequoteclose}\isanewline
{\isacharbar}\ {\isachardoublequoteopen}IN\ sA{\isadigit{8}}\ {\isacharequal}\ {\isacharbraceleft}\ data{\isadigit{1}}{\isadigit{7}}{\isacharcomma}\ data{\isadigit{1}}{\isadigit{8}}{\isacharcomma}\ data{\isadigit{1}}{\isadigit{9}}{\isacharcomma}\ data{\isadigit{2}}{\isadigit{2}}\ {\isacharbraceright}{\isachardoublequoteclose}\isanewline
{\isacharbar}\ {\isachardoublequoteopen}IN\ sA{\isadigit{9}}\ {\isacharequal}\ {\isacharbraceleft}\ data{\isadigit{2}}{\isadigit{0}}{\isacharcomma}\ data{\isadigit{2}}{\isadigit{1}}\ {\isacharbraceright}{\isachardoublequoteclose}\isanewline
{\isacharbar}\ {\isachardoublequoteopen}IN\ sA{\isadigit{1}}{\isadigit{1}}\ {\isacharequal}\ {\isacharbraceleft}\ data{\isadigit{1}}\ {\isacharbraceright}{\isachardoublequoteclose}\isanewline
{\isacharbar}\ {\isachardoublequoteopen}IN\ sA{\isadigit{1}}{\isadigit{2}}\ {\isacharequal}\ {\isacharbraceleft}\ data{\isadigit{1}}\ {\isacharbraceright}{\isachardoublequoteclose}\isanewline
{\isacharbar}\ {\isachardoublequoteopen}IN\ sA{\isadigit{2}}{\isadigit{1}}\ {\isacharequal}\ {\isacharbraceleft}\ data{\isadigit{2}}\ {\isacharbraceright}{\isachardoublequoteclose}\isanewline
{\isacharbar}\ {\isachardoublequoteopen}IN\ sA{\isadigit{2}}{\isadigit{2}}\ {\isacharequal}\ {\isacharbraceleft}\ data{\isadigit{2}}{\isacharcomma}\ data{\isadigit{3}}\ {\isacharbraceright}{\isachardoublequoteclose}\isanewline
{\isacharbar}\ {\isachardoublequoteopen}IN\ sA{\isadigit{2}}{\isadigit{3}}\ {\isacharequal}\ {\isacharbraceleft}\ data{\isadigit{2}}\ {\isacharbraceright}{\isachardoublequoteclose}\isanewline
{\isacharbar}\ {\isachardoublequoteopen}IN\ sA{\isadigit{3}}{\isadigit{1}}\ {\isacharequal}\ {\isacharbraceleft}\ data{\isadigit{4}}\ {\isacharbraceright}{\isachardoublequoteclose}\isanewline
{\isacharbar}\ {\isachardoublequoteopen}IN\ sA{\isadigit{3}}{\isadigit{2}}\ {\isacharequal}\ {\isacharbraceleft}\ data{\isadigit{5}}\ {\isacharbraceright}{\isachardoublequoteclose}\isanewline
{\isacharbar}\ {\isachardoublequoteopen}IN\ sA{\isadigit{4}}{\isadigit{1}}\ {\isacharequal}\ {\isacharbraceleft}\ data{\isadigit{6}}{\isacharcomma}\ data{\isadigit{7}}\ {\isacharbraceright}{\isachardoublequoteclose}\isanewline
{\isacharbar}\ {\isachardoublequoteopen}IN\ sA{\isadigit{4}}{\isadigit{2}}\ {\isacharequal}\ {\isacharbraceleft}\ data{\isadigit{1}}{\isadigit{3}}\ {\isacharbraceright}{\isachardoublequoteclose}\isanewline
{\isacharbar}\ {\isachardoublequoteopen}IN\ sA{\isadigit{7}}{\isadigit{1}}\ {\isacharequal}\ {\isacharbraceleft}\ data{\isadigit{1}}{\isadigit{5}}\ {\isacharbraceright}{\isachardoublequoteclose}\isanewline
{\isacharbar}\ {\isachardoublequoteopen}IN\ sA{\isadigit{7}}{\isadigit{2}}\ {\isacharequal}\ {\isacharbraceleft}\ data{\isadigit{1}}{\isadigit{6}}\ {\isacharbraceright}{\isachardoublequoteclose}\isanewline
{\isacharbar}\ {\isachardoublequoteopen}IN\ sA{\isadigit{8}}{\isadigit{1}}\ {\isacharequal}\ {\isacharbraceleft}\ data{\isadigit{1}}{\isadigit{7}}{\isacharcomma}\ data{\isadigit{2}}{\isadigit{2}}\ {\isacharbraceright}{\isachardoublequoteclose}\isanewline
{\isacharbar}\ {\isachardoublequoteopen}IN\ sA{\isadigit{8}}{\isadigit{2}}\ {\isacharequal}\ {\isacharbraceleft}\ data{\isadigit{1}}{\isadigit{8}}{\isacharcomma}\ data{\isadigit{1}}{\isadigit{9}}\ {\isacharbraceright}{\isachardoublequoteclose}\isanewline
{\isacharbar}\ {\isachardoublequoteopen}IN\ sA{\isadigit{9}}{\isadigit{1}}\ {\isacharequal}\ {\isacharbraceleft}\ data{\isadigit{2}}{\isadigit{0}}\ {\isacharbraceright}{\isachardoublequoteclose}\isanewline
{\isacharbar}\ {\isachardoublequoteopen}IN\ sA{\isadigit{9}}{\isadigit{2}}\ {\isacharequal}\ {\isacharbraceleft}\ data{\isadigit{2}}{\isadigit{0}}\ {\isacharbraceright}{\isachardoublequoteclose}\isanewline
{\isacharbar}\ {\isachardoublequoteopen}IN\ sA{\isadigit{9}}{\isadigit{3}}\ {\isacharequal}\ {\isacharbraceleft}\ data{\isadigit{2}}{\isadigit{1}}\ {\isacharbraceright}{\isachardoublequoteclose}\isanewline
{\isacharbar}\ {\isachardoublequoteopen}IN\ sS{\isadigit{1}}\ {\isacharequal}\ {\isacharbraceleft}\ data{\isadigit{1}}\ {\isacharbraceright}{\isachardoublequoteclose}\isanewline
{\isacharbar}\ {\isachardoublequoteopen}IN\ sS{\isadigit{2}}\ {\isacharequal}\ {\isacharbraceleft}\ data{\isadigit{1}}\ {\isacharbraceright}{\isachardoublequoteclose}\isanewline
{\isacharbar}\ {\isachardoublequoteopen}IN\ sS{\isadigit{3}}\ {\isacharequal}\ {\isacharbraceleft}\ data{\isadigit{2}}\ {\isacharbraceright}{\isachardoublequoteclose}\isanewline
{\isacharbar}\ {\isachardoublequoteopen}IN\ sS{\isadigit{4}}\ {\isacharequal}\ {\isacharbraceleft}\ data{\isadigit{2}}\ {\isacharbraceright}{\isachardoublequoteclose}\isanewline
{\isacharbar}\ {\isachardoublequoteopen}IN\ sS{\isadigit{5}}\ {\isacharequal}\ {\isacharbraceleft}\ data{\isadigit{5}}\ {\isacharbraceright}{\isachardoublequoteclose}\isanewline
{\isacharbar}\ {\isachardoublequoteopen}IN\ sS{\isadigit{6}}\ {\isacharequal}\ {\isacharbraceleft}\ data{\isadigit{2}}{\isacharcomma}\ data{\isadigit{7}}\ {\isacharbraceright}{\isachardoublequoteclose}\isanewline
{\isacharbar}\ {\isachardoublequoteopen}IN\ sS{\isadigit{7}}\ {\isacharequal}\ {\isacharbraceleft}\ data{\isadigit{1}}{\isadigit{3}}\ {\isacharbraceright}{\isachardoublequoteclose}\isanewline
{\isacharbar}\ {\isachardoublequoteopen}IN\ sS{\isadigit{8}}\ {\isacharequal}\ {\isacharbraceleft}\ data{\isadigit{8}}\ {\isacharbraceright}{\isachardoublequoteclose}\isanewline
{\isacharbar}\ {\isachardoublequoteopen}IN\ sS{\isadigit{9}}\ {\isacharequal}\ {\isacharbraceleft}\ data{\isadigit{1}}{\isadigit{4}}\ {\isacharbraceright}{\isachardoublequoteclose}\isanewline
{\isacharbar}\ {\isachardoublequoteopen}IN\ sS{\isadigit{1}}{\isadigit{0}}\ {\isacharequal}\ {\isacharbraceleft}\ data{\isadigit{1}}{\isadigit{5}}\ {\isacharbraceright}{\isachardoublequoteclose}\isanewline
{\isacharbar}\ {\isachardoublequoteopen}IN\ sS{\isadigit{1}}{\isadigit{1}}\ {\isacharequal}\ {\isacharbraceleft}\ data{\isadigit{1}}{\isadigit{6}}\ {\isacharbraceright}{\isachardoublequoteclose}\isanewline
{\isacharbar}\ {\isachardoublequoteopen}IN\ sS{\isadigit{1}}{\isadigit{2}}\ {\isacharequal}\ {\isacharbraceleft}\ data{\isadigit{1}}{\isadigit{7}}{\isacharbraceright}{\isachardoublequoteclose}\isanewline
{\isacharbar}\ {\isachardoublequoteopen}IN\ sS{\isadigit{1}}{\isadigit{3}}{\isacharequal}\ {\isacharbraceleft}\ data{\isadigit{2}}{\isadigit{0}}\ {\isacharbraceright}{\isachardoublequoteclose}\ \isanewline
{\isacharbar}\ {\isachardoublequoteopen}IN\ sS{\isadigit{1}}{\isadigit{4}}\ {\isacharequal}\ {\isacharbraceleft}\ data{\isadigit{1}}{\isadigit{8}}{\isacharcomma}\ data{\isadigit{1}}{\isadigit{9}}\ {\isacharbraceright}{\isachardoublequoteclose}\isanewline
{\isacharbar}\ {\isachardoublequoteopen}IN\ sS{\isadigit{1}}{\isadigit{5}}\ {\isacharequal}\ {\isacharbraceleft}\ data{\isadigit{2}}{\isadigit{1}}\ {\isacharbraceright}{\isachardoublequoteclose}\isanewline
{\isacharbar}\ {\isachardoublequoteopen}IN\ sS{\isadigit{1}}opt\ {\isacharequal}\ {\isacharbraceleft}\ data{\isadigit{1}}\ {\isacharbraceright}{\isachardoublequoteclose}\isanewline
{\isacharbar}\ {\isachardoublequoteopen}IN\ sS{\isadigit{4}}opt\ {\isacharequal}\ {\isacharbraceleft}\ data{\isadigit{2}}\ {\isacharbraceright}{\isachardoublequoteclose}\isanewline
{\isacharbar}\ {\isachardoublequoteopen}IN\ sS{\isadigit{7}}opt\ {\isacharequal}\ {\isacharbraceleft}\ data{\isadigit{1}}{\isadigit{3}}\ {\isacharbraceright}{\isachardoublequoteclose}\ \isanewline
{\isacharbar}\ {\isachardoublequoteopen}IN\ sS{\isadigit{1}}{\isadigit{1}}opt\ {\isacharequal}\ {\isacharbraceleft}\ data{\isadigit{1}}{\isadigit{6}}{\isacharcomma}\ data{\isadigit{1}}{\isadigit{9}}\ {\isacharbraceright}{\isachardoublequoteclose}\ \isanewline
\isanewline
\isanewline
\isanewline
\isanewline
\isanewline
\isamarkupcmt{function OUT maps component ID to the set of its output channels%
}
\isanewline
\isacommand{fun}\isamarkupfalse%
\ OUT\ {\isacharcolon}{\isacharcolon}\ \ {\isachardoublequoteopen}CSet\ {\isasymRightarrow}\ chanID\ set{\isachardoublequoteclose}\isanewline
\isakeyword{where}\ \isanewline
\ \ \ {\isachardoublequoteopen}OUT\ sA{\isadigit{1}}\ {\isacharequal}\ {\isacharbraceleft}\ data{\isadigit{2}}{\isacharcomma}\ data{\isadigit{1}}{\isadigit{0}}\ {\isacharbraceright}{\isachardoublequoteclose}\ \isanewline
{\isacharbar}\ {\isachardoublequoteopen}OUT\ sA{\isadigit{2}}\ {\isacharequal}\ {\isacharbraceleft}\ data{\isadigit{4}}{\isacharcomma}\ data{\isadigit{5}}{\isacharcomma}\ data{\isadigit{1}}{\isadigit{1}}{\isacharcomma}\ data{\isadigit{1}}{\isadigit{2}}\ {\isacharbraceright}{\isachardoublequoteclose}\isanewline
{\isacharbar}\ {\isachardoublequoteopen}OUT\ sA{\isadigit{3}}\ {\isacharequal}\ {\isacharbraceleft}\ data{\isadigit{6}}{\isacharcomma}\ data{\isadigit{7}}\ {\isacharbraceright}{\isachardoublequoteclose}\isanewline
{\isacharbar}\ {\isachardoublequoteopen}OUT\ sA{\isadigit{4}}\ {\isacharequal}\ {\isacharbraceleft}\ data{\isadigit{3}}{\isacharcomma}\ data{\isadigit{8}}\ {\isacharbraceright}{\isachardoublequoteclose}\isanewline
{\isacharbar}\ {\isachardoublequoteopen}OUT\ sA{\isadigit{5}}\ {\isacharequal}\ {\isacharbraceleft}\ data{\isadigit{9}}\ {\isacharbraceright}{\isachardoublequoteclose}\isanewline
{\isacharbar}\ {\isachardoublequoteopen}OUT\ sA{\isadigit{6}}\ {\isacharequal}\ {\isacharbraceleft}\ data{\isadigit{1}}{\isadigit{5}}{\isacharcomma}\ data{\isadigit{1}}{\isadigit{6}}\ {\isacharbraceright}{\isachardoublequoteclose}\isanewline
{\isacharbar}\ {\isachardoublequoteopen}OUT\ sA{\isadigit{7}}\ {\isacharequal}\ {\isacharbraceleft}\ data{\isadigit{1}}{\isadigit{7}}{\isacharcomma}\ data{\isadigit{1}}{\isadigit{8}}\ {\isacharbraceright}{\isachardoublequoteclose}\isanewline
{\isacharbar}\ {\isachardoublequoteopen}OUT\ sA{\isadigit{8}}\ {\isacharequal}\ {\isacharbraceleft}\ data{\isadigit{2}}{\isadigit{0}}{\isacharcomma}\ data{\isadigit{2}}{\isadigit{1}}\ {\isacharbraceright}{\isachardoublequoteclose}\isanewline
{\isacharbar}\ {\isachardoublequoteopen}OUT\ sA{\isadigit{9}}\ {\isacharequal}\ {\isacharbraceleft}\ data{\isadigit{2}}{\isadigit{2}}{\isacharcomma}\ data{\isadigit{2}}{\isadigit{3}}{\isacharcomma}\ data{\isadigit{2}}{\isadigit{4}}\ {\isacharbraceright}{\isachardoublequoteclose}\isanewline
{\isacharbar}\ {\isachardoublequoteopen}OUT\ sA{\isadigit{1}}{\isadigit{1}}\ {\isacharequal}\ {\isacharbraceleft}\ data{\isadigit{2}}\ {\isacharbraceright}{\isachardoublequoteclose}\isanewline
{\isacharbar}\ {\isachardoublequoteopen}OUT\ sA{\isadigit{1}}{\isadigit{2}}{\isacharequal}\ {\isacharbraceleft}\ data{\isadigit{1}}{\isadigit{0}}\ {\isacharbraceright}{\isachardoublequoteclose}\isanewline
{\isacharbar}\ {\isachardoublequoteopen}OUT\ sA{\isadigit{2}}{\isadigit{1}}\ {\isacharequal}\ {\isacharbraceleft}\ data{\isadigit{1}}{\isadigit{1}}\ {\isacharbraceright}{\isachardoublequoteclose}\isanewline
{\isacharbar}\ {\isachardoublequoteopen}OUT\ sA{\isadigit{2}}{\isadigit{2}}\ {\isacharequal}\ {\isacharbraceleft}\ data{\isadigit{4}}{\isacharcomma}\ data{\isadigit{1}}{\isadigit{2}}\ {\isacharbraceright}{\isachardoublequoteclose}\isanewline
{\isacharbar}\ {\isachardoublequoteopen}OUT\ sA{\isadigit{2}}{\isadigit{3}}\ {\isacharequal}\ {\isacharbraceleft}\ data{\isadigit{5}}\ {\isacharbraceright}{\isachardoublequoteclose}\isanewline
{\isacharbar}\ {\isachardoublequoteopen}OUT\ sA{\isadigit{3}}{\isadigit{1}}{\isacharequal}\ {\isacharbraceleft}\ data{\isadigit{6}}\ {\isacharbraceright}{\isachardoublequoteclose}\isanewline
{\isacharbar}\ {\isachardoublequoteopen}OUT\ sA{\isadigit{3}}{\isadigit{2}}\ {\isacharequal}\ {\isacharbraceleft}\ data{\isadigit{7}}\ {\isacharbraceright}{\isachardoublequoteclose}\isanewline
{\isacharbar}\ {\isachardoublequoteopen}OUT\ sA{\isadigit{4}}{\isadigit{1}}\ {\isacharequal}\ {\isacharbraceleft}\ data{\isadigit{3}}\ {\isacharbraceright}{\isachardoublequoteclose}\isanewline
{\isacharbar}\ {\isachardoublequoteopen}OUT\ sA{\isadigit{4}}{\isadigit{2}}\ {\isacharequal}\ {\isacharbraceleft}\ data{\isadigit{8}}\ {\isacharbraceright}{\isachardoublequoteclose}\isanewline
{\isacharbar}\ {\isachardoublequoteopen}OUT\ sA{\isadigit{7}}{\isadigit{1}}\ {\isacharequal}\ {\isacharbraceleft}\ data{\isadigit{1}}{\isadigit{7}}\ {\isacharbraceright}{\isachardoublequoteclose}\isanewline
{\isacharbar}\ {\isachardoublequoteopen}OUT\ sA{\isadigit{7}}{\isadigit{2}}\ {\isacharequal}\ {\isacharbraceleft}\ data{\isadigit{1}}{\isadigit{8}}\ {\isacharbraceright}{\isachardoublequoteclose}\isanewline
{\isacharbar}\ {\isachardoublequoteopen}OUT\ sA{\isadigit{8}}{\isadigit{1}}\ {\isacharequal}\ {\isacharbraceleft}\ data{\isadigit{2}}{\isadigit{0}}\ {\isacharbraceright}{\isachardoublequoteclose}\isanewline
{\isacharbar}\ {\isachardoublequoteopen}OUT\ sA{\isadigit{8}}{\isadigit{2}}\ {\isacharequal}\ {\isacharbraceleft}\ data{\isadigit{2}}{\isadigit{1}}\ {\isacharbraceright}{\isachardoublequoteclose}\isanewline
{\isacharbar}\ {\isachardoublequoteopen}OUT\ sA{\isadigit{9}}{\isadigit{1}}\ {\isacharequal}\ {\isacharbraceleft}\ data{\isadigit{2}}{\isadigit{2}}\ {\isacharbraceright}{\isachardoublequoteclose}\isanewline
{\isacharbar}\ {\isachardoublequoteopen}OUT\ sA{\isadigit{9}}{\isadigit{2}}\ {\isacharequal}\ {\isacharbraceleft}\ data{\isadigit{2}}{\isadigit{3}}\ {\isacharbraceright}{\isachardoublequoteclose}\isanewline
{\isacharbar}\ {\isachardoublequoteopen}OUT\ sA{\isadigit{9}}{\isadigit{3}}\ {\isacharequal}\ {\isacharbraceleft}\ data{\isadigit{2}}{\isadigit{4}}\ {\isacharbraceright}{\isachardoublequoteclose}\isanewline
{\isacharbar}\ {\isachardoublequoteopen}OUT\ sS{\isadigit{1}}\ {\isacharequal}\ {\isacharbraceleft}\ data{\isadigit{1}}{\isadigit{0}}\ {\isacharbraceright}{\isachardoublequoteclose}\isanewline
{\isacharbar}\ {\isachardoublequoteopen}OUT\ sS{\isadigit{2}}\ {\isacharequal}\ {\isacharbraceleft}\ data{\isadigit{2}}\ {\isacharbraceright}{\isachardoublequoteclose}\isanewline
{\isacharbar}\ {\isachardoublequoteopen}OUT\ sS{\isadigit{3}}\ {\isacharequal}\ {\isacharbraceleft}\ data{\isadigit{1}}{\isadigit{1}}\ {\isacharbraceright}{\isachardoublequoteclose}\isanewline
{\isacharbar}\ {\isachardoublequoteopen}OUT\ sS{\isadigit{4}}\ {\isacharequal}\ {\isacharbraceleft}\ data{\isadigit{5}}\ {\isacharbraceright}{\isachardoublequoteclose}\isanewline
{\isacharbar}\ {\isachardoublequoteopen}OUT\ sS{\isadigit{5}}\ {\isacharequal}\ {\isacharbraceleft}\ data{\isadigit{7}}\ {\isacharbraceright}{\isachardoublequoteclose}\isanewline
{\isacharbar}\ {\isachardoublequoteopen}OUT\ sS{\isadigit{6}}\ {\isacharequal}\ {\isacharbraceleft}\ data{\isadigit{1}}{\isadigit{2}}\ {\isacharbraceright}{\isachardoublequoteclose}\isanewline
{\isacharbar}\ {\isachardoublequoteopen}OUT\ sS{\isadigit{7}}\ {\isacharequal}\ {\isacharbraceleft}\ data{\isadigit{8}}\ {\isacharbraceright}{\isachardoublequoteclose}\isanewline
{\isacharbar}\ {\isachardoublequoteopen}OUT\ sS{\isadigit{8}}\ {\isacharequal}\ {\isacharbraceleft}\ data{\isadigit{9}}\ {\isacharbraceright}{\isachardoublequoteclose}\isanewline
{\isacharbar}\ {\isachardoublequoteopen}OUT\ sS{\isadigit{9}}\ {\isacharequal}\ {\isacharbraceleft}\ data{\isadigit{1}}{\isadigit{5}}{\isacharcomma}\ data{\isadigit{1}}{\isadigit{6}}\ {\isacharbraceright}{\isachardoublequoteclose}\isanewline
{\isacharbar}\ {\isachardoublequoteopen}OUT\ sS{\isadigit{1}}{\isadigit{0}}\ {\isacharequal}\ {\isacharbraceleft}\ data{\isadigit{1}}{\isadigit{7}}\ {\isacharbraceright}{\isachardoublequoteclose}\isanewline
{\isacharbar}\ {\isachardoublequoteopen}OUT\ sS{\isadigit{1}}{\isadigit{1}}\ {\isacharequal}\ {\isacharbraceleft}\ data{\isadigit{1}}{\isadigit{8}}\ {\isacharbraceright}{\isachardoublequoteclose}\isanewline
{\isacharbar}\ {\isachardoublequoteopen}OUT\ sS{\isadigit{1}}{\isadigit{2}}\ {\isacharequal}\ {\isacharbraceleft}\ data{\isadigit{2}}{\isadigit{0}}{\isacharbraceright}{\isachardoublequoteclose}\isanewline
{\isacharbar}\ {\isachardoublequoteopen}OUT\ sS{\isadigit{1}}{\isadigit{3}}{\isacharequal}\ {\isacharbraceleft}\ data{\isadigit{2}}{\isadigit{3}}\ {\isacharbraceright}{\isachardoublequoteclose}\ \isanewline
{\isacharbar}\ {\isachardoublequoteopen}OUT\ sS{\isadigit{1}}{\isadigit{4}}\ {\isacharequal}\ {\isacharbraceleft}\ data{\isadigit{2}}{\isadigit{1}}\ {\isacharbraceright}{\isachardoublequoteclose}\isanewline
{\isacharbar}\ {\isachardoublequoteopen}OUT\ sS{\isadigit{1}}{\isadigit{5}}\ {\isacharequal}\ {\isacharbraceleft}\ data{\isadigit{2}}{\isadigit{4}}\ {\isacharbraceright}{\isachardoublequoteclose}\isanewline
{\isacharbar}\ {\isachardoublequoteopen}OUT\ sS{\isadigit{1}}opt\ {\isacharequal}\ {\isacharbraceleft}\ data{\isadigit{2}}{\isacharcomma}\ data{\isadigit{1}}{\isadigit{0}}\ {\isacharbraceright}{\isachardoublequoteclose}\isanewline
{\isacharbar}\ {\isachardoublequoteopen}OUT\ sS{\isadigit{4}}opt\ {\isacharequal}\ {\isacharbraceleft}\ data{\isadigit{1}}{\isadigit{2}}\ {\isacharbraceright}{\isachardoublequoteclose}\isanewline
{\isacharbar}\ {\isachardoublequoteopen}OUT\ sS{\isadigit{7}}opt\ {\isacharequal}\ {\isacharbraceleft}\ data{\isadigit{9}}\ {\isacharbraceright}{\isachardoublequoteclose}\isanewline
{\isacharbar}\ {\isachardoublequoteopen}OUT\ sS{\isadigit{1}}{\isadigit{1}}opt\ {\isacharequal}\ {\isacharbraceleft}\ data{\isadigit{2}}{\isadigit{4}}\ {\isacharbraceright}{\isachardoublequoteclose}\isanewline
\isanewline
\isanewline
\isamarkupcmt{function VAR maps component IDs to the set of its local variables%
}
\isanewline
\isacommand{fun}\isamarkupfalse%
\ VAR\ {\isacharcolon}{\isacharcolon}\ \ {\isachardoublequoteopen}CSet\ {\isasymRightarrow}\ varID\ set{\isachardoublequoteclose}\isanewline
\isakeyword{where}\ \isanewline
\ \ \ {\isachardoublequoteopen}VAR\ sA{\isadigit{1}}\ {\isacharequal}\ {\isacharbraceleft}\ stA{\isadigit{1}}\ {\isacharbraceright}{\isachardoublequoteclose}\ \isanewline
{\isacharbar}\ {\isachardoublequoteopen}VAR\ sA{\isadigit{2}}\ {\isacharequal}\ {\isacharbraceleft}\ stA{\isadigit{2}}\ {\isacharbraceright}{\isachardoublequoteclose}\isanewline
{\isacharbar}\ {\isachardoublequoteopen}VAR\ sA{\isadigit{3}}\ {\isacharequal}\ {\isacharbraceleft}{\isacharbraceright}{\isachardoublequoteclose}\isanewline
{\isacharbar}\ {\isachardoublequoteopen}VAR\ sA{\isadigit{4}}\ {\isacharequal}\ {\isacharbraceleft}\ stA{\isadigit{4}}\ {\isacharbraceright}{\isachardoublequoteclose}\isanewline
{\isacharbar}\ {\isachardoublequoteopen}VAR\ sA{\isadigit{5}}\ {\isacharequal}\ {\isacharbraceleft}{\isacharbraceright}{\isachardoublequoteclose}\isanewline
{\isacharbar}\ {\isachardoublequoteopen}VAR\ sA{\isadigit{6}}\ {\isacharequal}\ {\isacharbraceleft}\ stA{\isadigit{6}}\ {\isacharbraceright}{\isachardoublequoteclose}\isanewline
{\isacharbar}\ {\isachardoublequoteopen}VAR\ sA{\isadigit{7}}\ {\isacharequal}\ {\isacharbraceleft}{\isacharbraceright}{\isachardoublequoteclose}\isanewline
{\isacharbar}\ {\isachardoublequoteopen}VAR\ sA{\isadigit{8}}\ {\isacharequal}\ {\isacharbraceleft}{\isacharbraceright}{\isachardoublequoteclose}\isanewline
{\isacharbar}\ {\isachardoublequoteopen}VAR\ sA{\isadigit{9}}\ {\isacharequal}\ {\isacharbraceleft}{\isacharbraceright}{\isachardoublequoteclose}\isanewline
{\isacharbar}\ {\isachardoublequoteopen}VAR\ sA{\isadigit{1}}{\isadigit{1}}\ {\isacharequal}\ {\isacharbraceleft}{\isacharbraceright}{\isachardoublequoteclose}\isanewline
{\isacharbar}\ {\isachardoublequoteopen}VAR\ sA{\isadigit{1}}{\isadigit{2}}\ {\isacharequal}\ {\isacharbraceleft}\ stA{\isadigit{1}}\ {\isacharbraceright}{\isachardoublequoteclose}\isanewline
{\isacharbar}\ {\isachardoublequoteopen}VAR\ sA{\isadigit{2}}{\isadigit{1}}\ {\isacharequal}\ {\isacharbraceleft}{\isacharbraceright}{\isachardoublequoteclose}\isanewline
{\isacharbar}\ {\isachardoublequoteopen}VAR\ sA{\isadigit{2}}{\isadigit{2}}\ {\isacharequal}\ {\isacharbraceleft}\ stA{\isadigit{2}}\ {\isacharbraceright}{\isachardoublequoteclose}\isanewline
{\isacharbar}\ {\isachardoublequoteopen}VAR\ sA{\isadigit{2}}{\isadigit{3}}\ {\isacharequal}\ {\isacharbraceleft}{\isacharbraceright}{\isachardoublequoteclose}\isanewline
{\isacharbar}\ {\isachardoublequoteopen}VAR\ sA{\isadigit{3}}{\isadigit{1}}\ {\isacharequal}\ {\isacharbraceleft}{\isacharbraceright}{\isachardoublequoteclose}\isanewline
{\isacharbar}\ {\isachardoublequoteopen}VAR\ sA{\isadigit{3}}{\isadigit{2}}\ {\isacharequal}\ {\isacharbraceleft}{\isacharbraceright}{\isachardoublequoteclose}\isanewline
{\isacharbar}\ {\isachardoublequoteopen}VAR\ sA{\isadigit{4}}{\isadigit{1}}\ {\isacharequal}\ {\isacharbraceleft}stA{\isadigit{4}}\ {\isacharbraceright}{\isachardoublequoteclose}\isanewline
{\isacharbar}\ {\isachardoublequoteopen}VAR\ sA{\isadigit{4}}{\isadigit{2}}\ {\isacharequal}\ {\isacharbraceleft}{\isacharbraceright}{\isachardoublequoteclose}\isanewline
{\isacharbar}\ {\isachardoublequoteopen}VAR\ sA{\isadigit{7}}{\isadigit{1}}\ {\isacharequal}\ {\isacharbraceleft}{\isacharbraceright}{\isachardoublequoteclose}\isanewline
{\isacharbar}\ {\isachardoublequoteopen}VAR\ sA{\isadigit{7}}{\isadigit{2}}\ {\isacharequal}\ {\isacharbraceleft}{\isacharbraceright}{\isachardoublequoteclose}\isanewline
{\isacharbar}\ {\isachardoublequoteopen}VAR\ sA{\isadigit{8}}{\isadigit{1}}\ {\isacharequal}\ {\isacharbraceleft}{\isacharbraceright}{\isachardoublequoteclose}\isanewline
{\isacharbar}\ {\isachardoublequoteopen}VAR\ sA{\isadigit{8}}{\isadigit{2}}\ {\isacharequal}\ {\isacharbraceleft}{\isacharbraceright}{\isachardoublequoteclose}\isanewline
{\isacharbar}\ {\isachardoublequoteopen}VAR\ sA{\isadigit{9}}{\isadigit{1}}\ {\isacharequal}\ {\isacharbraceleft}{\isacharbraceright}{\isachardoublequoteclose}\isanewline
{\isacharbar}\ {\isachardoublequoteopen}VAR\ sA{\isadigit{9}}{\isadigit{2}}\ {\isacharequal}\ {\isacharbraceleft}{\isacharbraceright}{\isachardoublequoteclose}\isanewline
{\isacharbar}\ {\isachardoublequoteopen}VAR\ sA{\isadigit{9}}{\isadigit{3}}\ {\isacharequal}\ {\isacharbraceleft}{\isacharbraceright}{\isachardoublequoteclose}\isanewline
{\isacharbar}\ {\isachardoublequoteopen}VAR\ sS{\isadigit{1}}\ {\isacharequal}\ {\isacharbraceleft}\ stA{\isadigit{1}}\ {\isacharbraceright}{\isachardoublequoteclose}\isanewline
{\isacharbar}\ {\isachardoublequoteopen}VAR\ sS{\isadigit{2}}\ {\isacharequal}\ {\isacharbraceleft}{\isacharbraceright}{\isachardoublequoteclose}\isanewline
{\isacharbar}\ {\isachardoublequoteopen}VAR\ sS{\isadigit{3}}\ {\isacharequal}\ {\isacharbraceleft}{\isacharbraceright}{\isachardoublequoteclose}\isanewline
{\isacharbar}\ {\isachardoublequoteopen}VAR\ sS{\isadigit{4}}\ {\isacharequal}\ {\isacharbraceleft}{\isacharbraceright}{\isachardoublequoteclose}\isanewline
{\isacharbar}\ {\isachardoublequoteopen}VAR\ sS{\isadigit{5}}\ {\isacharequal}\ {\isacharbraceleft}{\isacharbraceright}{\isachardoublequoteclose}\isanewline
{\isacharbar}\ {\isachardoublequoteopen}VAR\ sS{\isadigit{6}}\ {\isacharequal}\ {\isacharbraceleft}stA{\isadigit{2}}{\isacharcomma}\ stA{\isadigit{4}}{\isacharbraceright}{\isachardoublequoteclose}\isanewline
{\isacharbar}\ {\isachardoublequoteopen}VAR\ sS{\isadigit{7}}\ {\isacharequal}\ {\isacharbraceleft}{\isacharbraceright}{\isachardoublequoteclose}\isanewline
{\isacharbar}\ {\isachardoublequoteopen}VAR\ sS{\isadigit{8}}\ {\isacharequal}\ {\isacharbraceleft}{\isacharbraceright}{\isachardoublequoteclose}\isanewline
{\isacharbar}\ {\isachardoublequoteopen}VAR\ sS{\isadigit{9}}\ {\isacharequal}\ {\isacharbraceleft}stA{\isadigit{6}}{\isacharbraceright}{\isachardoublequoteclose}\isanewline
{\isacharbar}\ {\isachardoublequoteopen}VAR\ sS{\isadigit{1}}{\isadigit{0}}\ {\isacharequal}\ {\isacharbraceleft}{\isacharbraceright}{\isachardoublequoteclose}\isanewline
{\isacharbar}\ {\isachardoublequoteopen}VAR\ sS{\isadigit{1}}{\isadigit{1}}\ {\isacharequal}\ {\isacharbraceleft}{\isacharbraceright}{\isachardoublequoteclose}\isanewline
{\isacharbar}\ {\isachardoublequoteopen}VAR\ sS{\isadigit{1}}{\isadigit{2}}\ {\isacharequal}\ {\isacharbraceleft}{\isacharbraceright}{\isachardoublequoteclose}\isanewline
{\isacharbar}\ {\isachardoublequoteopen}VAR\ sS{\isadigit{1}}{\isadigit{3}}\ {\isacharequal}\ {\isacharbraceleft}{\isacharbraceright}{\isachardoublequoteclose}\isanewline
{\isacharbar}\ {\isachardoublequoteopen}VAR\ sS{\isadigit{1}}{\isadigit{4}}\ {\isacharequal}\ {\isacharbraceleft}{\isacharbraceright}{\isachardoublequoteclose}\isanewline
{\isacharbar}\ {\isachardoublequoteopen}VAR\ sS{\isadigit{1}}{\isadigit{5}}\ {\isacharequal}\ {\isacharbraceleft}{\isacharbraceright}{\isachardoublequoteclose}\isanewline
{\isacharbar}\ {\isachardoublequoteopen}VAR\ sS{\isadigit{1}}opt\ {\isacharequal}\ {\isacharbraceleft}\ stA{\isadigit{1}}\ {\isacharbraceright}{\isachardoublequoteclose}\isanewline
{\isacharbar}\ {\isachardoublequoteopen}VAR\ sS{\isadigit{4}}opt\ {\isacharequal}\ {\isacharbraceleft}\ stA{\isadigit{2}}{\isacharcomma}\ stA{\isadigit{4}}\ {\isacharbraceright}{\isachardoublequoteclose}\isanewline
{\isacharbar}\ {\isachardoublequoteopen}VAR\ sS{\isadigit{7}}opt\ {\isacharequal}\ {\isacharbraceleft}{\isacharbraceright}{\isachardoublequoteclose}\isanewline
{\isacharbar}\ {\isachardoublequoteopen}VAR\ sS{\isadigit{1}}{\isadigit{1}}opt\ {\isacharequal}\ {\isacharbraceleft}{\isacharbraceright}{\isachardoublequoteclose}\isanewline
\isanewline
\isanewline

\isamarkupcmt{function subcomp maps component ID to the set of its subcomponents%
}
\isanewline
\isanewline
\isacommand{fun}\isamarkupfalse%
\ subcomp\ {\isacharcolon}{\isacharcolon}\ \ {\isachardoublequoteopen}CSet\ {\isasymRightarrow}\ CSet\ set{\isachardoublequoteclose}\isanewline
\isakeyword{where}\ \isanewline
\ \ \ {\isachardoublequoteopen}subcomp\ sA{\isadigit{1}}\ {\isacharequal}\ {\isacharbraceleft}\ sA{\isadigit{1}}{\isadigit{1}}{\isacharcomma}\ sA{\isadigit{1}}{\isadigit{2}}\ {\isacharbraceright}{\isachardoublequoteclose}\ \isanewline
{\isacharbar}\ {\isachardoublequoteopen}subcomp\ sA{\isadigit{2}}\ {\isacharequal}\ {\isacharbraceleft}\ sA{\isadigit{2}}{\isadigit{1}}{\isacharcomma}\ sA{\isadigit{2}}{\isadigit{2}}{\isacharcomma}\ sA{\isadigit{2}}{\isadigit{3}}\ {\isacharbraceright}{\isachardoublequoteclose}\isanewline
{\isacharbar}\ {\isachardoublequoteopen}subcomp\ sA{\isadigit{3}}\ {\isacharequal}\ {\isacharbraceleft}\ sA{\isadigit{3}}{\isadigit{1}}{\isacharcomma}\ sA{\isadigit{3}}{\isadigit{2}}\ {\isacharbraceright}{\isachardoublequoteclose}\isanewline
{\isacharbar}\ {\isachardoublequoteopen}subcomp\ sA{\isadigit{4}}\ {\isacharequal}\ {\isacharbraceleft}\ sA{\isadigit{4}}{\isadigit{1}}{\isacharcomma}\ sA{\isadigit{4}}{\isadigit{2}}\ {\isacharbraceright}{\isachardoublequoteclose}\isanewline
{\isacharbar}\ {\isachardoublequoteopen}subcomp\ sA{\isadigit{5}}\ {\isacharequal}\ {\isacharbraceleft}{\isacharbraceright}{\isachardoublequoteclose}\isanewline
{\isacharbar}\ {\isachardoublequoteopen}subcomp\ sA{\isadigit{6}}\ {\isacharequal}\ {\isacharbraceleft}{\isacharbraceright}{\isachardoublequoteclose}\isanewline
{\isacharbar}\ {\isachardoublequoteopen}subcomp\ sA{\isadigit{7}}\ {\isacharequal}\ {\isacharbraceleft}\ sA{\isadigit{7}}{\isadigit{1}}{\isacharcomma}\ sA{\isadigit{7}}{\isadigit{2}}\ {\isacharbraceright}{\isachardoublequoteclose}\isanewline
{\isacharbar}\ {\isachardoublequoteopen}subcomp\ sA{\isadigit{8}}\ {\isacharequal}\ {\isacharbraceleft}\ sA{\isadigit{8}}{\isadigit{1}}{\isacharcomma}\ sA{\isadigit{8}}{\isadigit{2}}\ {\isacharbraceright}{\isachardoublequoteclose}\isanewline
{\isacharbar}\ {\isachardoublequoteopen}subcomp\ sA{\isadigit{9}}\ {\isacharequal}\ {\isacharbraceleft}\ sA{\isadigit{9}}{\isadigit{1}}{\isacharcomma}\ sA{\isadigit{9}}{\isadigit{2}}{\isacharcomma}\ sA{\isadigit{9}}{\isadigit{3}}\ {\isacharbraceright}{\isachardoublequoteclose}\ \isanewline
{\isacharbar}\ {\isachardoublequoteopen}subcomp\ sA{\isadigit{1}}{\isadigit{1}}\ {\isacharequal}\ {\isacharbraceleft}{\isacharbraceright}{\isachardoublequoteclose}\isanewline
{\isacharbar}\ {\isachardoublequoteopen}subcomp\ sA{\isadigit{1}}{\isadigit{2}}\ {\isacharequal}\ {\isacharbraceleft}{\isacharbraceright}{\isachardoublequoteclose}\isanewline
{\isacharbar}\ {\isachardoublequoteopen}subcomp\ sA{\isadigit{2}}{\isadigit{1}}\ {\isacharequal}\ {\isacharbraceleft}{\isacharbraceright}{\isachardoublequoteclose}\isanewline
{\isacharbar}\ {\isachardoublequoteopen}subcomp\ sA{\isadigit{2}}{\isadigit{2}}\ {\isacharequal}\ {\isacharbraceleft}{\isacharbraceright}{\isachardoublequoteclose}\isanewline
{\isacharbar}\ {\isachardoublequoteopen}subcomp\ sA{\isadigit{2}}{\isadigit{3}}\ {\isacharequal}\ {\isacharbraceleft}{\isacharbraceright}{\isachardoublequoteclose}\isanewline
{\isacharbar}\ {\isachardoublequoteopen}subcomp\ sA{\isadigit{3}}{\isadigit{1}}\ {\isacharequal}\ {\isacharbraceleft}{\isacharbraceright}{\isachardoublequoteclose}\isanewline
{\isacharbar}\ {\isachardoublequoteopen}subcomp\ sA{\isadigit{3}}{\isadigit{2}}\ {\isacharequal}\ {\isacharbraceleft}{\isacharbraceright}{\isachardoublequoteclose}\isanewline
{\isacharbar}\ {\isachardoublequoteopen}subcomp\ sA{\isadigit{4}}{\isadigit{1}}\ {\isacharequal}\ {\isacharbraceleft}{\isacharbraceright}{\isachardoublequoteclose}\isanewline
{\isacharbar}\ {\isachardoublequoteopen}subcomp\ sA{\isadigit{4}}{\isadigit{2}}\ {\isacharequal}\ {\isacharbraceleft}{\isacharbraceright}{\isachardoublequoteclose}\isanewline
{\isacharbar}\ {\isachardoublequoteopen}subcomp\ sA{\isadigit{7}}{\isadigit{1}}\ {\isacharequal}\ {\isacharbraceleft}{\isacharbraceright}{\isachardoublequoteclose}\isanewline
{\isacharbar}\ {\isachardoublequoteopen}subcomp\ sA{\isadigit{7}}{\isadigit{2}}\ {\isacharequal}\ {\isacharbraceleft}{\isacharbraceright}{\isachardoublequoteclose}\isanewline
{\isacharbar}\ {\isachardoublequoteopen}subcomp\ sA{\isadigit{8}}{\isadigit{1}}\ {\isacharequal}\ {\isacharbraceleft}{\isacharbraceright}{\isachardoublequoteclose}\isanewline
{\isacharbar}\ {\isachardoublequoteopen}subcomp\ sA{\isadigit{8}}{\isadigit{2}}\ {\isacharequal}\ {\isacharbraceleft}{\isacharbraceright}{\isachardoublequoteclose}\isanewline
{\isacharbar}\ {\isachardoublequoteopen}subcomp\ sA{\isadigit{9}}{\isadigit{1}}\ {\isacharequal}\ {\isacharbraceleft}{\isacharbraceright}{\isachardoublequoteclose}\isanewline
{\isacharbar}\ {\isachardoublequoteopen}subcomp\ sA{\isadigit{9}}{\isadigit{2}}\ {\isacharequal}\ {\isacharbraceleft}{\isacharbraceright}{\isachardoublequoteclose}\isanewline
{\isacharbar}\ {\isachardoublequoteopen}subcomp\ sA{\isadigit{9}}{\isadigit{3}}\ {\isacharequal}\ {\isacharbraceleft}{\isacharbraceright}{\isachardoublequoteclose}\isanewline
{\isacharbar}\ {\isachardoublequoteopen}subcomp\ sS{\isadigit{1}}\ {\isacharequal}\ {\isacharbraceleft}\ sA{\isadigit{1}}{\isadigit{2}}\ {\isacharbraceright}{\isachardoublequoteclose}\isanewline
{\isacharbar}\ {\isachardoublequoteopen}subcomp\ sS{\isadigit{2}}\ {\isacharequal}\ {\isacharbraceleft}\ sA{\isadigit{1}}{\isadigit{1}}\ {\isacharbraceright}{\isachardoublequoteclose}\isanewline
{\isacharbar}\ {\isachardoublequoteopen}subcomp\ sS{\isadigit{3}}\ {\isacharequal}\ {\isacharbraceleft}\ sA{\isadigit{2}}{\isadigit{1}}\ {\isacharbraceright}{\isachardoublequoteclose}\isanewline
{\isacharbar}\ {\isachardoublequoteopen}subcomp\ sS{\isadigit{4}}\ {\isacharequal}\ {\isacharbraceleft}\ sA{\isadigit{2}}{\isadigit{3}}\ {\isacharbraceright}{\isachardoublequoteclose}\isanewline
{\isacharbar}\ {\isachardoublequoteopen}subcomp\ sS{\isadigit{5}}\ {\isacharequal}\ {\isacharbraceleft}\ sA{\isadigit{3}}{\isadigit{2}}\ {\isacharbraceright}{\isachardoublequoteclose}\isanewline
{\isacharbar}\ {\isachardoublequoteopen}subcomp\ sS{\isadigit{6}}\ {\isacharequal}\ {\isacharbraceleft}\ sA{\isadigit{2}}{\isadigit{2}}{\isacharcomma}\ sA{\isadigit{3}}{\isadigit{1}}{\isacharcomma}\ sA{\isadigit{4}}{\isadigit{1}}\ {\isacharbraceright}{\isachardoublequoteclose}\isanewline
{\isacharbar}\ {\isachardoublequoteopen}subcomp\ sS{\isadigit{7}}\ {\isacharequal}\ {\isacharbraceleft}\ sA{\isadigit{4}}{\isadigit{2}}{\isacharbraceright}{\isachardoublequoteclose}\isanewline
{\isacharbar}\ {\isachardoublequoteopen}subcomp\ sS{\isadigit{8}}\ {\isacharequal}\ {\isacharbraceleft}\ sA{\isadigit{5}}\ {\isacharbraceright}{\isachardoublequoteclose}\isanewline
{\isacharbar}\ {\isachardoublequoteopen}subcomp\ sS{\isadigit{9}}\ {\isacharequal}\ {\isacharbraceleft}\ sA{\isadigit{6}}\ {\isacharbraceright}{\isachardoublequoteclose}\isanewline
{\isacharbar}\ {\isachardoublequoteopen}subcomp\ sS{\isadigit{1}}{\isadigit{0}}\ {\isacharequal}\ {\isacharbraceleft}\ sA{\isadigit{7}}{\isadigit{1}}\ {\isacharbraceright}{\isachardoublequoteclose}\isanewline
{\isacharbar}\ {\isachardoublequoteopen}subcomp\ sS{\isadigit{1}}{\isadigit{1}}\ {\isacharequal}\ {\isacharbraceleft}\ sA{\isadigit{7}}{\isadigit{2}}\ {\isacharbraceright}{\isachardoublequoteclose}\isanewline
{\isacharbar}\ {\isachardoublequoteopen}subcomp\ sS{\isadigit{1}}{\isadigit{2}}\ {\isacharequal}\ {\isacharbraceleft}\ sA{\isadigit{8}}{\isadigit{1}}{\isacharcomma}\ sA{\isadigit{9}}{\isadigit{1}}\ {\isacharbraceright}{\isachardoublequoteclose}\isanewline
{\isacharbar}\ {\isachardoublequoteopen}subcomp\ sS{\isadigit{1}}{\isadigit{3}}\ {\isacharequal}\ {\isacharbraceleft}\ sA{\isadigit{9}}{\isadigit{2}}\ {\isacharbraceright}{\isachardoublequoteclose}\isanewline
{\isacharbar}\ {\isachardoublequoteopen}subcomp\ sS{\isadigit{1}}{\isadigit{4}}\ {\isacharequal}\ {\isacharbraceleft}\ sA{\isadigit{8}}{\isadigit{2}}\ {\isacharbraceright}{\isachardoublequoteclose}\isanewline
{\isacharbar}\ {\isachardoublequoteopen}subcomp\ sS{\isadigit{1}}{\isadigit{5}}\ {\isacharequal}\ {\isacharbraceleft}\ sA{\isadigit{9}}{\isadigit{3}}\ {\isacharbraceright}{\isachardoublequoteclose}\isanewline
{\isacharbar}\ {\isachardoublequoteopen}subcomp\ sS{\isadigit{1}}opt\ {\isacharequal}\ {\isacharbraceleft}\ sA{\isadigit{1}}{\isadigit{1}}{\isacharcomma}\ sA{\isadigit{1}}{\isadigit{2}}\ {\isacharbraceright}{\isachardoublequoteclose}\isanewline
{\isacharbar}\ {\isachardoublequoteopen}subcomp\ sS{\isadigit{4}}opt\ {\isacharequal}\ {\isacharbraceleft}\ sA{\isadigit{2}}{\isadigit{2}}{\isacharcomma}\ sA{\isadigit{2}}{\isadigit{3}}{\isacharcomma}\ sA{\isadigit{3}}{\isadigit{1}}{\isacharcomma}\ sA{\isadigit{3}}{\isadigit{2}}{\isacharcomma}\ sA{\isadigit{4}}{\isadigit{1}}\ {\isacharbraceright}{\isachardoublequoteclose}\isanewline
{\isacharbar}\ {\isachardoublequoteopen}subcomp\ sS{\isadigit{7}}opt\ {\isacharequal}\ {\isacharbraceleft}\ sA{\isadigit{4}}{\isadigit{2}}{\isacharcomma}\ sA{\isadigit{5}}\ {\isacharbraceright}{\isachardoublequoteclose}\isanewline
{\isacharbar}\ {\isachardoublequoteopen}subcomp\ sS{\isadigit{1}}{\isadigit{1}}opt\ {\isacharequal}\ {\isacharbraceleft}\ sA{\isadigit{7}}{\isadigit{2}}{\isacharcomma}\ sA{\isadigit{8}}{\isadigit{2}}{\isacharcomma}\ sA{\isadigit{9}}{\isadigit{3}}\ {\isacharbraceright}{\isachardoublequoteclose}\isanewline
\isanewline
\isamarkupcmt{function AbstrLevel maps abstraction level ID to the corresponding set of components%
}
\isanewline

\isacommand{axiomatization}\isamarkupfalse%
\isanewline
\ \ AbstrLevel\ {\isacharcolon}{\isacharcolon}\ \ {\isachardoublequoteopen}AbstrLevelsID\ {\isasymRightarrow}\ CSet\ set{\isachardoublequoteclose}\isanewline
\isakeyword{where}\isanewline
AbstrLevel{\isadigit{0}}{\isacharcolon}\isanewline
{\isachardoublequoteopen}AbstrLevel\ level{\isadigit{0}}\ {\isacharequal}\ {\isacharbraceleft}sA{\isadigit{1}}{\isacharcomma}\ sA{\isadigit{2}}{\isacharcomma}\ sA{\isadigit{3}}{\isacharcomma}\ sA{\isadigit{4}}{\isacharcomma}\ sA{\isadigit{5}}{\isacharcomma}\ sA{\isadigit{6}}{\isacharcomma}\ sA{\isadigit{7}}{\isacharcomma}\ sA{\isadigit{8}}{\isacharcomma}\ sA{\isadigit{9}}{\isacharbraceright}{\isachardoublequoteclose}\isanewline
\isakeyword{and}\isanewline
AbstrLevel{\isadigit{1}}{\isacharcolon}\isanewline
{\isachardoublequoteopen}AbstrLevel\ level{\isadigit{1}}\ {\isacharequal}\ {\isacharbraceleft}sA{\isadigit{1}}{\isadigit{1}}{\isacharcomma}\ sA{\isadigit{1}}{\isadigit{2}}{\isacharcomma}\ sA{\isadigit{2}}{\isadigit{1}}{\isacharcomma}\ sA{\isadigit{2}}{\isadigit{2}}{\isacharcomma}\ sA{\isadigit{2}}{\isadigit{3}}{\isacharcomma}\ sA{\isadigit{3}}{\isadigit{1}}{\isacharcomma}\ sA{\isadigit{3}}{\isadigit{2}}{\isacharcomma}\ \isanewline
\ sA{\isadigit{4}}{\isadigit{1}}{\isacharcomma}\ sA{\isadigit{4}}{\isadigit{2}}{\isacharcomma}\ sA{\isadigit{5}}{\isacharcomma}\ sA{\isadigit{6}}{\isacharcomma}\ sA{\isadigit{7}}{\isadigit{1}}{\isacharcomma}\ sA{\isadigit{7}}{\isadigit{2}}{\isacharcomma}\ sA{\isadigit{8}}{\isadigit{1}}{\isacharcomma}\ sA{\isadigit{8}}{\isadigit{2}}{\isacharcomma}\ sA{\isadigit{9}}{\isadigit{1}}{\isacharcomma}\ sA{\isadigit{9}}{\isadigit{2}}{\isacharcomma}\ sA{\isadigit{9}}{\isadigit{3}}{\isacharbraceright}{\isachardoublequoteclose}\isanewline
\isakeyword{and}\isanewline
AbstrLevel{\isadigit{2}}{\isacharcolon}\isanewline
{\isachardoublequoteopen}AbstrLevel\ level{\isadigit{2}}\ {\isacharequal}\ {\isacharbraceleft}sS{\isadigit{1}}{\isacharcomma}\ sS{\isadigit{2}}{\isacharcomma}\ sS{\isadigit{3}}{\isacharcomma}\ sS{\isadigit{4}}{\isacharcomma}\ sS{\isadigit{5}}{\isacharcomma}\ sS{\isadigit{6}}{\isacharcomma}\ sS{\isadigit{7}}{\isacharcomma}\ sS{\isadigit{8}}{\isacharcomma}\ \isanewline
\ \ \ \ \ \ \ \ \ \ \ \ \ \ \ \ \ \ \ \ \ \ \ \ \ \ \ \ \ \ \ sS{\isadigit{9}}{\isacharcomma}\ sS{\isadigit{1}}{\isadigit{0}}{\isacharcomma}\ sS{\isadigit{1}}{\isadigit{1}}{\isacharcomma}\ sS{\isadigit{1}}{\isadigit{2}}{\isacharcomma}\ sS{\isadigit{1}}{\isadigit{3}}{\isacharcomma}\ sS{\isadigit{1}}{\isadigit{4}}{\isacharcomma}\ sS{\isadigit{1}}{\isadigit{5}}{\isacharbraceright}{\isachardoublequoteclose}\isanewline
\isakeyword{and}\isanewline
AbstrLevel{\isadigit{3}}{\isacharcolon}\isanewline
{\isachardoublequoteopen}AbstrLevel\ level{\isadigit{3}}\ {\isacharequal}\ {\isacharbraceleft}sS{\isadigit{1}}opt{\isacharcomma}\ sS{\isadigit{3}}{\isacharcomma}\ sS{\isadigit{4}}opt{\isacharcomma}\ sS{\isadigit{7}}opt{\isacharcomma}\ sS{\isadigit{9}}{\isacharcomma}\ sS{\isadigit{1}}{\isadigit{0}}{\isacharcomma}\ sS{\isadigit{1}}{\isadigit{1}}opt{\isacharcomma}\ sS{\isadigit{1}}{\isadigit{2}}{\isacharcomma}\ sS{\isadigit{1}}{\isadigit{3}}\ {\isacharbraceright}{\isachardoublequoteclose}\isanewline
\isanewline
\isamarkupcmt{function VARfrom maps variable ID to the set of input channels it depends from%
}
\isanewline
\isacommand{fun}\isamarkupfalse%
\ VARfrom\ {\isacharcolon}{\isacharcolon}\ {\isachardoublequoteopen}varID\ {\isasymRightarrow}\ chanID\ set{\isachardoublequoteclose}\isanewline
\isakeyword{where}\isanewline
\ \ \ {\isachardoublequoteopen}VARfrom\ stA{\isadigit{1}}\ {\isacharequal}\ {\isacharbraceleft}data{\isadigit{1}}{\isacharbraceright}{\isachardoublequoteclose}\isanewline
{\isacharbar}\ {\isachardoublequoteopen}VARfrom\ stA{\isadigit{2}}\ {\isacharequal}\ {\isacharbraceleft}data{\isadigit{3}}{\isacharbraceright}{\isachardoublequoteclose}\isanewline
{\isacharbar}\ {\isachardoublequoteopen}VARfrom\ stA{\isadigit{4}}\ {\isacharequal}\ {\isacharbraceleft}data{\isadigit{6}}{\isacharcomma}\ data{\isadigit{7}}{\isacharbraceright}{\isachardoublequoteclose}\ \isanewline
{\isacharbar}\ {\isachardoublequoteopen}VARfrom\ stA{\isadigit{6}}\ {\isacharequal}\ {\isacharbraceleft}data{\isadigit{1}}{\isadigit{4}}{\isacharbraceright}{\isachardoublequoteclose}\ \isanewline
\isanewline
\isamarkupcmt{function VARto maps variable ID to the set of output channels depending from this variable%
}
\isanewline
\isacommand{fun}\isamarkupfalse%
\ VARto\ {\isacharcolon}{\isacharcolon}\ {\isachardoublequoteopen}varID\ {\isasymRightarrow}\ chanID\ set{\isachardoublequoteclose}\isanewline
\isakeyword{where}\isanewline
\ \ \ {\isachardoublequoteopen}VARto\ stA{\isadigit{1}}\ {\isacharequal}\ {\isacharbraceleft}data{\isadigit{1}}{\isadigit{0}}{\isacharbraceright}{\isachardoublequoteclose}\isanewline
{\isacharbar}\ {\isachardoublequoteopen}VARto\ stA{\isadigit{2}}\ {\isacharequal}\ {\isacharbraceleft}data{\isadigit{4}}{\isacharcomma}\ data{\isadigit{1}}{\isadigit{2}}{\isacharbraceright}{\isachardoublequoteclose}\isanewline
{\isacharbar}\ {\isachardoublequoteopen}VARto\ stA{\isadigit{4}}\ {\isacharequal}\ {\isacharbraceleft}data{\isadigit{3}}{\isacharbraceright}{\isachardoublequoteclose}\ \isanewline
{\isacharbar}\ {\isachardoublequoteopen}VARto\ stA{\isadigit{6}}\ {\isacharequal}\ {\isacharbraceleft}data{\isadigit{1}}{\isadigit{5}}{\isacharcomma}\ data{\isadigit{1}}{\isadigit{6}}{\isacharbraceright}{\isachardoublequoteclose}\ \isanewline
\isanewline
\isamarkupcmt{function OUTfromCh maps channel ID to the set of input channels%
}
\isanewline
\isamarkupcmt{from which it depends derectly;%
}
\isanewline
\isamarkupcmt{an empty set means that the channel is either input of the system or%
}
\isanewline
\isamarkupcmt{its values are computed from local variables or are generated%
}
\isanewline
\isamarkupcmt{within some component independently%
}
\isanewline
\isacommand{fun}\isamarkupfalse%
\ OUTfromCh\ {\isacharcolon}{\isacharcolon}\ \ {\isachardoublequoteopen}chanID\ {\isasymRightarrow}\ chanID\ set{\isachardoublequoteclose}\isanewline
\isakeyword{where}\isanewline
\isanewline
\isanewline
\ \ \ {\isachardoublequoteopen}OUTfromCh\ data{\isadigit{1}}\ {\isacharequal}\ {\isacharbraceleft}{\isacharbraceright}{\isachardoublequoteclose}\isanewline
{\isacharbar}\ {\isachardoublequoteopen}OUTfromCh\ data{\isadigit{2}}\ {\isacharequal}\ {\isacharbraceleft}data{\isadigit{1}}{\isacharbraceright}{\isachardoublequoteclose}\isanewline
{\isacharbar}\ {\isachardoublequoteopen}OUTfromCh\ data{\isadigit{3}}\ {\isacharequal}\ {\isacharbraceleft}{\isacharbraceright}{\isachardoublequoteclose}\isanewline
{\isacharbar}\ {\isachardoublequoteopen}OUTfromCh\ data{\isadigit{4}}\ {\isacharequal}\ {\isacharbraceleft}data{\isadigit{2}}{\isacharbraceright}{\isachardoublequoteclose}\isanewline
{\isacharbar}\ {\isachardoublequoteopen}OUTfromCh\ data{\isadigit{5}}\ {\isacharequal}\ {\isacharbraceleft}data{\isadigit{2}}{\isacharbraceright}{\isachardoublequoteclose}\isanewline
{\isacharbar}\ {\isachardoublequoteopen}OUTfromCh\ data{\isadigit{6}}\ {\isacharequal}\ {\isacharbraceleft}data{\isadigit{4}}{\isacharbraceright}{\isachardoublequoteclose}\isanewline
{\isacharbar}\ {\isachardoublequoteopen}OUTfromCh\ data{\isadigit{7}}\ {\isacharequal}\ {\isacharbraceleft}data{\isadigit{5}}{\isacharbraceright}{\isachardoublequoteclose}\isanewline
{\isacharbar}\ {\isachardoublequoteopen}OUTfromCh\ data{\isadigit{8}}\ {\isacharequal}\ {\isacharbraceleft}data{\isadigit{1}}{\isadigit{3}}{\isacharbraceright}{\isachardoublequoteclose}\isanewline
{\isacharbar}\ {\isachardoublequoteopen}OUTfromCh\ data{\isadigit{9}}\ {\isacharequal}\ {\isacharbraceleft}data{\isadigit{8}}{\isacharbraceright}{\isachardoublequoteclose}\isanewline
{\isacharbar}\ {\isachardoublequoteopen}OUTfromCh\ data{\isadigit{1}}{\isadigit{0}}\ {\isacharequal}\ {\isacharbraceleft}{\isacharbraceright}{\isachardoublequoteclose}\isanewline
{\isacharbar}\ {\isachardoublequoteopen}OUTfromCh\ data{\isadigit{1}}{\isadigit{1}}\ {\isacharequal}\ {\isacharbraceleft}data{\isadigit{2}}{\isacharbraceright}{\isachardoublequoteclose}\isanewline
{\isacharbar}\ {\isachardoublequoteopen}OUTfromCh\ data{\isadigit{1}}{\isadigit{2}}\ {\isacharequal}\ {\isacharbraceleft}{\isacharbraceright}{\isachardoublequoteclose}\isanewline
{\isacharbar}\ {\isachardoublequoteopen}OUTfromCh\ data{\isadigit{1}}{\isadigit{3}}\ {\isacharequal}\ {\isacharbraceleft}{\isacharbraceright}{\isachardoublequoteclose}\isanewline
{\isacharbar}\ {\isachardoublequoteopen}OUTfromCh\ data{\isadigit{1}}{\isadigit{4}}\ {\isacharequal}\ {\isacharbraceleft}{\isacharbraceright}{\isachardoublequoteclose}\isanewline
{\isacharbar}\ {\isachardoublequoteopen}OUTfromCh\ data{\isadigit{1}}{\isadigit{5}}\ {\isacharequal}\ {\isacharbraceleft}{\isacharbraceright}{\isachardoublequoteclose}\isanewline
{\isacharbar}\ {\isachardoublequoteopen}OUTfromCh\ data{\isadigit{1}}{\isadigit{6}}\ {\isacharequal}\ {\isacharbraceleft}{\isacharbraceright}{\isachardoublequoteclose}\isanewline
{\isacharbar}\ {\isachardoublequoteopen}OUTfromCh\ data{\isadigit{1}}{\isadigit{7}}\ {\isacharequal}\ {\isacharbraceleft}data{\isadigit{1}}{\isadigit{5}}{\isacharbraceright}{\isachardoublequoteclose}\isanewline
{\isacharbar}\ {\isachardoublequoteopen}OUTfromCh\ data{\isadigit{1}}{\isadigit{8}}\ {\isacharequal}\ {\isacharbraceleft}data{\isadigit{1}}{\isadigit{6}}{\isacharbraceright}{\isachardoublequoteclose}\isanewline
{\isacharbar}\ {\isachardoublequoteopen}OUTfromCh\ data{\isadigit{1}}{\isadigit{9}}\ {\isacharequal}\ {\isacharbraceleft}{\isacharbraceright}{\isachardoublequoteclose}\isanewline
{\isacharbar}\ {\isachardoublequoteopen}OUTfromCh\ data{\isadigit{2}}{\isadigit{0}}\ {\isacharequal}\ {\isacharbraceleft}data{\isadigit{1}}{\isadigit{7}}{\isacharcomma}\ data{\isadigit{2}}{\isadigit{2}}{\isacharbraceright}{\isachardoublequoteclose}\isanewline
{\isacharbar}\ {\isachardoublequoteopen}OUTfromCh\ data{\isadigit{2}}{\isadigit{1}}\ {\isacharequal}\ {\isacharbraceleft}data{\isadigit{1}}{\isadigit{8}}{\isacharcomma}\ data{\isadigit{1}}{\isadigit{9}}{\isacharbraceright}{\isachardoublequoteclose}\isanewline
{\isacharbar}\ {\isachardoublequoteopen}OUTfromCh\ data{\isadigit{2}}{\isadigit{2}}\ {\isacharequal}\ {\isacharbraceleft}data{\isadigit{2}}{\isadigit{0}}{\isacharbraceright}{\isachardoublequoteclose}\isanewline
{\isacharbar}\ {\isachardoublequoteopen}OUTfromCh\ data{\isadigit{2}}{\isadigit{3}}\ {\isacharequal}\ {\isacharbraceleft}data{\isadigit{2}}{\isadigit{1}}{\isacharbraceright}{\isachardoublequoteclose}\isanewline
{\isacharbar}\ {\isachardoublequoteopen}OUTfromCh\ data{\isadigit{2}}{\isadigit{4}}\ {\isacharequal}\ {\isacharbraceleft}data{\isadigit{2}}{\isadigit{0}}{\isacharbraceright}{\isachardoublequoteclose}\isanewline
\isanewline
\isamarkupcmt{function OUTfromV maps channel ID to the set of local variables it depends from%
}
\isanewline
\isacommand{fun}\isamarkupfalse%
\ OUTfromV\ {\isacharcolon}{\isacharcolon}\ \ {\isachardoublequoteopen}chanID\ {\isasymRightarrow}\ varID\ set{\isachardoublequoteclose}\ \isanewline
\isakeyword{where}\isanewline
\ \ \ {\isachardoublequoteopen}OUTfromV\ data{\isadigit{1}}\ {\isacharequal}\ {\isacharbraceleft}{\isacharbraceright}{\isachardoublequoteclose}\isanewline
{\isacharbar}\ {\isachardoublequoteopen}OUTfromV\ data{\isadigit{2}}\ {\isacharequal}\ {\isacharbraceleft}{\isacharbraceright}{\isachardoublequoteclose}\isanewline
{\isacharbar}\ {\isachardoublequoteopen}OUTfromV\ data{\isadigit{3}}\ {\isacharequal}\ {\isacharbraceleft}stA{\isadigit{4}}{\isacharbraceright}{\isachardoublequoteclose}\isanewline
{\isacharbar}\ {\isachardoublequoteopen}OUTfromV\ data{\isadigit{4}}\ {\isacharequal}\ {\isacharbraceleft}stA{\isadigit{2}}{\isacharbraceright}{\isachardoublequoteclose}\isanewline
{\isacharbar}\ {\isachardoublequoteopen}OUTfromV\ data{\isadigit{5}}\ {\isacharequal}\ {\isacharbraceleft}{\isacharbraceright}{\isachardoublequoteclose}\isanewline
{\isacharbar}\ {\isachardoublequoteopen}OUTfromV\ data{\isadigit{6}}\ {\isacharequal}\ {\isacharbraceleft}{\isacharbraceright}{\isachardoublequoteclose}\isanewline
{\isacharbar}\ {\isachardoublequoteopen}OUTfromV\ data{\isadigit{7}}\ {\isacharequal}\ {\isacharbraceleft}{\isacharbraceright}{\isachardoublequoteclose}\isanewline
{\isacharbar}\ {\isachardoublequoteopen}OUTfromV\ data{\isadigit{8}}\ {\isacharequal}\ {\isacharbraceleft}{\isacharbraceright}{\isachardoublequoteclose}\isanewline
{\isacharbar}\ {\isachardoublequoteopen}OUTfromV\ data{\isadigit{9}}\ {\isacharequal}\ {\isacharbraceleft}{\isacharbraceright}{\isachardoublequoteclose}\isanewline
{\isacharbar}\ {\isachardoublequoteopen}OUTfromV\ data{\isadigit{1}}{\isadigit{0}}\ {\isacharequal}\ {\isacharbraceleft}stA{\isadigit{1}}{\isacharbraceright}{\isachardoublequoteclose}\isanewline
{\isacharbar}\ {\isachardoublequoteopen}OUTfromV\ data{\isadigit{1}}{\isadigit{1}}\ {\isacharequal}\ {\isacharbraceleft}{\isacharbraceright}{\isachardoublequoteclose}\isanewline
{\isacharbar}\ {\isachardoublequoteopen}OUTfromV\ data{\isadigit{1}}{\isadigit{2}}\ {\isacharequal}\ {\isacharbraceleft}stA{\isadigit{2}}{\isacharbraceright}{\isachardoublequoteclose}\isanewline
{\isacharbar}\ {\isachardoublequoteopen}OUTfromV\ data{\isadigit{1}}{\isadigit{3}}\ {\isacharequal}\ {\isacharbraceleft}{\isacharbraceright}{\isachardoublequoteclose}\isanewline
{\isacharbar}\ {\isachardoublequoteopen}OUTfromV\ data{\isadigit{1}}{\isadigit{4}}\ {\isacharequal}\ {\isacharbraceleft}{\isacharbraceright}{\isachardoublequoteclose}\isanewline
{\isacharbar}\ {\isachardoublequoteopen}OUTfromV\ data{\isadigit{1}}{\isadigit{5}}\ {\isacharequal}\ {\isacharbraceleft}stA{\isadigit{6}}{\isacharbraceright}{\isachardoublequoteclose}\isanewline
{\isacharbar}\ {\isachardoublequoteopen}OUTfromV\ data{\isadigit{1}}{\isadigit{6}}\ {\isacharequal}\ {\isacharbraceleft}stA{\isadigit{6}}{\isacharbraceright}{\isachardoublequoteclose}\isanewline
{\isacharbar}\ {\isachardoublequoteopen}OUTfromV\ data{\isadigit{1}}{\isadigit{7}}\ {\isacharequal}\ {\isacharbraceleft}{\isacharbraceright}{\isachardoublequoteclose}\isanewline
{\isacharbar}\ {\isachardoublequoteopen}OUTfromV\ data{\isadigit{1}}{\isadigit{8}}\ {\isacharequal}\ {\isacharbraceleft}{\isacharbraceright}{\isachardoublequoteclose}\isanewline
{\isacharbar}\ {\isachardoublequoteopen}OUTfromV\ data{\isadigit{1}}{\isadigit{9}}\ {\isacharequal}\ {\isacharbraceleft}{\isacharbraceright}{\isachardoublequoteclose}\isanewline
{\isacharbar}\ {\isachardoublequoteopen}OUTfromV\ data{\isadigit{2}}{\isadigit{0}}\ {\isacharequal}\ {\isacharbraceleft}{\isacharbraceright}{\isachardoublequoteclose}\isanewline
{\isacharbar}\ {\isachardoublequoteopen}OUTfromV\ data{\isadigit{2}}{\isadigit{1}}\ {\isacharequal}\ {\isacharbraceleft}{\isacharbraceright}{\isachardoublequoteclose}\isanewline
{\isacharbar}\ {\isachardoublequoteopen}OUTfromV\ data{\isadigit{2}}{\isadigit{2}}\ {\isacharequal}\ {\isacharbraceleft}{\isacharbraceright}{\isachardoublequoteclose}\isanewline
{\isacharbar}\ {\isachardoublequoteopen}OUTfromV\ data{\isadigit{2}}{\isadigit{3}}\ {\isacharequal}\ {\isacharbraceleft}{\isacharbraceright}{\isachardoublequoteclose}\isanewline
{\isacharbar}\ {\isachardoublequoteopen}OUTfromV\ data{\isadigit{2}}{\isadigit{4}}\ {\isacharequal}\ {\isacharbraceleft}{\isacharbraceright}{\isachardoublequoteclose}\isanewline
\isanewline
\isamarkupcmt{Set of channels channels which have  UplSize measure greather that the predifined value $HighLoad$%
}
\isanewline
\isacommand{definition}\isamarkupfalse%
\isanewline
\ \ \ UplSizeHighLoad\ {\isacharcolon}{\isacharcolon}\ \ {\isachardoublequoteopen}chanID\ set{\isachardoublequoteclose}\isanewline
\isakeyword{where}\isanewline
\ \ {\isachardoublequoteopen}UplSizeHighLoad\ {\isasymequiv}\ {\isacharbraceleft}data{\isadigit{1}}{\isacharcomma}\ data{\isadigit{4}}{\isacharcomma}\ data{\isadigit{5}}{\isacharcomma}\ data{\isadigit{6}}{\isacharcomma}\ data{\isadigit{7}}{\isacharcomma}\ data{\isadigit{8}}{\isacharcomma}\ data{\isadigit{1}}{\isadigit{8}}{\isacharcomma}\ data{\isadigit{2}}{\isadigit{1}}{\isacharbraceright}{\isachardoublequoteclose}\isanewline
\isanewline
\isamarkupcmt{Set of components from the abstraction level 1 for which the Perf measure is greather that the predifined value $HighPerf$%
}
\isanewline
\isacommand{definition}\isamarkupfalse%
\isanewline
\ \ \ HighPerfSet\ {\isacharcolon}{\isacharcolon}\ \ {\isachardoublequoteopen}CSet\ set{\isachardoublequoteclose}\isanewline
\isakeyword{where}\isanewline
\ \ {\isachardoublequoteopen}HighPerfSet\ {\isasymequiv}\ {\isacharbraceleft}sA{\isadigit{2}}{\isadigit{2}}{\isacharcomma}\ sA{\isadigit{2}}{\isadigit{3}}{\isacharcomma}\ sA{\isadigit{4}}{\isadigit{1}}{\isacharcomma}\ sA{\isadigit{4}}{\isadigit{2}}{\isacharcomma}\ sA{\isadigit{7}}{\isadigit{2}}{\isacharcomma}\ sA{\isadigit{9}}{\isadigit{3}}{\isacharbraceright}{\isachardoublequoteclose}\isanewline
\isadelimtheory
\isanewline
\endisadelimtheory
\isatagtheory
\isacommand{end}\isamarkupfalse%
\endisatagtheory
{\isafoldtheory}%
\isadelimtheory
\isanewline
\endisadelimtheory
\end{isabellebody}%

%
\begin{isabellebody}%
\def\isabellecontext{DataDependencies}%
\isamarkupheader{Inter-/Intracomponent dependencies%
}
\isamarkuptrue%
\isadelimtheory
\endisadelimtheory
\isatagtheory
\isacommand{theory}\isamarkupfalse%
\ DataDependencies\isanewline
\ \ \isakeyword{imports}\ DataDependenciesConcreteValues\isanewline
\isakeyword{begin}\isanewline
\isanewline
\isamarkupcmt{component and its subcomponents should be defined on different abstraction levels%
}
\endisatagtheory
{\isafoldtheory}%
\isadelimtheory
\endisadelimtheory
\isanewline
\isacommand{definition}\isamarkupfalse%
\isanewline
correctCompositionDiffLevels\ {\isacharcolon}{\isacharcolon}\ {\isachardoublequoteopen}CSet\ {\isasymRightarrow}\ bool{\isachardoublequoteclose}\isanewline
\isakeyword{where}\ \isanewline
\ \ {\isachardoublequoteopen}correctCompositionDiffLevels\ S\ {\isasymequiv}\ \isanewline
\ \ \ {\isasymforall}\ C\ {\isasymin}\ \ subcomp\ S{\isachardot}\ {\isasymforall}\ i{\isachardot}\ S\ {\isasymin}\ AbstrLevel\ i\ {\isasymlongrightarrow}\ C\ {\isasymnotin}\ AbstrLevel\ i{\isachardoublequoteclose}\isanewline
\isanewline
\isamarkupcmt{General system's property: for all abstraction levels and all components should hold%
}
\isanewline
\isamarkupcmt{component and its subcomponents should be defined on different abstraction levels%
}
\isanewline
\isacommand{definition}\isamarkupfalse%
\isanewline
correctCompositionDiffLevelsSYSTEM\ {\isacharcolon}{\isacharcolon}\ {\isachardoublequoteopen}bool{\isachardoublequoteclose}\isanewline
\isakeyword{where}\ \isanewline
\ \ {\isachardoublequoteopen}correctCompositionDiffLevelsSYSTEM\ {\isasymequiv}\ \isanewline
\ \ \ {\isacharparenleft}{\isasymforall}\ S{\isacharcolon}{\isacharcolon}CSet{\isachardot}\ {\isacharparenleft}correctCompositionDiffLevels\ S{\isacharparenright}{\isacharparenright}{\isachardoublequoteclose}\isanewline
\isanewline
\isamarkupcmt{if a local variable belongs to one of the subcomponents, it also belongs to the composed component%
}
\isanewline
\isacommand{definition}\isamarkupfalse%
\isanewline
correctCompositionVAR\ {\isacharcolon}{\isacharcolon}\ \ {\isachardoublequoteopen}CSet\ {\isasymRightarrow}\ bool{\isachardoublequoteclose}\isanewline
\isakeyword{where}\isanewline
\ \ {\isachardoublequoteopen}correctCompositionVAR\ S\ {\isasymequiv}\ \isanewline
\ \ \ {\isasymforall}\ C\ {\isasymin}\ \ subcomp\ S{\isachardot}\ {\isasymforall}\ v\ {\isasymin}\ VAR\ C{\isachardot}\ \ v\ {\isasymin}\ VAR\ S{\isachardoublequoteclose}\isanewline
\isanewline
\isamarkupcmt{General system's property: for all abstraction levels and all components should hold%
}
\isanewline
\isamarkupcmt{if a local variable belongs to one of the subcomponents, it also belongs to the composed component%
}
\isanewline
\isacommand{definition}\isamarkupfalse%
\isanewline
correctCompositionVARSYSTEM\ {\isacharcolon}{\isacharcolon}\ \ {\isachardoublequoteopen}bool{\isachardoublequoteclose}\isanewline
\isakeyword{where}\isanewline
\ \ {\isachardoublequoteopen}correctCompositionVARSYSTEM\ {\isasymequiv}\ \isanewline
\ \ \ {\isacharparenleft}{\isasymforall}\ S{\isacharcolon}{\isacharcolon}CSet{\isachardot}\ {\isacharparenleft}correctCompositionVAR\ S{\isacharparenright}{\isacharparenright}{\isachardoublequoteclose}\isanewline
\isanewline
\isamarkupcmt{after correct decomposition of a component each of its local variable can belong only to one of its subcomponents%
}
\isanewline
\isacommand{definition}\isamarkupfalse%
\isanewline
correctDeCompositionVAR\ {\isacharcolon}{\isacharcolon}\ \ {\isachardoublequoteopen}CSet\ {\isasymRightarrow}\ bool{\isachardoublequoteclose}\isanewline
\isakeyword{where}\isanewline
\ \ {\isachardoublequoteopen}correctDeCompositionVAR\ S\ {\isasymequiv}\ \isanewline
\ \ \ {\isasymforall}\ v\ {\isasymin}\ VAR\ S{\isachardot}\ \ {\isasymforall}\ C{\isadigit{1}}\ {\isasymin}\ \ subcomp\ S{\isachardot}\ {\isasymforall}\ C{\isadigit{2}}\ {\isasymin}\ \ subcomp\ S{\isachardot}\ v\ {\isasymin}\ VAR\ C{\isadigit{1}}\ {\isasymand}\ v\ {\isasymin}\ VAR\ C{\isadigit{2}}\ {\isasymlongrightarrow}\ C{\isadigit{1}}\ {\isacharequal}\ C{\isadigit{2}}{\isachardoublequoteclose}\isanewline
\isanewline
\isamarkupcmt{General system's property: for all abstraction levels and all components should hold%
}
\isanewline
\isamarkupcmt{after correct decomposition of a component each of its local variable can belong only to one of its subcomponents%
}
\isanewline
\isacommand{definition}\isamarkupfalse%
\isanewline
correctDeCompositionVARSYSTEM\ {\isacharcolon}{\isacharcolon}\ \ {\isachardoublequoteopen}bool{\isachardoublequoteclose}\isanewline
\isakeyword{where}\isanewline
\ \ {\isachardoublequoteopen}correctDeCompositionVARSYSTEM\ {\isasymequiv}\ \isanewline
\ \ {\isacharparenleft}{\isasymforall}\ S{\isacharcolon}{\isacharcolon}CSet{\isachardot}\ {\isacharparenleft}correctDeCompositionVAR\ S{\isacharparenright}{\isacharparenright}{\isachardoublequoteclose}\isanewline
\isanewline
\isamarkupcmt{if x is an output channel of a  component C on some anstraction level, 
it cannot be an output of another component on the same level%
}
\isanewline
\isacommand{definition}\isamarkupfalse%
\isanewline
correctCompositionOUT\ {\isacharcolon}{\isacharcolon}\ \ {\isachardoublequoteopen}chanID\ {\isasymRightarrow}\ bool{\isachardoublequoteclose}\isanewline
\isakeyword{where}\isanewline
\ \ {\isachardoublequoteopen}correctCompositionOUT\ x\ {\isasymequiv}\ \isanewline
\ \ \ {\isasymforall}\ C\ i{\isachardot}\ x\ {\isasymin}\ OUT\ C\ {\isasymand}\ C\ {\isasymin}\ AbstrLevel\ i\ {\isasymlongrightarrow}\ \ {\isacharparenleft}{\isasymforall}\ S\ {\isasymin}\ AbstrLevel\ i{\isachardot}\ x\ {\isasymnotin}\ OUT\ S{\isacharparenright}{\isachardoublequoteclose}\isanewline
\isanewline
\isamarkupcmt{General system's property: for all abstraction levels and all channels should hold%
}
\isanewline
\isacommand{definition}\isamarkupfalse%
\isanewline
correctCompositionOUTSYSTEM\ {\isacharcolon}{\isacharcolon}\ \ {\isachardoublequoteopen}bool{\isachardoublequoteclose}\isanewline
\isakeyword{where}\isanewline
\ \ {\isachardoublequoteopen}correctCompositionOUTSYSTEM\ {\isasymequiv}\ {\isacharparenleft}{\isasymforall}\ x{\isachardot}\ correctCompositionOUT\ x{\isacharparenright}{\isachardoublequoteclose}\isanewline
\isanewline
\isamarkupcmt{if X is a subcomponent of a  component C on some anstraction level, 
it cannot be a subcomponent of another component on the same level%
}
\isanewline
\isacommand{definition}\isamarkupfalse%
\isanewline
correctCompositionSubcomp\ {\isacharcolon}{\isacharcolon}\ \ {\isachardoublequoteopen}CSet\ {\isasymRightarrow}\ bool{\isachardoublequoteclose}\isanewline
\isakeyword{where}\isanewline
\ \ {\isachardoublequoteopen}correctCompositionSubcomp\ X\ {\isasymequiv}\ \isanewline
\ \ \ {\isasymforall}\ C\ i{\isachardot}\ X\ {\isasymin}\ subcomp\ C\ {\isasymand}\ C\ {\isasymin}\ AbstrLevel\ i\ {\isasymlongrightarrow}\ \ {\isacharparenleft}{\isasymforall}\ S\ {\isasymin}\ AbstrLevel\ i{\isachardot}\ {\isacharparenleft}S\ {\isasymnoteq}\ C\ {\isasymlongrightarrow}\ X\ {\isasymnotin}\ subcomp\ S{\isacharparenright}{\isacharparenright}{\isachardoublequoteclose}\isanewline
\isanewline
\isamarkupcmt{General system's property: for all abstraction levels and all components should hold%
}
\isanewline
\isacommand{definition}\isamarkupfalse%
\isanewline
correctCompositionSubcompSYSTEM\ {\isacharcolon}{\isacharcolon}\ \ {\isachardoublequoteopen}bool{\isachardoublequoteclose}\isanewline
\isakeyword{where}\isanewline
\ \ {\isachardoublequoteopen}correctCompositionSubcompSYSTEM\ {\isasymequiv}\ {\isacharparenleft}{\isasymforall}\ X{\isachardot}\ correctCompositionSubcomp\ X{\isacharparenright}{\isachardoublequoteclose}\isanewline
\isanewline
\isamarkupcmt{If a component belongs is defined in the set CSet, 
it should belong to at least one abstraction level%
}
\isanewline
\isacommand{definition}\isamarkupfalse%
\isanewline
allComponentsUsed\ {\isacharcolon}{\isacharcolon}\ \ {\isachardoublequoteopen}bool{\isachardoublequoteclose}\isanewline
\isakeyword{where}\isanewline
\ \ {\isachardoublequoteopen}allComponentsUsed\ {\isasymequiv}\ \ {\isasymforall}\ C{\isachardot}\ {\isasymexists}\ i{\isachardot}\ \ C\ {\isasymin}\ AbstrLevel\ i{\isachardoublequoteclose}\isanewline
\isanewline
\isamarkupcmt{if a component does not have any local variables, none of its subcomponents has any local variables%
}
\isanewline
\isacommand{lemma}\isamarkupfalse%
\ correctDeCompositionVARempty{\isacharcolon}\isanewline
\ \ \isakeyword{assumes}\ {\isachardoublequoteopen}correctCompositionVAR\ S{\isachardoublequoteclose}\ \isanewline
\ \ \ \ \ \ \ \ \ \ \isakeyword{and}\ {\isachardoublequoteopen}VAR\ S\ {\isacharequal}\ {\isacharbraceleft}{\isacharbraceright}{\isachardoublequoteclose}\isanewline
\ \ \isakeyword{shows}\ \ \ \ {\isachardoublequoteopen}{\isasymforall}\ C\ {\isasymin}\ \ subcomp\ S{\isachardot}\ VAR\ C\ {\isacharequal}\ {\isacharbraceleft}{\isacharbraceright}{\isachardoublequoteclose}\isanewline
\isadelimproof
\endisadelimproof
\isatagproof
\isacommand{using}\isamarkupfalse%
\ assms\ \isacommand{by}\isamarkupfalse%
\ {\isacharparenleft}metis\ all{\isacharunderscore}not{\isacharunderscore}in{\isacharunderscore}conv\ correctCompositionVAR{\isacharunderscore}def{\isacharparenright}\isanewline
\isanewline
\isanewline
\isamarkupcmt{function OUTfrom maps channel ID to the set of input channels it depends from,%
}
\isanewline
\isamarkupcmt{directly (OUTfromCh) or via local variables (VARfrom)%
}
\ \isanewline
\isamarkupcmt{an empty set means that the channel is either input of the system or%
}
\isanewline
\isamarkupcmt{its values are generated within some component independently%
}
\endisatagproof
{\isafoldproof}%
\isadelimproof
\isanewline
\endisadelimproof
\isacommand{definition}\isamarkupfalse%
\ OUTfrom\ {\isacharcolon}{\isacharcolon}\ \ {\isachardoublequoteopen}chanID\ {\isasymRightarrow}\ chanID\ set{\isachardoublequoteclose}\isanewline
\isakeyword{where}\isanewline
\ {\isachardoublequoteopen}OUTfrom\ x\ {\isasymequiv}\ {\isacharparenleft}OUTfromCh\ x{\isacharparenright}\ {\isasymunion}\ {\isacharbraceleft}y{\isachardot}\ {\isasymexists}\ v{\isachardot}\ v\ {\isasymin}\ {\isacharparenleft}OUTfromV\ x{\isacharparenright}\ {\isasymand}\ y\ {\isasymin}\ {\isacharparenleft}VARfrom\ v{\isacharparenright}{\isacharbraceright}{\isachardoublequoteclose}\isanewline
\ \isanewline
\isamarkupcmt{if x depends from some input channel(s) directly, then exists%
}
\isanewline
\isamarkupcmt{a component which has them as input channels and x as an output channel%
}
\isanewline
\isacommand{definition}\isamarkupfalse%
\isanewline
\ \ OUTfromChCorrect\ {\isacharcolon}{\isacharcolon}\ {\isachardoublequoteopen}chanID\ {\isasymRightarrow}\ bool{\isachardoublequoteclose}\isanewline
\isakeyword{where}\isanewline
\ \ {\isachardoublequoteopen}OUTfromChCorrect\ x\ {\isasymequiv}\isanewline
\ \ \ {\isacharparenleft}OUTfromCh\ x\ {\isasymnoteq}\ {\isacharbraceleft}{\isacharbraceright}\ {\isasymlongrightarrow}\ \isanewline
\ \ \ \ \ \ {\isacharparenleft}{\isasymexists}\ Z\ {\isachardot}\ {\isacharparenleft}x\ {\isasymin}\ {\isacharparenleft}OUT\ Z{\isacharparenright}\ {\isasymand}\ {\isacharparenleft}{\isasymforall}\ y\ {\isasymin}\ {\isacharparenleft}OUTfromCh\ x{\isacharparenright}{\isachardot}\ y\ {\isasymin}\ IN\ Z{\isacharparenright}\ {\isacharparenright}{\isacharparenright}{\isacharparenright}{\isachardoublequoteclose}\isanewline
\isanewline
\isamarkupcmt{General system's property: for channels in the system should hold:%
}
\isanewline
\isamarkupcmt{if x depends from some input channel(s) directly, then exists%
}
\isanewline
\isamarkupcmt{a component which has them as input channels and x as an output channel%
}
\isanewline
\isacommand{definition}\isamarkupfalse%
\isanewline
\ \ OUTfromChCorrectSYSTEM\ {\isacharcolon}{\isacharcolon}\ {\isachardoublequoteopen}bool{\isachardoublequoteclose}\isanewline
\isakeyword{where}\isanewline
\ \ {\isachardoublequoteopen}OUTfromChCorrectSYSTEM\ {\isasymequiv}\ {\isacharparenleft}{\isasymforall}\ x{\isacharcolon}{\isacharcolon}chanID{\isachardot}\ {\isacharparenleft}OUTfromChCorrect\ x{\isacharparenright}{\isacharparenright}{\isachardoublequoteclose}\isanewline
\isanewline
\isanewline
\isamarkupcmt{if x depends from some local variables, then exists a component%
}
\isanewline
\isamarkupcmt{to which these variables belong and which has  x as an output channel%
}
\isanewline
\isacommand{definition}\isamarkupfalse%
\isanewline
\ \ OUTfromVCorrect{\isadigit{1}}\ {\isacharcolon}{\isacharcolon}\ {\isachardoublequoteopen}chanID\ {\isasymRightarrow}\ bool{\isachardoublequoteclose}\isanewline
\isakeyword{where}\isanewline
\ \ {\isachardoublequoteopen}OUTfromVCorrect{\isadigit{1}}\ x\ {\isasymequiv}\isanewline
\ \ \ {\isacharparenleft}OUTfromV\ x\ {\isasymnoteq}\ {\isacharbraceleft}{\isacharbraceright}\ {\isasymlongrightarrow}\ \isanewline
\ \ \ \ \ \ {\isacharparenleft}{\isasymexists}\ Z\ {\isachardot}\ {\isacharparenleft}x\ {\isasymin}\ {\isacharparenleft}OUT\ Z{\isacharparenright}\ {\isasymand}\ {\isacharparenleft}{\isasymforall}\ v\ {\isasymin}\ {\isacharparenleft}OUTfromV\ x{\isacharparenright}{\isachardot}\ v\ {\isasymin}\ VAR\ Z{\isacharparenright}\ {\isacharparenright}{\isacharparenright}{\isacharparenright}{\isachardoublequoteclose}\isanewline
\isanewline
\isamarkupcmt{General system's property: for channels in the system should hold the above property:%
}
\isanewline
\isacommand{definition}\isamarkupfalse%
\isanewline
\ \ OUTfromVCorrect{\isadigit{1}}SYSTEM\ {\isacharcolon}{\isacharcolon}\ {\isachardoublequoteopen}bool{\isachardoublequoteclose}\isanewline
\isakeyword{where}\isanewline
\ \ {\isachardoublequoteopen}OUTfromVCorrect{\isadigit{1}}SYSTEM\ {\isasymequiv}\ {\isacharparenleft}{\isasymforall}\ x{\isacharcolon}{\isacharcolon}chanID{\isachardot}\ {\isacharparenleft}OUTfromVCorrect{\isadigit{1}}\ x{\isacharparenright}{\isacharparenright}{\isachardoublequoteclose}\isanewline
\isanewline
\isamarkupcmt{if x does not depend from any local variables, then it does not belong to any set VARfrom%
}
\isanewline
\isacommand{definition}\isamarkupfalse%
\isanewline
\ \ OUTfromVCorrect{\isadigit{2}}\ {\isacharcolon}{\isacharcolon}\ {\isachardoublequoteopen}chanID\ {\isasymRightarrow}\ bool{\isachardoublequoteclose}\isanewline
\isakeyword{where}\isanewline
\ \ {\isachardoublequoteopen}OUTfromVCorrect{\isadigit{2}}\ x\ {\isasymequiv}\isanewline
\ \ \ {\isacharparenleft}OUTfromV\ x\ {\isacharequal}\ {\isacharbraceleft}{\isacharbraceright}\ {\isasymlongrightarrow}\ {\isacharparenleft}{\isasymforall}\ v{\isacharcolon}{\isacharcolon}varID{\isachardot}\ x\ {\isasymnotin}\ {\isacharparenleft}VARto\ v{\isacharparenright}{\isacharparenright}\ {\isacharparenright}{\isachardoublequoteclose}\isanewline
\isanewline
\isamarkupcmt{General system's property: for channels in the system should hold the above property:%
}
\ \isanewline
\isacommand{definition}\isamarkupfalse%
\isanewline
\ \ OUTfromVCorrect{\isadigit{2}}SYSTEM\ {\isacharcolon}{\isacharcolon}\ {\isachardoublequoteopen}bool{\isachardoublequoteclose}\isanewline
\isakeyword{where}\isanewline
\ \ {\isachardoublequoteopen}OUTfromVCorrect{\isadigit{2}}SYSTEM\ {\isasymequiv}\ \ {\isacharparenleft}{\isasymforall}\ x{\isacharcolon}{\isacharcolon}chanID{\isachardot}\ {\isacharparenleft}OUTfromVCorrect{\isadigit{2}}\ x{\isacharparenright}{\isacharparenright}{\isachardoublequoteclose}\isanewline
\isanewline
\isamarkupcmt{General system's property:%
}
\isanewline
\isamarkupcmt{definitions OUTfromV and VARto should give equivalent mappings%
}
\isanewline
\isacommand{definition}\isamarkupfalse%
\isanewline
\ \ OUTfromV{\isacharunderscore}VARto\ {\isacharcolon}{\isacharcolon}\ {\isachardoublequoteopen}bool{\isachardoublequoteclose}\isanewline
\isakeyword{where}\isanewline
\ \ {\isachardoublequoteopen}OUTfromV{\isacharunderscore}VARto\ {\isasymequiv}\isanewline
\ \ \ {\isacharparenleft}{\isasymforall}\ x{\isacharcolon}{\isacharcolon}chanID{\isachardot}\ {\isasymforall}\ v{\isacharcolon}{\isacharcolon}varID{\isachardot}\ {\isacharparenleft}v\ {\isasymin}\ OUTfromV\ x\ {\isasymlongleftrightarrow}\ x\ {\isasymin}\ {\isacharparenleft}VARto\ v{\isacharparenright}{\isacharparenright}\ {\isacharparenright}{\isachardoublequoteclose}\isanewline
\isanewline
\isamarkupcmt{General system's property for abstraction levels 0 and 1%
}
\isanewline
\isamarkupcmt{if a variable v belongs to a component, then all the channels v%
}
\isanewline
\isamarkupcmt{depends from should be input channels of this component%
}
\isanewline
\isacommand{definition}\isamarkupfalse%
\isanewline
\ \ VARfromCorrectSYSTEM\ {\isacharcolon}{\isacharcolon}\ {\isachardoublequoteopen}bool{\isachardoublequoteclose}\isanewline
\isakeyword{where}\isanewline
\ \ {\isachardoublequoteopen}VARfromCorrectSYSTEM\ {\isasymequiv}\isanewline
\ \ \ {\isacharparenleft}{\isasymforall}\ v{\isacharcolon}{\isacharcolon}varID{\isachardot}\ {\isasymforall}\ Z{\isasymin}\ {\isacharparenleft}{\isacharparenleft}AbstrLevel\ level{\isadigit{0}}{\isacharparenright}\ {\isasymunion}\ {\isacharparenleft}AbstrLevel\ level{\isadigit{1}}{\isacharparenright}{\isacharparenright}{\isachardot}\ \isanewline
\ \ \ \ \ {\isacharparenleft}\ {\isacharparenleft}v\ {\isasymin}\ VAR\ Z{\isacharparenright}\ {\isasymlongrightarrow}\ \ {\isacharparenleft}{\isasymforall}\ x\ {\isasymin}\ VARfrom\ v{\isachardot}\ x\ {\isasymin}\ IN\ Z{\isacharparenright}\ {\isacharparenright}{\isacharparenright}{\isachardoublequoteclose}\isanewline
\isanewline
\isamarkupcmt{General system's property for abstraction levels 0 and 1%
}
\isanewline
\isamarkupcmt{if a variable v belongs to a component, then all the channels v%
}
\isanewline
\isamarkupcmt{provides value to should be input channels of this component%
}
\isanewline
\isacommand{definition}\isamarkupfalse%
\isanewline
\ \ VARtoCorrectSYSTEM\ {\isacharcolon}{\isacharcolon}\ {\isachardoublequoteopen}bool{\isachardoublequoteclose}\isanewline
\isakeyword{where}\isanewline
\ \ {\isachardoublequoteopen}VARtoCorrectSYSTEM\ {\isasymequiv}\isanewline
\ \ \ {\isacharparenleft}{\isasymforall}\ v{\isacharcolon}{\isacharcolon}varID{\isachardot}\ {\isasymforall}\ Z\ {\isasymin}\ {\isacharparenleft}{\isacharparenleft}AbstrLevel\ level{\isadigit{0}}{\isacharparenright}\ {\isasymunion}\ {\isacharparenleft}AbstrLevel\ level{\isadigit{1}}{\isacharparenright}{\isacharparenright}{\isachardot}\ \isanewline
\ \ \ \ \ {\isacharparenleft}\ {\isacharparenleft}v\ {\isasymin}\ VAR\ Z{\isacharparenright}\ {\isasymlongrightarrow}\ \ \ {\isacharparenleft}{\isasymforall}\ x\ {\isasymin}\ VARto\ v{\isachardot}\ x\ {\isasymin}\ OUT\ Z{\isacharparenright}{\isacharparenright}{\isacharparenright}{\isachardoublequoteclose}\isanewline
\isanewline
\isamarkupcmt{to detect local variables, unused for computation of any output%
}
\ \isanewline
\isacommand{definition}\isamarkupfalse%
\isanewline
\ \ VARusefulSYSTEM\ {\isacharcolon}{\isacharcolon}\ {\isachardoublequoteopen}bool{\isachardoublequoteclose}\isanewline
\isakeyword{where}\isanewline
\ \ {\isachardoublequoteopen}VARusefulSYSTEM\ {\isasymequiv}\ {\isacharparenleft}{\isasymforall}\ v{\isacharcolon}{\isacharcolon}varID{\isachardot}\ {\isacharparenleft}VARto\ v\ {\isasymnoteq}\ {\isacharbraceleft}{\isacharbraceright}{\isacharparenright}{\isacharparenright}{\isachardoublequoteclose}\isanewline
\isanewline
\isacommand{lemma}\isamarkupfalse%
\isanewline
\ \ OUTfromV{\isacharunderscore}VARto{\isacharunderscore}lemma{\isacharcolon}\ \isanewline
\ \isakeyword{assumes}\ {\isachardoublequoteopen}OUTfromV\ x\ {\isasymnoteq}\ {\isacharbraceleft}{\isacharbraceright}{\isachardoublequoteclose}\ \ \isakeyword{and}\ {\isachardoublequoteopen}OUTfromV{\isacharunderscore}VARto{\isachardoublequoteclose}\isanewline
\ \isakeyword{shows}\ \ \ \ {\isachardoublequoteopen}{\isasymexists}\ v{\isacharcolon}{\isacharcolon}varID{\isachardot}\ x\ {\isasymin}\ {\isacharparenleft}VARto\ v{\isacharparenright}{\isachardoublequoteclose}\ \isanewline
\isadelimproof
\ %
\endisadelimproof
\isatagproof
\isacommand{using}\isamarkupfalse%
\ assms\ \isacommand{by}\isamarkupfalse%
\ {\isacharparenleft}simp\ add{\isacharcolon}\ OUTfromV{\isacharunderscore}VARto{\isacharunderscore}def{\isacharcomma}\ auto{\isacharparenright}\isanewline
\endisatagproof
{\isafoldproof}%
\isadelimproof
\endisadelimproof
\isamarkupsubsection{Direct and indirect data dependencies between components%
}
\isamarkuptrue%
\isamarkupcmt{The component C should be defined on the same abstraction%
}
\isanewline
\isamarkupcmt{level we are seaching for its direct or indirect sources,%
}
\isanewline
\isamarkupcmt{otherwise we get an empty set as result%
}
\isanewline
\isacommand{definition}\isamarkupfalse%
\isanewline
\ \ DSources\ {\isacharcolon}{\isacharcolon}\ {\isachardoublequoteopen}AbstrLevelsID\ {\isasymRightarrow}\ CSet\ {\isasymRightarrow}\ CSet\ set{\isachardoublequoteclose}\isanewline
\isakeyword{where}\isanewline
\ {\isachardoublequoteopen}DSources\ i\ C\ {\isasymequiv}\ {\isacharbraceleft}Z{\isachardot}\ \ {\isasymexists}\ x{\isachardot}\ x\ {\isasymin}\ {\isacharparenleft}IN\ C{\isacharparenright}\ {\isasymand}\ x\ {\isasymin}\ {\isacharparenleft}OUT\ Z{\isacharparenright}\ {\isasymand}\ Z\ {\isasymin}\ {\isacharparenleft}AbstrLevel\ i{\isacharparenright}\ {\isasymand}\ C\ {\isasymin}\ {\isacharparenleft}AbstrLevel\ i{\isacharparenright}{\isacharbraceright}{\isachardoublequoteclose}\isanewline
\isanewline
\isacommand{lemma}\isamarkupfalse%
\ DSourcesLevelX{\isacharcolon}\isanewline
{\isachardoublequoteopen}{\isacharparenleft}DSources\ i\ X{\isacharparenright}\ \ {\isasymsubseteq}\ {\isacharparenleft}AbstrLevel\ i{\isacharparenright}{\isachardoublequoteclose}\ \isanewline
\isadelimproof
\endisadelimproof
\isatagproof
\isacommand{by}\isamarkupfalse%
\ {\isacharparenleft}simp\ add{\isacharcolon}\ DSources{\isacharunderscore}def{\isacharcomma}\ auto{\isacharparenright}\isanewline
\isanewline
\isamarkupcmt{The component C should be defined on the same abstraction level we are%
}
\ \isanewline
\isamarkupcmt{seaching for its direct or indirect acceptors (coponents, for which C is a source),%
}
\isanewline
\isamarkupcmt{otherwise we get an empty set as result%
}
\endisatagproof
{\isafoldproof}%
\isadelimproof
\isanewline
\endisadelimproof
\isacommand{definition}\isamarkupfalse%
\isanewline
\ \ DAcc\ {\isacharcolon}{\isacharcolon}\ {\isachardoublequoteopen}AbstrLevelsID\ {\isasymRightarrow}\ CSet\ {\isasymRightarrow}\ CSet\ set{\isachardoublequoteclose}\isanewline
\isakeyword{where}\isanewline
\ {\isachardoublequoteopen}DAcc\ i\ C\ {\isasymequiv}\ {\isacharbraceleft}Z{\isachardot}\ \ {\isasymexists}\ x{\isachardot}\ x\ {\isasymin}\ {\isacharparenleft}OUT\ C{\isacharparenright}\ {\isasymand}\ x\ {\isasymin}\ {\isacharparenleft}IN\ Z{\isacharparenright}\ {\isasymand}\ Z\ {\isasymin}\ {\isacharparenleft}AbstrLevel\ i{\isacharparenright}\ {\isasymand}\ C\ {\isasymin}\ {\isacharparenleft}AbstrLevel\ i{\isacharparenright}{\isacharbraceright}{\isachardoublequoteclose}\isanewline
\isanewline
\isacommand{axiomatization}\isamarkupfalse%
\isanewline
\ \ Sources\ {\isacharcolon}{\isacharcolon}\ {\isachardoublequoteopen}AbstrLevelsID\ {\isasymRightarrow}\ CSet\ {\isasymRightarrow}\ CSet\ set{\isachardoublequoteclose}\isanewline
\isakeyword{where}\ \isanewline
SourcesDef{\isacharcolon}\isanewline
{\isachardoublequoteopen}{\isacharparenleft}Sources\ i\ C{\isacharparenright}\ {\isacharequal}\ {\isacharparenleft}DSources\ i\ C{\isacharparenright}\ {\isasymunion}\ {\isacharparenleft}{\isasymUnion}\ S\ {\isasymin}\ {\isacharparenleft}DSources\ i\ C{\isacharparenright}{\isachardot}\ {\isacharparenleft}Sources\ i\ S{\isacharparenright}{\isacharparenright}{\isachardoublequoteclose}\ \isanewline
\isakeyword{and}\isanewline
SourceExistsDSource{\isacharcolon}\isanewline
{\isachardoublequoteopen}S\ {\isasymin}\ {\isacharparenleft}Sources\ i\ C{\isacharparenright}\ {\isasymlongrightarrow}\ {\isacharparenleft}{\isasymexists}\ Z{\isachardot}\ S\ {\isasymin}\ {\isacharparenleft}DSources\ i\ Z{\isacharparenright}{\isacharparenright}{\isachardoublequoteclose}\isanewline
\isakeyword{and}\isanewline
NDSourceExistsDSource{\isacharcolon}\isanewline
{\isachardoublequoteopen}S\ {\isasymin}\ {\isacharparenleft}Sources\ i\ C{\isacharparenright}\ {\isasymand}\ S\ {\isasymnotin}\ {\isacharparenleft}DSources\ i\ C{\isacharparenright}\ {\isasymlongrightarrow}\ \isanewline
\ {\isacharparenleft}{\isasymexists}\ Z{\isachardot}\ S\ {\isasymin}\ {\isacharparenleft}DSources\ i\ Z{\isacharparenright}\ {\isasymand}\ Z\ {\isasymin}\ {\isacharparenleft}Sources\ i\ C{\isacharparenright}{\isacharparenright}{\isachardoublequoteclose}\isanewline
\isakeyword{and}\isanewline
SourcesTrans{\isacharcolon}\isanewline
{\isachardoublequoteopen}{\isacharparenleft}C\ {\isasymin}\ Sources\ i\ S\ {\isasymand}\ S\ {\isasymin}\ Sources\ i\ Z{\isacharparenright}\ {\isasymlongrightarrow}\ C\ {\isasymin}\ Sources\ i\ Z{\isachardoublequoteclose}\isanewline
\isakeyword{and}\ \isanewline
SourcesLevelX{\isacharcolon}\isanewline
{\isachardoublequoteopen}{\isacharparenleft}Sources\ i\ X{\isacharparenright}\ \ {\isasymsubseteq}\ {\isacharparenleft}AbstrLevel\ i{\isacharparenright}{\isachardoublequoteclose}\ \isanewline
\isakeyword{and}\isanewline
SourcesLoop{\isacharcolon}\isanewline
{\isachardoublequoteopen}{\isacharparenleft}Sources\ i\ C{\isacharparenright}\ {\isacharequal}\ {\isacharparenleft}XS\ {\isasymunion}\ {\isacharparenleft}Sources\ i\ S{\isacharparenright}{\isacharparenright}\ {\isasymand}\ {\isacharparenleft}Sources\ i\ S{\isacharparenright}\ {\isacharequal}\ {\isacharparenleft}ZS\ {\isasymunion}\ {\isacharparenleft}Sources\ i\ C{\isacharparenright}{\isacharparenright}\ \isanewline
{\isasymlongrightarrow}\ {\isacharparenleft}Sources\ i\ C{\isacharparenright}\ {\isacharequal}\ XS\ \ {\isasymunion}\ ZS\ {\isasymunion}\ {\isacharbraceleft}\ C{\isacharcomma}\ S{\isacharbraceright}{\isachardoublequoteclose}\ \isanewline
\isamarkupcmt{if we have a loop in the dependencies we need to cut it for counting the sources%
}
\isanewline
\isanewline
\isacommand{axiomatization}\isamarkupfalse%
\isanewline
\ \ Acc\ {\isacharcolon}{\isacharcolon}\ {\isachardoublequoteopen}AbstrLevelsID\ {\isasymRightarrow}\ CSet\ {\isasymRightarrow}\ CSet\ set{\isachardoublequoteclose}\isanewline
\isakeyword{where}\ \isanewline
AccDef{\isacharcolon}\isanewline
{\isachardoublequoteopen}{\isacharparenleft}Acc\ i\ C{\isacharparenright}\ {\isacharequal}\ {\isacharparenleft}DAcc\ i\ C{\isacharparenright}\ {\isasymunion}\ {\isacharparenleft}{\isasymUnion}\ S\ {\isasymin}\ {\isacharparenleft}DAcc\ i\ C{\isacharparenright}{\isachardot}\ {\isacharparenleft}Acc\ i\ S{\isacharparenright}{\isacharparenright}{\isachardoublequoteclose}\ \isanewline
\isakeyword{and}\isanewline
Acc{\isacharunderscore}Sources{\isacharcolon}\isanewline
{\isachardoublequoteopen}{\isacharparenleft}X\ {\isasymin}\ Acc\ i\ C{\isacharparenright}\ {\isacharequal}\ {\isacharparenleft}C\ {\isasymin}\ Sources\ i\ X{\isacharparenright}{\isachardoublequoteclose}\isanewline
\isakeyword{and}\isanewline
AccSigleLoop{\isacharcolon}\isanewline
{\isachardoublequoteopen}DAcc\ i\ C\ {\isacharequal}\ {\isacharbraceleft}S{\isacharbraceright}\ {\isasymand}\ DAcc\ i\ S\ {\isacharequal}\ {\isacharbraceleft}C{\isacharbraceright}\ {\isasymlongrightarrow}\ Acc\ i\ C\ {\isacharequal}\ {\isacharbraceleft}C{\isacharcomma}\ S{\isacharbraceright}{\isachardoublequoteclose}\ \isanewline
\isakeyword{and}\isanewline
AccLoop{\isacharcolon}\isanewline
{\isachardoublequoteopen}{\isacharparenleft}Acc\ i\ C{\isacharparenright}\ {\isacharequal}\ {\isacharparenleft}XS\ {\isasymunion}\ {\isacharparenleft}Acc\ i\ S{\isacharparenright}{\isacharparenright}\ {\isasymand}\ {\isacharparenleft}Acc\ i\ S{\isacharparenright}\ {\isacharequal}\ {\isacharparenleft}ZS\ {\isasymunion}\ {\isacharparenleft}Acc\ i\ C{\isacharparenright}{\isacharparenright}\ \isanewline
{\isasymlongrightarrow}\ {\isacharparenleft}Acc\ i\ C{\isacharparenright}\ {\isacharequal}\ XS\ \ {\isasymunion}\ ZS\ {\isasymunion}\ {\isacharbraceleft}\ C{\isacharcomma}\ S{\isacharbraceright}{\isachardoublequoteclose}\ \isanewline
\isamarkupcmt{if we have a loop in the dependencies we need to cut it for counting the accessors%
}
\isanewline
\isanewline
\isacommand{lemma}\isamarkupfalse%
\ Acc{\isacharunderscore}SourcesNOT{\isacharcolon}\ {\isachardoublequoteopen}{\isacharparenleft}X\ {\isasymnotin}\ Acc\ i\ C{\isacharparenright}\ {\isacharequal}\ {\isacharparenleft}C\ {\isasymnotin}\ Sources\ i\ X{\isacharparenright}{\isachardoublequoteclose}\isanewline
\isadelimproof
\endisadelimproof
\isatagproof
\isacommand{by}\isamarkupfalse%
\ {\isacharparenleft}metis\ Acc{\isacharunderscore}Sources{\isacharparenright}\isanewline
\isanewline
\isamarkupcmt{component S is not a source for any component on the abstraction level i%
}
\endisatagproof
{\isafoldproof}%
\isadelimproof
\isanewline
\endisadelimproof
\isacommand{definition}\isamarkupfalse%
\isanewline
\ \ isNotDSource\ {\isacharcolon}{\isacharcolon}\ {\isachardoublequoteopen}AbstrLevelsID\ {\isasymRightarrow}\ CSet\ {\isasymRightarrow}\ bool{\isachardoublequoteclose}\isanewline
\isakeyword{where}\isanewline
\ {\isachardoublequoteopen}isNotDSource\ i\ S\ {\isasymequiv}\ {\isacharparenleft}{\isasymforall}\ x\ {\isasymin}\ {\isacharparenleft}OUT\ S{\isacharparenright}{\isachardot}\ {\isacharparenleft}{\isasymforall}\ Z\ {\isasymin}\ {\isacharparenleft}AbstrLevel\ i{\isacharparenright}{\isachardot}\ {\isacharparenleft}x\ {\isasymnotin}\ {\isacharparenleft}IN\ Z{\isacharparenright}{\isacharparenright}{\isacharparenright}{\isacharparenright}{\isachardoublequoteclose}\isanewline
\isanewline
\isamarkupcmt{component S is not a source for a component Z on the abstraction level i%
}
\isanewline
\isacommand{definition}\isamarkupfalse%
\isanewline
\ \ isNotDSourceX\ {\isacharcolon}{\isacharcolon}\ {\isachardoublequoteopen}AbstrLevelsID\ {\isasymRightarrow}\ CSet\ {\isasymRightarrow}\ CSet\ {\isasymRightarrow}\ bool{\isachardoublequoteclose}\isanewline
\isakeyword{where}\isanewline
\ {\isachardoublequoteopen}isNotDSourceX\ i\ S\ C\ {\isasymequiv}\ {\isacharparenleft}{\isasymforall}\ x\ {\isasymin}\ {\isacharparenleft}OUT\ S{\isacharparenright}{\isachardot}\ {\isacharparenleft}C\ {\isasymnotin}\ {\isacharparenleft}AbstrLevel\ i{\isacharparenright}\ {\isasymor}\ {\isacharparenleft}x\ {\isasymnotin}\ {\isacharparenleft}IN\ C{\isacharparenright}{\isacharparenright}{\isacharparenright}{\isacharparenright}{\isachardoublequoteclose}\isanewline
\isanewline
\isacommand{lemma}\isamarkupfalse%
\ isNotSource{\isacharunderscore}isNotSourceX{\isacharcolon}\isanewline
{\isachardoublequoteopen}isNotDSource\ i\ S\ {\isacharequal}\ {\isacharparenleft}{\isasymforall}\ C{\isachardot}\ isNotDSourceX\ i\ S\ C{\isacharparenright}{\isachardoublequoteclose}\ \isanewline
\isadelimproof
\endisadelimproof
\isatagproof
\isacommand{by}\isamarkupfalse%
\ {\isacharparenleft}auto{\isacharcomma}\ {\isacharparenleft}simp\ add{\isacharcolon}\ isNotDSource{\isacharunderscore}def\ isNotDSourceX{\isacharunderscore}def{\isacharparenright}{\isacharplus}{\isacharparenright}%
\endisatagproof
{\isafoldproof}%
\isadelimproof
\isanewline
\endisadelimproof
\isanewline
\isacommand{lemma}\isamarkupfalse%
\ DAcc{\isacharunderscore}DSources{\isacharcolon}\isanewline
{\isachardoublequoteopen}{\isacharparenleft}X\ {\isasymin}\ DAcc\ i\ C{\isacharparenright}\ {\isacharequal}\ {\isacharparenleft}C\ {\isasymin}\ DSources\ i\ X{\isacharparenright}{\isachardoublequoteclose}\isanewline
\isadelimproof
\endisadelimproof
\isatagproof
\isacommand{by}\isamarkupfalse%
\ {\isacharparenleft}auto{\isacharcomma}\ {\isacharparenleft}simp\ add{\isacharcolon}\ DAcc{\isacharunderscore}def\ DSources{\isacharunderscore}def{\isacharcomma}\ auto{\isacharparenright}{\isacharplus}{\isacharparenright}%
\endisatagproof
{\isafoldproof}%
\isadelimproof
\isanewline
\endisadelimproof
\ \isanewline
\isacommand{lemma}\isamarkupfalse%
\ DAcc{\isacharunderscore}DSourcesNOT{\isacharcolon}\isanewline
{\isachardoublequoteopen}{\isacharparenleft}X\ {\isasymnotin}\ DAcc\ i\ C{\isacharparenright}\ {\isacharequal}\ {\isacharparenleft}C\ {\isasymnotin}\ DSources\ i\ X{\isacharparenright}{\isachardoublequoteclose}\isanewline
\isadelimproof
\endisadelimproof
\isatagproof
\isacommand{by}\isamarkupfalse%
\ {\isacharparenleft}auto{\isacharcomma}\ {\isacharparenleft}simp\ add{\isacharcolon}\ DAcc{\isacharunderscore}def\ DSources{\isacharunderscore}def{\isacharcomma}\ auto{\isacharparenright}{\isacharplus}{\isacharparenright}%
\endisatagproof
{\isafoldproof}%
\isadelimproof
\isanewline
\endisadelimproof
\isanewline
\isacommand{lemma}\isamarkupfalse%
\ DSource{\isacharunderscore}level{\isacharcolon}\isanewline
\ \ \isakeyword{assumes}\ {\isachardoublequoteopen}S\ {\isasymin}\ {\isacharparenleft}DSources\ i\ C{\isacharparenright}{\isachardoublequoteclose}\isanewline
\ \ \isakeyword{shows}\ \ \ \ {\isachardoublequoteopen}C\ \ {\isasymin}\ {\isacharparenleft}AbstrLevel\ i{\isacharparenright}{\isachardoublequoteclose}\isanewline
\isadelimproof
\endisadelimproof
\isatagproof
\isacommand{using}\isamarkupfalse%
\ assms\ \isacommand{by}\isamarkupfalse%
\ {\isacharparenleft}simp\ add{\isacharcolon}\ DSources{\isacharunderscore}def{\isacharcomma}\ auto{\isacharparenright}%
\endisatagproof
{\isafoldproof}%
\isadelimproof
\isanewline
\endisadelimproof
\isanewline
\isacommand{lemma}\isamarkupfalse%
\ SourceExistsDSource{\isacharunderscore}level{\isacharcolon}\isanewline
\ \ \isakeyword{assumes}\ {\isachardoublequoteopen}S\ {\isasymin}\ {\isacharparenleft}Sources\ i\ C{\isacharparenright}{\isachardoublequoteclose}\isanewline
\ \ \isakeyword{shows}\ \ \ \ {\isachardoublequoteopen}{\isasymexists}\ Z\ \ {\isasymin}\ {\isacharparenleft}AbstrLevel\ i{\isacharparenright}{\isachardot}\ {\isacharparenleft}S\ {\isasymin}\ {\isacharparenleft}DSources\ i\ Z{\isacharparenright}{\isacharparenright}{\isachardoublequoteclose}\isanewline
\isadelimproof
\endisadelimproof
\isatagproof
\isacommand{using}\isamarkupfalse%
\ assms\ \isacommand{by}\isamarkupfalse%
\ {\isacharparenleft}metis\ DSource{\isacharunderscore}level\ SourceExistsDSource{\isacharparenright}%
\endisatagproof
{\isafoldproof}%
\isadelimproof
\ \isanewline
\endisadelimproof
\ \ \isanewline
\isacommand{lemma}\isamarkupfalse%
\ Sources{\isacharunderscore}DSources{\isacharcolon}\isanewline
\ {\isachardoublequoteopen}{\isacharparenleft}DSources\ i\ C{\isacharparenright}\ {\isasymsubseteq}\ {\isacharparenleft}Sources\ i\ C{\isacharparenright}{\isachardoublequoteclose}\ \ \ \isanewline
\isadelimproof
\endisadelimproof
\isatagproof
\isacommand{proof}\isamarkupfalse%
\ {\isacharminus}\ \ \isanewline
\ \ \isacommand{have}\isamarkupfalse%
\ {\isachardoublequoteopen}{\isacharparenleft}Sources\ i\ C{\isacharparenright}\ {\isacharequal}\ {\isacharparenleft}DSources\ i\ C{\isacharparenright}\ {\isasymunion}\ {\isacharparenleft}{\isasymUnion}\ S\ {\isasymin}\ {\isacharparenleft}DSources\ i\ C{\isacharparenright}{\isachardot}\ {\isacharparenleft}Sources\ i\ S{\isacharparenright}{\isacharparenright}{\isachardoublequoteclose}\ \isanewline
\ \ \ \ \isacommand{by}\isamarkupfalse%
\ {\isacharparenleft}rule\ SourcesDef{\isacharparenright}\isanewline
\ \ \isacommand{thus}\isamarkupfalse%
\ {\isacharquery}thesis\ \isacommand{by}\isamarkupfalse%
\ auto\isanewline
\isacommand{qed}\isamarkupfalse%
\endisatagproof
{\isafoldproof}%
\isadelimproof
\isanewline
\endisadelimproof
\isanewline
\isacommand{lemma}\isamarkupfalse%
\ NoDSourceNoSource{\isacharcolon}\isanewline
\ \ \isakeyword{assumes}\ {\isachardoublequoteopen}S\ {\isasymnotin}\ {\isacharparenleft}Sources\ i\ C{\isacharparenright}{\isachardoublequoteclose}\isanewline
\ \ \isakeyword{shows}\ \ \ \ \ {\isachardoublequoteopen}S\ {\isasymnotin}\ {\isacharparenleft}DSources\ i\ C{\isacharparenright}{\isachardoublequoteclose}\isanewline
\isadelimproof
\endisadelimproof
\isatagproof
\isacommand{using}\isamarkupfalse%
\ assms\ \isacommand{by}\isamarkupfalse%
\ {\isacharparenleft}metis\ {\isacharparenleft}full{\isacharunderscore}types{\isacharparenright}\ Sources{\isacharunderscore}DSources\ set{\isacharunderscore}rev{\isacharunderscore}mp{\isacharparenright}%
\endisatagproof
{\isafoldproof}%
\isadelimproof
\isanewline
\endisadelimproof
\isanewline
\isacommand{lemma}\isamarkupfalse%
\ DSourcesEmptySources{\isacharcolon}\isanewline
\ \ \isakeyword{assumes}\ {\isachardoublequoteopen}DSources\ i\ C\ {\isacharequal}\ {\isacharbraceleft}{\isacharbraceright}{\isachardoublequoteclose}\isanewline
\ \ \isakeyword{shows}\ \ \ \ {\isachardoublequoteopen}Sources\ i\ C\ {\isacharequal}\ {\isacharbraceleft}{\isacharbraceright}{\isachardoublequoteclose}\ \isanewline
\isadelimproof
\endisadelimproof
\isatagproof
\isacommand{proof}\isamarkupfalse%
\ {\isacharminus}\ \isanewline
\ \ \isacommand{have}\isamarkupfalse%
\ {\isachardoublequoteopen}{\isacharparenleft}Sources\ i\ C{\isacharparenright}\ {\isacharequal}\ {\isacharparenleft}DSources\ i\ C{\isacharparenright}\ {\isasymunion}\ {\isacharparenleft}{\isasymUnion}\ S\ {\isasymin}\ {\isacharparenleft}DSources\ i\ C{\isacharparenright}{\isachardot}\ {\isacharparenleft}Sources\ i\ S{\isacharparenright}{\isacharparenright}{\isachardoublequoteclose}\ \ \isanewline
\ \ \ \ \isacommand{by}\isamarkupfalse%
\ {\isacharparenleft}rule\ SourcesDef{\isacharparenright}\ \isanewline
\ \ \isacommand{with}\isamarkupfalse%
\ assms\ \isacommand{show}\isamarkupfalse%
\ {\isacharquery}thesis\ \isacommand{by}\isamarkupfalse%
\ auto\isanewline
\isacommand{qed}\isamarkupfalse%
\endisatagproof
{\isafoldproof}%
\isadelimproof
\isanewline
\endisadelimproof
\isanewline
\isacommand{lemma}\isamarkupfalse%
\ DSource{\isacharunderscore}Sources{\isacharcolon}\isanewline
\ \ \isakeyword{assumes}\ \ {\isachardoublequoteopen}S\ {\isasymin}\ {\isacharparenleft}DSources\ i\ C{\isacharparenright}{\isachardoublequoteclose}\isanewline
\ \ \isakeyword{shows}\ \ \ \ \ {\isachardoublequoteopen}{\isacharparenleft}Sources\ i\ S{\isacharparenright}\ \ {\isasymsubseteq}\ {\isacharparenleft}Sources\ i\ C{\isacharparenright}{\isachardoublequoteclose}\isanewline
\isadelimproof
\endisadelimproof
\isatagproof
\isacommand{proof}\isamarkupfalse%
\ {\isacharminus}\ \isanewline
\ \isacommand{have}\isamarkupfalse%
\ \ {\isachardoublequoteopen}{\isacharparenleft}Sources\ i\ C{\isacharparenright}\ {\isacharequal}\ {\isacharparenleft}DSources\ i\ C{\isacharparenright}\ {\isasymunion}\ {\isacharparenleft}{\isasymUnion}\ S\ {\isasymin}\ {\isacharparenleft}DSources\ i\ C{\isacharparenright}{\isachardot}\ {\isacharparenleft}Sources\ i\ S{\isacharparenright}{\isacharparenright}{\isachardoublequoteclose}\isanewline
\ \ \isacommand{by}\isamarkupfalse%
\ {\isacharparenleft}rule\ SourcesDef{\isacharparenright}\isanewline
\ \ \isacommand{with}\isamarkupfalse%
\ assms\ \isacommand{show}\isamarkupfalse%
\ {\isacharquery}thesis\ \isacommand{by}\isamarkupfalse%
\ auto\isanewline
\isacommand{qed}\isamarkupfalse%
\endisatagproof
{\isafoldproof}%
\isadelimproof
\isanewline
\endisadelimproof
\isanewline
\isacommand{lemma}\isamarkupfalse%
\ SourcesOnlyDSources{\isacharcolon}\isanewline
\ \ \isakeyword{assumes}\ {\isachardoublequoteopen}{\isasymforall}\ X{\isachardot}\ {\isacharparenleft}X\ {\isasymin}\ {\isacharparenleft}DSources\ i\ C{\isacharparenright}\ {\isasymlongrightarrow}\ {\isacharparenleft}DSources\ i\ X{\isacharparenright}\ {\isacharequal}\ {\isacharbraceleft}{\isacharbraceright}{\isacharparenright}{\isachardoublequoteclose}\isanewline
\ \ \isakeyword{shows}\ \ \ \ {\isachardoublequoteopen}Sources\ i\ C\ {\isacharequal}\ DSources\ i\ C{\isachardoublequoteclose}\isanewline
\isadelimproof
\endisadelimproof
\isatagproof
\isacommand{proof}\isamarkupfalse%
\ {\isacharminus}\ \isanewline
\ \isacommand{have}\isamarkupfalse%
\ sDef{\isacharcolon}\ \ {\isachardoublequoteopen}{\isacharparenleft}Sources\ i\ C{\isacharparenright}\ {\isacharequal}\ {\isacharparenleft}DSources\ i\ C{\isacharparenright}\ {\isasymunion}\ {\isacharparenleft}{\isasymUnion}\ S\ {\isasymin}\ {\isacharparenleft}DSources\ i\ C{\isacharparenright}{\isachardot}\ {\isacharparenleft}Sources\ i\ S{\isacharparenright}{\isacharparenright}{\isachardoublequoteclose}\isanewline
\ \ \isacommand{by}\isamarkupfalse%
\ {\isacharparenleft}rule\ SourcesDef{\isacharparenright}\isanewline
\ \isacommand{from}\isamarkupfalse%
\ assms\ \isacommand{have}\isamarkupfalse%
\ \ {\isachardoublequoteopen}{\isasymforall}\ X{\isachardot}\ {\isacharparenleft}X\ {\isasymin}\ {\isacharparenleft}DSources\ i\ C{\isacharparenright}\ {\isasymlongrightarrow}\ {\isacharparenleft}Sources\ i\ X{\isacharparenright}\ {\isacharequal}\ {\isacharbraceleft}{\isacharbraceright}{\isacharparenright}{\isachardoublequoteclose}\isanewline
\ \ \ \isacommand{by}\isamarkupfalse%
\ {\isacharparenleft}simp\ add{\isacharcolon}\ DSourcesEmptySources{\isacharparenright}\isanewline
\ \isacommand{hence}\isamarkupfalse%
\ {\isachardoublequoteopen}{\isacharparenleft}{\isasymUnion}\ S\ {\isasymin}\ {\isacharparenleft}DSources\ i\ C{\isacharparenright}{\isachardot}\ {\isacharparenleft}Sources\ i\ S{\isacharparenright}{\isacharparenright}\ {\isacharequal}\ {\isacharbraceleft}{\isacharbraceright}{\isachardoublequoteclose}\ \ \isacommand{by}\isamarkupfalse%
\ auto\isanewline
\ \isacommand{with}\isamarkupfalse%
\ sDef\ \ \isacommand{show}\isamarkupfalse%
\ {\isacharquery}thesis\ \isacommand{by}\isamarkupfalse%
\ simp\isanewline
\isacommand{qed}\isamarkupfalse%
\endisatagproof
{\isafoldproof}%
\isadelimproof
\isanewline
\endisadelimproof
\isanewline
\isacommand{lemma}\isamarkupfalse%
\ SourcesEmptyDSources{\isacharcolon}\isanewline
\ \isakeyword{assumes}\ {\isachardoublequoteopen}Sources\ i\ C\ {\isacharequal}\ {\isacharbraceleft}{\isacharbraceright}{\isachardoublequoteclose}\isanewline
\ \isakeyword{shows}\ {\isachardoublequoteopen}DSources\ i\ C\ {\isacharequal}\ {\isacharbraceleft}{\isacharbraceright}{\isachardoublequoteclose}\isanewline
\isadelimproof
\endisadelimproof
\isatagproof
\isacommand{using}\isamarkupfalse%
\ assms\ \ \isacommand{by}\isamarkupfalse%
\ {\isacharparenleft}metis\ Sources{\isacharunderscore}DSources\ bot{\isachardot}extremum{\isacharunderscore}uniqueI{\isacharparenright}%
\endisatagproof
{\isafoldproof}%
\isadelimproof
\isanewline
\endisadelimproof
\isanewline
\isacommand{lemma}\isamarkupfalse%
\ NotDSource{\isacharcolon}\isanewline
\ \isakeyword{assumes}\ {\isachardoublequoteopen}{\isasymforall}\ x\ {\isasymin}\ {\isacharparenleft}OUT\ S{\isacharparenright}{\isachardot}\ {\isacharparenleft}{\isasymforall}\ Z\ {\isasymin}\ {\isacharparenleft}AbstrLevel\ i{\isacharparenright}{\isachardot}\ {\isacharparenleft}x\ {\isasymnotin}\ {\isacharparenleft}IN\ Z{\isacharparenright}{\isacharparenright}{\isacharparenright}{\isachardoublequoteclose}\isanewline
\ \isakeyword{shows}\ \ \ \ {\isachardoublequoteopen}{\isasymforall}\ C\ {\isasymin}\ {\isacharparenleft}AbstrLevel\ i{\isacharparenright}\ {\isachardot}\ S\ {\isasymnotin}\ {\isacharparenleft}DSources\ i\ C{\isacharparenright}{\isachardoublequoteclose}\ \isanewline
\isadelimproof
\endisadelimproof
\isatagproof
\isacommand{using}\isamarkupfalse%
\ assms\ \ \isacommand{by}\isamarkupfalse%
\ {\isacharparenleft}simp\ add{\isacharcolon}\ AbstrLevel{\isadigit{0}}\ DSources{\isacharunderscore}def{\isacharparenright}%
\endisatagproof
{\isafoldproof}%
\isadelimproof
\ \isanewline
\endisadelimproof
\isanewline
\isacommand{lemma}\isamarkupfalse%
\ allNotDSource{\isacharunderscore}NotSource{\isacharcolon}\isanewline
\ \isakeyword{assumes}\ {\isachardoublequoteopen}{\isasymforall}\ C\ {\isachardot}\ S\ {\isasymnotin}\ {\isacharparenleft}DSources\ i\ C{\isacharparenright}{\isachardoublequoteclose}\ \isanewline
\ \isakeyword{shows}\ \ \ \ {\isachardoublequoteopen}{\isasymforall}\ Z{\isachardot}\ S\ {\isasymnotin}\ {\isacharparenleft}Sources\ i\ Z{\isacharparenright}{\isachardoublequoteclose}\ \isanewline
\isadelimproof
\endisadelimproof
\isatagproof
\isacommand{using}\isamarkupfalse%
\ assms\ \ \isacommand{by}\isamarkupfalse%
\ {\isacharparenleft}metis\ SourceExistsDSource{\isacharparenright}%
\endisatagproof
{\isafoldproof}%
\isadelimproof
\ \isanewline
\endisadelimproof
\isanewline
\isacommand{lemma}\isamarkupfalse%
\ NotDSource{\isacharunderscore}NotSource{\isacharcolon}\isanewline
\ \isakeyword{assumes}\ {\isachardoublequoteopen}{\isasymforall}\ C\ {\isasymin}\ {\isacharparenleft}AbstrLevel\ i{\isacharparenright}{\isachardot}\ S\ {\isasymnotin}\ {\isacharparenleft}DSources\ i\ C{\isacharparenright}{\isachardoublequoteclose}\ \isanewline
\ \isakeyword{shows}\ \ \ \ {\isachardoublequoteopen}{\isasymforall}\ Z\ {\isasymin}\ {\isacharparenleft}AbstrLevel\ i{\isacharparenright}{\isachardot}\ S\ {\isasymnotin}\ {\isacharparenleft}Sources\ i\ Z{\isacharparenright}{\isachardoublequoteclose}\ \isanewline
\isadelimproof
\endisadelimproof
\isatagproof
\isacommand{using}\isamarkupfalse%
\ assms\ \isacommand{by}\isamarkupfalse%
\ {\isacharparenleft}metis\ SourceExistsDSource{\isacharunderscore}level{\isacharparenright}%
\endisatagproof
{\isafoldproof}%
\isadelimproof
\ \ \isanewline
\endisadelimproof
\isanewline
\isacommand{lemma}\isamarkupfalse%
\ isNotSource{\isacharunderscore}Sources{\isacharcolon}\ \isanewline
\ \isakeyword{assumes}\ {\isachardoublequoteopen}isNotDSource\ i\ S{\isachardoublequoteclose}\isanewline
\ \isakeyword{shows}\ {\isachardoublequoteopen}{\isasymforall}\ C\ \ {\isasymin}\ {\isacharparenleft}AbstrLevel\ i{\isacharparenright}{\isachardot}\ S\ {\isasymnotin}\ {\isacharparenleft}Sources\ i\ C{\isacharparenright}{\isachardoublequoteclose}\ \isanewline
\isadelimproof
\endisadelimproof
\isatagproof
\isacommand{using}\isamarkupfalse%
\ assms\ \ \isanewline
\isacommand{by}\isamarkupfalse%
\ {\isacharparenleft}simp\ add{\isacharcolon}\ isNotDSource{\isacharunderscore}def{\isacharcomma}\ metis\ {\isacharparenleft}full{\isacharunderscore}types{\isacharparenright}\ NotDSource\ NotDSource{\isacharunderscore}NotSource{\isacharparenright}%
\endisatagproof
{\isafoldproof}%
\isadelimproof
\isanewline
\endisadelimproof
\isanewline
\isacommand{lemma}\isamarkupfalse%
\ SourcesAbstrLevel{\isacharcolon}\isanewline
\isakeyword{assumes}\ {\isachardoublequoteopen}x\ {\isasymin}\ Sources\ i\ S{\isachardoublequoteclose}\isanewline
\isakeyword{shows}\ {\isachardoublequoteopen}x\ {\isasymin}\ AbstrLevel\ i{\isachardoublequoteclose}\isanewline
\isadelimproof
\endisadelimproof
\isatagproof
\isacommand{using}\isamarkupfalse%
\ assms\isanewline
\isacommand{by}\isamarkupfalse%
\ {\isacharparenleft}metis\ SourcesLevelX\ in{\isacharunderscore}mono{\isacharparenright}%
\endisatagproof
{\isafoldproof}%
\isadelimproof
\isanewline
\endisadelimproof
\isanewline
\isacommand{lemma}\isamarkupfalse%
\ DSourceIsSource{\isacharcolon}\isanewline
\ \ \isakeyword{assumes}\ \ {\isachardoublequoteopen}C\ {\isasymin}\ DSources\ i\ S{\isachardoublequoteclose}\ \isanewline
\ \ \ \ \ \isakeyword{shows}\ \ {\isachardoublequoteopen}C\ {\isasymin}\ Sources\ i\ S{\isachardoublequoteclose}\ \isanewline
\isadelimproof
\endisadelimproof
\isatagproof
\isacommand{proof}\isamarkupfalse%
\ {\isacharminus}\isanewline
\ \ \isacommand{have}\isamarkupfalse%
\ {\isachardoublequoteopen}{\isacharparenleft}Sources\ i\ S{\isacharparenright}\ {\isacharequal}\ {\isacharparenleft}DSources\ i\ S{\isacharparenright}\ {\isasymunion}\ {\isacharparenleft}{\isasymUnion}\ Z\ {\isasymin}\ {\isacharparenleft}DSources\ i\ S{\isacharparenright}{\isachardot}\ {\isacharparenleft}Sources\ i\ Z{\isacharparenright}{\isacharparenright}{\isachardoublequoteclose}\ \isanewline
\ \ \ \ \isacommand{by}\isamarkupfalse%
\ {\isacharparenleft}rule\ SourcesDef{\isacharparenright}\isanewline
\ \ \isacommand{with}\isamarkupfalse%
\ assms\ \isacommand{show}\isamarkupfalse%
\ {\isacharquery}thesis\ \ \isacommand{by}\isamarkupfalse%
\ simp\isanewline
\isacommand{qed}\isamarkupfalse%
\endisatagproof
{\isafoldproof}%
\isadelimproof
\isanewline
\endisadelimproof
\isanewline
\isacommand{lemma}\isamarkupfalse%
\ DSourceOfDSource{\isacharcolon}\isanewline
\ \ \isakeyword{assumes}\ \ {\isachardoublequoteopen}Z\ {\isasymin}\ DSources\ i\ S{\isachardoublequoteclose}\ \isanewline
\ \ \ \ \ \ \ \ \ \isakeyword{and}\ \ {\isachardoublequoteopen}S\ {\isasymin}\ DSources\ i\ C{\isachardoublequoteclose}\isanewline
\ \ \isakeyword{shows}\ \ \ \ \ {\isachardoublequoteopen}Z\ {\isasymin}\ Sources\ i\ C{\isachardoublequoteclose}\isanewline
\isadelimproof
\endisadelimproof
\isatagproof
\isacommand{using}\isamarkupfalse%
\ assms\isanewline
\isacommand{proof}\isamarkupfalse%
\ {\isacharminus}\isanewline
\ \ \isacommand{from}\isamarkupfalse%
\ assms\ \isacommand{have}\isamarkupfalse%
\ src{\isacharcolon}{\isachardoublequoteopen}Sources\ i\ S\ {\isasymsubseteq}\ Sources\ i\ C{\isachardoublequoteclose}\ \isacommand{by}\isamarkupfalse%
\ {\isacharparenleft}simp\ add{\isacharcolon}\ DSource{\isacharunderscore}Sources{\isacharparenright}\isanewline
\ \ \isacommand{from}\isamarkupfalse%
\ assms\ \isacommand{have}\isamarkupfalse%
\ \ {\isachardoublequoteopen}Z\ {\isasymin}\ Sources\ i\ S{\isachardoublequoteclose}\ \ \isacommand{by}\isamarkupfalse%
\ {\isacharparenleft}simp\ add{\isacharcolon}\ DSourceIsSource{\isacharparenright}\isanewline
\ \ \isacommand{with}\isamarkupfalse%
\ src\ \ \ \isacommand{show}\isamarkupfalse%
\ {\isacharquery}thesis\ \ \isacommand{by}\isamarkupfalse%
\ auto\isanewline
\isacommand{qed}\isamarkupfalse%
\endisatagproof
{\isafoldproof}%
\isadelimproof
\isanewline
\endisadelimproof
\isanewline
\isacommand{lemma}\isamarkupfalse%
\ SourceOfDSource{\isacharcolon}\isanewline
\ \ \isakeyword{assumes}\ \ {\isachardoublequoteopen}Z\ {\isasymin}\ Sources\ i\ S{\isachardoublequoteclose}\ \isanewline
\ \ \ \ \ \ \ \ \ \isakeyword{and}\ \ {\isachardoublequoteopen}S\ {\isasymin}\ DSources\ i\ C{\isachardoublequoteclose}\isanewline
\ \ \isakeyword{shows}\ \ \ \ \ {\isachardoublequoteopen}Z\ {\isasymin}\ Sources\ i\ C{\isachardoublequoteclose}\isanewline
\isadelimproof
\endisadelimproof
\isatagproof
\isacommand{using}\isamarkupfalse%
\ assms\isanewline
\isacommand{proof}\isamarkupfalse%
\ {\isacharminus}\isanewline
\ \ \isacommand{from}\isamarkupfalse%
\ assms\ \isacommand{have}\isamarkupfalse%
\ {\isachardoublequoteopen}Sources\ i\ S\ {\isasymsubseteq}\ Sources\ i\ C{\isachardoublequoteclose}\ \isacommand{by}\isamarkupfalse%
\ {\isacharparenleft}simp\ add{\isacharcolon}\ DSource{\isacharunderscore}Sources{\isacharparenright}\ \isanewline
\ \ \isacommand{thus}\isamarkupfalse%
\ {\isacharquery}thesis\ \isacommand{by}\isamarkupfalse%
\ {\isacharparenleft}metis\ {\isacharparenleft}full{\isacharunderscore}types{\isacharparenright}\ assms{\isacharparenleft}{\isadigit{1}}{\isacharparenright}\ set{\isacharunderscore}rev{\isacharunderscore}mp{\isacharparenright}\ \ \isanewline
\isacommand{qed}\isamarkupfalse%
\endisatagproof
{\isafoldproof}%
\isadelimproof
\isanewline
\endisadelimproof
\isanewline
\isacommand{lemma}\isamarkupfalse%
\ DSourceOfSource{\isacharcolon}\isanewline
\ \ \isakeyword{assumes}\ \ cDS{\isacharcolon}{\isachardoublequoteopen}C\ {\isasymin}\ DSources\ i\ S{\isachardoublequoteclose}\ \isanewline
\ \ \ \ \ \ \ \ \ \isakeyword{and}\ \ sS{\isacharcolon}{\isachardoublequoteopen}S\ {\isasymin}\ Sources\ i\ Z{\isachardoublequoteclose}\ \ \isanewline
\ \ \isakeyword{shows}\ \ \ \ \ {\isachardoublequoteopen}C\ {\isasymin}\ Sources\ i\ Z{\isachardoublequoteclose}\isanewline
\isadelimproof
\endisadelimproof
\isatagproof
\isacommand{proof}\isamarkupfalse%
\ {\isacharminus}\isanewline
\ \ \isacommand{from}\isamarkupfalse%
\ cDS\ \isacommand{have}\isamarkupfalse%
\ \ {\isachardoublequoteopen}C\ {\isasymin}\ Sources\ i\ S{\isachardoublequoteclose}\ \ \isacommand{by}\isamarkupfalse%
\ {\isacharparenleft}simp\ add{\isacharcolon}\ DSourceIsSource{\isacharparenright}\isanewline
\ \ \isacommand{from}\isamarkupfalse%
\ this\ \isakeyword{and}\ sS\ \isacommand{show}\isamarkupfalse%
\ {\isacharquery}thesis\ \isacommand{by}\isamarkupfalse%
\ {\isacharparenleft}metis\ {\isacharparenleft}full{\isacharunderscore}types{\isacharparenright}\ SourcesTrans{\isacharparenright}\ \ \isanewline
\isacommand{qed}\isamarkupfalse%
\endisatagproof
{\isafoldproof}%
\isadelimproof
\isanewline
\endisadelimproof
\isanewline
\isacommand{lemma}\isamarkupfalse%
\ Sources{\isacharunderscore}singleDSource{\isacharcolon}\isanewline
\ \ \isakeyword{assumes}\ {\isachardoublequoteopen}DSources\ i\ S\ {\isacharequal}\ {\isacharbraceleft}C{\isacharbraceright}{\isachardoublequoteclose}\ \isanewline
\ \ \isakeyword{shows}\ \ \ \ {\isachardoublequoteopen}Sources\ i\ S\ {\isacharequal}\ {\isacharbraceleft}C{\isacharbraceright}\ {\isasymunion}\ Sources\ i\ C{\isachardoublequoteclose}\isanewline
\isadelimproof
\endisadelimproof
\isatagproof
\isacommand{proof}\isamarkupfalse%
\ {\isacharminus}\ \isanewline
\ \isacommand{have}\isamarkupfalse%
\ sDef{\isacharcolon}\ \ {\isachardoublequoteopen}{\isacharparenleft}Sources\ i\ S{\isacharparenright}\ {\isacharequal}\ {\isacharparenleft}DSources\ i\ S{\isacharparenright}\ {\isasymunion}\ {\isacharparenleft}{\isasymUnion}\ Z\ {\isasymin}\ {\isacharparenleft}DSources\ i\ S{\isacharparenright}{\isachardot}\ {\isacharparenleft}Sources\ i\ Z{\isacharparenright}{\isacharparenright}{\isachardoublequoteclose}\isanewline
\ \ \ \ \ \isacommand{by}\isamarkupfalse%
\ {\isacharparenleft}rule\ SourcesDef{\isacharparenright}\isanewline
\ \ \isacommand{from}\isamarkupfalse%
\ assms\ \isacommand{have}\isamarkupfalse%
\ {\isachardoublequoteopen}{\isacharparenleft}{\isasymUnion}\ Z\ {\isasymin}\ {\isacharparenleft}DSources\ i\ S{\isacharparenright}{\isachardot}\ {\isacharparenleft}Sources\ i\ Z{\isacharparenright}{\isacharparenright}\ {\isacharequal}\ Sources\ i\ C{\isachardoublequoteclose}\isanewline
\ \ \ \ \ \isacommand{by}\isamarkupfalse%
\ auto\isanewline
\ \ \isacommand{with}\isamarkupfalse%
\ sDef\ assms\ \isacommand{show}\isamarkupfalse%
\ {\isacharquery}thesis\ \isacommand{by}\isamarkupfalse%
\ simp\isanewline
\isacommand{qed}\isamarkupfalse%
\endisatagproof
{\isafoldproof}%
\isadelimproof
\isanewline
\endisadelimproof
\isanewline
\isacommand{lemma}\isamarkupfalse%
\ Sources{\isacharunderscore}{\isadigit{2}}DSources{\isacharcolon}\isanewline
\ \ \isakeyword{assumes}\ {\isachardoublequoteopen}DSources\ i\ S\ {\isacharequal}\ {\isacharbraceleft}C{\isadigit{1}}{\isacharcomma}\ C{\isadigit{2}}{\isacharbraceright}{\isachardoublequoteclose}\ \isanewline
\ \ \isakeyword{shows}\ \ \ \ {\isachardoublequoteopen}Sources\ i\ S\ {\isacharequal}\ {\isacharbraceleft}C{\isadigit{1}}{\isacharcomma}\ C{\isadigit{2}}{\isacharbraceright}\ {\isasymunion}\ Sources\ i\ C{\isadigit{1}}\ \ {\isasymunion}\ Sources\ i\ C{\isadigit{2}}{\isachardoublequoteclose}\isanewline
\isadelimproof
\endisadelimproof
\isatagproof
\isacommand{proof}\isamarkupfalse%
\ {\isacharminus}\ \isanewline
\ \ \isacommand{have}\isamarkupfalse%
\ sDef{\isacharcolon}\ \ {\isachardoublequoteopen}{\isacharparenleft}Sources\ i\ S{\isacharparenright}\ {\isacharequal}\ {\isacharparenleft}DSources\ i\ S{\isacharparenright}\ {\isasymunion}\ {\isacharparenleft}{\isasymUnion}\ Z\ {\isasymin}\ {\isacharparenleft}DSources\ i\ S{\isacharparenright}{\isachardot}\ {\isacharparenleft}Sources\ i\ Z{\isacharparenright}{\isacharparenright}{\isachardoublequoteclose}\isanewline
\ \ \ \ \ \isacommand{by}\isamarkupfalse%
\ {\isacharparenleft}rule\ SourcesDef{\isacharparenright}\isanewline
\ \ \isacommand{from}\isamarkupfalse%
\ assms\ \isacommand{have}\isamarkupfalse%
\ {\isachardoublequoteopen}{\isacharparenleft}{\isasymUnion}\ Z\ {\isasymin}\ {\isacharparenleft}DSources\ i\ S{\isacharparenright}{\isachardot}\ {\isacharparenleft}Sources\ i\ Z{\isacharparenright}{\isacharparenright}\ {\isacharequal}\ Sources\ i\ C{\isadigit{1}}\ \ {\isasymunion}\ Sources\ i\ C{\isadigit{2}}{\isachardoublequoteclose}\isanewline
\ \ \ \ \ \isacommand{by}\isamarkupfalse%
\ auto\isanewline
\ \ \isacommand{with}\isamarkupfalse%
\ sDef\ \isakeyword{and}\ assms\ \isacommand{show}\isamarkupfalse%
\ {\isacharquery}thesis\ \isacommand{by}\isamarkupfalse%
\ simp\isanewline
\isacommand{qed}\isamarkupfalse%
\endisatagproof
{\isafoldproof}%
\isadelimproof
\isanewline
\endisadelimproof
\isanewline
\isacommand{lemma}\isamarkupfalse%
\ Sources{\isacharunderscore}{\isadigit{3}}DSources{\isacharcolon}\isanewline
\ \ \isakeyword{assumes}\ {\isachardoublequoteopen}DSources\ i\ S\ {\isacharequal}\ {\isacharbraceleft}C{\isadigit{1}}{\isacharcomma}\ C{\isadigit{2}}{\isacharcomma}\ C{\isadigit{3}}{\isacharbraceright}{\isachardoublequoteclose}\ \isanewline
\ \ \isakeyword{shows}\ \ \ \ {\isachardoublequoteopen}Sources\ i\ S\ {\isacharequal}\ {\isacharbraceleft}C{\isadigit{1}}{\isacharcomma}\ C{\isadigit{2}}{\isacharcomma}\ C{\isadigit{3}}{\isacharbraceright}\ {\isasymunion}\ Sources\ i\ C{\isadigit{1}}\ \ {\isasymunion}\ Sources\ i\ C{\isadigit{2}}\ \ {\isasymunion}\ Sources\ i\ C{\isadigit{3}}{\isachardoublequoteclose}\isanewline
\isadelimproof
\endisadelimproof
\isatagproof
\isacommand{proof}\isamarkupfalse%
\ {\isacharminus}\ \isanewline
\ \ \isacommand{have}\isamarkupfalse%
\ sDef{\isacharcolon}\ {\isachardoublequoteopen}{\isacharparenleft}Sources\ i\ S{\isacharparenright}\ {\isacharequal}\ {\isacharparenleft}DSources\ i\ S{\isacharparenright}\ {\isasymunion}\ {\isacharparenleft}{\isasymUnion}\ Z\ {\isasymin}\ {\isacharparenleft}DSources\ i\ S{\isacharparenright}{\isachardot}\ {\isacharparenleft}Sources\ i\ Z{\isacharparenright}{\isacharparenright}{\isachardoublequoteclose}\ \isanewline
\ \ \ \ \ \isacommand{by}\isamarkupfalse%
\ {\isacharparenleft}rule\ SourcesDef{\isacharparenright}\isanewline
\ \ \isacommand{from}\isamarkupfalse%
\ assms\ \isacommand{have}\isamarkupfalse%
\ {\isachardoublequoteopen}{\isacharparenleft}{\isasymUnion}\ Z\ {\isasymin}\ {\isacharparenleft}DSources\ i\ S{\isacharparenright}{\isachardot}\ {\isacharparenleft}Sources\ i\ Z{\isacharparenright}{\isacharparenright}\ {\isacharequal}\ Sources\ i\ C{\isadigit{1}}\ \ {\isasymunion}\ Sources\ i\ C{\isadigit{2}}\ \ {\isasymunion}\ Sources\ i\ C{\isadigit{3}}{\isachardoublequoteclose}\isanewline
\ \ \ \ \ \isacommand{by}\isamarkupfalse%
\ auto\isanewline
\ \ \isacommand{with}\isamarkupfalse%
\ sDef\ \isakeyword{and}\ assms\ \isacommand{show}\isamarkupfalse%
\ {\isacharquery}thesis\ \isacommand{by}\isamarkupfalse%
\ simp\isanewline
\isacommand{qed}\isamarkupfalse%
\endisatagproof
{\isafoldproof}%
\isadelimproof
\isanewline
\endisadelimproof
\isanewline
\isacommand{lemma}\isamarkupfalse%
\ singleDSourceEmpty{\isadigit{4}}isNotDSource{\isacharcolon}\isanewline
\ \ \isakeyword{assumes}\ {\isachardoublequoteopen}DAcc\ i\ C\ {\isacharequal}\ {\isacharbraceleft}S{\isacharbraceright}{\isachardoublequoteclose}\ \isanewline
\ \ \ \ \ \ \ \ \ \isakeyword{and}\ {\isachardoublequoteopen}Z\ {\isasymnoteq}\ S{\isachardoublequoteclose}\isanewline
\ \ \isakeyword{shows}\ {\isachardoublequoteopen}C\ {\isasymnotin}\ {\isacharparenleft}DSources\ i\ Z{\isacharparenright}{\isachardoublequoteclose}\ \isanewline
\isadelimproof
\endisadelimproof
\isatagproof
\isacommand{proof}\isamarkupfalse%
\ {\isacharminus}\isanewline
\ \ \isacommand{from}\isamarkupfalse%
\ assms\ \isacommand{have}\isamarkupfalse%
\ {\isachardoublequoteopen}{\isacharparenleft}Z\ {\isasymnotin}\ DAcc\ i\ C{\isacharparenright}{\isachardoublequoteclose}\ \ \isacommand{by}\isamarkupfalse%
\ simp\isanewline
\ \ \isacommand{thus}\isamarkupfalse%
\ {\isacharquery}thesis\ \isacommand{by}\isamarkupfalse%
\ {\isacharparenleft}simp\ add{\isacharcolon}\ DAcc{\isacharunderscore}DSourcesNOT{\isacharparenright}\isanewline
\isacommand{qed}\isamarkupfalse%
\endisatagproof
{\isafoldproof}%
\isadelimproof
\isanewline
\endisadelimproof
\isanewline
\isacommand{lemma}\isamarkupfalse%
\ singleDSourceEmpty{\isadigit{4}}isNotDSourceLevel{\isacharcolon}\isanewline
\ \ \isakeyword{assumes}\ {\isachardoublequoteopen}DAcc\ i\ C\ {\isacharequal}\ {\isacharbraceleft}S{\isacharbraceright}{\isachardoublequoteclose}\isanewline
\ \ \isakeyword{shows}\ {\isachardoublequoteopen}{\isasymforall}\ Z\ {\isasymin}\ {\isacharparenleft}AbstrLevel\ i{\isacharparenright}{\isachardot}\ Z\ {\isasymnoteq}\ S\ {\isasymlongrightarrow}\ C\ {\isasymnotin}\ {\isacharparenleft}DSources\ i\ Z{\isacharparenright}{\isachardoublequoteclose}\ \isanewline
\isadelimproof
\endisadelimproof
\isatagproof
\isacommand{using}\isamarkupfalse%
\ assms\ \ \isacommand{by}\isamarkupfalse%
\ {\isacharparenleft}metis\ singleDSourceEmpty{\isadigit{4}}isNotDSource{\isacharparenright}%
\endisatagproof
{\isafoldproof}%
\isadelimproof
\isanewline
\endisadelimproof
\isanewline
\isanewline
\isacommand{lemma}\isamarkupfalse%
\ {\isachardoublequoteopen}isNotDSource{\isacharunderscore}EmptyDAcc{\isachardoublequoteclose}{\isacharcolon}\isanewline
\ \ \isakeyword{assumes}\ {\isachardoublequoteopen}isNotDSource\ i\ S{\isachardoublequoteclose}\ \isanewline
\ \ \isakeyword{shows}\ \ \ \ {\isachardoublequoteopen}DAcc\ i\ S\ {\isacharequal}{\isacharbraceleft}{\isacharbraceright}{\isachardoublequoteclose}\isanewline
\isadelimproof
\endisadelimproof
\isatagproof
\isacommand{using}\isamarkupfalse%
\ assms\ \ \isacommand{by}\isamarkupfalse%
\ {\isacharparenleft}simp\ add{\isacharcolon}\ DAcc{\isacharunderscore}def\ isNotDSource{\isacharunderscore}def{\isacharcomma}\ auto{\isacharparenright}%
\endisatagproof
{\isafoldproof}%
\isadelimproof
\isanewline
\endisadelimproof
\isanewline
\isacommand{lemma}\isamarkupfalse%
\ {\isachardoublequoteopen}isNotDSource{\isacharunderscore}EmptyAcc{\isachardoublequoteclose}{\isacharcolon}\isanewline
\ \ \isakeyword{assumes}\ {\isachardoublequoteopen}isNotDSource\ i\ S{\isachardoublequoteclose}\ \isanewline
\ \ \isakeyword{shows}\ \ \ \ {\isachardoublequoteopen}Acc\ i\ S\ {\isacharequal}\ {\isacharbraceleft}{\isacharbraceright}{\isachardoublequoteclose}\isanewline
\isadelimproof
\endisadelimproof
\isatagproof
\isacommand{proof}\isamarkupfalse%
\ {\isacharminus}\isanewline
\ \ \isacommand{have}\isamarkupfalse%
\ {\isachardoublequoteopen}{\isacharparenleft}Acc\ i\ S{\isacharparenright}\ {\isacharequal}\ {\isacharparenleft}DAcc\ i\ S{\isacharparenright}\ {\isasymunion}\ {\isacharparenleft}{\isasymUnion}\ X\ {\isasymin}\ {\isacharparenleft}DAcc\ i\ S{\isacharparenright}{\isachardot}\ {\isacharparenleft}Acc\ i\ X{\isacharparenright}{\isacharparenright}{\isachardoublequoteclose}\ \ \isanewline
\ \ \ \ \ \isacommand{by}\isamarkupfalse%
\ {\isacharparenleft}rule\ AccDef{\isacharparenright}\isanewline
\ \ \isacommand{thus}\isamarkupfalse%
\ {\isacharquery}thesis\ \isacommand{by}\isamarkupfalse%
\ {\isacharparenleft}metis\ SUP{\isacharunderscore}empty\ Un{\isacharunderscore}absorb\ assms\ isNotDSource{\isacharunderscore}EmptyDAcc{\isacharparenright}\ \isanewline
\isacommand{qed}\isamarkupfalse%
\endisatagproof
{\isafoldproof}%
\isadelimproof
\isanewline
\endisadelimproof
\isanewline
\isacommand{lemma}\isamarkupfalse%
\ singleDSourceEmpty{\isacharunderscore}Acc{\isacharcolon}\isanewline
\ \ \isakeyword{assumes}\ {\isachardoublequoteopen}DAcc\ i\ C\ {\isacharequal}\ {\isacharbraceleft}S{\isacharbraceright}{\isachardoublequoteclose}\ \isanewline
\ \ \ \ \ \ \ \ \ \isakeyword{and}\ {\isachardoublequoteopen}isNotDSource\ i\ S{\isachardoublequoteclose}\ \isanewline
\ \ \isakeyword{shows}\ \ {\isachardoublequoteopen}Acc\ i\ C\ {\isacharequal}\ {\isacharbraceleft}S{\isacharbraceright}{\isachardoublequoteclose}\ \ \isanewline
\isadelimproof
\endisadelimproof
\isatagproof
\isacommand{proof}\isamarkupfalse%
\ {\isacharminus}\ \isanewline
\ \ \isacommand{have}\isamarkupfalse%
\ AccC{\isacharcolon}{\isachardoublequoteopen}{\isacharparenleft}Acc\ i\ C{\isacharparenright}\ {\isacharequal}\ {\isacharparenleft}DAcc\ i\ C{\isacharparenright}\ {\isasymunion}\ {\isacharparenleft}{\isasymUnion}\ S\ {\isasymin}\ {\isacharparenleft}DAcc\ i\ C{\isacharparenright}{\isachardot}\ {\isacharparenleft}Acc\ i\ S{\isacharparenright}{\isacharparenright}{\isachardoublequoteclose}\ \ \isanewline
\ \ \ \ \ \isacommand{by}\isamarkupfalse%
\ {\isacharparenleft}rule\ AccDef{\isacharparenright}\isanewline
\ \ \isacommand{from}\isamarkupfalse%
\ assms\ \isacommand{have}\isamarkupfalse%
\ {\isachardoublequoteopen}Acc\ i\ S\ {\isacharequal}\ {\isacharbraceleft}{\isacharbraceright}{\isachardoublequoteclose}\ \isacommand{by}\isamarkupfalse%
\ {\isacharparenleft}simp\ add{\isacharcolon}\ isNotDSource{\isacharunderscore}EmptyAcc{\isacharparenright}\isanewline
\ \ \isacommand{with}\isamarkupfalse%
\ AccC\ \isacommand{show}\isamarkupfalse%
\ {\isacharquery}thesis\isanewline
\ \ \ \ \ \isacommand{by}\isamarkupfalse%
\ {\isacharparenleft}metis\ SUP{\isacharunderscore}empty\ UN{\isacharunderscore}insert\ Un{\isacharunderscore}commute\ Un{\isacharunderscore}empty{\isacharunderscore}left\ assms{\isacharparenleft}{\isadigit{1}}{\isacharparenright}{\isacharparenright}\ \isanewline
\isacommand{qed}\isamarkupfalse%
\endisatagproof
{\isafoldproof}%
\isadelimproof
\isanewline
\endisadelimproof
\isanewline
\isacommand{lemma}\isamarkupfalse%
\ singleDSourceEmpty{\isadigit{4}}isNotSource{\isacharcolon}\isanewline
\ \ \isakeyword{assumes}\ {\isachardoublequoteopen}DAcc\ i\ C\ {\isacharequal}\ {\isacharbraceleft}S{\isacharbraceright}{\isachardoublequoteclose}\isanewline
\ \ \ \ \ \ \ \ \ \isakeyword{and}\ nSourcS{\isacharcolon}{\isachardoublequoteopen}isNotDSource\ i\ S{\isachardoublequoteclose}\isanewline
\ \ \ \ \ \ \ \ \ \isakeyword{and}\ {\isachardoublequoteopen}Z\ {\isasymnoteq}\ S{\isachardoublequoteclose}\isanewline
\ \ \isakeyword{shows}\ {\isachardoublequoteopen}C\ {\isasymnotin}\ {\isacharparenleft}Sources\ i\ Z{\isacharparenright}{\isachardoublequoteclose}\ \isanewline
\isadelimproof
\endisadelimproof
\isatagproof
\isacommand{proof}\isamarkupfalse%
\ {\isacharminus}\ \isanewline
\ \ \isacommand{from}\isamarkupfalse%
\ assms\ \isacommand{have}\isamarkupfalse%
\ \ {\isachardoublequoteopen}Acc\ i\ C\ {\isacharequal}\ {\isacharbraceleft}S{\isacharbraceright}{\isachardoublequoteclose}\ \ \isacommand{by}\isamarkupfalse%
\ {\isacharparenleft}simp\ add{\isacharcolon}\ singleDSourceEmpty{\isacharunderscore}Acc{\isacharparenright}\isanewline
\ \ \isacommand{with}\isamarkupfalse%
\ assms\ \isacommand{have}\isamarkupfalse%
\ {\isachardoublequoteopen}Z\ {\isasymnotin}\ Acc\ i\ C{\isachardoublequoteclose}\ \isacommand{by}\isamarkupfalse%
\ simp\isanewline
\ \ \isacommand{thus}\isamarkupfalse%
\ {\isacharquery}thesis\ \isacommand{by}\isamarkupfalse%
\ {\isacharparenleft}simp\ add{\isacharcolon}\ Acc{\isacharunderscore}SourcesNOT{\isacharparenright}\isanewline
\isacommand{qed}\isamarkupfalse%
\endisatagproof
{\isafoldproof}%
\isadelimproof
\isanewline
\endisadelimproof
\isanewline
\isacommand{lemma}\isamarkupfalse%
\ singleDSourceEmpty{\isadigit{4}}isNotSourceLevel{\isacharcolon}\isanewline
\ \ \isakeyword{assumes}\ {\isachardoublequoteopen}DAcc\ i\ C\ {\isacharequal}\ {\isacharbraceleft}S{\isacharbraceright}{\isachardoublequoteclose}\isanewline
\ \ \ \ \ \ \ \ \ \isakeyword{and}\ nSourcS{\isacharcolon}{\isachardoublequoteopen}isNotDSource\ i\ S{\isachardoublequoteclose}\ \isanewline
\ \ \isakeyword{shows}\ {\isachardoublequoteopen}{\isasymforall}\ Z\ {\isasymin}\ {\isacharparenleft}AbstrLevel\ i{\isacharparenright}{\isachardot}\ Z\ {\isasymnoteq}\ S\ {\isasymlongrightarrow}\ C\ {\isasymnotin}\ {\isacharparenleft}Sources\ i\ Z{\isacharparenright}{\isachardoublequoteclose}\ \isanewline
\isadelimproof
\endisadelimproof
\isatagproof
\isacommand{using}\isamarkupfalse%
\ assms\isanewline
\isacommand{by}\isamarkupfalse%
\ {\isacharparenleft}metis\ singleDSourceEmpty{\isadigit{4}}isNotSource{\isacharparenright}%
\endisatagproof
{\isafoldproof}%
\isadelimproof
\isanewline
\endisadelimproof
\isanewline
\isanewline
\isacommand{lemma}\isamarkupfalse%
\ singleDSourceLoop{\isacharcolon}\isanewline
\ \ \isakeyword{assumes}\ {\isachardoublequoteopen}DAcc\ i\ C\ {\isacharequal}\ {\isacharbraceleft}S{\isacharbraceright}{\isachardoublequoteclose}\isanewline
\ \ \ \ \ \ \ \ \ \isakeyword{and}\ {\isachardoublequoteopen}DAcc\ i\ S\ {\isacharequal}\ {\isacharbraceleft}C{\isacharbraceright}{\isachardoublequoteclose}\isanewline
\ \ \isakeyword{shows}\ {\isachardoublequoteopen}{\isasymforall}\ Z\ {\isasymin}\ {\isacharparenleft}AbstrLevel\ i{\isacharparenright}{\isachardot}\ {\isacharparenleft}Z\ {\isasymnoteq}\ S\ {\isasymand}\ Z\ {\isasymnoteq}\ C\ {\isasymlongrightarrow}\ C\ {\isasymnotin}\ {\isacharparenleft}Sources\ i\ Z{\isacharparenright}{\isacharparenright}{\isachardoublequoteclose}\ \isanewline
\isadelimproof
\endisadelimproof
\isatagproof
\isacommand{using}\isamarkupfalse%
\ assms\isanewline
\isacommand{by}\isamarkupfalse%
\ {\isacharparenleft}metis\ AccSigleLoop\ Acc{\isacharunderscore}SourcesNOT\ empty{\isacharunderscore}iff\ insert{\isacharunderscore}iff{\isacharparenright}\isanewline
\isanewline
\endisatagproof
{\isafoldproof}%
\isadelimproof
\endisadelimproof
\isamarkupsubsection{Components that are elementary wrt. data dependencies%
}
\isamarkuptrue%
\isamarkupcmt{two output channels of a component C are corelated, if they mutually depend on the same local variable(s)%
}
\isanewline
\isacommand{definition}\isamarkupfalse%
\isanewline
\ \ \ outPairCorelated\ {\isacharcolon}{\isacharcolon}\ {\isachardoublequoteopen}CSet\ {\isasymRightarrow}\ chanID\ {\isasymRightarrow}\ chanID\ {\isasymRightarrow}\ bool{\isachardoublequoteclose}\isanewline
\isakeyword{where}\isanewline
\ \ {\isachardoublequoteopen}outPairCorelated\ C\ x\ y\ {\isasymequiv}\isanewline
\ \ {\isacharparenleft}x\ {\isasymin}\ OUT\ C{\isacharparenright}\ {\isasymand}\ \ \ {\isacharparenleft}y\ {\isasymin}\ OUT\ C{\isacharparenright}\ {\isasymand}\ \isanewline
\ \ {\isacharparenleft}OUTfromV\ x{\isacharparenright}\ {\isasyminter}\ {\isacharparenleft}OUTfromV\ y{\isacharparenright}\ {\isasymnoteq}\ {\isacharbraceleft}{\isacharbraceright}{\isachardoublequoteclose}\isanewline
\isanewline
\isamarkupcmt{We call a set of output channels of a conponent correlated to it output channel x,%
}
\isanewline
\isamarkupcmt{if they mutually depend on the same local variable(s)%
}
\isanewline
\isacommand{definition}\isamarkupfalse%
\isanewline
\ \ \ outSetCorelated\ {\isacharcolon}{\isacharcolon}\ {\isachardoublequoteopen}chanID\ {\isasymRightarrow}\ chanID\ set{\isachardoublequoteclose}\isanewline
\isakeyword{where}\isanewline
\ \ {\isachardoublequoteopen}outSetCorelated\ x\ \ {\isasymequiv}\ \isanewline
\ \ {\isacharbraceleft}\ y{\isacharcolon}{\isacharcolon}chanID\ {\isachardot}\ {\isasymexists}\ v{\isacharcolon}{\isacharcolon}varID{\isachardot}\ {\isacharparenleft}v\ {\isasymin}\ {\isacharparenleft}OUTfromV\ x{\isacharparenright}\ {\isasymand}\ {\isacharparenleft}y\ {\isasymin}\ VARto\ v{\isacharparenright}{\isacharparenright}\ {\isacharbraceright}{\isachardoublequoteclose}\isanewline
\isanewline
\isamarkupcmt{Elementary component according to the data dependencies.%
}
\isanewline
\isamarkupcmt{This constraint should hold for all components on the abstraction level 1%
}
\isanewline
\isacommand{definition}\isamarkupfalse%
\isanewline
elementaryCompDD\ {\isacharcolon}{\isacharcolon}\ {\isachardoublequoteopen}CSet\ {\isasymRightarrow}\ bool{\isachardoublequoteclose}\isanewline
\isakeyword{where}\isanewline
\ \ {\isachardoublequoteopen}elementaryCompDD\ C\ {\isasymequiv}\ \isanewline
\ \ {\isacharparenleft}{\isacharparenleft}{\isasymexists}\ x{\isachardot}\ {\isacharparenleft}OUT\ C{\isacharparenright}\ {\isacharequal}\ {\isacharbraceleft}x{\isacharbraceright}\ {\isacharparenright}\ {\isasymor}\ \isanewline
\ \ \ {\isacharparenleft}{\isasymforall}\ x\ {\isasymin}\ {\isacharparenleft}OUT\ C{\isacharparenright}{\isachardot}\ {\isasymforall}\ y\ {\isasymin}\ {\isacharparenleft}OUT\ C{\isacharparenright}{\isachardot}\ {\isacharparenleft}{\isacharparenleft}outSetCorelated\ x{\isacharparenright}\ {\isasyminter}\ {\isacharparenleft}outSetCorelated\ y{\isacharparenright}\ {\isasymnoteq}\ {\isacharbraceleft}{\isacharbraceright}{\isacharparenright}\ {\isacharparenright}{\isacharparenright}{\isachardoublequoteclose}\isanewline
\isanewline
\isamarkupcmt{the set (outSetCorelated x) is empty if x does not depend from any variable%
}
\isanewline
\isacommand{lemma}\isamarkupfalse%
\ outSetCorelatedEmpty{\isadigit{1}}{\isacharcolon}\isanewline
\ \isakeyword{assumes}\ {\isachardoublequoteopen}OUTfromV\ x\ {\isacharequal}\ {\isacharbraceleft}{\isacharbraceright}{\isachardoublequoteclose}\isanewline
\ \isakeyword{shows}\ {\isachardoublequoteopen}outSetCorelated\ x\ {\isacharequal}\ {\isacharbraceleft}{\isacharbraceright}{\isachardoublequoteclose}\isanewline
\isadelimproof
\endisadelimproof
\isatagproof
\isacommand{using}\isamarkupfalse%
\ assms\ \isacommand{by}\isamarkupfalse%
\ {\isacharparenleft}simp\ add{\isacharcolon}\ outSetCorelated{\isacharunderscore}def{\isacharparenright}\isanewline
\isanewline
\isamarkupcmt{if x depends from at least one variable and the predicates OUTfromV and VARto are defined correctly,%
}
\isanewline
\isamarkupcmt{the set (outSetCorelated x) contains x itself%
}
\endisatagproof
{\isafoldproof}%
\isadelimproof
\isanewline
\endisadelimproof
\isacommand{lemma}\isamarkupfalse%
\ outSetCorelatedNonemptyX{\isacharcolon}\isanewline
\ \isakeyword{assumes}\ {\isachardoublequoteopen}OUTfromV\ x\ \ {\isasymnoteq}\ {\isacharbraceleft}{\isacharbraceright}{\isachardoublequoteclose}\ \isakeyword{and}\ correct{\isadigit{3}}{\isacharcolon}{\isachardoublequoteopen}OUTfromV{\isacharunderscore}VARto{\isachardoublequoteclose}\isanewline
\ \isakeyword{shows}\ {\isachardoublequoteopen}x\ {\isasymin}\ outSetCorelated\ x{\isachardoublequoteclose}\isanewline
\isadelimproof
\endisadelimproof
\isatagproof
\isacommand{proof}\isamarkupfalse%
\ {\isacharminus}\isanewline
\ \ \isacommand{from}\isamarkupfalse%
\ assms\ \isacommand{have}\isamarkupfalse%
\ {\isachardoublequoteopen}{\isasymexists}\ v{\isacharcolon}{\isacharcolon}varID{\isachardot}\ x\ {\isasymin}\ {\isacharparenleft}VARto\ v{\isacharparenright}{\isachardoublequoteclose}\ \isanewline
\ \ \ \ \isacommand{by}\isamarkupfalse%
\ {\isacharparenleft}rule\ OUTfromV{\isacharunderscore}VARto{\isacharunderscore}lemma{\isacharparenright}\isanewline
\ \ \isacommand{from}\isamarkupfalse%
\ this\ \isakeyword{and}\ assms\ \isacommand{show}\isamarkupfalse%
\ {\isacharquery}thesis\isanewline
\ \ \ \ \isacommand{by}\isamarkupfalse%
\ {\isacharparenleft}simp\ add{\isacharcolon}\ \ outSetCorelated{\isacharunderscore}def\ OUTfromV{\isacharunderscore}VARto{\isacharunderscore}def{\isacharparenright}\isanewline
\isacommand{qed}\isamarkupfalse%
\isanewline
\isanewline
\isamarkupcmt{if the set (outSetCorelated x) is empty, this means that x does not depend from any variable%
}
\endisatagproof
{\isafoldproof}%
\isadelimproof
\isanewline
\endisadelimproof
\isacommand{lemma}\isamarkupfalse%
\ outSetCorelatedEmpty{\isadigit{2}}{\isacharcolon}\isanewline
\ \isakeyword{assumes}\ {\isachardoublequoteopen}outSetCorelated\ x\ {\isacharequal}\ {\isacharbraceleft}{\isacharbraceright}{\isachardoublequoteclose}\ \ \ \isakeyword{and}\ correct{\isadigit{3}}{\isacharcolon}{\isachardoublequoteopen}OUTfromV{\isacharunderscore}VARto{\isachardoublequoteclose}\isanewline
\ \isakeyword{shows}\ \ {\isachardoublequoteopen}OUTfromV\ x\ {\isacharequal}\ {\isacharbraceleft}{\isacharbraceright}{\isachardoublequoteclose}\isanewline
\isadelimproof
\endisadelimproof
\isatagproof
\isacommand{proof}\isamarkupfalse%
\ {\isacharparenleft}rule\ ccontr{\isacharparenright}\isanewline
\ \ \isacommand{assume}\isamarkupfalse%
\ OUTfromVNonempty{\isacharcolon}{\isachardoublequoteopen}OUTfromV\ x\ {\isasymnoteq}\ {\isacharbraceleft}{\isacharbraceright}{\isachardoublequoteclose}\isanewline
\ \ \isacommand{from}\isamarkupfalse%
\ this\ \isakeyword{and}\ correct{\isadigit{3}}\ \isacommand{have}\isamarkupfalse%
\ {\isachardoublequoteopen}x\ {\isasymin}\ outSetCorelated\ x{\isachardoublequoteclose}\ \isanewline
\ \ \ \ \isacommand{by}\isamarkupfalse%
\ {\isacharparenleft}rule\ outSetCorelatedNonemptyX{\isacharparenright}\isanewline
\ \ \isacommand{from}\isamarkupfalse%
\ this\ \isakeyword{and}\ assms\ \isacommand{show}\isamarkupfalse%
\ False\ \isacommand{by}\isamarkupfalse%
\ simp\isanewline
\isacommand{qed}\isamarkupfalse%
\isanewline
\endisatagproof
{\isafoldproof}%
\isadelimproof
\endisadelimproof
\isamarkupsubsection{Set of components needed to check a specific property%
}
\isamarkuptrue%
\isamarkupcmt{set of components specified on abstreaction level i, which input channels belong to the set chSet%
}
\isanewline
\isacommand{definition}\isamarkupfalse%
\isanewline
\ \ inSetOfComponents\ {\isacharcolon}{\isacharcolon}\ {\isachardoublequoteopen}AbstrLevelsID\ {\isasymRightarrow}\ chanID\ set\ {\isasymRightarrow}\ CSet\ set{\isachardoublequoteclose}\isanewline
\isakeyword{where}\isanewline
\ {\isachardoublequoteopen}inSetOfComponents\ i\ chSet\ {\isasymequiv}\isanewline
\ \ {\isacharbraceleft}X{\isachardot}\ {\isacharparenleft}{\isacharparenleft}{\isacharparenleft}IN\ X{\isacharparenright}\ {\isasyminter}\ chSet\ {\isasymnoteq}\ {\isacharbraceleft}{\isacharbraceright}{\isacharparenright}\ \ {\isasymand}\ X\ {\isasymin}\ {\isacharparenleft}AbstrLevel\ i{\isacharparenright}{\isacharparenright}{\isacharbraceright}{\isachardoublequoteclose}\isanewline
\isanewline
\isamarkupcmt{Set of components from the abstraction level i, which output channels belong to the set chSet%
}
\isanewline
\isacommand{definition}\isamarkupfalse%
\isanewline
\ \ outSetOfComponents\ {\isacharcolon}{\isacharcolon}\ {\isachardoublequoteopen}AbstrLevelsID\ {\isasymRightarrow}\ chanID\ set\ {\isasymRightarrow}\ CSet\ set{\isachardoublequoteclose}\isanewline
\isakeyword{where}\isanewline
\ {\isachardoublequoteopen}outSetOfComponents\ i\ chSet\ {\isasymequiv}\isanewline
\ \ {\isacharbraceleft}Y{\isachardot}\ {\isacharparenleft}{\isacharparenleft}{\isacharparenleft}OUT\ Y{\isacharparenright}\ {\isasyminter}\ chSet\ {\isasymnoteq}\ {\isacharbraceleft}{\isacharbraceright}{\isacharparenright}\ {\isasymand}\ Y\ {\isasymin}\ {\isacharparenleft}AbstrLevel\ i{\isacharparenright}{\isacharparenright}{\isacharbraceright}{\isachardoublequoteclose}\isanewline
\isanewline
\isamarkupcmt{Set of components from the abstraction level i,%
}
\isanewline
\isamarkupcmt{which have output channels from the set chSet or are sources for such components%
}
\isanewline
\isacommand{definition}\isamarkupfalse%
\isanewline
\ \ minSetOfComponents\ {\isacharcolon}{\isacharcolon}\ \ {\isachardoublequoteopen}AbstrLevelsID\ {\isasymRightarrow}\ chanID\ set\ {\isasymRightarrow}\ CSet\ set{\isachardoublequoteclose}\isanewline
\isakeyword{where}\isanewline
\ {\isachardoublequoteopen}minSetOfComponents\ i\ chSet\ {\isasymequiv}\isanewline
\ \ {\isacharparenleft}outSetOfComponents\ i\ chSet{\isacharparenright}\ {\isasymunion}\isanewline
\ \ {\isacharparenleft}{\isasymUnion}\ S\ {\isasymin}\ {\isacharparenleft}outSetOfComponents\ i\ chSet{\isacharparenright}{\isachardot}\ {\isacharparenleft}Sources\ i\ S{\isacharparenright}{\isacharparenright}{\isachardoublequoteclose}\isanewline
\isanewline
\isamarkupcmt{Please note that a system output cannot beat the same time a local chanel.%
}
\isanewline
\isanewline
\isamarkupcmt{channel x is a system input on an abstraction level i%
}
\isanewline
\isacommand{definition}\isamarkupfalse%
\ systemIN\ {\isacharcolon}{\isacharcolon}{\isachardoublequoteopen}chanID\ {\isasymRightarrow}\ AbstrLevelsID\ {\isasymRightarrow}\ bool{\isachardoublequoteclose}\isanewline
\isakeyword{where}\isanewline
\ \ {\isachardoublequoteopen}systemIN\ x\ i\ {\isasymequiv}\ {\isacharparenleft}{\isasymexists}\ C{\isadigit{1}}\ {\isasymin}\ {\isacharparenleft}AbstrLevel\ i{\isacharparenright}{\isachardot}\ x\ {\isasymin}\ {\isacharparenleft}IN\ C{\isadigit{1}}{\isacharparenright}{\isacharparenright}\ {\isasymand}\ {\isacharparenleft}{\isasymforall}\ C{\isadigit{2}}\ {\isasymin}\ {\isacharparenleft}AbstrLevel\ i{\isacharparenright}{\isachardot}\ x\ {\isasymnotin}\ {\isacharparenleft}OUT\ C{\isadigit{2}}{\isacharparenright}{\isacharparenright}{\isachardoublequoteclose}\isanewline
\isanewline
\isamarkupcmt{channel x is a system input on an abstraction level i%
}
\isanewline
\isacommand{definition}\isamarkupfalse%
\ systemOUT\ {\isacharcolon}{\isacharcolon}{\isachardoublequoteopen}chanID\ {\isasymRightarrow}\ AbstrLevelsID\ {\isasymRightarrow}\ bool{\isachardoublequoteclose}\isanewline
\isakeyword{where}\isanewline
\ \ {\isachardoublequoteopen}systemOUT\ x\ i\ {\isasymequiv}\ {\isacharparenleft}{\isasymforall}\ C{\isadigit{1}}\ {\isasymin}\ {\isacharparenleft}AbstrLevel\ i{\isacharparenright}{\isachardot}\ x\ {\isasymnotin}\ {\isacharparenleft}IN\ C{\isadigit{1}}{\isacharparenright}{\isacharparenright}\ {\isasymand}\ {\isacharparenleft}{\isasymexists}\ C{\isadigit{2}}\ {\isasymin}\ {\isacharparenleft}AbstrLevel\ i{\isacharparenright}{\isachardot}\ x\ {\isasymin}\ {\isacharparenleft}OUT\ C{\isadigit{2}}{\isacharparenright}{\isacharparenright}{\isachardoublequoteclose}\isanewline
\isanewline
\isamarkupcmt{channel x is a system local channel on an abstraction level i%
}
\isanewline
\isacommand{definition}\isamarkupfalse%
\ systemLOC\ {\isacharcolon}{\isacharcolon}{\isachardoublequoteopen}chanID\ {\isasymRightarrow}\ AbstrLevelsID\ {\isasymRightarrow}\ bool{\isachardoublequoteclose}\isanewline
\isakeyword{where}\isanewline
\ \ {\isachardoublequoteopen}systemLOC\ x\ i\ {\isasymequiv}\ {\isacharparenleft}{\isasymexists}\ C{\isadigit{1}}\ {\isasymin}\ {\isacharparenleft}AbstrLevel\ i{\isacharparenright}{\isachardot}\ x\ {\isasymin}\ {\isacharparenleft}IN\ C{\isadigit{1}}{\isacharparenright}{\isacharparenright}\ {\isasymand}\ {\isacharparenleft}{\isasymexists}\ C{\isadigit{2}}\ {\isasymin}\ {\isacharparenleft}AbstrLevel\ i{\isacharparenright}{\isachardot}\ x\ {\isasymin}\ {\isacharparenleft}OUT\ C{\isadigit{2}}{\isacharparenright}{\isacharparenright}{\isachardoublequoteclose}\isanewline
\isanewline
\isacommand{lemma}\isamarkupfalse%
\ systemIN{\isacharunderscore}noOUT{\isacharcolon}\isanewline
\ \ \isakeyword{assumes}\ {\isachardoublequoteopen}systemIN\ x\ i{\isachardoublequoteclose}\isanewline
\ \ \isakeyword{shows}\ \ \ \ {\isachardoublequoteopen}{\isasymnot}\ systemOUT\ x\ i{\isachardoublequoteclose}\isanewline
\isadelimproof
\endisadelimproof
\isatagproof
\isacommand{using}\isamarkupfalse%
\ assms\ \isacommand{by}\isamarkupfalse%
\ {\isacharparenleft}simp\ add{\isacharcolon}\ systemIN{\isacharunderscore}def\ systemOUT{\isacharunderscore}def{\isacharparenright}%
\endisatagproof
{\isafoldproof}%
\isadelimproof
\isanewline
\endisadelimproof
\isanewline
\isacommand{lemma}\isamarkupfalse%
\ systemOUT{\isacharunderscore}noIN{\isacharcolon}\isanewline
\ \ \isakeyword{assumes}\ {\isachardoublequoteopen}systemOUT\ x\ i{\isachardoublequoteclose}\isanewline
\ \ \isakeyword{shows}\ \ \ \ {\isachardoublequoteopen}{\isasymnot}\ systemIN\ x\ i{\isachardoublequoteclose}\isanewline
\isadelimproof
\endisadelimproof
\isatagproof
\isacommand{using}\isamarkupfalse%
\ assms\ \isacommand{by}\isamarkupfalse%
\ {\isacharparenleft}simp\ add{\isacharcolon}\ systemIN{\isacharunderscore}def\ systemOUT{\isacharunderscore}def{\isacharparenright}%
\endisatagproof
{\isafoldproof}%
\isadelimproof
\isanewline
\endisadelimproof
\isanewline
\isacommand{lemma}\isamarkupfalse%
\ systemIN{\isacharunderscore}noLOC{\isacharcolon}\isanewline
\ \ \isakeyword{assumes}\ {\isachardoublequoteopen}systemIN\ x\ i{\isachardoublequoteclose}\isanewline
\ \ \isakeyword{shows}\ \ \ \ {\isachardoublequoteopen}{\isasymnot}\ systemLOC\ x\ i{\isachardoublequoteclose}\isanewline
\isadelimproof
\endisadelimproof
\isatagproof
\isacommand{using}\isamarkupfalse%
\ assms\ \isacommand{by}\isamarkupfalse%
\ {\isacharparenleft}simp\ add{\isacharcolon}\ systemIN{\isacharunderscore}def\ systemLOC{\isacharunderscore}def{\isacharparenright}%
\endisatagproof
{\isafoldproof}%
\isadelimproof
\isanewline
\endisadelimproof
\isanewline
\isacommand{lemma}\isamarkupfalse%
\ systemLOC{\isacharunderscore}noIN{\isacharcolon}\isanewline
\ \ \isakeyword{assumes}\ {\isachardoublequoteopen}systemLOC\ x\ i{\isachardoublequoteclose}\isanewline
\ \ \isakeyword{shows}\ \ \ \ {\isachardoublequoteopen}{\isasymnot}\ systemIN\ x\ i{\isachardoublequoteclose}\isanewline
\isadelimproof
\endisadelimproof
\isatagproof
\isacommand{using}\isamarkupfalse%
\ assms\ \isacommand{by}\isamarkupfalse%
\ {\isacharparenleft}simp\ add{\isacharcolon}\ systemIN{\isacharunderscore}def\ systemLOC{\isacharunderscore}def{\isacharparenright}%
\endisatagproof
{\isafoldproof}%
\isadelimproof
\isanewline
\endisadelimproof
\isanewline
\isacommand{lemma}\isamarkupfalse%
\ systemOUT{\isacharunderscore}noLOC{\isacharcolon}\isanewline
\ \ \isakeyword{assumes}\ {\isachardoublequoteopen}systemOUT\ x\ i{\isachardoublequoteclose}\isanewline
\ \ \isakeyword{shows}\ \ \ \ {\isachardoublequoteopen}{\isasymnot}\ systemLOC\ x\ i{\isachardoublequoteclose}\isanewline
\isadelimproof
\endisadelimproof
\isatagproof
\isacommand{using}\isamarkupfalse%
\ assms\ \isacommand{by}\isamarkupfalse%
\ {\isacharparenleft}simp\ add{\isacharcolon}\ systemOUT{\isacharunderscore}def\ systemLOC{\isacharunderscore}def{\isacharparenright}%
\endisatagproof
{\isafoldproof}%
\isadelimproof
\isanewline
\endisadelimproof
\isanewline
\isacommand{lemma}\isamarkupfalse%
\ systemLOC{\isacharunderscore}noOUT{\isacharcolon}\isanewline
\ \ \isakeyword{assumes}\ {\isachardoublequoteopen}systemLOC\ x\ i{\isachardoublequoteclose}\isanewline
\ \ \isakeyword{shows}\ \ \ \ {\isachardoublequoteopen}{\isasymnot}\ systemOUT\ x\ i{\isachardoublequoteclose}\isanewline
\isadelimproof
\endisadelimproof
\isatagproof
\isacommand{using}\isamarkupfalse%
\ assms\ \isacommand{by}\isamarkupfalse%
\ {\isacharparenleft}simp\ add{\isacharcolon}\ systemLOC{\isacharunderscore}def\ systemOUT{\isacharunderscore}def{\isacharparenright}%
\endisatagproof
{\isafoldproof}%
\isadelimproof
\isanewline
\endisadelimproof
\isanewline
\isacommand{definition}\isamarkupfalse%
\isanewline
\ \ noIrrelevantChannels\ {\isacharcolon}{\isacharcolon}\ \ {\isachardoublequoteopen}AbstrLevelsID\ {\isasymRightarrow}\ chanID\ set\ {\isasymRightarrow}\ bool{\isachardoublequoteclose}\isanewline
\isakeyword{where}\isanewline
\ {\isachardoublequoteopen}noIrrelevantChannels\ i\ chSet\ {\isasymequiv}\isanewline
\ \ {\isasymforall}\ x\ {\isasymin}\ chSet{\isachardot}\ {\isacharparenleft}{\isacharparenleft}systemIN\ x\ i{\isacharparenright}\ {\isasymlongrightarrow}\isanewline
\ \ \ {\isacharparenleft}{\isasymexists}\ Z\ {\isasymin}\ {\isacharparenleft}minSetOfComponents\ i\ chSet{\isacharparenright}{\isachardot}\ x\ {\isasymin}\ {\isacharparenleft}IN\ Z{\isacharparenright}{\isacharparenright}{\isacharparenright}{\isachardoublequoteclose}\isanewline
\isanewline
\isanewline
\isacommand{definition}\isamarkupfalse%
\isanewline
\ \ allNeededINChannels\ {\isacharcolon}{\isacharcolon}\ \ {\isachardoublequoteopen}AbstrLevelsID\ {\isasymRightarrow}\ chanID\ set\ {\isasymRightarrow}\ bool{\isachardoublequoteclose}\isanewline
\isakeyword{where}\isanewline
\ {\isachardoublequoteopen}allNeededINChannels\ i\ chSet\ {\isasymequiv}\isanewline
\ \ {\isacharparenleft}{\isasymforall}\ Z\ {\isasymin}\ {\isacharparenleft}minSetOfComponents\ i\ chSet{\isacharparenright}{\isachardot}\ {\isasymexists}\ x\ {\isasymin}\ {\isacharparenleft}IN\ Z{\isacharparenright}{\isachardot}\ {\isacharparenleft}{\isacharparenleft}systemIN\ x\ i{\isacharparenright}\ {\isasymlongrightarrow}\ {\isacharparenleft}x\ {\isasymin}\ chSet{\isacharparenright}{\isacharparenright}{\isacharparenright}{\isachardoublequoteclose}\isanewline
\isanewline
\isamarkupcmt{the set (outSetOfComponents i chSet) should be a subset of all components specified on the abstraction level i%
}
\isanewline
\isacommand{lemma}\isamarkupfalse%
\ outSetOfComponentsLimit{\isacharcolon}\isanewline
{\isachardoublequoteopen}outSetOfComponents\ i\ chSet\ {\isasymsubseteq}\ AbstrLevel\ i{\isachardoublequoteclose}\isanewline
\isadelimproof
\endisadelimproof
\isatagproof
\isacommand{by}\isamarkupfalse%
\ {\isacharparenleft}metis\ {\isacharparenleft}lifting{\isacharparenright}\ mem{\isacharunderscore}Collect{\isacharunderscore}eq\ outSetOfComponents{\isacharunderscore}def\ subsetI{\isacharparenright}\isanewline
\isanewline
\isamarkupcmt{the set (inSetOfComponents i chSet) should be a subset of all components specified on the abstraction level i%
}
\endisatagproof
{\isafoldproof}%
\isadelimproof
\isanewline
\endisadelimproof
\isacommand{lemma}\isamarkupfalse%
\ inSetOfComponentsLimit{\isacharcolon}\isanewline
{\isachardoublequoteopen}inSetOfComponents\ i\ chSet\ {\isasymsubseteq}\ AbstrLevel\ i{\isachardoublequoteclose}\isanewline
\isadelimproof
\endisadelimproof
\isatagproof
\isacommand{by}\isamarkupfalse%
\ {\isacharparenleft}metis\ {\isacharparenleft}lifting{\isacharparenright}\ inSetOfComponents{\isacharunderscore}def\ mem{\isacharunderscore}Collect{\isacharunderscore}eq\ subsetI{\isacharparenright}\isanewline
\isanewline
\isamarkupcmt{the set of components, which are sources for the components%
}
\isanewline
\isamarkupcmt{out of (inSetOfComponents i chSet), should be a subset of%
}
\ \isanewline
\isamarkupcmt{all components specified on the abstraction level i%
}
\endisatagproof
{\isafoldproof}%
\isadelimproof
\isanewline
\endisadelimproof
\isacommand{lemma}\isamarkupfalse%
\ SourcesLevelLimit{\isacharcolon}\isanewline
{\isachardoublequoteopen}{\isacharparenleft}{\isasymUnion}\ S\ {\isasymin}\ {\isacharparenleft}outSetOfComponents\ i\ chSet{\isacharparenright}{\isachardot}\ {\isacharparenleft}Sources\ i\ S{\isacharparenright}{\isacharparenright}\ {\isasymsubseteq}\ AbstrLevel\ i{\isachardoublequoteclose}\isanewline
\isadelimproof
\endisadelimproof
\isatagproof
\isacommand{proof}\isamarkupfalse%
\ {\isacharminus}\isanewline
\ \ \isacommand{have}\isamarkupfalse%
\ sg{\isadigit{1}}{\isacharcolon}{\isachardoublequoteopen}outSetOfComponents\ i\ chSet\ {\isasymsubseteq}\ AbstrLevel\ i{\isachardoublequoteclose}\isanewline
\ \ \ \ \isacommand{by}\isamarkupfalse%
\ {\isacharparenleft}simp\ add{\isacharcolon}\ outSetOfComponentsLimit{\isacharparenright}\isanewline
\ \ \isacommand{have}\isamarkupfalse%
\ {\isachardoublequoteopen}{\isasymforall}\ S{\isachardot}\ S\ {\isasymin}\ {\isacharparenleft}outSetOfComponents\ i\ chSet{\isacharparenright}\ {\isasymlongrightarrow}\ Sources\ i\ S\ {\isasymsubseteq}\ AbstrLevel\ i{\isachardoublequoteclose}\isanewline
\ \ \ \ \isacommand{by}\isamarkupfalse%
\ {\isacharparenleft}metis\ SourcesLevelX{\isacharparenright}\isanewline
\ \ \isacommand{from}\isamarkupfalse%
\ this\ \isakeyword{and}\ sg{\isadigit{1}}\ \isacommand{show}\isamarkupfalse%
\ {\isacharquery}thesis\ \isacommand{by}\isamarkupfalse%
\ auto\isanewline
\isacommand{qed}\isamarkupfalse%
\endisatagproof
{\isafoldproof}%
\isadelimproof
\isanewline
\endisadelimproof
\isanewline
\isacommand{lemma}\isamarkupfalse%
\ minSetOfComponentsLimit{\isacharcolon}\isanewline
{\isachardoublequoteopen}minSetOfComponents\ i\ chSet\ {\isasymsubseteq}\ AbstrLevel\ i{\isachardoublequoteclose}\isanewline
\isadelimproof
\endisadelimproof
\isatagproof
\isacommand{proof}\isamarkupfalse%
\ {\isacharminus}\isanewline
\ \ \isacommand{have}\isamarkupfalse%
\ sg{\isadigit{1}}{\isacharcolon}\ {\isachardoublequoteopen}outSetOfComponents\ i\ chSet\ {\isasymsubseteq}\ AbstrLevel\ i{\isachardoublequoteclose}\isanewline
\ \ \ \ \isacommand{by}\isamarkupfalse%
\ {\isacharparenleft}simp\ add{\isacharcolon}\ outSetOfComponentsLimit{\isacharparenright}\isanewline
\ \ \isacommand{have}\isamarkupfalse%
\ {\isachardoublequoteopen}{\isacharparenleft}{\isasymUnion}\ S\ {\isasymin}\ {\isacharparenleft}outSetOfComponents\ i\ chSet{\isacharparenright}{\isachardot}\ {\isacharparenleft}Sources\ i\ S{\isacharparenright}{\isacharparenright}\ {\isasymsubseteq}\ AbstrLevel\ i{\isachardoublequoteclose}\isanewline
\ \ \ \ \isacommand{by}\isamarkupfalse%
\ {\isacharparenleft}simp\ add{\isacharcolon}\ \ SourcesLevelLimit{\isacharparenright}\isanewline
\ \ \isacommand{with}\isamarkupfalse%
\ sg{\isadigit{1}}\ \isacommand{show}\isamarkupfalse%
\ {\isacharquery}thesis\ \ \isacommand{by}\isamarkupfalse%
\ {\isacharparenleft}simp\ add{\isacharcolon}\ minSetOfComponents{\isacharunderscore}def{\isacharparenright}\isanewline
\isacommand{qed}\isamarkupfalse%
\ \isanewline
\endisatagproof
{\isafoldproof}%
\isadelimproof
\endisadelimproof
\isamarkupsubsection{Additional properties: Remote Computation%
}
\isamarkuptrue%
\isamarkupcmt{The value of  $UplSizeHighLoad$ $x$ is True if its $UplSize$ measure is greather that a predifined value%
}
\isanewline
\isacommand{definition}\isamarkupfalse%
\ UplSizeHighLoadCh\ {\isacharcolon}{\isacharcolon}\ \ {\isachardoublequoteopen}chanID\ {\isasymRightarrow}\ bool{\isachardoublequoteclose}\isanewline
\isakeyword{where}\isanewline
\ \ \ {\isachardoublequoteopen}UplSizeHighLoadCh\ x\ {\isasymequiv}\ {\isacharparenleft}x\ {\isasymin}\ UplSizeHighLoad{\isacharparenright}{\isachardoublequoteclose}\isanewline
\isanewline
\isamarkupcmt{if the $Perf$ measure of at least one subcomponent is greather than a predifined value,%
}
\isanewline
\isamarkupcmt{the $Perf$ measure of this component is greather than $HighPerf$ too%
}
\isanewline
\isacommand{axiomatization}\isamarkupfalse%
\ HighPerfComp\ {\isacharcolon}{\isacharcolon}\ \ {\isachardoublequoteopen}CSet\ {\isasymRightarrow}\ bool{\isachardoublequoteclose}\isanewline
\isakeyword{where}\isanewline
HighPerfComDef{\isacharcolon}\isanewline
\ \ \ {\isachardoublequoteopen}HighPerfComp\ C\ {\isacharequal}\isanewline
\ \ \ {\isacharparenleft}{\isacharparenleft}C\ {\isasymin}\ HighPerfSet{\isacharparenright}\ {\isasymor}\ {\isacharparenleft}{\isasymexists}\ Z\ {\isasymin}\ subcomp\ C{\isachardot}\ {\isacharparenleft}HighPerfComp\ Z{\isacharparenright}{\isacharparenright}{\isacharparenright}{\isachardoublequoteclose}\isanewline
\isadelimtheory
\isanewline
\endisadelimtheory
\isatagtheory
\isacommand{end}\isamarkupfalse%
\endisatagtheory
{\isafoldtheory}%
\isadelimtheory
\isanewline
\endisadelimtheory
\end{isabellebody}%

%
\begin{isabellebody}%
\def\isabellecontext{DataDependenciesCaseStudy}%
\isamarkupheader{Case Study: Verification of Properties%
}
\isamarkuptrue%
\isadelimtheory
\endisadelimtheory
\isatagtheory
\isacommand{theory}\isamarkupfalse%
\ DataDependenciesCaseStudy\isanewline
\ \ \isakeyword{imports}\ DataDependencies\isanewline
\isakeyword{begin}%
\endisatagtheory
{\isafoldtheory}%
\isadelimtheory
\endisadelimtheory
\isamarkupsubsection{Correct composition of components%
}
\isamarkuptrue%
\isamarkupcmt{the lemmas  AbstrLevels X Y with corresponding proofs can be composend%
}
\isanewline
\isamarkupcmt{and proven automatically, their proofs are identical%
}
\isanewline
\isacommand{lemma}\isamarkupfalse%
\ AbstrLevels{\isacharunderscore}A{\isadigit{1}}{\isacharunderscore}A{\isadigit{1}}{\isadigit{1}}{\isacharcolon}\isanewline
\ \ \isakeyword{assumes}\ {\isachardoublequoteopen}sA{\isadigit{1}}\ {\isasymin}\ AbstrLevel\ i{\isachardoublequoteclose}\isanewline
\ \ \isakeyword{shows}\ {\isachardoublequoteopen}sA{\isadigit{1}}{\isadigit{1}}\ {\isasymnotin}\ AbstrLevel\ i{\isachardoublequoteclose}\isanewline
\isadelimproof
\endisadelimproof
\isatagproof
\isacommand{using}\isamarkupfalse%
\ assms\ \isanewline
\isacommand{by}\isamarkupfalse%
\ {\isacharparenleft}induct\ i{\isacharcomma}\ simp\ add{\isacharcolon}\ AbstrLevel{\isadigit{0}}{\isacharcomma}\ simp\ add{\isacharcolon}\ \ AbstrLevel{\isadigit{1}}{\isacharcomma}\ simp\ add{\isacharcolon}\ \ AbstrLevel{\isadigit{2}}{\isacharcomma}\ \ simp\ add{\isacharcolon}\ \ AbstrLevel{\isadigit{3}}{\isacharparenright}%
\endisatagproof
{\isafoldproof}%
\isadelimproof
\isanewline
\endisadelimproof
\isanewline
\isacommand{lemma}\isamarkupfalse%
\ AbstrLevels{\isacharunderscore}A{\isadigit{1}}{\isacharunderscore}A{\isadigit{1}}{\isadigit{2}}{\isacharcolon}\isanewline
\ \ \isakeyword{assumes}\ {\isachardoublequoteopen}sA{\isadigit{1}}\ {\isasymin}\ AbstrLevel\ i{\isachardoublequoteclose}
\ \ \isakeyword{shows}\ {\isachardoublequoteopen}sA{\isadigit{1}}{\isadigit{2}}\ {\isasymnotin}\ AbstrLevel\ i{\isachardoublequoteclose}%
\isadelimproof
\endisadelimproof
\isatagproof
\endisatagproof
{\isafoldproof}%
\isadelimproof
\isanewline
\endisadelimproof
\isanewline
\isacommand{lemma}\isamarkupfalse%
\ AbstrLevels{\isacharunderscore}A{\isadigit{2}}{\isacharunderscore}A{\isadigit{2}}{\isadigit{1}}{\isacharcolon}\isanewline
\ \ \isakeyword{assumes}\ {\isachardoublequoteopen}sA{\isadigit{2}}\ {\isasymin}\ AbstrLevel\ i{\isachardoublequoteclose} 
\ \ \isakeyword{shows}\ {\isachardoublequoteopen}sA{\isadigit{2}}{\isadigit{1}}\ {\isasymnotin}\ AbstrLevel\ i{\isachardoublequoteclose}%
\isadelimproof
\endisadelimproof
\isatagproof
\endisatagproof
{\isafoldproof}%
\isadelimproof
\isanewline
\endisadelimproof
\isanewline
\isacommand{lemma}\isamarkupfalse%
\ AbstrLevels{\isacharunderscore}A{\isadigit{2}}{\isacharunderscore}A{\isadigit{2}}{\isadigit{2}}{\isacharcolon}\isanewline
\ \ \isakeyword{assumes}\ {\isachardoublequoteopen}sA{\isadigit{2}}\ {\isasymin}\ AbstrLevel\ i{\isachardoublequoteclose} 
\ \ \isakeyword{shows}\ {\isachardoublequoteopen}sA{\isadigit{2}}{\isadigit{2}}\ {\isasymnotin}\ AbstrLevel\ i{\isachardoublequoteclose}%
\isadelimproof
\endisadelimproof
\isatagproof
\endisatagproof
{\isafoldproof}%
\isadelimproof
\isanewline
\endisadelimproof
\isanewline
\isacommand{lemma}\isamarkupfalse%
\ AbstrLevels{\isacharunderscore}A{\isadigit{2}}{\isacharunderscore}A{\isadigit{2}}{\isadigit{3}}{\isacharcolon}\isanewline
\ \ \isakeyword{assumes}\ {\isachardoublequoteopen}sA{\isadigit{2}}\ {\isasymin}\ AbstrLevel\ i{\isachardoublequoteclose} 
\ \ \isakeyword{shows}\ {\isachardoublequoteopen}sA{\isadigit{2}}{\isadigit{3}}\ {\isasymnotin}\ AbstrLevel\ i{\isachardoublequoteclose}%
\isadelimproof
\endisadelimproof
\isatagproof
\endisatagproof
{\isafoldproof}%
\isadelimproof
\isanewline
\endisadelimproof
\isanewline
\isacommand{lemma}\isamarkupfalse%
\ AbstrLevels{\isacharunderscore}A{\isadigit{3}}{\isacharunderscore}A{\isadigit{3}}{\isadigit{1}}{\isacharcolon}\isanewline
\ \ \isakeyword{assumes}\ {\isachardoublequoteopen}sA{\isadigit{3}}\ {\isasymin}\ AbstrLevel\ i{\isachardoublequoteclose} 
\ \ \isakeyword{shows}\ {\isachardoublequoteopen}sA{\isadigit{3}}{\isadigit{1}}\ {\isasymnotin}\ AbstrLevel\ i{\isachardoublequoteclose}%
\isadelimproof
\endisadelimproof
\isatagproof
\endisatagproof
{\isafoldproof}%
\isadelimproof
\isanewline
\endisadelimproof
\isanewline
\isacommand{lemma}\isamarkupfalse%
\ AbstrLevels{\isacharunderscore}A{\isadigit{3}}{\isacharunderscore}A{\isadigit{3}}{\isadigit{2}}{\isacharcolon}\isanewline
\ \ \isakeyword{assumes}\ {\isachardoublequoteopen}sA{\isadigit{3}}\ {\isasymin}\ AbstrLevel\ i{\isachardoublequoteclose} 
\ \ \isakeyword{shows}\ {\isachardoublequoteopen}sA{\isadigit{3}}{\isadigit{2}}\ {\isasymnotin}\ AbstrLevel\ i{\isachardoublequoteclose}%
\isadelimproof
\endisadelimproof
\isatagproof
\endisatagproof
{\isafoldproof}%
\isadelimproof
\isanewline
\endisadelimproof
\isanewline
\isacommand{lemma}\isamarkupfalse%
\ AbstrLevels{\isacharunderscore}A{\isadigit{4}}{\isacharunderscore}A{\isadigit{4}}{\isadigit{1}}{\isacharcolon}\isanewline
\ \ \isakeyword{assumes}\ {\isachardoublequoteopen}sA{\isadigit{4}}\ {\isasymin}\ AbstrLevel\ i{\isachardoublequoteclose} 
\ \ \isakeyword{shows}\ {\isachardoublequoteopen}sA{\isadigit{4}}{\isadigit{1}}\ {\isasymnotin}\ AbstrLevel\ i{\isachardoublequoteclose}%
\isadelimproof
\endisadelimproof
\isatagproof
\endisatagproof
{\isafoldproof}%
\isadelimproof
\isanewline
\endisadelimproof
\isanewline
\isacommand{lemma}\isamarkupfalse%
\ AbstrLevels{\isacharunderscore}A{\isadigit{4}}{\isacharunderscore}A{\isadigit{4}}{\isadigit{2}}{\isacharcolon}\isanewline
\ \ \isakeyword{assumes}\ {\isachardoublequoteopen}sA{\isadigit{4}}\ {\isasymin}\ AbstrLevel\ i{\isachardoublequoteclose} 
\ \ \isakeyword{shows}\ {\isachardoublequoteopen}sA{\isadigit{4}}{\isadigit{2}}\ {\isasymnotin}\ AbstrLevel\ i{\isachardoublequoteclose}%
\isadelimproof
\endisadelimproof
\isatagproof
\endisatagproof
{\isafoldproof}%
\isadelimproof
\isanewline
\endisadelimproof
\isanewline
\isacommand{lemma}\isamarkupfalse%
\ AbstrLevels{\isacharunderscore}A{\isadigit{7}}{\isacharunderscore}A{\isadigit{7}}{\isadigit{1}}{\isacharcolon}\isanewline
\ \ \isakeyword{assumes}\ {\isachardoublequoteopen}sA{\isadigit{7}}\ {\isasymin}\ AbstrLevel\ i{\isachardoublequoteclose} 
\ \ \isakeyword{shows}\ {\isachardoublequoteopen}sA{\isadigit{7}}{\isadigit{1}}\ {\isasymnotin}\ AbstrLevel\ i{\isachardoublequoteclose}%
\isadelimproof
\endisadelimproof
\isatagproof
\endisatagproof
{\isafoldproof}%
\isadelimproof
\isanewline
\endisadelimproof
\isanewline
\isacommand{lemma}\isamarkupfalse%
\ AbstrLevels{\isacharunderscore}A{\isadigit{7}}{\isacharunderscore}A{\isadigit{7}}{\isadigit{2}}{\isacharcolon}\isanewline
\ \ \isakeyword{assumes}\ {\isachardoublequoteopen}sA{\isadigit{7}}\ {\isasymin}\ AbstrLevel\ i{\isachardoublequoteclose} 
\ \ \isakeyword{shows}\ {\isachardoublequoteopen}sA{\isadigit{7}}{\isadigit{2}}\ {\isasymnotin}\ AbstrLevel\ i{\isachardoublequoteclose}%
\isadelimproof
\endisadelimproof
\isatagproof
\endisatagproof
{\isafoldproof}%
\isadelimproof
\isanewline \isanewline
\endisadelimproof
\isacommand{lemma}\isamarkupfalse%
\ AbstrLevels{\isacharunderscore}A{\isadigit{8}}{\isacharunderscore}A{\isadigit{8}}{\isadigit{1}}{\isacharcolon}\isanewline
\ \ \isakeyword{assumes}\ {\isachardoublequoteopen}sA{\isadigit{8}}\ {\isasymin}\ AbstrLevel\ i{\isachardoublequoteclose}
\ \ \isakeyword{shows}\ {\isachardoublequoteopen}sA{\isadigit{8}}{\isadigit{1}}\ {\isasymnotin}\ AbstrLevel\ i{\isachardoublequoteclose}%
\isadelimproof
\endisadelimproof
\isatagproof
\endisatagproof
{\isafoldproof}%
\isadelimproof
\isanewline \isanewline
\endisadelimproof
\isacommand{lemma}\isamarkupfalse%
\ AbstrLevels{\isacharunderscore}A{\isadigit{8}}{\isacharunderscore}A{\isadigit{8}}{\isadigit{2}}{\isacharcolon}\isanewline
\ \ \isakeyword{assumes}\ {\isachardoublequoteopen}sA{\isadigit{8}}\ {\isasymin}\ AbstrLevel\ i{\isachardoublequoteclose} 
\ \ \isakeyword{shows}\ {\isachardoublequoteopen}sA{\isadigit{8}}{\isadigit{2}}\ {\isasymnotin}\ AbstrLevel\ i{\isachardoublequoteclose}%
\isadelimproof
\endisadelimproof
\isatagproof
\endisatagproof
{\isafoldproof}%
\isadelimproof
\isanewline
\endisadelimproof
\isanewline
\isacommand{lemma}\isamarkupfalse%
\ AbstrLevels{\isacharunderscore}A{\isadigit{9}}{\isacharunderscore}A{\isadigit{9}}{\isadigit{1}}{\isacharcolon}\isanewline
\ \ \isakeyword{assumes}\ {\isachardoublequoteopen}sA{\isadigit{9}}\ {\isasymin}\ AbstrLevel\ i{\isachardoublequoteclose} 
\ \ \isakeyword{shows}\ {\isachardoublequoteopen}sA{\isadigit{9}}{\isadigit{1}}\ {\isasymnotin}\ AbstrLevel\ i{\isachardoublequoteclose}%
\isadelimproof
\endisadelimproof
\isatagproof
\endisatagproof
{\isafoldproof}%
\isadelimproof
\isanewline
\endisadelimproof
\isanewline
\isacommand{lemma}\isamarkupfalse%
\ AbstrLevels{\isacharunderscore}A{\isadigit{9}}{\isacharunderscore}A{\isadigit{9}}{\isadigit{2}}{\isacharcolon}\isanewline
\ \ \isakeyword{assumes}\ {\isachardoublequoteopen}sA{\isadigit{9}}\ {\isasymin}\ AbstrLevel\ i{\isachardoublequoteclose} 
\ \ \isakeyword{shows}\ {\isachardoublequoteopen}sA{\isadigit{9}}{\isadigit{2}}\ {\isasymnotin}\ AbstrLevel\ i{\isachardoublequoteclose}%
\isadelimproof
\endisadelimproof
\isatagproof
\endisatagproof
{\isafoldproof}%
\isadelimproof
\isanewline
\endisadelimproof
\isanewline
\isacommand{lemma}\isamarkupfalse%
\ AbstrLevels{\isacharunderscore}A{\isadigit{9}}{\isacharunderscore}A{\isadigit{9}}{\isadigit{3}}{\isacharcolon}\isanewline
\ \ \isakeyword{assumes}\ {\isachardoublequoteopen}sA{\isadigit{9}}\ {\isasymin}\ AbstrLevel\ i{\isachardoublequoteclose} 
\ \ \isakeyword{shows}\ {\isachardoublequoteopen}sA{\isadigit{9}}{\isadigit{3}}\ {\isasymnotin}\ AbstrLevel\ i{\isachardoublequoteclose}%
\isadelimproof
\endisadelimproof
\isatagproof
\endisatagproof
{\isafoldproof}%
\isadelimproof
\isanewline
\endisadelimproof
\isanewline
\isacommand{lemma}\isamarkupfalse%
\ AbstrLevels{\isacharunderscore}S{\isadigit{1}}{\isacharunderscore}A{\isadigit{1}}{\isadigit{2}}{\isacharcolon}\isanewline
\ \ \isakeyword{assumes}\ {\isachardoublequoteopen}sS{\isadigit{1}}\ {\isasymin}\ AbstrLevel\ i{\isachardoublequoteclose} 
\ \ \isakeyword{shows}\ {\isachardoublequoteopen}sA{\isadigit{1}}{\isadigit{2}}\ {\isasymnotin}\ AbstrLevel\ i{\isachardoublequoteclose}%
\isadelimproof
\endisadelimproof
\isatagproof
\endisatagproof
{\isafoldproof}%
\isadelimproof
\isanewline
\endisadelimproof
\isanewline
\isacommand{lemma}\isamarkupfalse%
\ AbstrLevels{\isacharunderscore}S{\isadigit{2}}{\isacharunderscore}A{\isadigit{1}}{\isadigit{1}}{\isacharcolon}\isanewline
\ \ \isakeyword{assumes}\ {\isachardoublequoteopen}sS{\isadigit{2}}\ {\isasymin}\ AbstrLevel\ i{\isachardoublequoteclose} 
\ \ \isakeyword{shows}\ {\isachardoublequoteopen}sA{\isadigit{1}}{\isadigit{1}}\ {\isasymnotin}\ AbstrLevel\ i{\isachardoublequoteclose}%
\isadelimproof
\endisadelimproof
\isatagproof
\endisatagproof
{\isafoldproof}%
\isadelimproof
\isanewline
\endisadelimproof
\isanewline
\isacommand{lemma}\isamarkupfalse%
\ AbstrLevels{\isacharunderscore}S{\isadigit{3}}{\isacharunderscore}A{\isadigit{2}}{\isadigit{1}}{\isacharcolon}\isanewline
\ \ \isakeyword{assumes}\ {\isachardoublequoteopen}sS{\isadigit{3}}\ {\isasymin}\ AbstrLevel\ i{\isachardoublequoteclose} 
\ \ \isakeyword{shows}\ {\isachardoublequoteopen}sA{\isadigit{2}}{\isadigit{1}}\ {\isasymnotin}\ AbstrLevel\ i{\isachardoublequoteclose}%
\isadelimproof
\endisadelimproof
\isatagproof
\endisatagproof
{\isafoldproof}%
\isadelimproof
\isanewline
\endisadelimproof
\isanewline
\isacommand{lemma}\isamarkupfalse%
\ AbstrLevels{\isacharunderscore}S{\isadigit{4}}{\isacharunderscore}A{\isadigit{2}}{\isadigit{3}}{\isacharcolon}\isanewline
\ \ \isakeyword{assumes}\ {\isachardoublequoteopen}sS{\isadigit{4}}\ {\isasymin}\ AbstrLevel\ i{\isachardoublequoteclose} 
\ \ \isakeyword{shows}\ {\isachardoublequoteopen}sA{\isadigit{2}}{\isadigit{3}}\ {\isasymnotin}\ AbstrLevel\ i{\isachardoublequoteclose}%
\isadelimproof
\endisadelimproof
\isatagproof
\endisatagproof
{\isafoldproof}%
\isadelimproof
\isanewline
\endisadelimproof
\isanewline
\isacommand{lemma}\isamarkupfalse%
\ AbstrLevels{\isacharunderscore}S{\isadigit{5}}{\isacharunderscore}A{\isadigit{3}}{\isadigit{2}}{\isacharcolon}\isanewline
\ \ \isakeyword{assumes}\ {\isachardoublequoteopen}sS{\isadigit{5}}\ {\isasymin}\ AbstrLevel\ i{\isachardoublequoteclose} 
\ \ \isakeyword{shows}\ {\isachardoublequoteopen}sA{\isadigit{3}}{\isadigit{2}}\ {\isasymnotin}\ AbstrLevel\ i{\isachardoublequoteclose}%
\isadelimproof
\endisadelimproof
\isatagproof
\endisatagproof
{\isafoldproof}%
\isadelimproof
\isanewline
\endisadelimproof
\isanewline
\isacommand{lemma}\isamarkupfalse%
\ AbstrLevels{\isacharunderscore}S{\isadigit{6}}{\isacharunderscore}A{\isadigit{2}}{\isadigit{2}}{\isacharcolon}\isanewline
\ \ \isakeyword{assumes}\ {\isachardoublequoteopen}sS{\isadigit{6}}\ {\isasymin}\ AbstrLevel\ i{\isachardoublequoteclose} 
\ \ \isakeyword{shows}\ {\isachardoublequoteopen}sA{\isadigit{2}}{\isadigit{2}}\ {\isasymnotin}\ AbstrLevel\ i{\isachardoublequoteclose}%
\isadelimproof
\endisadelimproof
\isatagproof
\endisatagproof
{\isafoldproof}%
\isadelimproof
\isanewline
\endisadelimproof
\isanewline
\isacommand{lemma}\isamarkupfalse%
\ AbstrLevels{\isacharunderscore}S{\isadigit{6}}{\isacharunderscore}A{\isadigit{3}}{\isadigit{1}}{\isacharcolon}\isanewline
\ \ \isakeyword{assumes}\ {\isachardoublequoteopen}sS{\isadigit{6}}\ {\isasymin}\ AbstrLevel\ i{\isachardoublequoteclose} 
\ \ \isakeyword{shows}\ {\isachardoublequoteopen}sA{\isadigit{3}}{\isadigit{1}}\ {\isasymnotin}\ AbstrLevel\ i{\isachardoublequoteclose}%
\isadelimproof
\endisadelimproof
\isatagproof
\endisatagproof
{\isafoldproof}%
\isadelimproof
\isanewline
\endisadelimproof
\isanewline
\isacommand{lemma}\isamarkupfalse%
\ AbstrLevels{\isacharunderscore}S{\isadigit{6}}{\isacharunderscore}A{\isadigit{4}}{\isadigit{1}}{\isacharcolon}\isanewline
\ \ \isakeyword{assumes}\ {\isachardoublequoteopen}sS{\isadigit{6}}\ {\isasymin}\ AbstrLevel\ i{\isachardoublequoteclose} 
\ \ \isakeyword{shows}\ {\isachardoublequoteopen}sA{\isadigit{4}}{\isadigit{1}}\ {\isasymnotin}\ AbstrLevel\ i{\isachardoublequoteclose}%
\isadelimproof
\endisadelimproof
\isatagproof
\endisatagproof
{\isafoldproof}%
\isadelimproof
\isanewline
\endisadelimproof
\isanewline
\isacommand{lemma}\isamarkupfalse%
\ AbstrLevels{\isacharunderscore}S{\isadigit{7}}{\isacharunderscore}A{\isadigit{4}}{\isadigit{2}}{\isacharcolon}\isanewline
\ \ \isakeyword{assumes}\ {\isachardoublequoteopen}sS{\isadigit{7}}\ {\isasymin}\ AbstrLevel\ i{\isachardoublequoteclose} 
\ \ \isakeyword{shows}\ {\isachardoublequoteopen}sA{\isadigit{4}}{\isadigit{2}}\ {\isasymnotin}\ AbstrLevel\ i{\isachardoublequoteclose}%
\isadelimproof
\endisadelimproof
\isatagproof
\endisatagproof
{\isafoldproof}%
\isadelimproof
\isanewline
\endisadelimproof
\isanewline
\isacommand{lemma}\isamarkupfalse%
\ AbstrLevels{\isacharunderscore}S{\isadigit{8}}{\isacharunderscore}A{\isadigit{5}}{\isacharcolon}\isanewline
\ \ \isakeyword{assumes}\ {\isachardoublequoteopen}sS{\isadigit{8}}\ {\isasymin}\ AbstrLevel\ i{\isachardoublequoteclose} 
\ \ \isakeyword{shows}\ {\isachardoublequoteopen}sA{\isadigit{5}}\ {\isasymnotin}\ AbstrLevel\ i{\isachardoublequoteclose}%
\isadelimproof
\endisadelimproof
\isatagproof
\endisatagproof
{\isafoldproof}%
\isadelimproof
\isanewline
\endisadelimproof
\isanewline
\isacommand{lemma}\isamarkupfalse%
\ AbstrLevels{\isacharunderscore}S{\isadigit{9}}{\isacharunderscore}A{\isadigit{6}}{\isacharcolon}\isanewline
\ \ \isakeyword{assumes}\ {\isachardoublequoteopen}sS{\isadigit{9}}\ {\isasymin}\ AbstrLevel\ i{\isachardoublequoteclose} 
\ \ \isakeyword{shows}\ {\isachardoublequoteopen}sA{\isadigit{6}}\ {\isasymnotin}\ AbstrLevel\ i{\isachardoublequoteclose}%
\isadelimproof
\endisadelimproof
\isatagproof
\endisatagproof
{\isafoldproof}%
\isadelimproof
\isanewline
\endisadelimproof
\isanewline
\isacommand{lemma}\isamarkupfalse%
\ AbstrLevels{\isacharunderscore}S{\isadigit{1}}{\isadigit{0}}{\isacharunderscore}A{\isadigit{7}}{\isadigit{1}}{\isacharcolon}\isanewline
\ \ \isakeyword{assumes}\ {\isachardoublequoteopen}sS{\isadigit{1}}{\isadigit{0}}\ {\isasymin}\ AbstrLevel\ i{\isachardoublequoteclose} 
\ \ \isakeyword{shows}\ {\isachardoublequoteopen}sA{\isadigit{7}}{\isadigit{1}}\ {\isasymnotin}\ AbstrLevel\ i{\isachardoublequoteclose}%
\isadelimproof
\endisadelimproof
\isatagproof
\endisatagproof
{\isafoldproof}%
\isadelimproof
\isanewline
\endisadelimproof
\isanewline
\isacommand{lemma}\isamarkupfalse%
\ AbstrLevels{\isacharunderscore}S{\isadigit{1}}{\isadigit{1}}{\isacharunderscore}A{\isadigit{7}}{\isadigit{2}}{\isacharcolon}\isanewline
\ \ \isakeyword{assumes}\ {\isachardoublequoteopen}sS{\isadigit{1}}{\isadigit{1}}\ {\isasymin}\ AbstrLevel\ i{\isachardoublequoteclose} 
\ \ \isakeyword{shows}\ {\isachardoublequoteopen}sA{\isadigit{7}}{\isadigit{2}}\ {\isasymnotin}\ AbstrLevel\ i{\isachardoublequoteclose}%
\isadelimproof
\endisadelimproof
\isatagproof
\endisatagproof
{\isafoldproof}%
\isadelimproof
\isanewline
\endisadelimproof
\isanewline
\isacommand{lemma}\isamarkupfalse%
\ AbstrLevels{\isacharunderscore}S{\isadigit{1}}{\isadigit{2}}{\isacharunderscore}A{\isadigit{8}}{\isadigit{1}}{\isacharcolon}\isanewline
\ \ \isakeyword{assumes}\ {\isachardoublequoteopen}sS{\isadigit{1}}{\isadigit{2}}\ {\isasymin}\ AbstrLevel\ i{\isachardoublequoteclose} 
\ \ \isakeyword{shows}\ {\isachardoublequoteopen}sA{\isadigit{8}}{\isadigit{1}}\ {\isasymnotin}\ AbstrLevel\ i{\isachardoublequoteclose}%
\isadelimproof
\endisadelimproof
\isatagproof
\endisatagproof
{\isafoldproof}%
\isadelimproof
\isanewline
\endisadelimproof
\isanewline
\isacommand{lemma}\isamarkupfalse%
\ AbstrLevels{\isacharunderscore}S{\isadigit{1}}{\isadigit{2}}{\isacharunderscore}A{\isadigit{9}}{\isadigit{1}}{\isacharcolon}\isanewline
\ \ \isakeyword{assumes}\ {\isachardoublequoteopen}sS{\isadigit{1}}{\isadigit{2}}\ {\isasymin}\ AbstrLevel\ i{\isachardoublequoteclose} 
\ \ \isakeyword{shows}\ {\isachardoublequoteopen}sA{\isadigit{9}}{\isadigit{1}}\ {\isasymnotin}\ AbstrLevel\ i{\isachardoublequoteclose}%
\isadelimproof
\endisadelimproof
\isatagproof
\endisatagproof
{\isafoldproof}%
\isadelimproof
\isanewline
\endisadelimproof
\isanewline
\isacommand{lemma}\isamarkupfalse%
\ AbstrLevels{\isacharunderscore}S{\isadigit{1}}{\isadigit{3}}{\isacharunderscore}A{\isadigit{9}}{\isadigit{2}}{\isacharcolon}\isanewline
\ \ \isakeyword{assumes}\ {\isachardoublequoteopen}sS{\isadigit{1}}{\isadigit{3}}\ {\isasymin}\ AbstrLevel\ i{\isachardoublequoteclose} 
\ \ \isakeyword{shows}\ {\isachardoublequoteopen}sA{\isadigit{9}}{\isadigit{2}}\ {\isasymnotin}\ AbstrLevel\ i{\isachardoublequoteclose}%
\isadelimproof
\endisadelimproof
\isatagproof
\endisatagproof
{\isafoldproof}%
\isadelimproof
\isanewline
\endisadelimproof
\isanewline
\isacommand{lemma}\isamarkupfalse%
\ AbstrLevels{\isacharunderscore}S{\isadigit{1}}{\isadigit{4}}{\isacharunderscore}A{\isadigit{8}}{\isadigit{2}}{\isacharcolon}\isanewline
\ \ \isakeyword{assumes}\ {\isachardoublequoteopen}sS{\isadigit{1}}{\isadigit{4}}\ {\isasymin}\ AbstrLevel\ i{\isachardoublequoteclose} 
\ \ \isakeyword{shows}\ {\isachardoublequoteopen}sA{\isadigit{8}}{\isadigit{2}}\ {\isasymnotin}\ AbstrLevel\ i{\isachardoublequoteclose}%
\isadelimproof
\endisadelimproof
\isatagproof
\endisatagproof
{\isafoldproof}%
\isadelimproof
\isanewline
\endisadelimproof
\isanewline
\isacommand{lemma}\isamarkupfalse%
\ AbstrLevels{\isacharunderscore}S{\isadigit{1}}{\isadigit{5}}{\isacharunderscore}A{\isadigit{9}}{\isadigit{3}}{\isacharcolon}\isanewline
\ \ \isakeyword{assumes}\ {\isachardoublequoteopen}sS{\isadigit{1}}{\isadigit{5}}\ {\isasymin}\ AbstrLevel\ i{\isachardoublequoteclose} 
\ \ \isakeyword{shows}\ {\isachardoublequoteopen}sA{\isadigit{9}}{\isadigit{3}}\ {\isasymnotin}\ AbstrLevel\ i{\isachardoublequoteclose}%
\isadelimproof
\endisadelimproof
\isatagproof
\endisatagproof
{\isafoldproof}%
\isadelimproof
\isanewline
\endisadelimproof
\isanewline
\isacommand{lemma}\isamarkupfalse%
\ AbstrLevels{\isacharunderscore}S{\isadigit{1}}opt{\isacharunderscore}A{\isadigit{1}}{\isadigit{1}}{\isacharcolon}\isanewline
\ \ \isakeyword{assumes}\ {\isachardoublequoteopen}sS{\isadigit{1}}opt\ {\isasymin}\ AbstrLevel\ i{\isachardoublequoteclose} 
\ \ \isakeyword{shows}\ {\isachardoublequoteopen}sA{\isadigit{1}}{\isadigit{1}}\ {\isasymnotin}\ AbstrLevel\ i{\isachardoublequoteclose}%
\isadelimproof
\endisadelimproof
\isatagproof
\endisatagproof
{\isafoldproof}%
\isadelimproof
\isanewline
\endisadelimproof
\isanewline
\isacommand{lemma}\isamarkupfalse%
\ AbstrLevels{\isacharunderscore}S{\isadigit{1}}opt{\isacharunderscore}A{\isadigit{1}}{\isadigit{2}}{\isacharcolon}\isanewline
\ \ \isakeyword{assumes}\ {\isachardoublequoteopen}sS{\isadigit{1}}opt\ {\isasymin}\ AbstrLevel\ i{\isachardoublequoteclose}
\ \ \isakeyword{shows}\ {\isachardoublequoteopen}sA{\isadigit{1}}{\isadigit{2}}\ {\isasymnotin}\ AbstrLevel\ i{\isachardoublequoteclose}%
\isadelimproof
\endisadelimproof
\isatagproof
\endisatagproof
{\isafoldproof}%
\isadelimproof
\isanewline
\endisadelimproof
\isanewline
\isacommand{lemma}\isamarkupfalse%
\ AbstrLevels{\isacharunderscore}S{\isadigit{4}}opt{\isacharunderscore}A{\isadigit{2}}{\isadigit{3}}{\isacharcolon}\isanewline
\ \ \isakeyword{assumes}\ {\isachardoublequoteopen}sS{\isadigit{4}}opt\ {\isasymin}\ AbstrLevel\ i{\isachardoublequoteclose} 
\ \ \isakeyword{shows}\ {\isachardoublequoteopen}sA{\isadigit{2}}{\isadigit{3}}\ {\isasymnotin}\ AbstrLevel\ i{\isachardoublequoteclose}%
\isadelimproof
\endisadelimproof
\isatagproof
\endisatagproof
{\isafoldproof}%
\isadelimproof
\isanewline
\endisadelimproof
\isanewline
\isacommand{lemma}\isamarkupfalse%
\ AbstrLevels{\isacharunderscore}S{\isadigit{4}}opt{\isacharunderscore}A{\isadigit{3}}{\isadigit{2}}{\isacharcolon}\isanewline
\ \ \isakeyword{assumes}\ {\isachardoublequoteopen}sS{\isadigit{4}}opt\ {\isasymin}\ AbstrLevel\ i{\isachardoublequoteclose} 
\ \ \isakeyword{shows}\ {\isachardoublequoteopen}sA{\isadigit{3}}{\isadigit{2}}\ {\isasymnotin}\ AbstrLevel\ i{\isachardoublequoteclose}%
\isadelimproof
\endisadelimproof
\isatagproof
\endisatagproof
{\isafoldproof}%
\isadelimproof
\isanewline
\endisadelimproof
\isanewline
\isacommand{lemma}\isamarkupfalse%
\ AbstrLevels{\isacharunderscore}S{\isadigit{4}}opt{\isacharunderscore}A{\isadigit{2}}{\isadigit{2}}{\isacharcolon}\isanewline
\ \ \isakeyword{assumes}\ {\isachardoublequoteopen}sS{\isadigit{4}}opt\ {\isasymin}\ AbstrLevel\ i{\isachardoublequoteclose} 
\ \ \isakeyword{shows}\ {\isachardoublequoteopen}sA{\isadigit{2}}{\isadigit{2}}\ {\isasymnotin}\ AbstrLevel\ i{\isachardoublequoteclose}%
\isadelimproof
\endisadelimproof
\isatagproof
\endisatagproof
{\isafoldproof}%
\isadelimproof
\isanewline
\endisadelimproof
\isanewline
\isacommand{lemma}\isamarkupfalse%
\ AbstrLevels{\isacharunderscore}S{\isadigit{4}}opt{\isacharunderscore}A{\isadigit{3}}{\isadigit{1}}{\isacharcolon}\isanewline
\ \ \isakeyword{assumes}\ {\isachardoublequoteopen}sS{\isadigit{4}}opt\ {\isasymin}\ AbstrLevel\ i{\isachardoublequoteclose} 
\ \ \isakeyword{shows}\ {\isachardoublequoteopen}sA{\isadigit{3}}{\isadigit{1}}\ {\isasymnotin}\ AbstrLevel\ i{\isachardoublequoteclose}%
\isadelimproof
\endisadelimproof
\isatagproof
\endisatagproof
{\isafoldproof}%
\isadelimproof
\isanewline
\endisadelimproof
\isanewline
\isacommand{lemma}\isamarkupfalse%
\ AbstrLevels{\isacharunderscore}S{\isadigit{4}}opt{\isacharunderscore}A{\isadigit{4}}{\isadigit{1}}{\isacharcolon}\isanewline
\ \ \isakeyword{assumes}\ {\isachardoublequoteopen}sS{\isadigit{4}}opt\ {\isasymin}\ AbstrLevel\ i{\isachardoublequoteclose} 
\ \ \isakeyword{shows}\ {\isachardoublequoteopen}sA{\isadigit{4}}{\isadigit{1}}\ {\isasymnotin}\ AbstrLevel\ i{\isachardoublequoteclose}%
\isadelimproof
\endisadelimproof
\isatagproof
\endisatagproof
{\isafoldproof}%
\isadelimproof
\isanewline
\endisadelimproof
\isanewline
\isacommand{lemma}\isamarkupfalse%
\ AbstrLevels{\isacharunderscore}S{\isadigit{7}}opt{\isacharunderscore}A{\isadigit{4}}{\isadigit{2}}{\isacharcolon}\isanewline
\ \ \isakeyword{assumes}\ {\isachardoublequoteopen}sS{\isadigit{7}}opt\ {\isasymin}\ AbstrLevel\ i{\isachardoublequoteclose} 
\ \ \isakeyword{shows}\ {\isachardoublequoteopen}sA{\isadigit{4}}{\isadigit{2}}\ {\isasymnotin}\ AbstrLevel\ i{\isachardoublequoteclose}%
\isadelimproof
\endisadelimproof
\isatagproof
\endisatagproof
{\isafoldproof}%
\isadelimproof
\isanewline
\endisadelimproof
\isanewline
\isacommand{lemma}\isamarkupfalse%
\ AbstrLevels{\isacharunderscore}S{\isadigit{7}}opt{\isacharunderscore}A{\isadigit{5}}{\isacharcolon}\isanewline
\ \ \isakeyword{assumes}\ {\isachardoublequoteopen}sS{\isadigit{7}}opt\ {\isasymin}\ AbstrLevel\ i{\isachardoublequoteclose} 
\ \ \isakeyword{shows}\ {\isachardoublequoteopen}sA{\isadigit{5}}\ {\isasymnotin}\ AbstrLevel\ i{\isachardoublequoteclose}%
\isadelimproof
\endisadelimproof
\isatagproof
\endisatagproof
{\isafoldproof}%
\isadelimproof
\isanewline
\endisadelimproof
\isanewline
\isacommand{lemma}\isamarkupfalse%
\ AbstrLevels{\isacharunderscore}S{\isadigit{1}}{\isadigit{1}}opt{\isacharunderscore}A{\isadigit{7}}{\isadigit{2}}{\isacharcolon}\isanewline
\ \ \isakeyword{assumes}\ {\isachardoublequoteopen}sS{\isadigit{1}}{\isadigit{1}}opt\ {\isasymin}\ AbstrLevel\ i{\isachardoublequoteclose} 
\ \ \isakeyword{shows}\ {\isachardoublequoteopen}sA{\isadigit{7}}{\isadigit{2}}\ {\isasymnotin}\ AbstrLevel\ i{\isachardoublequoteclose}%
\isadelimproof
\endisadelimproof
\isatagproof
\endisatagproof
{\isafoldproof}%
\isadelimproof
\isanewline
\endisadelimproof
\isanewline
\isacommand{lemma}\isamarkupfalse%
\ AbstrLevels{\isacharunderscore}S{\isadigit{1}}{\isadigit{1}}opt{\isacharunderscore}A{\isadigit{8}}{\isadigit{2}}{\isacharcolon}\isanewline
\ \ \isakeyword{assumes}\ {\isachardoublequoteopen}sS{\isadigit{1}}{\isadigit{1}}opt\ {\isasymin}\ AbstrLevel\ i{\isachardoublequoteclose} 
\ \ \isakeyword{shows}\ {\isachardoublequoteopen}sA{\isadigit{8}}{\isadigit{2}}\ {\isasymnotin}\ AbstrLevel\ i{\isachardoublequoteclose}%
\isadelimproof
\endisadelimproof
\isatagproof
\endisatagproof
{\isafoldproof}%
\isadelimproof
\isanewline
\endisadelimproof
\isanewline
\isacommand{lemma}\isamarkupfalse%
\ AbstrLevels{\isacharunderscore}S{\isadigit{1}}{\isadigit{1}}opt{\isacharunderscore}A{\isadigit{9}}{\isadigit{3}}{\isacharcolon}\isanewline
\ \ \isakeyword{assumes}\ {\isachardoublequoteopen}sS{\isadigit{1}}{\isadigit{1}}opt\ {\isasymin}\ AbstrLevel\ i{\isachardoublequoteclose} 
\ \ \isakeyword{shows}\ {\isachardoublequoteopen}sA{\isadigit{9}}{\isadigit{3}}\ {\isasymnotin}\ AbstrLevel\ i{\isachardoublequoteclose}%
\isadelimproof
\endisadelimproof
\isatagproof
\endisatagproof
{\isafoldproof}%
\isadelimproof
\isanewline
\endisadelimproof
\isanewline
\isacommand{lemma}\isamarkupfalse%
\ correctCompositionDiffLevelsA{\isadigit{1}}{\isacharcolon}\ {\isachardoublequoteopen}correctCompositionDiffLevels\ sA{\isadigit{1}}{\isachardoublequoteclose}%
\isadelimproof
\endisadelimproof
\isatagproof
\endisatagproof
{\isafoldproof}%
\isadelimproof
\endisadelimproof
\isanewline
\isanewline
\isacommand{lemma}\isamarkupfalse%
\ correctCompositionDiffLevelsA{\isadigit{2}}{\isacharcolon}\ {\isachardoublequoteopen}correctCompositionDiffLevels\ sA{\isadigit{2}}{\isachardoublequoteclose}%
\isadelimproof
\endisadelimproof
\isatagproof
\endisatagproof
{\isafoldproof}%
\isadelimproof
\endisadelimproof
\isanewline
\isanewline
\isacommand{lemma}\isamarkupfalse%
\ correctCompositionDiffLevelsA{\isadigit{3}}{\isacharcolon}\ {\isachardoublequoteopen}correctCompositionDiffLevels\ sA{\isadigit{3}}{\isachardoublequoteclose}%
\isadelimproof
\endisadelimproof
\isatagproof
\endisatagproof
{\isafoldproof}%
\isadelimproof
\endisadelimproof
\isanewline
\isanewline
\isacommand{lemma}\isamarkupfalse%
\ correctCompositionDiffLevelsA{\isadigit{4}}{\isacharcolon}\ {\isachardoublequoteopen}correctCompositionDiffLevels\ sA{\isadigit{4}}{\isachardoublequoteclose}%
\isadelimproof
\endisadelimproof
\isatagproof
\isanewline
\isanewline
\isamarkupcmt{lemmas  correctCompositionDiffLevelsX and corresponding proofs%
}
\isanewline
\isamarkupcmt{are identical for all elementary components, they can be constructed automatically%
}
\endisatagproof
{\isafoldproof}%
\isadelimproof
\endisadelimproof
\ \isanewline
\isacommand{lemma}\isamarkupfalse%
\ correctCompositionDiffLevelsA{\isadigit{5}}{\isacharcolon}\ {\isachardoublequoteopen}correctCompositionDiffLevels\ sA{\isadigit{5}}{\isachardoublequoteclose}%
\isadelimproof
\endisadelimproof
\isatagproof
\endisatagproof
{\isafoldproof}%
\isadelimproof
\endisadelimproof
\isanewline
\isacommand{lemma}\isamarkupfalse%
\ correctCompositionDiffLevelsA{\isadigit{6}}{\isacharcolon}\ {\isachardoublequoteopen}correctCompositionDiffLevels\ sA{\isadigit{6}}{\isachardoublequoteclose}%
\isadelimproof
\endisadelimproof
\isatagproof
\endisatagproof
{\isafoldproof}%
\isadelimproof
\endisadelimproof
\isanewline
\isacommand{lemma}\isamarkupfalse%
\ correctCompositionDiffLevelsA{\isadigit{7}}{\isacharcolon}\ {\isachardoublequoteopen}correctCompositionDiffLevels\ sA{\isadigit{7}}{\isachardoublequoteclose}%
\isadelimproof
\endisadelimproof
\isatagproof
\endisatagproof
{\isafoldproof}%
\isadelimproof
\endisadelimproof
\isanewline
\isacommand{lemma}\isamarkupfalse%
\ correctCompositionDiffLevelsA{\isadigit{8}}{\isacharcolon}\ {\isachardoublequoteopen}correctCompositionDiffLevels\ sA{\isadigit{8}}{\isachardoublequoteclose}%
\isadelimproof
\endisadelimproof
\isatagproof
\endisatagproof
{\isafoldproof}%
\isadelimproof
\endisadelimproof
\isanewline
\isacommand{lemma}\isamarkupfalse%
\ correctCompositionDiffLevelsA{\isadigit{9}}{\isacharcolon}\ {\isachardoublequoteopen}correctCompositionDiffLevels\ sA{\isadigit{9}}{\isachardoublequoteclose}%
\isadelimproof
\endisadelimproof
\isatagproof
\endisatagproof
{\isafoldproof}%
\isadelimproof
\endisadelimproof
\isanewline
\isacommand{lemma}\isamarkupfalse%
\ correctCompositionDiffLevelsA{\isadigit{1}}{\isadigit{1}}{\isacharcolon}\ {\isachardoublequoteopen}correctCompositionDiffLevels\ sA{\isadigit{1}}{\isadigit{1}}{\isachardoublequoteclose}%
\isadelimproof
\endisadelimproof
\isatagproof
\endisatagproof
{\isafoldproof}%
\isadelimproof
\endisadelimproof
\isanewline
\isacommand{lemma}\isamarkupfalse%
\ correctCompositionDiffLevelsA{\isadigit{1}}{\isadigit{2}}{\isacharcolon}\ {\isachardoublequoteopen}correctCompositionDiffLevels\ sA{\isadigit{1}}{\isadigit{2}}{\isachardoublequoteclose}%
\isadelimproof
\endisadelimproof
\isatagproof
\endisatagproof
{\isafoldproof}%
\isadelimproof
\endisadelimproof
\isanewline
\isacommand{lemma}\isamarkupfalse%
\ correctCompositionDiffLevelsA{\isadigit{2}}{\isadigit{1}}{\isacharcolon}\ {\isachardoublequoteopen}correctCompositionDiffLevels\ sA{\isadigit{2}}{\isadigit{1}}{\isachardoublequoteclose}%
\isadelimproof
\endisadelimproof
\isatagproof
\endisatagproof
{\isafoldproof}%
\isadelimproof
\endisadelimproof
\isanewline
\isacommand{lemma}\isamarkupfalse%
\ correctCompositionDiffLevelsA{\isadigit{2}}{\isadigit{2}}{\isacharcolon}\ {\isachardoublequoteopen}correctCompositionDiffLevels\ sA{\isadigit{2}}{\isadigit{2}}{\isachardoublequoteclose}%
\isadelimproof
\endisadelimproof
\isatagproof
\endisatagproof
{\isafoldproof}%
\isadelimproof
\endisadelimproof
\isanewline
\isacommand{lemma}\isamarkupfalse%
\ correctCompositionDiffLevelsA{\isadigit{2}}{\isadigit{3}}{\isacharcolon}\ {\isachardoublequoteopen}correctCompositionDiffLevels\ sA{\isadigit{2}}{\isadigit{3}}{\isachardoublequoteclose}%
\isadelimproof
\endisadelimproof
\isatagproof
\endisatagproof
{\isafoldproof}%
\isadelimproof
\endisadelimproof
\isanewline
\isacommand{lemma}\isamarkupfalse%
\ correctCompositionDiffLevelsA{\isadigit{3}}{\isadigit{1}}{\isacharcolon}\ {\isachardoublequoteopen}correctCompositionDiffLevels\ sA{\isadigit{3}}{\isadigit{1}}{\isachardoublequoteclose}%
\isadelimproof
\endisadelimproof
\isatagproof
\endisatagproof
{\isafoldproof}%
\isadelimproof
\endisadelimproof
\isanewline
\isacommand{lemma}\isamarkupfalse%
\ correctCompositionDiffLevelsA{\isadigit{3}}{\isadigit{2}}{\isacharcolon}\ {\isachardoublequoteopen}correctCompositionDiffLevels\ sA{\isadigit{3}}{\isadigit{2}}{\isachardoublequoteclose}%
\isadelimproof
\endisadelimproof
\isatagproof
\endisatagproof
{\isafoldproof}%
\isadelimproof
\endisadelimproof
\isanewline
\isacommand{lemma}\isamarkupfalse%
\ correctCompositionDiffLevelsA{\isadigit{4}}{\isadigit{1}}{\isacharcolon}\ {\isachardoublequoteopen}correctCompositionDiffLevels\ sA{\isadigit{4}}{\isadigit{1}}{\isachardoublequoteclose}%
\isadelimproof
\endisadelimproof
\isatagproof
\endisatagproof
{\isafoldproof}%
\isadelimproof
\endisadelimproof
\isanewline
\isacommand{lemma}\isamarkupfalse%
\ correctCompositionDiffLevelsA{\isadigit{4}}{\isadigit{2}}{\isacharcolon}\ {\isachardoublequoteopen}correctCompositionDiffLevels\ sA{\isadigit{4}}{\isadigit{2}}{\isachardoublequoteclose}%
\isadelimproof
\endisadelimproof
\isatagproof
\endisatagproof
{\isafoldproof}%
\isadelimproof
\endisadelimproof
\isanewline
\isacommand{lemma}\isamarkupfalse%
\ correctCompositionDiffLevelsA{\isadigit{7}}{\isadigit{1}}{\isacharcolon}\ {\isachardoublequoteopen}correctCompositionDiffLevels\ sA{\isadigit{7}}{\isadigit{1}}{\isachardoublequoteclose}%
\isadelimproof
\endisadelimproof
\isatagproof
\endisatagproof
{\isafoldproof}%
\isadelimproof
\endisadelimproof
\isanewline
\isacommand{lemma}\isamarkupfalse%
\ correctCompositionDiffLevelsA{\isadigit{7}}{\isadigit{2}}{\isacharcolon}\ {\isachardoublequoteopen}correctCompositionDiffLevels\ sA{\isadigit{7}}{\isadigit{2}}{\isachardoublequoteclose}%
\isadelimproof
\endisadelimproof
\isatagproof
\endisatagproof
{\isafoldproof}%
\isadelimproof
\endisadelimproof
\isanewline
\isacommand{lemma}\isamarkupfalse%
\ correctCompositionDiffLevelsA{\isadigit{8}}{\isadigit{1}}{\isacharcolon}\ {\isachardoublequoteopen}correctCompositionDiffLevels\ sA{\isadigit{8}}{\isadigit{1}}{\isachardoublequoteclose}%
\isadelimproof
\endisadelimproof
\isatagproof
\endisatagproof
{\isafoldproof}%
\isadelimproof
\endisadelimproof
\isanewline
\isacommand{lemma}\isamarkupfalse%
\ correctCompositionDiffLevelsA{\isadigit{8}}{\isadigit{2}}{\isacharcolon}\ {\isachardoublequoteopen}correctCompositionDiffLevels\ sA{\isadigit{8}}{\isadigit{2}}{\isachardoublequoteclose}%
\isadelimproof
\endisadelimproof
\isatagproof
\endisatagproof
{\isafoldproof}%
\isadelimproof
\endisadelimproof
\isanewline
\isacommand{lemma}\isamarkupfalse%
\ correctCompositionDiffLevelsA{\isadigit{9}}{\isadigit{1}}{\isacharcolon}\ {\isachardoublequoteopen}correctCompositionDiffLevels\ sA{\isadigit{9}}{\isadigit{1}}{\isachardoublequoteclose}%
\isadelimproof
\endisadelimproof
\isatagproof
\endisatagproof
{\isafoldproof}%
\isadelimproof
\endisadelimproof
\isanewline
\isacommand{lemma}\isamarkupfalse%
\ correctCompositionDiffLevelsA{\isadigit{9}}{\isadigit{2}}{\isacharcolon}\ {\isachardoublequoteopen}correctCompositionDiffLevels\ sA{\isadigit{9}}{\isadigit{2}}{\isachardoublequoteclose}%
\isadelimproof
\endisadelimproof
\isatagproof
\endisatagproof
{\isafoldproof}%
\isadelimproof
\endisadelimproof
\isanewline
\isacommand{lemma}\isamarkupfalse%
\ correctCompositionDiffLevelsA{\isadigit{9}}{\isadigit{3}}{\isacharcolon}\ {\isachardoublequoteopen}correctCompositionDiffLevels\ sA{\isadigit{9}}{\isadigit{3}}{\isachardoublequoteclose}%
\isadelimproof
\endisadelimproof
\isatagproof
\endisatagproof
{\isafoldproof}%
\isadelimproof
\endisadelimproof
\isanewline
\isacommand{lemma}\isamarkupfalse%
\ correctCompositionDiffLevelsS{\isadigit{1}}{\isacharcolon}\ {\isachardoublequoteopen}correctCompositionDiffLevels\ sS{\isadigit{1}}{\isachardoublequoteclose}%
\isadelimproof
\endisadelimproof
\isatagproof
\endisatagproof
{\isafoldproof}%
\isadelimproof
\endisadelimproof
\isanewline
\isacommand{lemma}\isamarkupfalse%
\ correctCompositionDiffLevelsS{\isadigit{2}}{\isacharcolon}\ {\isachardoublequoteopen}correctCompositionDiffLevels\ sS{\isadigit{2}}{\isachardoublequoteclose}%
\isadelimproof
\endisadelimproof
\isatagproof
\endisatagproof
{\isafoldproof}%
\isadelimproof
\endisadelimproof
\isanewline
\isacommand{lemma}\isamarkupfalse%
\ correctCompositionDiffLevelsS{\isadigit{3}}{\isacharcolon}\ {\isachardoublequoteopen}correctCompositionDiffLevels\ sS{\isadigit{3}}{\isachardoublequoteclose}%
\isadelimproof
\endisadelimproof
\isatagproof
\endisatagproof
{\isafoldproof}%
\isadelimproof
\endisadelimproof
\isanewline
\isacommand{lemma}\isamarkupfalse%
\ correctCompositionDiffLevelsS{\isadigit{4}}{\isacharcolon}\ {\isachardoublequoteopen}correctCompositionDiffLevels\ sS{\isadigit{4}}{\isachardoublequoteclose}%
\isadelimproof
\endisadelimproof
\isatagproof
\endisatagproof
{\isafoldproof}%
\isadelimproof
\endisadelimproof
\isanewline
\isacommand{lemma}\isamarkupfalse%
\ correctCompositionDiffLevelsS{\isadigit{5}}{\isacharcolon}\ {\isachardoublequoteopen}correctCompositionDiffLevels\ sS{\isadigit{5}}{\isachardoublequoteclose}%
\isadelimproof
\endisadelimproof
\isatagproof
\endisatagproof
{\isafoldproof}%
\isadelimproof
\endisadelimproof
\isanewline
\isacommand{lemma}\isamarkupfalse%
\ correctCompositionDiffLevelsS{\isadigit{6}}{\isacharcolon}\ {\isachardoublequoteopen}correctCompositionDiffLevels\ sS{\isadigit{6}}{\isachardoublequoteclose}%
\isadelimproof
\endisadelimproof
\isatagproof
\endisatagproof
{\isafoldproof}%
\isadelimproof
\endisadelimproof
\isanewline
\isacommand{lemma}\isamarkupfalse%
\ correctCompositionDiffLevelsS{\isadigit{7}}{\isacharcolon}\ {\isachardoublequoteopen}correctCompositionDiffLevels\ sS{\isadigit{7}}{\isachardoublequoteclose}%
\isadelimproof
\endisadelimproof
\isatagproof
\endisatagproof
{\isafoldproof}%
\isadelimproof
\endisadelimproof
\isanewline
\isacommand{lemma}\isamarkupfalse%
\ correctCompositionDiffLevelsS{\isadigit{8}}{\isacharcolon}\ {\isachardoublequoteopen}correctCompositionDiffLevels\ sS{\isadigit{8}}{\isachardoublequoteclose}%
\isadelimproof
\endisadelimproof
\isatagproof
\endisatagproof
{\isafoldproof}%
\isadelimproof
\endisadelimproof
\isanewline
\isacommand{lemma}\isamarkupfalse%
\ correctCompositionDiffLevelsS{\isadigit{9}}{\isacharcolon}\ {\isachardoublequoteopen}correctCompositionDiffLevels\ sS{\isadigit{9}}{\isachardoublequoteclose}%
\isadelimproof
\endisadelimproof
\isatagproof
\endisatagproof
{\isafoldproof}%
\isadelimproof
\endisadelimproof
\isanewline
\isacommand{lemma}\isamarkupfalse%
\ correctCompositionDiffLevelsS{\isadigit{1}}{\isadigit{0}}{\isacharcolon}\ {\isachardoublequoteopen}correctCompositionDiffLevels\ sS{\isadigit{1}}{\isadigit{0}}{\isachardoublequoteclose}%
\isadelimproof
\endisadelimproof
\isatagproof
\endisatagproof
{\isafoldproof}%
\isadelimproof
\endisadelimproof
\isanewline
\isacommand{lemma}\isamarkupfalse%
\ correctCompositionDiffLevelsS{\isadigit{1}}{\isadigit{1}}{\isacharcolon}\ {\isachardoublequoteopen}correctCompositionDiffLevels\ sS{\isadigit{1}}{\isadigit{1}}{\isachardoublequoteclose}%
\isadelimproof
\endisadelimproof
\isatagproof
\endisatagproof
{\isafoldproof}%
\isadelimproof
\endisadelimproof
\isanewline
\isacommand{lemma}\isamarkupfalse%
\ correctCompositionDiffLevelsS{\isadigit{1}}{\isadigit{2}}{\isacharcolon}\ {\isachardoublequoteopen}correctCompositionDiffLevels\ sS{\isadigit{1}}{\isadigit{2}}{\isachardoublequoteclose}%
\isadelimproof
\endisadelimproof
\isatagproof
\endisatagproof
{\isafoldproof}%
\isadelimproof
\endisadelimproof
\isanewline
\isacommand{lemma}\isamarkupfalse%
\ correctCompositionDiffLevelsS{\isadigit{1}}{\isadigit{3}}{\isacharcolon}\ {\isachardoublequoteopen}correctCompositionDiffLevels\ sS{\isadigit{1}}{\isadigit{3}}{\isachardoublequoteclose}%
\isadelimproof
\endisadelimproof
\isatagproof
\endisatagproof
{\isafoldproof}%
\isadelimproof
\endisadelimproof
\isanewline
\isacommand{lemma}\isamarkupfalse%
\ correctCompositionDiffLevelsS{\isadigit{1}}{\isadigit{4}}{\isacharcolon}\ {\isachardoublequoteopen}correctCompositionDiffLevels\ sS{\isadigit{1}}{\isadigit{4}}{\isachardoublequoteclose}%
\isadelimproof
\endisadelimproof
\isatagproof
\endisatagproof
{\isafoldproof}%
\isadelimproof
\endisadelimproof
\isanewline
\isacommand{lemma}\isamarkupfalse%
\ correctCompositionDiffLevelsS{\isadigit{1}}{\isadigit{5}}{\isacharcolon}\ {\isachardoublequoteopen}correctCompositionDiffLevels\ sS{\isadigit{1}}{\isadigit{5}}{\isachardoublequoteclose}%
\isadelimproof
\endisadelimproof
\isatagproof
\endisatagproof
{\isafoldproof}%
\isadelimproof
\endisadelimproof
\isanewline
\isacommand{lemma}\isamarkupfalse%
\ correctCompositionDiffLevelsS{\isadigit{1}}opt{\isacharcolon}\ {\isachardoublequoteopen}correctCompositionDiffLevels\ sS{\isadigit{1}}opt{\isachardoublequoteclose}%
\isadelimproof
\endisadelimproof
\isatagproof
\endisatagproof
{\isafoldproof}%
\isadelimproof
\endisadelimproof
\isanewline
\isacommand{lemma}\isamarkupfalse%
\ correctCompositionDiffLevelsS{\isadigit{4}}opt{\isacharcolon}\ {\isachardoublequoteopen}correctCompositionDiffLevels\ sS{\isadigit{4}}opt{\isachardoublequoteclose}%
\isadelimproof
\endisadelimproof
\isatagproof
\endisatagproof
{\isafoldproof}%
\isadelimproof
\endisadelimproof
\isanewline
\isacommand{lemma}\isamarkupfalse%
\ correctCompositionDiffLevelsS{\isadigit{7}}opt{\isacharcolon}\ {\isachardoublequoteopen}correctCompositionDiffLevels\ sS{\isadigit{7}}opt{\isachardoublequoteclose}%
\isadelimproof
\endisadelimproof
\isatagproof
\endisatagproof
{\isafoldproof}%
\isadelimproof
\endisadelimproof
\isanewline
\isacommand{lemma}\isamarkupfalse%
\ correctCompositionDiffLevelsS{\isadigit{1}}{\isadigit{1}}opt{\isacharcolon}\ {\isachardoublequoteopen}correctCompositionDiffLevels\ sS{\isadigit{1}}{\isadigit{1}}opt{\isachardoublequoteclose}%
\isadelimproof
\endisadelimproof
\isatagproof
\endisatagproof
{\isafoldproof}%
\isadelimproof
\endisadelimproof
\isanewline
\isacommand{lemma}\isamarkupfalse%
\ correctCompositionDiffLevelsSYSTEM{\isacharunderscore}holds{\isacharcolon}\isanewline
{\isachardoublequoteopen}correctCompositionDiffLevelsSYSTEM{\isachardoublequoteclose}%
\isadelimproof
\endisadelimproof
\isatagproof
\endisatagproof
{\isafoldproof}%
\isadelimproof
\endisadelimproof
\isanewline
\isacommand{lemma}\isamarkupfalse%
\ correctCompositionVARSYSTEM{\isacharunderscore}holds{\isacharcolon}\isanewline
{\isachardoublequoteopen}correctCompositionVARSYSTEM{\isachardoublequoteclose}\isanewline
\isadelimproof
\endisadelimproof
\isatagproof
\isacommand{by}\isamarkupfalse%
\ {\isacharparenleft}simp\ add{\isacharcolon}\ correctCompositionVARSYSTEM{\isacharunderscore}def{\isacharcomma}\ clarify{\isacharcomma}\ case{\isacharunderscore}tac\ S{\isacharcomma}\ {\isacharparenleft}simp\ add{\isacharcolon}\ correctCompositionVAR{\isacharunderscore}def{\isacharparenright}{\isacharplus}{\isacharparenright}%
\endisatagproof
{\isafoldproof}%
\isadelimproof
\isanewline
\endisadelimproof
\isanewline
\isacommand{lemma}\isamarkupfalse%
\ correctDeCompositionVARSYSTEM{\isacharunderscore}holds{\isacharcolon}\isanewline
{\isachardoublequoteopen}correctDeCompositionVARSYSTEM{\isachardoublequoteclose}\isanewline
\isadelimproof
\endisadelimproof
\isatagproof
\isacommand{by}\isamarkupfalse%
\ {\isacharparenleft}simp\ add{\isacharcolon}\ correctDeCompositionVARSYSTEM{\isacharunderscore}def{\isacharcomma}\ clarify{\isacharcomma}\ case{\isacharunderscore}tac\ S{\isacharcomma}\ {\isacharparenleft}simp\ add{\isacharcolon}\ correctDeCompositionVAR{\isacharunderscore}def{\isacharparenright}{\isacharplus}{\isacharparenright}%
\endisatagproof
{\isafoldproof}%
\isadelimproof
\endisadelimproof
\isamarkupsubsection{Correct specification of the relations between channels%
}
\isamarkuptrue%
\isacommand{lemma}\isamarkupfalse%
\ OUTfromChCorrect{\isacharunderscore}data{\isadigit{1}}{\isacharcolon}\ {\isachardoublequoteopen}OUTfromChCorrect\ data{\isadigit{1}}{\isachardoublequoteclose}\isanewline
\isadelimproof
\endisadelimproof
\isatagproof
\isacommand{by}\isamarkupfalse%
\ {\isacharparenleft}simp\ add{\isacharcolon}\ OUTfromChCorrect{\isacharunderscore}def{\isacharparenright}%
\endisatagproof
{\isafoldproof}%
\isadelimproof
\isanewline
\endisadelimproof
\isanewline
\isacommand{lemma}\isamarkupfalse%
\ OUTfromChCorrect{\isacharunderscore}data{\isadigit{2}}{\isacharcolon}\ {\isachardoublequoteopen}OUTfromChCorrect\ data{\isadigit{2}}{\isachardoublequoteclose}\isanewline
\isadelimproof
\endisadelimproof
\isatagproof
\isacommand{by}\isamarkupfalse%
\ {\isacharparenleft}metis\ IN{\isachardot}simps{\isacharparenleft}{\isadigit{2}}{\isadigit{7}}{\isacharparenright}\ OUT{\isachardot}simps{\isacharparenleft}{\isadigit{2}}{\isadigit{7}}{\isacharparenright}\ OUTfromCh{\isachardot}simps{\isacharparenleft}{\isadigit{2}}{\isacharparenright}\ OUTfromChCorrect{\isacharunderscore}def\ insertI{\isadigit{1}}{\isacharparenright}%
\endisatagproof
{\isafoldproof}%
\isadelimproof
\isanewline
\endisadelimproof
\isanewline
\isacommand{lemma}\isamarkupfalse%
\ OUTfromChCorrect{\isacharunderscore}data{\isadigit{3}}{\isacharcolon}\ {\isachardoublequoteopen}OUTfromChCorrect\ data{\isadigit{3}}{\isachardoublequoteclose}\isanewline
\isadelimproof
\endisadelimproof
\isatagproof
\isacommand{by}\isamarkupfalse%
\ {\isacharparenleft}metis\ OUTfromCh{\isachardot}simps{\isacharparenleft}{\isadigit{3}}{\isacharparenright}\ OUTfromChCorrect{\isacharunderscore}def{\isacharparenright}%
\endisatagproof
{\isafoldproof}%
\isadelimproof
\isanewline
\endisadelimproof
\isanewline
\isacommand{lemma}\isamarkupfalse%
\ OUTfromChCorrect{\isacharunderscore}data{\isadigit{4}}{\isacharcolon}\ {\isachardoublequoteopen}OUTfromChCorrect\ data{\isadigit{4}}{\isachardoublequoteclose}\isanewline
\isadelimproof
\endisadelimproof
\isatagproof
\isacommand{by}\isamarkupfalse%
\ {\isacharparenleft}metis\ IN{\isachardot}simps{\isacharparenleft}{\isadigit{2}}{\isacharparenright}\ OUT{\isachardot}simps{\isacharparenleft}{\isadigit{2}}{\isacharparenright}\ OUTfromCh{\isachardot}simps{\isacharparenleft}{\isadigit{4}}{\isacharparenright}\ OUTfromChCorrect{\isacharunderscore}def\ insertI{\isadigit{1}}\ singleton{\isacharunderscore}iff{\isacharparenright}%
\endisatagproof
{\isafoldproof}%
\isadelimproof
\isanewline
\endisadelimproof
\isanewline
\isacommand{lemma}\isamarkupfalse%
\ OUTfromChCorrect{\isacharunderscore}data{\isadigit{5}}{\isacharcolon}\ {\isachardoublequoteopen}OUTfromChCorrect\ data{\isadigit{5}}{\isachardoublequoteclose}\isanewline
\isadelimproof
\endisadelimproof
\isatagproof
\isacommand{by}\isamarkupfalse%
\ \ {\isacharparenleft}simp\ add{\isacharcolon}\ OUTfromChCorrect{\isacharunderscore}def{\isacharcomma}\ metis\ IN{\isachardot}simps{\isacharparenleft}{\isadigit{1}}{\isadigit{4}}{\isacharparenright}\ OUT{\isachardot}simps{\isacharparenleft}{\isadigit{1}}{\isadigit{4}}{\isacharparenright}\ insertI{\isadigit{1}}{\isacharparenright}%
\endisatagproof
{\isafoldproof}%
\isadelimproof
\isanewline
\endisadelimproof
\isanewline
\isacommand{lemma}\isamarkupfalse%
\ OUTfromChCorrect{\isacharunderscore}data{\isadigit{6}}{\isacharcolon}\ {\isachardoublequoteopen}OUTfromChCorrect\ data{\isadigit{6}}{\isachardoublequoteclose}\isanewline
\isadelimproof
\endisadelimproof
\isatagproof
\isacommand{by}\isamarkupfalse%
\ \ {\isacharparenleft}simp\ add{\isacharcolon}\ OUTfromChCorrect{\isacharunderscore}def{\isacharcomma}\ metis\ IN{\isachardot}simps{\isacharparenleft}{\isadigit{1}}{\isadigit{5}}{\isacharparenright}\ OUT{\isachardot}simps{\isacharparenleft}{\isadigit{1}}{\isadigit{5}}{\isacharparenright}\ insertI{\isadigit{1}}{\isacharparenright}%
\endisatagproof
{\isafoldproof}%
\isadelimproof
\isanewline
\endisadelimproof
\isanewline
\isacommand{lemma}\isamarkupfalse%
\ OUTfromChCorrect{\isacharunderscore}data{\isadigit{7}}{\isacharcolon}\ {\isachardoublequoteopen}OUTfromChCorrect\ data{\isadigit{7}}{\isachardoublequoteclose}\isanewline
\isadelimproof
\endisadelimproof
\isatagproof
\isacommand{by}\isamarkupfalse%
\ {\isacharparenleft}simp\ add{\isacharcolon}\ OUTfromChCorrect{\isacharunderscore}def{\isacharcomma}\ metis\ IN{\isachardot}simps{\isacharparenleft}{\isadigit{1}}{\isadigit{6}}{\isacharparenright}\ OUT{\isachardot}simps{\isacharparenleft}{\isadigit{1}}{\isadigit{6}}{\isacharparenright}\ insertI{\isadigit{1}}{\isacharparenright}%
\endisatagproof
{\isafoldproof}%
\isadelimproof
\isanewline
\endisadelimproof
\isanewline
\isacommand{lemma}\isamarkupfalse%
\ OUTfromChCorrect{\isacharunderscore}data{\isadigit{8}}{\isacharcolon}\ {\isachardoublequoteopen}OUTfromChCorrect\ data{\isadigit{8}}{\isachardoublequoteclose}\isanewline
\isadelimproof
\endisadelimproof
\isatagproof
\isacommand{by}\isamarkupfalse%
\ {\isacharparenleft}simp\ add{\isacharcolon}\ OUTfromChCorrect{\isacharunderscore}def{\isacharcomma}\ metis\ IN{\isachardot}simps{\isacharparenleft}{\isadigit{1}}{\isadigit{8}}{\isacharparenright}\ OUT{\isachardot}simps{\isacharparenleft}{\isadigit{1}}{\isadigit{8}}{\isacharparenright}\ insertI{\isadigit{1}}{\isacharparenright}%
\endisatagproof
{\isafoldproof}%
\isadelimproof
\ \isanewline
\endisadelimproof
\isanewline
\ \isanewline
\isacommand{lemma}\isamarkupfalse%
\ OUTfromChCorrect{\isacharunderscore}data{\isadigit{9}}{\isacharcolon}\ {\isachardoublequoteopen}OUTfromChCorrect\ data{\isadigit{9}}{\isachardoublequoteclose}\isanewline
\isadelimproof
\endisadelimproof
\isatagproof
\isacommand{by}\isamarkupfalse%
\ {\isacharparenleft}simp\ add{\isacharcolon}\ OUTfromChCorrect{\isacharunderscore}def\ {\isacharcomma}\ metis\ IN{\isachardot}simps{\isacharparenleft}{\isadigit{3}}{\isadigit{3}}{\isacharparenright}\ OUT{\isachardot}simps{\isacharparenleft}{\isadigit{3}}{\isadigit{3}}{\isacharparenright}\ singleton{\isacharunderscore}iff{\isacharparenright}%
\endisatagproof
{\isafoldproof}%
\isadelimproof
\isanewline
\endisadelimproof
\isanewline
\isacommand{lemma}\isamarkupfalse%
\ OUTfromChCorrect{\isacharunderscore}data{\isadigit{1}}{\isadigit{0}}{\isacharcolon}\ {\isachardoublequoteopen}OUTfromChCorrect\ data{\isadigit{1}}{\isadigit{0}}{\isachardoublequoteclose}\isanewline
\isadelimproof
\endisadelimproof
\isatagproof
\isacommand{by}\isamarkupfalse%
\ {\isacharparenleft}simp\ add{\isacharcolon}\ OUTfromChCorrect{\isacharunderscore}def{\isacharparenright}%
\endisatagproof
{\isafoldproof}%
\isadelimproof
\isanewline
\endisadelimproof
\isanewline
\isacommand{lemma}\isamarkupfalse%
\ OUTfromChCorrect{\isacharunderscore}data{\isadigit{1}}{\isadigit{1}}{\isacharcolon}\ {\isachardoublequoteopen}OUTfromChCorrect\ data{\isadigit{1}}{\isadigit{1}}{\isachardoublequoteclose}\isanewline
\isadelimproof
\endisadelimproof
\isatagproof
\isacommand{by}\isamarkupfalse%
\ {\isacharparenleft}simp\ add{\isacharcolon}\ OUTfromChCorrect{\isacharunderscore}def{\isacharcomma}\ metis\ {\isacharparenleft}full{\isacharunderscore}types{\isacharparenright}\ IN{\isachardot}simps{\isacharparenleft}{\isadigit{2}}{\isacharparenright}\ \isanewline
OUT{\isachardot}simps{\isacharparenleft}{\isadigit{2}}{\isacharparenright}\ OUT{\isachardot}simps{\isacharparenleft}{\isadigit{3}}{\isadigit{1}}{\isacharparenright}\ Un{\isacharunderscore}empty{\isacharunderscore}right\ Un{\isacharunderscore}insert{\isacharunderscore}left\ Un{\isacharunderscore}insert{\isacharunderscore}right\ insertI{\isadigit{1}}{\isacharparenright}%
\endisatagproof
{\isafoldproof}%
\isadelimproof
\isanewline
\endisadelimproof
\isanewline
\isacommand{lemma}\isamarkupfalse%
\ OUTfromChCorrect{\isacharunderscore}data{\isadigit{1}}{\isadigit{2}}{\isacharcolon}\ {\isachardoublequoteopen}OUTfromChCorrect\ data{\isadigit{1}}{\isadigit{2}}{\isachardoublequoteclose}\isanewline
\isadelimproof
\endisadelimproof
\isatagproof
\isacommand{by}\isamarkupfalse%
\ {\isacharparenleft}simp\ add{\isacharcolon}\ OUTfromChCorrect{\isacharunderscore}def{\isacharparenright}%
\endisatagproof
{\isafoldproof}%
\isadelimproof
\isanewline
\endisadelimproof
\isanewline
\isacommand{lemma}\isamarkupfalse%
\ OUTfromChCorrect{\isacharunderscore}data{\isadigit{1}}{\isadigit{3}}{\isacharcolon}\ {\isachardoublequoteopen}OUTfromChCorrect\ data{\isadigit{1}}{\isadigit{3}}{\isachardoublequoteclose}\isanewline
\isadelimproof
\endisadelimproof
\isatagproof
\isacommand{by}\isamarkupfalse%
\ {\isacharparenleft}simp\ add{\isacharcolon}\ OUTfromChCorrect{\isacharunderscore}def{\isacharparenright}%
\endisatagproof
{\isafoldproof}%
\isadelimproof
\isanewline
\endisadelimproof
\isanewline
\isacommand{lemma}\isamarkupfalse%
\ OUTfromChCorrect{\isacharunderscore}data{\isadigit{1}}{\isadigit{4}}{\isacharcolon}\ {\isachardoublequoteopen}OUTfromChCorrect\ data{\isadigit{1}}{\isadigit{4}}{\isachardoublequoteclose}\isanewline
\isadelimproof
\endisadelimproof
\isatagproof
\isacommand{by}\isamarkupfalse%
\ {\isacharparenleft}metis\ OUTfromCh{\isachardot}simps{\isacharparenleft}{\isadigit{1}}{\isadigit{4}}{\isacharparenright}\ OUTfromChCorrect{\isacharunderscore}def{\isacharparenright}%
\endisatagproof
{\isafoldproof}%
\isadelimproof
\isanewline
\endisadelimproof
\isanewline
\isacommand{lemma}\isamarkupfalse%
\ OUTfromChCorrect{\isacharunderscore}data{\isadigit{1}}{\isadigit{5}}{\isacharcolon}\ {\isachardoublequoteopen}OUTfromChCorrect\ data{\isadigit{1}}{\isadigit{5}}{\isachardoublequoteclose}\isanewline
\isadelimproof
\endisadelimproof
\isatagproof
\isacommand{by}\isamarkupfalse%
\ {\isacharparenleft}metis\ OUTfromCh{\isachardot}simps{\isacharparenleft}{\isadigit{1}}{\isadigit{5}}{\isacharparenright}\ OUTfromChCorrect{\isacharunderscore}def{\isacharparenright}%
\endisatagproof
{\isafoldproof}%
\isadelimproof
\isanewline
\endisadelimproof
\isanewline
\isacommand{lemma}\isamarkupfalse%
\ OUTfromChCorrect{\isacharunderscore}data{\isadigit{1}}{\isadigit{6}}{\isacharcolon}\ {\isachardoublequoteopen}OUTfromChCorrect\ data{\isadigit{1}}{\isadigit{6}}{\isachardoublequoteclose}\isanewline
\isadelimproof
\endisadelimproof
\isatagproof
\isacommand{by}\isamarkupfalse%
\ {\isacharparenleft}metis\ OUTfromCh{\isachardot}simps{\isacharparenleft}{\isadigit{1}}{\isadigit{6}}{\isacharparenright}\ OUTfromChCorrect{\isacharunderscore}def{\isacharparenright}%
\endisatagproof
{\isafoldproof}%
\isadelimproof
\isanewline
\endisadelimproof
\isanewline
\isacommand{lemma}\isamarkupfalse%
\ OUTfromChCorrect{\isacharunderscore}data{\isadigit{1}}{\isadigit{7}}{\isacharcolon}\ {\isachardoublequoteopen}OUTfromChCorrect\ data{\isadigit{1}}{\isadigit{7}}{\isachardoublequoteclose}\isanewline
\isadelimproof
\endisadelimproof
\isatagproof
\isacommand{proof}\isamarkupfalse%
\ {\isacharminus}\ \isanewline
\ \ \isacommand{have}\isamarkupfalse%
\ {\isachardoublequoteopen}data{\isadigit{1}}{\isadigit{7}}\ {\isasymin}\ OUT\ sA{\isadigit{7}}{\isadigit{1}}\ {\isasymand}\ data{\isadigit{1}}{\isadigit{5}}\ {\isasymin}\ IN\ sA{\isadigit{7}}{\isadigit{1}}{\isachardoublequoteclose}\isanewline
\ \ \ \ \isacommand{by}\isamarkupfalse%
\ {\isacharparenleft}metis\ IN{\isachardot}simps{\isacharparenleft}{\isadigit{1}}{\isadigit{9}}{\isacharparenright}\ OUT{\isachardot}simps{\isacharparenleft}{\isadigit{1}}{\isadigit{9}}{\isacharparenright}\ insertI{\isadigit{1}}{\isacharparenright}\ \ \isanewline
\ \ \isacommand{thus}\isamarkupfalse%
\ {\isacharquery}thesis\ \isacommand{by}\isamarkupfalse%
\ {\isacharparenleft}metis\ IN{\isachardot}simps{\isacharparenleft}{\isadigit{1}}{\isadigit{9}}{\isacharparenright}\ OUTfromCh{\isachardot}simps{\isacharparenleft}{\isadigit{1}}{\isadigit{7}}{\isacharparenright}\ OUTfromChCorrect{\isacharunderscore}def{\isacharparenright}\ \isanewline
\isacommand{qed}\isamarkupfalse%
\endisatagproof
{\isafoldproof}%
\isadelimproof
\isanewline
\endisadelimproof
\isanewline
\isacommand{lemma}\isamarkupfalse%
\ OUTfromChCorrect{\isacharunderscore}data{\isadigit{1}}{\isadigit{8}}{\isacharcolon}\ {\isachardoublequoteopen}OUTfromChCorrect\ data{\isadigit{1}}{\isadigit{8}}{\isachardoublequoteclose}\isanewline
\isadelimproof
\endisadelimproof
\isatagproof
\isacommand{by}\isamarkupfalse%
\ {\isacharparenleft}simp\ add{\isacharcolon}\ OUTfromChCorrect{\isacharunderscore}def{\isacharcomma}\ metis\ IN{\isachardot}simps{\isacharparenleft}{\isadigit{2}}{\isadigit{0}}{\isacharparenright}\ OUT{\isachardot}simps{\isacharparenleft}{\isadigit{2}}{\isadigit{0}}{\isacharparenright}\ insertI{\isadigit{1}}{\isacharparenright}%
\endisatagproof
{\isafoldproof}%
\isadelimproof
\isanewline
\endisadelimproof
\isanewline
\isacommand{lemma}\isamarkupfalse%
\ OUTfromChCorrect{\isacharunderscore}data{\isadigit{1}}{\isadigit{9}}{\isacharcolon}\ {\isachardoublequoteopen}OUTfromChCorrect\ data{\isadigit{1}}{\isadigit{9}}{\isachardoublequoteclose}\isanewline
\isadelimproof
\endisadelimproof
\isatagproof
\isacommand{by}\isamarkupfalse%
\ {\isacharparenleft}metis\ OUTfromCh{\isachardot}simps{\isacharparenleft}{\isadigit{1}}{\isadigit{9}}{\isacharparenright}\ OUTfromChCorrect{\isacharunderscore}def{\isacharparenright}%
\endisatagproof
{\isafoldproof}%
\isadelimproof
\isanewline
\endisadelimproof
\isanewline
\isacommand{lemma}\isamarkupfalse%
\ OUTfromChCorrect{\isacharunderscore}data{\isadigit{2}}{\isadigit{0}}{\isacharcolon}\ {\isachardoublequoteopen}OUTfromChCorrect\ data{\isadigit{2}}{\isadigit{0}}{\isachardoublequoteclose}\isanewline
\isadelimproof
\endisadelimproof
\isatagproof
\isacommand{by}\isamarkupfalse%
\ \ {\isacharparenleft}simp\ add{\isacharcolon}\ OUTfromChCorrect{\isacharunderscore}def{\isacharcomma}\ metis\ IN{\isachardot}simps{\isacharparenleft}{\isadigit{2}}{\isadigit{1}}{\isacharparenright}\ OUT{\isachardot}simps{\isacharparenleft}{\isadigit{2}}{\isadigit{1}}{\isacharparenright}\ insertI{\isadigit{1}}\ insert{\isacharunderscore}subset\ subset{\isacharunderscore}insertI{\isacharparenright}%
\endisatagproof
{\isafoldproof}%
\isadelimproof
\isanewline
\endisadelimproof
\isanewline
\isacommand{lemma}\isamarkupfalse%
\ OUTfromChCorrect{\isacharunderscore}data{\isadigit{2}}{\isadigit{1}}{\isacharcolon}\ {\isachardoublequoteopen}OUTfromChCorrect\ data{\isadigit{2}}{\isadigit{1}}{\isachardoublequoteclose}\isanewline
\isadelimproof
\endisadelimproof
\isatagproof
\isacommand{by}\isamarkupfalse%
\ {\isacharparenleft}simp\ add{\isacharcolon}\ OUTfromChCorrect{\isacharunderscore}def{\isacharcomma}\ metis\ {\isacharparenleft}full{\isacharunderscore}types{\isacharparenright}\ \isanewline
IN{\isachardot}simps{\isacharparenleft}{\isadigit{2}}{\isadigit{2}}{\isacharparenright}\ OUT{\isachardot}simps{\isacharparenleft}{\isadigit{2}}{\isadigit{2}}{\isacharparenright}\ insertI{\isadigit{1}}\ insert{\isacharunderscore}subset\ subset{\isacharunderscore}insertI{\isacharparenright}%
\endisatagproof
{\isafoldproof}%
\isadelimproof
\isanewline
\endisadelimproof
\isanewline
\isacommand{lemma}\isamarkupfalse%
\ OUTfromChCorrect{\isacharunderscore}data{\isadigit{2}}{\isadigit{2}}{\isacharcolon}\ {\isachardoublequoteopen}OUTfromChCorrect\ data{\isadigit{2}}{\isadigit{2}}{\isachardoublequoteclose}\isanewline
\isadelimproof
\endisadelimproof
\isatagproof
\isacommand{by}\isamarkupfalse%
\ {\isacharparenleft}simp\ add{\isacharcolon}\ OUTfromChCorrect{\isacharunderscore}def{\isacharcomma}\ metis\ {\isacharparenleft}full{\isacharunderscore}types{\isacharparenright}\ IN{\isachardot}simps{\isacharparenleft}{\isadigit{2}}{\isadigit{3}}{\isacharparenright}\ OUT{\isachardot}simps{\isacharparenleft}{\isadigit{2}}{\isadigit{3}}{\isacharparenright}\ insertI{\isadigit{1}}{\isacharparenright}%
\endisatagproof
{\isafoldproof}%
\isadelimproof
\isanewline
\endisadelimproof
\isanewline
\isacommand{lemma}\isamarkupfalse%
\ OUTfromChCorrect{\isacharunderscore}data{\isadigit{2}}{\isadigit{3}}{\isacharcolon}\ {\isachardoublequoteopen}OUTfromChCorrect\ data{\isadigit{2}}{\isadigit{3}}{\isachardoublequoteclose}\isanewline
\isadelimproof
\endisadelimproof
\isatagproof
\isacommand{by}\isamarkupfalse%
\ {\isacharparenleft}simp\ add{\isacharcolon}\ OUTfromChCorrect{\isacharunderscore}def{\isacharcomma}\ metis\ {\isacharparenleft}full{\isacharunderscore}types{\isacharparenright}\ IN{\isachardot}simps{\isacharparenleft}{\isadigit{9}}{\isacharparenright}\ OUT{\isachardot}simps{\isacharparenleft}{\isadigit{9}}{\isacharparenright}\ insert{\isacharunderscore}subset\ subset{\isacharunderscore}insertI{\isacharparenright}%
\endisatagproof
{\isafoldproof}%
\isadelimproof
\isanewline
\endisadelimproof
\isanewline
\isacommand{lemma}\isamarkupfalse%
\ OUTfromChCorrect{\isacharunderscore}data{\isadigit{2}}{\isadigit{4}}{\isacharcolon}\ {\isachardoublequoteopen}OUTfromChCorrect\ data{\isadigit{2}}{\isadigit{4}}{\isachardoublequoteclose}\isanewline
\isadelimproof
\endisadelimproof
\isatagproof
\isacommand{by}\isamarkupfalse%
\ {\isacharparenleft}simp\ add{\isacharcolon}\ OUTfromChCorrect{\isacharunderscore}def{\isacharcomma}\ metis\ IN{\isachardot}simps{\isacharparenleft}{\isadigit{9}}{\isacharparenright}\ OUT{\isachardot}simps{\isacharparenleft}{\isadigit{9}}{\isacharparenright}\ insertI{\isadigit{1}}\ insert{\isacharunderscore}subset\ subset{\isacharunderscore}insertI{\isacharparenright}%
\endisatagproof
{\isafoldproof}%
\isadelimproof
\isanewline
\endisadelimproof
\isanewline
\isacommand{lemma}\isamarkupfalse%
\ OUTfromChCorrectSYSTEM{\isacharunderscore}holds{\isacharcolon}\ {\isachardoublequoteopen}OUTfromChCorrectSYSTEM{\isachardoublequoteclose}\isanewline
\isadelimproof
\endisadelimproof
\isatagproof
\isacommand{by}\isamarkupfalse%
\ {\isacharparenleft}simp\ add{\isacharcolon}\ OUTfromChCorrectSYSTEM{\isacharunderscore}def{\isacharcomma}\ \ clarify{\isacharcomma}\ case{\isacharunderscore}tac\ x{\isacharcomma}\isanewline
simp\ add{\isacharcolon}\ OUTfromChCorrect{\isacharunderscore}data{\isadigit{1}}{\isacharcomma}\ simp\ add{\isacharcolon}\ OUTfromChCorrect{\isacharunderscore}data{\isadigit{2}}{\isacharcomma}\ \isanewline
simp\ add{\isacharcolon}\ OUTfromChCorrect{\isacharunderscore}data{\isadigit{3}}{\isacharcomma}\ simp\ add{\isacharcolon}\ OUTfromChCorrect{\isacharunderscore}data{\isadigit{4}}{\isacharcomma}\ \ \isanewline
simp\ add{\isacharcolon}\ OUTfromChCorrect{\isacharunderscore}data{\isadigit{5}}{\isacharcomma}\ simp\ add{\isacharcolon}\ OUTfromChCorrect{\isacharunderscore}data{\isadigit{6}}{\isacharcomma}\ \isanewline
simp\ add{\isacharcolon}\ OUTfromChCorrect{\isacharunderscore}data{\isadigit{7}}{\isacharcomma}\ simp\ add{\isacharcolon}\ OUTfromChCorrect{\isacharunderscore}data{\isadigit{8}}{\isacharcomma}\isanewline
simp\ add{\isacharcolon}\ OUTfromChCorrect{\isacharunderscore}data{\isadigit{9}}{\isacharcomma}\ simp\ add{\isacharcolon}\ OUTfromChCorrect{\isacharunderscore}data{\isadigit{1}}{\isadigit{0}}{\isacharcomma}\isanewline
simp\ add{\isacharcolon}\ OUTfromChCorrect{\isacharunderscore}data{\isadigit{1}}{\isadigit{1}}{\isacharcomma}\ simp\ add{\isacharcolon}\ OUTfromChCorrect{\isacharunderscore}data{\isadigit{1}}{\isadigit{2}}{\isacharcomma}\isanewline
simp\ add{\isacharcolon}\ OUTfromChCorrect{\isacharunderscore}data{\isadigit{1}}{\isadigit{3}}{\isacharcomma}\ simp\ add{\isacharcolon}\ OUTfromChCorrect{\isacharunderscore}data{\isadigit{1}}{\isadigit{4}}{\isacharcomma}\ \isanewline
simp\ add{\isacharcolon}\ OUTfromChCorrect{\isacharunderscore}data{\isadigit{1}}{\isadigit{5}}{\isacharcomma}\ simp\ add{\isacharcolon}\ OUTfromChCorrect{\isacharunderscore}data{\isadigit{1}}{\isadigit{6}}{\isacharcomma}\isanewline
simp\ add{\isacharcolon}\ OUTfromChCorrect{\isacharunderscore}data{\isadigit{1}}{\isadigit{7}}{\isacharcomma}\ simp\ add{\isacharcolon}\ OUTfromChCorrect{\isacharunderscore}data{\isadigit{1}}{\isadigit{8}}{\isacharcomma}\isanewline
simp\ add{\isacharcolon}\ OUTfromChCorrect{\isacharunderscore}data{\isadigit{1}}{\isadigit{9}}{\isacharcomma}\ simp\ add{\isacharcolon}\ OUTfromChCorrect{\isacharunderscore}data{\isadigit{2}}{\isadigit{0}}{\isacharcomma}\isanewline
simp\ add{\isacharcolon}\ OUTfromChCorrect{\isacharunderscore}data{\isadigit{2}}{\isadigit{1}}{\isacharcomma}\ simp\ add{\isacharcolon}\ OUTfromChCorrect{\isacharunderscore}data{\isadigit{2}}{\isadigit{2}}{\isacharcomma}\ \isanewline
simp\ add{\isacharcolon}\ OUTfromChCorrect{\isacharunderscore}data{\isadigit{2}}{\isadigit{3}}{\isacharcomma}\ simp\ add{\isacharcolon}\ OUTfromChCorrect{\isacharunderscore}data{\isadigit{2}}{\isadigit{4}}{\isacharparenright}%
\endisatagproof
{\isafoldproof}%
\isadelimproof
\isanewline
\endisadelimproof
\isanewline
\isacommand{lemma}\isamarkupfalse%
\ OUTfromVCorrect{\isadigit{1}}{\isacharunderscore}data{\isadigit{1}}{\isacharcolon}\ {\isachardoublequoteopen}OUTfromVCorrect{\isadigit{1}}\ data{\isadigit{1}}{\isachardoublequoteclose}\isanewline
\isadelimproof
\endisadelimproof
\isatagproof
\isacommand{by}\isamarkupfalse%
\ {\isacharparenleft}simp\ add{\isacharcolon}\ OUTfromVCorrect{\isadigit{1}}{\isacharunderscore}def{\isacharparenright}%
\endisatagproof
{\isafoldproof}%
\isadelimproof
\isanewline
\endisadelimproof
\isanewline
\isacommand{lemma}\isamarkupfalse%
\ OUTfromVCorrect{\isadigit{1}}{\isacharunderscore}data{\isadigit{2}}{\isacharcolon}\ {\isachardoublequoteopen}OUTfromVCorrect{\isadigit{1}}\ data{\isadigit{2}}{\isachardoublequoteclose}\isanewline
\isadelimproof
\endisadelimproof
\isatagproof
\isacommand{by}\isamarkupfalse%
\ {\isacharparenleft}simp\ add{\isacharcolon}\ OUTfromVCorrect{\isadigit{1}}{\isacharunderscore}def{\isacharparenright}%
\endisatagproof
{\isafoldproof}%
\isadelimproof
\isanewline
\endisadelimproof
\isanewline
\isacommand{lemma}\isamarkupfalse%
\ OUTfromVCorrect{\isadigit{1}}{\isacharunderscore}data{\isadigit{3}}{\isacharcolon}\ {\isachardoublequoteopen}OUTfromVCorrect{\isadigit{1}}\ data{\isadigit{3}}{\isachardoublequoteclose}\isanewline
\isadelimproof
\endisadelimproof
\isatagproof
\isacommand{proof}\isamarkupfalse%
\ {\isacharminus}\ \isanewline
\ \ \isacommand{have}\isamarkupfalse%
\ {\isachardoublequoteopen}data{\isadigit{3}}\ {\isasymin}\ OUT\ sA{\isadigit{4}}{\isadigit{1}}\ {\isasymand}\ stA{\isadigit{4}}\ {\isasymin}\ VAR\ sA{\isadigit{4}}{\isadigit{1}}{\isachardoublequoteclose}\isanewline
\ \ \ \ \isacommand{by}\isamarkupfalse%
\ {\isacharparenleft}metis\ OUT{\isachardot}simps{\isacharparenleft}{\isadigit{1}}{\isadigit{7}}{\isacharparenright}\ VAR{\isachardot}simps{\isacharparenleft}{\isadigit{1}}{\isadigit{7}}{\isacharparenright}\ insertI{\isadigit{1}}{\isacharparenright}\ \isanewline
\ \ \isacommand{thus}\isamarkupfalse%
\ {\isacharquery}thesis\ \isacommand{by}\isamarkupfalse%
\ {\isacharparenleft}metis\ OUTfromV{\isachardot}simps{\isacharparenleft}{\isadigit{3}}{\isacharparenright}\ OUTfromVCorrect{\isadigit{1}}{\isacharunderscore}def\ VAR{\isachardot}simps{\isacharparenleft}{\isadigit{1}}{\isadigit{7}}{\isacharparenright}{\isacharparenright}\ \isanewline
\isacommand{qed}\isamarkupfalse%
\endisatagproof
{\isafoldproof}%
\isadelimproof
\isanewline
\endisadelimproof
\isanewline
\isacommand{lemma}\isamarkupfalse%
\ OUTfromVCorrect{\isadigit{1}}{\isacharunderscore}data{\isadigit{4}}{\isacharcolon}\ {\isachardoublequoteopen}OUTfromVCorrect{\isadigit{1}}\ data{\isadigit{4}}{\isachardoublequoteclose}\isanewline
\isadelimproof
\endisadelimproof
\isatagproof
\isacommand{by}\isamarkupfalse%
\ {\isacharparenleft}simp\ add{\isacharcolon}\ OUTfromVCorrect{\isadigit{1}}{\isacharunderscore}def{\isacharcomma}\ metis\ {\isacharparenleft}full{\isacharunderscore}types{\isacharparenright}\ OUT{\isachardot}simps{\isacharparenleft}{\isadigit{2}}{\isacharparenright}\ VAR{\isachardot}simps{\isacharparenleft}{\isadigit{2}}{\isacharparenright}\ insertI{\isadigit{1}}{\isacharparenright}%
\endisatagproof
{\isafoldproof}%
\isadelimproof
\ \isanewline
\endisadelimproof
\isanewline
\isacommand{lemma}\isamarkupfalse%
\ OUTfromVCorrect{\isadigit{1}}{\isacharunderscore}data{\isadigit{5}}{\isacharcolon}\ {\isachardoublequoteopen}OUTfromVCorrect{\isadigit{1}}\ data{\isadigit{5}}{\isachardoublequoteclose}\isanewline
\isadelimproof
\endisadelimproof
\isatagproof
\isacommand{by}\isamarkupfalse%
\ {\isacharparenleft}simp\ add{\isacharcolon}\ OUTfromVCorrect{\isadigit{1}}{\isacharunderscore}def{\isacharparenright}%
\endisatagproof
{\isafoldproof}%
\isadelimproof
\isanewline
\endisadelimproof
\isanewline
\isacommand{lemma}\isamarkupfalse%
\ OUTfromVCorrect{\isadigit{1}}{\isacharunderscore}data{\isadigit{6}}{\isacharcolon}\ {\isachardoublequoteopen}OUTfromVCorrect{\isadigit{1}}\ data{\isadigit{6}}{\isachardoublequoteclose}\isanewline
\isadelimproof
\endisadelimproof
\isatagproof
\isacommand{by}\isamarkupfalse%
\ {\isacharparenleft}simp\ add{\isacharcolon}\ OUTfromVCorrect{\isadigit{1}}{\isacharunderscore}def{\isacharparenright}%
\endisatagproof
{\isafoldproof}%
\isadelimproof
\isanewline
\endisadelimproof
\isanewline
\isacommand{lemma}\isamarkupfalse%
\ OUTfromVCorrect{\isadigit{1}}{\isacharunderscore}data{\isadigit{7}}{\isacharcolon}\ {\isachardoublequoteopen}OUTfromVCorrect{\isadigit{1}}\ data{\isadigit{7}}{\isachardoublequoteclose}\isanewline
\isadelimproof
\endisadelimproof
\isatagproof
\isacommand{by}\isamarkupfalse%
\ {\isacharparenleft}simp\ add{\isacharcolon}\ OUTfromVCorrect{\isadigit{1}}{\isacharunderscore}def{\isacharparenright}%
\endisatagproof
{\isafoldproof}%
\isadelimproof
\isanewline
\endisadelimproof
\isanewline
\isacommand{lemma}\isamarkupfalse%
\ OUTfromVCorrect{\isadigit{1}}{\isacharunderscore}data{\isadigit{8}}{\isacharcolon}\ {\isachardoublequoteopen}OUTfromVCorrect{\isadigit{1}}\ data{\isadigit{8}}{\isachardoublequoteclose}\isanewline
\isadelimproof
\endisadelimproof
\isatagproof
\isacommand{by}\isamarkupfalse%
\ {\isacharparenleft}simp\ add{\isacharcolon}\ OUTfromVCorrect{\isadigit{1}}{\isacharunderscore}def{\isacharparenright}%
\endisatagproof
{\isafoldproof}%
\isadelimproof
\isanewline
\endisadelimproof
\isanewline
\isacommand{lemma}\isamarkupfalse%
\ OUTfromVCorrect{\isadigit{1}}{\isacharunderscore}data{\isadigit{9}}{\isacharcolon}\ {\isachardoublequoteopen}OUTfromVCorrect{\isadigit{1}}\ data{\isadigit{9}}{\isachardoublequoteclose}\isanewline
\isadelimproof
\endisadelimproof
\isatagproof
\isacommand{by}\isamarkupfalse%
\ {\isacharparenleft}simp\ add{\isacharcolon}\ OUTfromVCorrect{\isadigit{1}}{\isacharunderscore}def{\isacharparenright}%
\endisatagproof
{\isafoldproof}%
\isadelimproof
\isanewline
\endisadelimproof
\isanewline
\isacommand{lemma}\isamarkupfalse%
\ OUTfromVCorrect{\isadigit{1}}{\isacharunderscore}data{\isadigit{1}}{\isadigit{0}}{\isacharcolon}\ {\isachardoublequoteopen}OUTfromVCorrect{\isadigit{1}}\ data{\isadigit{1}}{\isadigit{0}}{\isachardoublequoteclose}\isanewline
\isadelimproof
\endisadelimproof
\isatagproof
\isacommand{proof}\isamarkupfalse%
\ {\isacharminus}\isanewline
\ \ \isacommand{have}\isamarkupfalse%
\ {\isachardoublequoteopen}data{\isadigit{1}}{\isadigit{0}}\ {\isasymin}\ OUT\ sA{\isadigit{1}}{\isadigit{2}}\ {\isasymand}\ stA{\isadigit{1}}\ {\isasymin}\ VAR\ sA{\isadigit{1}}{\isadigit{2}}{\isachardoublequoteclose}\isanewline
\ \ \ \ \isacommand{by}\isamarkupfalse%
\ {\isacharparenleft}metis\ OUT{\isachardot}simps{\isacharparenleft}{\isadigit{1}}{\isadigit{1}}{\isacharparenright}\ VAR{\isachardot}simps{\isacharparenleft}{\isadigit{1}}{\isadigit{1}}{\isacharparenright}\ insertI{\isadigit{1}}{\isacharparenright}\ \isanewline
\ \ \isacommand{thus}\isamarkupfalse%
\ {\isacharquery}thesis\ \isacommand{by}\isamarkupfalse%
\ {\isacharparenleft}metis\ OUT{\isachardot}simps{\isacharparenleft}{\isadigit{2}}{\isadigit{6}}{\isacharparenright}\ OUTfromV{\isachardot}simps{\isacharparenleft}{\isadigit{1}}{\isadigit{0}}{\isacharparenright}\ OUTfromVCorrect{\isadigit{1}}{\isacharunderscore}def\ VAR{\isachardot}simps{\isacharparenleft}{\isadigit{2}}{\isadigit{6}}{\isacharparenright}\ insertI{\isadigit{1}}{\isacharparenright}\ \isanewline
\isacommand{qed}\isamarkupfalse%
\endisatagproof
{\isafoldproof}%
\isadelimproof
\ \isanewline
\endisadelimproof
\isanewline
\isacommand{lemma}\isamarkupfalse%
\ OUTfromVCorrect{\isadigit{1}}{\isacharunderscore}data{\isadigit{1}}{\isadigit{1}}{\isacharcolon}\ {\isachardoublequoteopen}OUTfromVCorrect{\isadigit{1}}\ data{\isadigit{1}}{\isadigit{1}}{\isachardoublequoteclose}\isanewline
\isadelimproof
\endisadelimproof
\isatagproof
\isacommand{by}\isamarkupfalse%
\ {\isacharparenleft}simp\ add{\isacharcolon}\ OUTfromVCorrect{\isadigit{1}}{\isacharunderscore}def{\isacharparenright}%
\endisatagproof
{\isafoldproof}%
\isadelimproof
\isanewline
\endisadelimproof
\isanewline
\isacommand{lemma}\isamarkupfalse%
\ OUTfromVCorrect{\isadigit{1}}{\isacharunderscore}data{\isadigit{1}}{\isadigit{2}}{\isacharcolon}\ {\isachardoublequoteopen}OUTfromVCorrect{\isadigit{1}}\ data{\isadigit{1}}{\isadigit{2}}{\isachardoublequoteclose}\isanewline
\isadelimproof
\endisadelimproof
\isatagproof
\isacommand{proof}\isamarkupfalse%
\ {\isacharminus}\ \isanewline
\ \ \isacommand{have}\isamarkupfalse%
\ {\isachardoublequoteopen}data{\isadigit{1}}{\isadigit{2}}\ {\isasymin}\ OUT\ sA{\isadigit{2}}{\isadigit{2}}\ {\isasymand}\ stA{\isadigit{2}}\ {\isasymin}\ VAR\ sA{\isadigit{2}}{\isadigit{2}}{\isachardoublequoteclose}\isanewline
\ \ \ \ \isacommand{by}\isamarkupfalse%
\ {\isacharparenleft}metis\ {\isacharparenleft}full{\isacharunderscore}types{\isacharparenright}\ OUT{\isachardot}simps{\isacharparenleft}{\isadigit{1}}{\isadigit{3}}{\isacharparenright}\ VAR{\isachardot}simps{\isacharparenleft}{\isadigit{1}}{\isadigit{3}}{\isacharparenright}\ insertCI{\isacharparenright}\ \isanewline
\ \ \isacommand{thus}\isamarkupfalse%
\ {\isacharquery}thesis\ \isacommand{by}\isamarkupfalse%
\ {\isacharparenleft}metis\ OUTfromV{\isachardot}simps{\isacharparenleft}{\isadigit{1}}{\isadigit{2}}{\isacharparenright}\ OUTfromVCorrect{\isadigit{1}}{\isacharunderscore}def\ VAR{\isachardot}simps{\isacharparenleft}{\isadigit{1}}{\isadigit{3}}{\isacharparenright}{\isacharparenright}\ \isanewline
\isacommand{qed}\isamarkupfalse%
\endisatagproof
{\isafoldproof}%
\isadelimproof
\isanewline
\endisadelimproof
\isanewline
\isacommand{lemma}\isamarkupfalse%
\ OUTfromVCorrect{\isadigit{1}}{\isacharunderscore}data{\isadigit{1}}{\isadigit{3}}{\isacharcolon}\ {\isachardoublequoteopen}OUTfromVCorrect{\isadigit{1}}\ data{\isadigit{1}}{\isadigit{3}}{\isachardoublequoteclose}\isanewline
\isadelimproof
\endisadelimproof
\isatagproof
\isacommand{by}\isamarkupfalse%
\ {\isacharparenleft}simp\ add{\isacharcolon}\ OUTfromVCorrect{\isadigit{1}}{\isacharunderscore}def{\isacharparenright}%
\endisatagproof
{\isafoldproof}%
\isadelimproof
\isanewline
\endisadelimproof
\isanewline
\isacommand{lemma}\isamarkupfalse%
\ OUTfromVCorrect{\isadigit{1}}{\isacharunderscore}data{\isadigit{1}}{\isadigit{4}}{\isacharcolon}\ {\isachardoublequoteopen}OUTfromVCorrect{\isadigit{1}}\ data{\isadigit{1}}{\isadigit{4}}{\isachardoublequoteclose}\isanewline
\isadelimproof
\endisadelimproof
\isatagproof
\isacommand{by}\isamarkupfalse%
\ {\isacharparenleft}simp\ add{\isacharcolon}\ OUTfromVCorrect{\isadigit{1}}{\isacharunderscore}def{\isacharparenright}%
\endisatagproof
{\isafoldproof}%
\isadelimproof
\isanewline
\endisadelimproof
\isanewline
\isacommand{lemma}\isamarkupfalse%
\ OUTfromVCorrect{\isadigit{1}}{\isacharunderscore}data{\isadigit{1}}{\isadigit{5}}{\isacharcolon}\ {\isachardoublequoteopen}OUTfromVCorrect{\isadigit{1}}\ data{\isadigit{1}}{\isadigit{5}}{\isachardoublequoteclose}\isanewline
\isadelimproof
\endisadelimproof
\isatagproof
\isacommand{proof}\isamarkupfalse%
\ {\isacharminus}\isanewline
\ \ \isacommand{have}\isamarkupfalse%
\ A{\isadigit{6}}ch{\isacharcolon}{\isachardoublequoteopen}data{\isadigit{1}}{\isadigit{5}}\ {\isasymin}\ OUT\ sA{\isadigit{6}}\ {\isasymand}\ stA{\isadigit{6}}\ {\isasymin}\ VAR\ sA{\isadigit{6}}{\isachardoublequoteclose}\isanewline
\ \ \ \ \isacommand{by}\isamarkupfalse%
\ {\isacharparenleft}metis\ OUT{\isachardot}simps{\isacharparenleft}{\isadigit{6}}{\isacharparenright}\ VAR{\isachardot}simps{\isacharparenleft}{\isadigit{6}}{\isacharparenright}\ insertI{\isadigit{1}}{\isacharparenright}\ \isanewline
\ \ \isacommand{thus}\isamarkupfalse%
\ {\isacharquery}thesis\ \isacommand{by}\isamarkupfalse%
\ {\isacharparenleft}simp\ add{\isacharcolon}\ OUTfromVCorrect{\isadigit{1}}{\isacharunderscore}def{\isacharcomma}\ metis\ A{\isadigit{6}}ch{\isacharparenright}\ \isanewline
\isacommand{qed}\isamarkupfalse%
\endisatagproof
{\isafoldproof}%
\isadelimproof
\isanewline
\endisadelimproof
\isanewline
\isacommand{lemma}\isamarkupfalse%
\ OUTfromVCorrect{\isadigit{1}}{\isacharunderscore}data{\isadigit{1}}{\isadigit{6}}{\isacharcolon}\ {\isachardoublequoteopen}OUTfromVCorrect{\isadigit{1}}\ data{\isadigit{1}}{\isadigit{6}}{\isachardoublequoteclose}\isanewline
\isadelimproof
\endisadelimproof
\isatagproof
\isacommand{proof}\isamarkupfalse%
\ {\isacharminus}\isanewline
\ \ \isacommand{have}\isamarkupfalse%
\ A{\isadigit{6}}ch{\isacharcolon}{\isachardoublequoteopen}data{\isadigit{1}}{\isadigit{6}}\ {\isasymin}\ OUT\ sA{\isadigit{6}}\ {\isasymand}\ stA{\isadigit{6}}\ {\isasymin}\ VAR\ sA{\isadigit{6}}{\isachardoublequoteclose}\isanewline
\ \ \ \ \isacommand{by}\isamarkupfalse%
\ {\isacharparenleft}metis\ {\isacharparenleft}full{\isacharunderscore}types{\isacharparenright}\ OUT{\isachardot}simps{\isacharparenleft}{\isadigit{6}}{\isacharparenright}\ VAR{\isachardot}simps{\isacharparenleft}{\isadigit{6}}{\isacharparenright}\ insertCI{\isacharparenright}\isanewline
\ \ \isacommand{thus}\isamarkupfalse%
\ {\isacharquery}thesis\ \isacommand{by}\isamarkupfalse%
\ {\isacharparenleft}simp\ add{\isacharcolon}\ OUTfromVCorrect{\isadigit{1}}{\isacharunderscore}def{\isacharcomma}\ metis\ A{\isadigit{6}}ch{\isacharparenright}\ \isanewline
\isacommand{qed}\isamarkupfalse%
\endisatagproof
{\isafoldproof}%
\isadelimproof
\isanewline
\endisadelimproof
\isanewline
\isacommand{lemma}\isamarkupfalse%
\ OUTfromVCorrect{\isadigit{1}}{\isacharunderscore}data{\isadigit{1}}{\isadigit{7}}{\isacharcolon}\ {\isachardoublequoteopen}OUTfromVCorrect{\isadigit{1}}\ data{\isadigit{1}}{\isadigit{7}}{\isachardoublequoteclose}\isanewline
\isadelimproof
\endisadelimproof
\isatagproof
\isacommand{by}\isamarkupfalse%
\ {\isacharparenleft}simp\ add{\isacharcolon}\ OUTfromVCorrect{\isadigit{1}}{\isacharunderscore}def{\isacharparenright}%
\endisatagproof
{\isafoldproof}%
\isadelimproof
\isanewline
\endisadelimproof
\isanewline
\isacommand{lemma}\isamarkupfalse%
\ OUTfromVCorrect{\isadigit{1}}{\isacharunderscore}data{\isadigit{1}}{\isadigit{8}}{\isacharcolon}\ {\isachardoublequoteopen}OUTfromVCorrect{\isadigit{1}}\ data{\isadigit{1}}{\isadigit{8}}{\isachardoublequoteclose}\isanewline
\isadelimproof
\endisadelimproof
\isatagproof
\isacommand{by}\isamarkupfalse%
\ {\isacharparenleft}simp\ add{\isacharcolon}\ OUTfromVCorrect{\isadigit{1}}{\isacharunderscore}def{\isacharparenright}%
\endisatagproof
{\isafoldproof}%
\isadelimproof
\isanewline
\endisadelimproof
\isanewline
\isacommand{lemma}\isamarkupfalse%
\ OUTfromVCorrect{\isadigit{1}}{\isacharunderscore}data{\isadigit{1}}{\isadigit{9}}{\isacharcolon}\ {\isachardoublequoteopen}OUTfromVCorrect{\isadigit{1}}\ data{\isadigit{1}}{\isadigit{9}}{\isachardoublequoteclose}\isanewline
\isadelimproof
\endisadelimproof
\isatagproof
\isacommand{by}\isamarkupfalse%
\ {\isacharparenleft}simp\ add{\isacharcolon}\ OUTfromVCorrect{\isadigit{1}}{\isacharunderscore}def{\isacharparenright}%
\endisatagproof
{\isafoldproof}%
\isadelimproof
\isanewline
\endisadelimproof
\isanewline
\isacommand{lemma}\isamarkupfalse%
\ OUTfromVCorrect{\isadigit{1}}{\isacharunderscore}data{\isadigit{2}}{\isadigit{0}}{\isacharcolon}\ {\isachardoublequoteopen}OUTfromVCorrect{\isadigit{1}}\ data{\isadigit{2}}{\isadigit{0}}{\isachardoublequoteclose}\isanewline
\isadelimproof
\endisadelimproof
\isatagproof
\isacommand{by}\isamarkupfalse%
\ {\isacharparenleft}simp\ add{\isacharcolon}\ OUTfromVCorrect{\isadigit{1}}{\isacharunderscore}def{\isacharparenright}%
\endisatagproof
{\isafoldproof}%
\isadelimproof
\isanewline
\endisadelimproof
\isanewline
\isacommand{lemma}\isamarkupfalse%
\ OUTfromVCorrect{\isadigit{1}}{\isacharunderscore}data{\isadigit{2}}{\isadigit{1}}{\isacharcolon}\ {\isachardoublequoteopen}OUTfromVCorrect{\isadigit{1}}\ data{\isadigit{2}}{\isadigit{1}}{\isachardoublequoteclose}\isanewline
\isadelimproof
\endisadelimproof
\isatagproof
\isacommand{by}\isamarkupfalse%
\ {\isacharparenleft}simp\ add{\isacharcolon}\ OUTfromVCorrect{\isadigit{1}}{\isacharunderscore}def{\isacharparenright}%
\endisatagproof
{\isafoldproof}%
\isadelimproof
\isanewline
\endisadelimproof
\isanewline
\isacommand{lemma}\isamarkupfalse%
\ OUTfromVCorrect{\isadigit{1}}{\isacharunderscore}data{\isadigit{2}}{\isadigit{2}}{\isacharcolon}\ {\isachardoublequoteopen}OUTfromVCorrect{\isadigit{1}}\ data{\isadigit{2}}{\isadigit{2}}{\isachardoublequoteclose}\isanewline
\isadelimproof
\endisadelimproof
\isatagproof
\isacommand{by}\isamarkupfalse%
\ {\isacharparenleft}simp\ add{\isacharcolon}\ OUTfromVCorrect{\isadigit{1}}{\isacharunderscore}def{\isacharparenright}%
\endisatagproof
{\isafoldproof}%
\isadelimproof
\isanewline
\endisadelimproof
\isanewline
\isacommand{lemma}\isamarkupfalse%
\ OUTfromVCorrect{\isadigit{1}}{\isacharunderscore}data{\isadigit{2}}{\isadigit{3}}{\isacharcolon}\ {\isachardoublequoteopen}OUTfromVCorrect{\isadigit{1}}\ data{\isadigit{2}}{\isadigit{3}}{\isachardoublequoteclose}\isanewline
\isadelimproof
\endisadelimproof
\isatagproof
\isacommand{by}\isamarkupfalse%
\ {\isacharparenleft}simp\ add{\isacharcolon}\ OUTfromVCorrect{\isadigit{1}}{\isacharunderscore}def{\isacharparenright}%
\endisatagproof
{\isafoldproof}%
\isadelimproof
\isanewline
\endisadelimproof
\isanewline
\isacommand{lemma}\isamarkupfalse%
\ OUTfromVCorrect{\isadigit{1}}{\isacharunderscore}data{\isadigit{2}}{\isadigit{4}}{\isacharcolon}\ {\isachardoublequoteopen}OUTfromVCorrect{\isadigit{1}}\ data{\isadigit{2}}{\isadigit{4}}{\isachardoublequoteclose}\isanewline
\isadelimproof
\endisadelimproof
\isatagproof
\isacommand{by}\isamarkupfalse%
\ {\isacharparenleft}simp\ add{\isacharcolon}\ OUTfromVCorrect{\isadigit{1}}{\isacharunderscore}def{\isacharparenright}%
\endisatagproof
{\isafoldproof}%
\isadelimproof
\isanewline
\endisadelimproof
\isanewline
\isacommand{lemma}\isamarkupfalse%
\ OUTfromVCorrect{\isadigit{1}}SYSTEM{\isacharunderscore}holds{\isacharcolon}\ {\isachardoublequoteopen}OUTfromVCorrect{\isadigit{1}}SYSTEM{\isachardoublequoteclose}\isanewline
\isadelimproof
\endisadelimproof
\isatagproof
\isacommand{by}\isamarkupfalse%
\ {\isacharparenleft}simp\ add{\isacharcolon}\ OUTfromVCorrect{\isadigit{1}}SYSTEM{\isacharunderscore}def{\isacharcomma}\ clarify{\isacharcomma}\ case{\isacharunderscore}tac\ x{\isacharcomma}\ \isanewline
simp\ add{\isacharcolon}\ OUTfromVCorrect{\isadigit{1}}{\isacharunderscore}data{\isadigit{1}}{\isacharcomma}\ simp\ add{\isacharcolon}\ OUTfromVCorrect{\isadigit{1}}{\isacharunderscore}data{\isadigit{2}}{\isacharcomma}\isanewline
simp\ add{\isacharcolon}\ OUTfromVCorrect{\isadigit{1}}{\isacharunderscore}data{\isadigit{3}}{\isacharcomma}\ simp\ add{\isacharcolon}\ OUTfromVCorrect{\isadigit{1}}{\isacharunderscore}data{\isadigit{4}}{\isacharcomma}\ \isanewline
simp\ add{\isacharcolon}\ OUTfromVCorrect{\isadigit{1}}{\isacharunderscore}data{\isadigit{5}}{\isacharcomma}\ simp\ add{\isacharcolon}\ OUTfromVCorrect{\isadigit{1}}{\isacharunderscore}data{\isadigit{6}}{\isacharcomma}\ \isanewline
simp\ add{\isacharcolon}\ OUTfromVCorrect{\isadigit{1}}{\isacharunderscore}data{\isadigit{7}}{\isacharcomma}\ simp\ add{\isacharcolon}\ OUTfromVCorrect{\isadigit{1}}{\isacharunderscore}data{\isadigit{8}}{\isacharcomma}\ \isanewline
simp\ add{\isacharcolon}\ OUTfromVCorrect{\isadigit{1}}{\isacharunderscore}data{\isadigit{9}}{\isacharcomma}\ simp\ add{\isacharcolon}\ OUTfromVCorrect{\isadigit{1}}{\isacharunderscore}data{\isadigit{1}}{\isadigit{0}}{\isacharcomma}\ \isanewline
simp\ add{\isacharcolon}\ OUTfromVCorrect{\isadigit{1}}{\isacharunderscore}data{\isadigit{1}}{\isadigit{1}}{\isacharcomma}\ simp\ add{\isacharcolon}\ OUTfromVCorrect{\isadigit{1}}{\isacharunderscore}data{\isadigit{1}}{\isadigit{2}}{\isacharcomma}\ \isanewline
simp\ add{\isacharcolon}\ OUTfromVCorrect{\isadigit{1}}{\isacharunderscore}data{\isadigit{1}}{\isadigit{3}}{\isacharcomma}\ simp\ add{\isacharcolon}\ OUTfromVCorrect{\isadigit{1}}{\isacharunderscore}data{\isadigit{1}}{\isadigit{4}}{\isacharcomma}\ \isanewline
simp\ add{\isacharcolon}\ OUTfromVCorrect{\isadigit{1}}{\isacharunderscore}data{\isadigit{1}}{\isadigit{5}}{\isacharcomma}\ simp\ add{\isacharcolon}\ OUTfromVCorrect{\isadigit{1}}{\isacharunderscore}data{\isadigit{1}}{\isadigit{6}}{\isacharcomma}\ \isanewline
simp\ add{\isacharcolon}\ OUTfromVCorrect{\isadigit{1}}{\isacharunderscore}data{\isadigit{1}}{\isadigit{7}}{\isacharcomma}\ simp\ add{\isacharcolon}\ OUTfromVCorrect{\isadigit{1}}{\isacharunderscore}data{\isadigit{1}}{\isadigit{8}}{\isacharcomma}\isanewline
simp\ add{\isacharcolon}\ OUTfromVCorrect{\isadigit{1}}{\isacharunderscore}data{\isadigit{1}}{\isadigit{9}}{\isacharcomma}\ simp\ add{\isacharcolon}\ OUTfromVCorrect{\isadigit{1}}{\isacharunderscore}data{\isadigit{2}}{\isadigit{0}}{\isacharcomma}\isanewline
simp\ add{\isacharcolon}\ OUTfromVCorrect{\isadigit{1}}{\isacharunderscore}data{\isadigit{2}}{\isadigit{1}}{\isacharcomma}\ simp\ add{\isacharcolon}\ OUTfromVCorrect{\isadigit{1}}{\isacharunderscore}data{\isadigit{2}}{\isadigit{2}}{\isacharcomma}\isanewline
simp\ add{\isacharcolon}\ OUTfromVCorrect{\isadigit{1}}{\isacharunderscore}data{\isadigit{2}}{\isadigit{3}}{\isacharcomma}\ simp\ add{\isacharcolon}\ OUTfromVCorrect{\isadigit{1}}{\isacharunderscore}data{\isadigit{2}}{\isadigit{4}}{\isacharparenright}%
\endisatagproof
{\isafoldproof}%
\isadelimproof
\isanewline
\endisadelimproof
\isanewline
\isacommand{lemma}\isamarkupfalse%
\ OUTfromVCorrect{\isadigit{2}}SYSTEM{\isacharcolon}\ {\isachardoublequoteopen}OUTfromVCorrect{\isadigit{2}}SYSTEM{\isachardoublequoteclose}\isanewline
\isadelimproof
\endisadelimproof
\isatagproof
\isacommand{by}\isamarkupfalse%
\ {\isacharparenleft}simp\ add{\isacharcolon}\ OUTfromVCorrect{\isadigit{2}}SYSTEM{\isacharunderscore}def{\isacharcomma}\ auto{\isacharcomma}\ case{\isacharunderscore}tac\ x{\isacharcomma}\isanewline
\ \ \ \ \ \ {\isacharparenleft}{\isacharparenleft}simp\ add{\isacharcolon}\ OUTfromVCorrect{\isadigit{2}}{\isacharunderscore}def{\isacharcomma}\ auto{\isacharcomma}\ case{\isacharunderscore}tac\ v{\isacharcomma}\ auto{\isacharparenright}\ {\isacharbar}\ \isanewline
\ \ \ \ \ \ \ {\isacharparenleft}simp\ add{\isacharcolon}\ OUTfromVCorrect{\isadigit{2}}{\isacharunderscore}def{\isacharparenright}\ {\isacharparenright}{\isacharplus}{\isacharparenright}%
\endisatagproof
{\isafoldproof}%
\isadelimproof
\isanewline
\endisadelimproof
\isanewline
\isacommand{lemma}\isamarkupfalse%
\ OUTfromV{\isacharunderscore}VARto{\isacharunderscore}holds{\isacharcolon}\isanewline
{\isachardoublequoteopen}OUTfromV{\isacharunderscore}VARto{\isachardoublequoteclose}\isanewline
\isadelimproof
\endisadelimproof
\isatagproof
\isacommand{by}\isamarkupfalse%
\ {\isacharparenleft}simp\ add{\isacharcolon}\ OUTfromV{\isacharunderscore}VARto{\isacharunderscore}def{\isacharcomma}\ auto{\isacharcomma}\ {\isacharparenleft}case{\isacharunderscore}tac\ x{\isacharcomma}\ auto{\isacharparenright}{\isacharcomma}\ {\isacharparenleft}case{\isacharunderscore}tac\ v{\isacharcomma}\ auto{\isacharparenright}{\isacharparenright}%
\endisatagproof
{\isafoldproof}%
\isadelimproof
\isanewline
\endisadelimproof
\isanewline
\isacommand{lemma}\isamarkupfalse%
\ VARfromCorrectSYSTEM{\isacharunderscore}holds{\isacharcolon}\isanewline
{\isachardoublequoteopen}VARfromCorrectSYSTEM{\isachardoublequoteclose}\isanewline
\isadelimproof
\endisadelimproof
\isatagproof
\isacommand{by}\isamarkupfalse%
\ {\isacharparenleft}simp\ add{\isacharcolon}\ VARfromCorrectSYSTEM{\isacharunderscore}def\ AbstrLevel{\isadigit{0}}\ AbstrLevel{\isadigit{1}}{\isacharparenright}%
\endisatagproof
{\isafoldproof}%
\isadelimproof
\isanewline
\endisadelimproof
\isanewline
\isacommand{lemma}\isamarkupfalse%
\ VARtoCorrectSYSTEM{\isacharunderscore}holds{\isacharcolon}\isanewline
{\isachardoublequoteopen}VARtoCorrectSYSTEM{\isachardoublequoteclose}\isanewline
\isadelimproof
\endisadelimproof
\isatagproof
\isacommand{by}\isamarkupfalse%
\ {\isacharparenleft}simp\ add{\isacharcolon}\ VARtoCorrectSYSTEM{\isacharunderscore}def\ AbstrLevel{\isadigit{0}}\ AbstrLevel{\isadigit{1}}{\isacharparenright}%
\endisatagproof
{\isafoldproof}%
\isadelimproof
\isanewline
\endisadelimproof
\isanewline
\isacommand{lemma}\isamarkupfalse%
\ VARusefulSYSTEM{\isacharunderscore}holds{\isacharcolon}\isanewline
{\isachardoublequoteopen}VARusefulSYSTEM{\isachardoublequoteclose}\isanewline
\isadelimproof
\endisadelimproof
\isatagproof
\isacommand{by}\isamarkupfalse%
\ {\isacharparenleft}simp\ add{\isacharcolon}\ VARusefulSYSTEM{\isacharunderscore}def{\isacharcomma}\ auto{\isacharcomma}\ case{\isacharunderscore}tac\ v{\isacharcomma}\ auto{\isacharparenright}%
\endisatagproof
{\isafoldproof}%
\isadelimproof
\endisadelimproof
\isamarkupsubsection{Elementary components%
}
\isamarkuptrue%
\isamarkupcmt{On the abstraction level 0 only the components sA5 and sA6 are elementary%
}
\isanewline
\isanewline
\isacommand{lemma}\isamarkupfalse%
\ NOT{\isacharunderscore}elementaryCompDD{\isacharunderscore}sA{\isadigit{1}}{\isacharcolon}\ \ {\isachardoublequoteopen}{\isasymnot}\ elementaryCompDD\ sA{\isadigit{1}}{\isachardoublequoteclose}\ \isanewline
\isadelimproof
\endisadelimproof
\isatagproof
\isacommand{proof}\isamarkupfalse%
\ {\isacharminus}\isanewline
\ \ \isacommand{have}\isamarkupfalse%
\ {\isachardoublequoteopen}outSetCorelated\ data{\isadigit{2}}\ {\isasyminter}\ outSetCorelated\ data{\isadigit{1}}{\isadigit{0}}\ {\isacharequal}\ {\isacharbraceleft}{\isacharbraceright}{\isachardoublequoteclose}\isanewline
\ \ \isacommand{by}\isamarkupfalse%
\ {\isacharparenleft}metis\ OUTfromV{\isachardot}simps{\isacharparenleft}{\isadigit{2}}{\isacharparenright}\ inf{\isacharunderscore}bot{\isacharunderscore}left\ outSetCorelatedEmpty{\isadigit{1}}{\isacharparenright}\ \isanewline
\ \ \isacommand{thus}\isamarkupfalse%
\ {\isacharquery}thesis\ \isacommand{by}\isamarkupfalse%
\ {\isacharparenleft}simp\ add{\isacharcolon}\ elementaryCompDD{\isacharunderscore}def{\isacharparenright}\isanewline
\isacommand{qed}\isamarkupfalse%
\endisatagproof
{\isafoldproof}%
\isadelimproof
\isanewline
\endisadelimproof
\isanewline
\isacommand{lemma}\isamarkupfalse%
\ NOT{\isacharunderscore}elementaryCompDD{\isacharunderscore}sA{\isadigit{2}}{\isacharcolon}\ {\isachardoublequoteopen}{\isasymnot}\ elementaryCompDD\ sA{\isadigit{2}}{\isachardoublequoteclose}\ \isanewline
\isadelimproof
\endisadelimproof
\isatagproof
\isacommand{proof}\isamarkupfalse%
\ {\isacharminus}\isanewline
\ \ \isacommand{have}\isamarkupfalse%
\ {\isachardoublequoteopen}outSetCorelated\ data{\isadigit{5}}\ {\isasyminter}\ outSetCorelated\ data{\isadigit{1}}{\isadigit{1}}\ {\isacharequal}\ {\isacharbraceleft}{\isacharbraceright}{\isachardoublequoteclose}\isanewline
\ \ \isacommand{by}\isamarkupfalse%
\ {\isacharparenleft}metis\ OUTfromV{\isachardot}simps{\isacharparenleft}{\isadigit{5}}{\isacharparenright}\ inf{\isacharunderscore}bot{\isacharunderscore}right\ inf{\isacharunderscore}commute\ outSetCorelatedEmpty{\isadigit{1}}{\isacharparenright}\isanewline
\ \ \isacommand{thus}\isamarkupfalse%
\ {\isacharquery}thesis\ \isacommand{by}\isamarkupfalse%
\ {\isacharparenleft}simp\ add{\isacharcolon}\ elementaryCompDD{\isacharunderscore}def{\isacharparenright}\isanewline
\isacommand{qed}\isamarkupfalse%
\endisatagproof
{\isafoldproof}%
\isadelimproof
\ \isanewline
\endisadelimproof
\isanewline
\isacommand{lemma}\isamarkupfalse%
\ NOT{\isacharunderscore}elementaryCompDD{\isacharunderscore}sA{\isadigit{3}}{\isacharcolon}\ \ {\isachardoublequoteopen}{\isasymnot}\ elementaryCompDD\ sA{\isadigit{3}}{\isachardoublequoteclose}\ \isanewline
\isadelimproof
\endisadelimproof
\isatagproof
\isacommand{proof}\isamarkupfalse%
\ {\isacharminus}\isanewline
\ \ \isacommand{have}\isamarkupfalse%
\ {\isachardoublequoteopen}outSetCorelated\ data{\isadigit{6}}\ {\isasyminter}\ outSetCorelated\ data{\isadigit{7}}\ {\isacharequal}\ {\isacharbraceleft}{\isacharbraceright}{\isachardoublequoteclose}\isanewline
\ \ \isacommand{by}\isamarkupfalse%
\ {\isacharparenleft}metis\ OUTfromV{\isachardot}simps{\isacharparenleft}{\isadigit{7}}{\isacharparenright}\ inf{\isacharunderscore}bot{\isacharunderscore}right\ outSetCorelatedEmpty{\isadigit{1}}{\isacharparenright}\ \isanewline
\ \ \isacommand{thus}\isamarkupfalse%
\ {\isacharquery}thesis\ \isacommand{by}\isamarkupfalse%
\ {\isacharparenleft}simp\ add{\isacharcolon}\ elementaryCompDD{\isacharunderscore}def{\isacharparenright}\isanewline
\isacommand{qed}\isamarkupfalse%
\endisatagproof
{\isafoldproof}%
\isadelimproof
\isanewline
\endisadelimproof
\isanewline
\isacommand{lemma}\isamarkupfalse%
\ NOT{\isacharunderscore}elementaryCompDD{\isacharunderscore}sA{\isadigit{4}}{\isacharcolon}\ \ {\isachardoublequoteopen}{\isasymnot}\ elementaryCompDD\ sA{\isadigit{4}}{\isachardoublequoteclose}\ \isanewline
\isadelimproof
\endisadelimproof
\isatagproof
\isacommand{proof}\isamarkupfalse%
\ {\isacharminus}\isanewline
\ \ \isacommand{have}\isamarkupfalse%
\ {\isachardoublequoteopen}outSetCorelated\ data{\isadigit{3}}\ {\isasyminter}\ outSetCorelated\ data{\isadigit{8}}\ {\isacharequal}\ {\isacharbraceleft}{\isacharbraceright}{\isachardoublequoteclose}\isanewline
\ \ \isacommand{by}\isamarkupfalse%
\ {\isacharparenleft}metis\ OUTfromV{\isachardot}simps{\isacharparenleft}{\isadigit{8}}{\isacharparenright}\ inf{\isacharunderscore}bot{\isacharunderscore}left\ inf{\isacharunderscore}commute\ outSetCorelatedEmpty{\isadigit{1}}{\isacharparenright}\isanewline
\ \ \isacommand{thus}\isamarkupfalse%
\ {\isacharquery}thesis\ \isacommand{by}\isamarkupfalse%
\ {\isacharparenleft}simp\ add{\isacharcolon}\ elementaryCompDD{\isacharunderscore}def{\isacharparenright}\ \ \isanewline
\isacommand{qed}\isamarkupfalse%
\endisatagproof
{\isafoldproof}%
\isadelimproof
\isanewline
\endisadelimproof
\isanewline
\isacommand{lemma}\isamarkupfalse%
\ elementaryCompDD{\isacharunderscore}sA{\isadigit{5}}{\isacharcolon}\ \ {\isachardoublequoteopen}elementaryCompDD\ sA{\isadigit{5}}{\isachardoublequoteclose}\ \isanewline
\isadelimproof
\endisadelimproof
\isatagproof
\isacommand{by}\isamarkupfalse%
\ \ {\isacharparenleft}simp\ add{\isacharcolon}\ elementaryCompDD{\isacharunderscore}def{\isacharparenright}%
\endisatagproof
{\isafoldproof}%
\isadelimproof
\isanewline
\endisadelimproof
\isanewline
\isacommand{lemma}\isamarkupfalse%
\ elementaryCompDD{\isacharunderscore}sA{\isadigit{6}}{\isacharcolon}\ \ {\isachardoublequoteopen}elementaryCompDD\ sA{\isadigit{6}}{\isachardoublequoteclose}\ \isanewline
\isadelimproof
\endisadelimproof
\isatagproof
\isacommand{proof}\isamarkupfalse%
\ {\isacharminus}\isanewline
\ \ \isacommand{have}\isamarkupfalse%
\ oSet{\isadigit{1}}{\isadigit{5}}{\isacharcolon}{\isachardoublequoteopen}outSetCorelated\ data{\isadigit{1}}{\isadigit{5}}\ {\isasymnoteq}\ {\isacharbraceleft}{\isacharbraceright}{\isachardoublequoteclose}\ \isanewline
\ \ \ \ \isacommand{by}\isamarkupfalse%
\ {\isacharparenleft}simp\ add{\isacharcolon}\ outSetCorelated{\isacharunderscore}def{\isacharcomma}\ auto{\isacharparenright}\isanewline
\ \ \isacommand{have}\isamarkupfalse%
\ oSet{\isadigit{1}}{\isadigit{6}}{\isacharcolon}{\isachardoublequoteopen}outSetCorelated\ data{\isadigit{1}}{\isadigit{6}}\ {\isasymnoteq}\ {\isacharbraceleft}{\isacharbraceright}{\isachardoublequoteclose}\isanewline
\ \ \ \ \isacommand{by}\isamarkupfalse%
\ {\isacharparenleft}simp\ add{\isacharcolon}\ outSetCorelated{\isacharunderscore}def{\isacharcomma}\ auto{\isacharparenright}\isanewline
\ \ \isacommand{have}\isamarkupfalse%
\ {\isachardoublequoteopen}outSetCorelated\ data{\isadigit{1}}{\isadigit{5}}\ {\isasyminter}\ outSetCorelated\ data{\isadigit{1}}{\isadigit{6}}\ {\isasymnoteq}\ {\isacharbraceleft}{\isacharbraceright}{\isachardoublequoteclose}\isanewline
\ \ \ \ \isacommand{by}\isamarkupfalse%
\ {\isacharparenleft}simp\ add{\isacharcolon}\ outSetCorelated{\isacharunderscore}def{\isacharcomma}\ auto{\isacharparenright}\isanewline
\ \ \isacommand{with}\isamarkupfalse%
\ oSet{\isadigit{1}}{\isadigit{5}}\ oSet{\isadigit{1}}{\isadigit{6}}\ \isacommand{show}\isamarkupfalse%
\ {\isacharquery}thesis\ \isacommand{by}\isamarkupfalse%
\ {\isacharparenleft}simp\ add{\isacharcolon}\ elementaryCompDD{\isacharunderscore}def{\isacharcomma}\ auto{\isacharparenright}\ \isanewline
\isacommand{qed}\isamarkupfalse%
\endisatagproof
{\isafoldproof}%
\isadelimproof
\isanewline
\endisadelimproof
\isanewline
\isacommand{lemma}\isamarkupfalse%
\ NOT{\isacharunderscore}elementaryCompDD{\isacharunderscore}sA{\isadigit{7}}{\isacharcolon}\ \ {\isachardoublequoteopen}{\isasymnot}\ elementaryCompDD\ sA{\isadigit{7}}{\isachardoublequoteclose}\ \isanewline
\isadelimproof
\endisadelimproof
\isatagproof
\isacommand{proof}\isamarkupfalse%
\ {\isacharminus}\ \isanewline
\ \ \isacommand{have}\isamarkupfalse%
\ {\isachardoublequoteopen}outSetCorelated\ data{\isadigit{1}}{\isadigit{7}}\ {\isasyminter}\ outSetCorelated\ data{\isadigit{1}}{\isadigit{8}}\ {\isacharequal}\ {\isacharbraceleft}{\isacharbraceright}{\isachardoublequoteclose}\isanewline
\ \ \isacommand{by}\isamarkupfalse%
\ {\isacharparenleft}metis\ {\isacharparenleft}full{\isacharunderscore}types{\isacharparenright}\ OUTfromV{\isachardot}simps{\isacharparenleft}{\isadigit{1}}{\isadigit{7}}{\isacharparenright}\ disjoint{\isacharunderscore}iff{\isacharunderscore}not{\isacharunderscore}equal\ empty{\isacharunderscore}iff\ outSetCorelatedEmpty{\isadigit{1}}{\isacharparenright}\ \isanewline
\ \ \isacommand{thus}\isamarkupfalse%
\ {\isacharquery}thesis\ \isacommand{by}\isamarkupfalse%
\ \ {\isacharparenleft}simp\ add{\isacharcolon}\ elementaryCompDD{\isacharunderscore}def{\isacharparenright}\isanewline
\isacommand{qed}\isamarkupfalse%
\endisatagproof
{\isafoldproof}%
\isadelimproof
\isanewline
\endisadelimproof
\isanewline
\isacommand{lemma}\isamarkupfalse%
\ NOT{\isacharunderscore}elementaryCompDD{\isacharunderscore}sA{\isadigit{8}}{\isacharcolon}\ \ {\isachardoublequoteopen}{\isasymnot}\ elementaryCompDD\ sA{\isadigit{8}}{\isachardoublequoteclose}\ \isanewline
\isadelimproof
\endisadelimproof
\isatagproof
\isacommand{proof}\isamarkupfalse%
\ {\isacharminus}\ \isanewline
\ \ \isacommand{have}\isamarkupfalse%
\ {\isachardoublequoteopen}outSetCorelated\ data{\isadigit{2}}{\isadigit{0}}\ {\isasyminter}\ outSetCorelated\ data{\isadigit{2}}{\isadigit{1}}\ {\isacharequal}\ {\isacharbraceleft}{\isacharbraceright}{\isachardoublequoteclose}\isanewline
\ \ \isacommand{by}\isamarkupfalse%
\ {\isacharparenleft}metis\ OUTfromV{\isachardot}simps{\isacharparenleft}{\isadigit{2}}{\isadigit{1}}{\isacharparenright}\ inf{\isacharunderscore}bot{\isacharunderscore}right\ outSetCorelatedEmpty{\isadigit{1}}{\isacharparenright}\isanewline
\ \ \isacommand{thus}\isamarkupfalse%
\ {\isacharquery}thesis\ \isacommand{by}\isamarkupfalse%
\ \ {\isacharparenleft}simp\ add{\isacharcolon}\ elementaryCompDD{\isacharunderscore}def{\isacharparenright}\isanewline
\isacommand{qed}\isamarkupfalse%
\endisatagproof
{\isafoldproof}%
\isadelimproof
\isanewline
\endisadelimproof
\isanewline
\isacommand{lemma}\isamarkupfalse%
\ NOT{\isacharunderscore}elementaryCompDD{\isacharunderscore}sA{\isadigit{9}}{\isacharcolon}\ \ {\isachardoublequoteopen}{\isasymnot}\ elementaryCompDD\ sA{\isadigit{9}}{\isachardoublequoteclose}\ \isanewline
\isadelimproof
\endisadelimproof
\isatagproof
\isacommand{proof}\isamarkupfalse%
\ {\isacharminus}\ \isanewline
\ \ \isacommand{have}\isamarkupfalse%
\ {\isachardoublequoteopen}outSetCorelated\ data{\isadigit{2}}{\isadigit{3}}\ {\isasyminter}\ outSetCorelated\ data{\isadigit{2}}{\isadigit{4}}\ {\isacharequal}\ {\isacharbraceleft}{\isacharbraceright}{\isachardoublequoteclose}\isanewline
\ \ \isacommand{by}\isamarkupfalse%
\ {\isacharparenleft}metis\ {\isacharparenleft}full{\isacharunderscore}types{\isacharparenright}\ OUTfromV{\isachardot}simps{\isacharparenleft}{\isadigit{2}}{\isadigit{3}}{\isacharparenright}\ disjoint{\isacharunderscore}iff{\isacharunderscore}not{\isacharunderscore}equal\ empty{\isacharunderscore}iff\ outSetCorelatedEmpty{\isadigit{1}}{\isacharparenright}\isanewline
\ \ \isacommand{thus}\isamarkupfalse%
\ {\isacharquery}thesis\ \isacommand{by}\isamarkupfalse%
\ \ {\isacharparenleft}simp\ add{\isacharcolon}\ elementaryCompDD{\isacharunderscore}def{\isacharparenright}\ \ \isanewline
\isacommand{qed}\isamarkupfalse%
\isanewline
\isanewline
\isamarkupcmt{On the abstraction level 1 all components are elementary%
}
\endisatagproof
{\isafoldproof}%
\isadelimproof
\isanewline
\endisadelimproof
\isanewline
\isacommand{lemma}\isamarkupfalse%
\ elementaryCompDD{\isacharunderscore}sA{\isadigit{1}}{\isadigit{1}}{\isacharcolon}\ \ {\isachardoublequoteopen}elementaryCompDD\ sA{\isadigit{1}}{\isadigit{1}}{\isachardoublequoteclose}\ \isanewline
\isadelimproof
\endisadelimproof
\isatagproof
\isacommand{by}\isamarkupfalse%
\ \ {\isacharparenleft}simp\ add{\isacharcolon}\ elementaryCompDD{\isacharunderscore}def{\isacharparenright}%
\endisatagproof
{\isafoldproof}%
\isadelimproof
\isanewline
\endisadelimproof
\isanewline
\isacommand{lemma}\isamarkupfalse%
\ elementaryCompDD{\isacharunderscore}sA{\isadigit{1}}{\isadigit{2}}{\isacharcolon}\ \ {\isachardoublequoteopen}elementaryCompDD\ sA{\isadigit{1}}{\isadigit{2}}{\isachardoublequoteclose}\ \isanewline
\isadelimproof
\endisadelimproof
\isatagproof
\isacommand{by}\isamarkupfalse%
\ \ {\isacharparenleft}simp\ add{\isacharcolon}\ elementaryCompDD{\isacharunderscore}def{\isacharparenright}%
\endisatagproof
{\isafoldproof}%
\isadelimproof
\isanewline
\endisadelimproof
\isanewline
\isacommand{lemma}\isamarkupfalse%
\ elementaryCompDD{\isacharunderscore}sA{\isadigit{2}}{\isadigit{1}}{\isacharcolon}\ {\isachardoublequoteopen}elementaryCompDD\ sA{\isadigit{2}}{\isadigit{1}}{\isachardoublequoteclose}\ \isanewline
\isadelimproof
\endisadelimproof
\isatagproof
\isacommand{by}\isamarkupfalse%
\ \ {\isacharparenleft}simp\ add{\isacharcolon}\ elementaryCompDD{\isacharunderscore}def{\isacharparenright}%
\endisatagproof
{\isafoldproof}%
\isadelimproof
\isanewline
\endisadelimproof
\isanewline
\isacommand{lemma}\isamarkupfalse%
\ elementaryCompDD{\isacharunderscore}sA{\isadigit{2}}{\isadigit{2}}{\isacharcolon}\ {\isachardoublequoteopen}elementaryCompDD\ sA{\isadigit{2}}{\isadigit{2}}{\isachardoublequoteclose}\ \isanewline
\isadelimproof
\endisadelimproof
\isatagproof
\isacommand{proof}\isamarkupfalse%
\ {\isacharminus}\ \isanewline
\ \ \isacommand{have}\isamarkupfalse%
\ oSet{\isadigit{4}}{\isacharcolon}{\isachardoublequoteopen}outSetCorelated\ data{\isadigit{4}}\ {\isasymnoteq}\ {\isacharbraceleft}{\isacharbraceright}{\isachardoublequoteclose}\ \ \isanewline
\ \ \ \ \isacommand{by}\isamarkupfalse%
\ {\isacharparenleft}simp\ add{\isacharcolon}\ outSetCorelated{\isacharunderscore}def{\isacharcomma}\ auto{\isacharparenright}\isanewline
\ \ \isacommand{have}\isamarkupfalse%
\ oSet{\isadigit{1}}{\isadigit{2}}{\isacharcolon}{\isachardoublequoteopen}outSetCorelated\ data{\isadigit{1}}{\isadigit{2}}\ {\isasymnoteq}\ {\isacharbraceleft}{\isacharbraceright}{\isachardoublequoteclose}\ \ \isanewline
\ \ \ \ \isacommand{by}\isamarkupfalse%
\ {\isacharparenleft}simp\ add{\isacharcolon}\ outSetCorelated{\isacharunderscore}def{\isacharcomma}\ auto{\isacharparenright}\isanewline
\ \ \isacommand{have}\isamarkupfalse%
\ {\isachardoublequoteopen}outSetCorelated\ data{\isadigit{4}}\ {\isasyminter}\ outSetCorelated\ data{\isadigit{1}}{\isadigit{2}}\ {\isasymnoteq}\ {\isacharbraceleft}{\isacharbraceright}{\isachardoublequoteclose}\isanewline
\ \ \ \ \isacommand{by}\isamarkupfalse%
\ {\isacharparenleft}simp\ add{\isacharcolon}\ outSetCorelated{\isacharunderscore}def{\isacharcomma}\ auto{\isacharparenright}\isanewline
\ \ \isacommand{with}\isamarkupfalse%
\ oSet{\isadigit{4}}\ oSet{\isadigit{1}}{\isadigit{2}}\ \isacommand{show}\isamarkupfalse%
\ {\isacharquery}thesis\ \isanewline
\ \ \ \ \isacommand{by}\isamarkupfalse%
\ \ {\isacharparenleft}simp\ add{\isacharcolon}\ elementaryCompDD{\isacharunderscore}def{\isacharcomma}\ auto{\isacharparenright}\isanewline
\isacommand{qed}\isamarkupfalse%
\endisatagproof
{\isafoldproof}%
\isadelimproof
\ \isanewline
\endisadelimproof
\isanewline
\isacommand{lemma}\isamarkupfalse%
\ elementaryCompDD{\isacharunderscore}sA{\isadigit{2}}{\isadigit{3}}{\isacharcolon}\ {\isachardoublequoteopen}elementaryCompDD\ sA{\isadigit{2}}{\isadigit{3}}{\isachardoublequoteclose}\ \isanewline
\isadelimproof
\endisadelimproof
\isatagproof
\isacommand{by}\isamarkupfalse%
\ \ {\isacharparenleft}simp\ add{\isacharcolon}\ elementaryCompDD{\isacharunderscore}def{\isacharparenright}%
\endisatagproof
{\isafoldproof}%
\isadelimproof
\isanewline
\endisadelimproof
\isanewline
\isacommand{lemma}\isamarkupfalse%
\ elementaryCompDD{\isacharunderscore}sA{\isadigit{3}}{\isadigit{1}}{\isacharcolon}\ {\isachardoublequoteopen}elementaryCompDD\ sA{\isadigit{3}}{\isadigit{1}}{\isachardoublequoteclose}\ \isanewline
\isadelimproof
\endisadelimproof
\isatagproof
\isacommand{by}\isamarkupfalse%
\ \ {\isacharparenleft}simp\ add{\isacharcolon}\ elementaryCompDD{\isacharunderscore}def{\isacharparenright}%
\endisatagproof
{\isafoldproof}%
\isadelimproof
\isanewline
\endisadelimproof
\isanewline
\isacommand{lemma}\isamarkupfalse%
\ elementaryCompDD{\isacharunderscore}sA{\isadigit{3}}{\isadigit{2}}{\isacharcolon}\ {\isachardoublequoteopen}elementaryCompDD\ sA{\isadigit{3}}{\isadigit{2}}{\isachardoublequoteclose}\ \isanewline
\isadelimproof
\endisadelimproof
\isatagproof
\isacommand{by}\isamarkupfalse%
\ \ {\isacharparenleft}simp\ add{\isacharcolon}\ elementaryCompDD{\isacharunderscore}def{\isacharparenright}%
\endisatagproof
{\isafoldproof}%
\isadelimproof
\isanewline
\endisadelimproof
\isanewline
\isanewline
\isacommand{lemma}\isamarkupfalse%
\ elementaryCompDD{\isacharunderscore}sA{\isadigit{4}}{\isadigit{1}}{\isacharcolon}\ {\isachardoublequoteopen}elementaryCompDD\ sA{\isadigit{4}}{\isadigit{1}}{\isachardoublequoteclose}\ \isanewline
\isadelimproof
\endisadelimproof
\isatagproof
\isacommand{by}\isamarkupfalse%
\ \ {\isacharparenleft}simp\ add{\isacharcolon}\ elementaryCompDD{\isacharunderscore}def{\isacharparenright}%
\endisatagproof
{\isafoldproof}%
\isadelimproof
\ \isanewline
\endisadelimproof
\isanewline
\isacommand{lemma}\isamarkupfalse%
\ elementaryCompDD{\isacharunderscore}sA{\isadigit{4}}{\isadigit{2}}{\isacharcolon}\ {\isachardoublequoteopen}elementaryCompDD\ sA{\isadigit{4}}{\isadigit{2}}{\isachardoublequoteclose}\ \isanewline
\isadelimproof
\endisadelimproof
\isatagproof
\isacommand{by}\isamarkupfalse%
\ \ {\isacharparenleft}simp\ add{\isacharcolon}\ elementaryCompDD{\isacharunderscore}def{\isacharparenright}%
\endisatagproof
{\isafoldproof}%
\isadelimproof
\isanewline
\endisadelimproof
\isanewline
\isacommand{lemma}\isamarkupfalse%
\ elementaryCompDD{\isacharunderscore}sA{\isadigit{7}}{\isadigit{1}}{\isacharcolon}\ {\isachardoublequoteopen}elementaryCompDD\ sA{\isadigit{7}}{\isadigit{1}}{\isachardoublequoteclose}\ \isanewline
\isadelimproof
\endisadelimproof
\isatagproof
\isacommand{by}\isamarkupfalse%
\ \ {\isacharparenleft}simp\ add{\isacharcolon}\ elementaryCompDD{\isacharunderscore}def{\isacharparenright}%
\endisatagproof
{\isafoldproof}%
\isadelimproof
\isanewline
\endisadelimproof
\isanewline
\isacommand{lemma}\isamarkupfalse%
\ elementaryCompDD{\isacharunderscore}sA{\isadigit{7}}{\isadigit{2}}{\isacharcolon}\ {\isachardoublequoteopen}elementaryCompDD\ sA{\isadigit{7}}{\isadigit{2}}{\isachardoublequoteclose}\ \isanewline
\isadelimproof
\endisadelimproof
\isatagproof
\isacommand{by}\isamarkupfalse%
\ \ {\isacharparenleft}simp\ add{\isacharcolon}\ elementaryCompDD{\isacharunderscore}def{\isacharparenright}%
\endisatagproof
{\isafoldproof}%
\isadelimproof
\isanewline
\endisadelimproof
\isanewline
\isacommand{lemma}\isamarkupfalse%
\ elementaryCompDD{\isacharunderscore}sA{\isadigit{8}}{\isadigit{1}}{\isacharcolon}\ {\isachardoublequoteopen}elementaryCompDD\ sA{\isadigit{8}}{\isadigit{1}}{\isachardoublequoteclose}\ \isanewline
\isadelimproof
\endisadelimproof
\isatagproof
\isacommand{by}\isamarkupfalse%
\ \ {\isacharparenleft}simp\ add{\isacharcolon}\ elementaryCompDD{\isacharunderscore}def{\isacharparenright}%
\endisatagproof
{\isafoldproof}%
\isadelimproof
\isanewline
\endisadelimproof
\isanewline
\isacommand{lemma}\isamarkupfalse%
\ elementaryCompDD{\isacharunderscore}sA{\isadigit{8}}{\isadigit{2}}{\isacharcolon}\ {\isachardoublequoteopen}elementaryCompDD\ sA{\isadigit{8}}{\isadigit{2}}{\isachardoublequoteclose}\ \isanewline
\isadelimproof
\endisadelimproof
\isatagproof
\isacommand{by}\isamarkupfalse%
\ \ {\isacharparenleft}simp\ add{\isacharcolon}\ elementaryCompDD{\isacharunderscore}def{\isacharparenright}%
\endisatagproof
{\isafoldproof}%
\isadelimproof
\isanewline
\endisadelimproof
\isanewline
\isacommand{lemma}\isamarkupfalse%
\ elementaryCompDD{\isacharunderscore}sA{\isadigit{9}}{\isadigit{1}}{\isacharcolon}\ {\isachardoublequoteopen}elementaryCompDD\ sA{\isadigit{9}}{\isadigit{1}}{\isachardoublequoteclose}\ \isanewline
\isadelimproof
\endisadelimproof
\isatagproof
\isacommand{by}\isamarkupfalse%
\ \ {\isacharparenleft}simp\ add{\isacharcolon}\ elementaryCompDD{\isacharunderscore}def{\isacharparenright}%
\endisatagproof
{\isafoldproof}%
\isadelimproof
\isanewline
\endisadelimproof
\isanewline
\isacommand{lemma}\isamarkupfalse%
\ elementaryCompDD{\isacharunderscore}sA{\isadigit{9}}{\isadigit{2}}{\isacharcolon}\ {\isachardoublequoteopen}elementaryCompDD\ sA{\isadigit{9}}{\isadigit{2}}{\isachardoublequoteclose}\ \isanewline
\isadelimproof
\endisadelimproof
\isatagproof
\isacommand{by}\isamarkupfalse%
\ \ {\isacharparenleft}simp\ add{\isacharcolon}\ elementaryCompDD{\isacharunderscore}def{\isacharparenright}%
\endisatagproof
{\isafoldproof}%
\isadelimproof
\isanewline
\endisadelimproof
\isanewline
\isacommand{lemma}\isamarkupfalse%
\ elementaryCompDD{\isacharunderscore}sA{\isadigit{9}}{\isadigit{3}}{\isacharcolon}\ {\isachardoublequoteopen}elementaryCompDD\ sA{\isadigit{9}}{\isadigit{3}}{\isachardoublequoteclose}\ \isanewline
\isadelimproof
\endisadelimproof
\isatagproof
\isacommand{by}\isamarkupfalse%
\ \ {\isacharparenleft}simp\ add{\isacharcolon}\ elementaryCompDD{\isacharunderscore}def{\isacharparenright}%
\endisatagproof
{\isafoldproof}%
\isadelimproof
\endisadelimproof
\isamarkupsubsection{Source components%
}
\isamarkuptrue%
\isamarkupcmt{Abstraction level 0%
}
\isanewline
\isanewline
\isacommand{lemma}\isamarkupfalse%
\ A{\isadigit{5}}{\isacharunderscore}NotDSource{\isacharunderscore}level{\isadigit{0}}{\isacharcolon}\ {\isachardoublequoteopen}isNotDSource\ level{\isadigit{0}}\ sA{\isadigit{5}}{\isachardoublequoteclose}\isanewline
\isadelimproof
\endisadelimproof
\isatagproof
\isacommand{by}\isamarkupfalse%
\ {\isacharparenleft}simp\ add{\isacharcolon}\ isNotDSource{\isacharunderscore}def{\isacharcomma}\ auto{\isacharcomma}\ \ case{\isacharunderscore}tac\ {\isachardoublequoteopen}Z{\isachardoublequoteclose}{\isacharcomma}\ auto{\isacharparenright}%
\endisatagproof
{\isafoldproof}%
\isadelimproof
\isanewline
\endisadelimproof
\isanewline
\isacommand{lemma}\isamarkupfalse%
\ DSourcesA{\isadigit{1}}{\isacharunderscore}L{\isadigit{0}}{\isacharcolon}\ {\isachardoublequoteopen}DSources\ level{\isadigit{0}}\ sA{\isadigit{1}}\ {\isacharequal}\ {\isacharbraceleft}{\isacharbraceright}{\isachardoublequoteclose}\isanewline
\isadelimproof
\endisadelimproof
\isatagproof
\isacommand{by}\isamarkupfalse%
\ {\isacharparenleft}simp\ add{\isacharcolon}\ DSources{\isacharunderscore}def{\isacharcomma}\ auto{\isacharcomma}\ case{\isacharunderscore}tac\ {\isachardoublequoteopen}x{\isachardoublequoteclose}{\isacharcomma}\ auto{\isacharparenright}%
\endisatagproof
{\isafoldproof}%
\isadelimproof
\ \isanewline
\endisadelimproof
\isanewline
\isacommand{lemma}\isamarkupfalse%
\ DSourcesA{\isadigit{2}}{\isacharunderscore}L{\isadigit{0}}{\isacharcolon}\ {\isachardoublequoteopen}DSources\ level{\isadigit{0}}\ sA{\isadigit{2}}\ {\isacharequal}\ {\isacharbraceleft}\ sA{\isadigit{1}}{\isacharcomma}\ sA{\isadigit{4}}{\isacharbraceright}{\isachardoublequoteclose}\isanewline
\isadelimproof
\endisadelimproof
\isatagproof
\isacommand{by}\isamarkupfalse%
\ {\isacharparenleft}simp\ add{\isacharcolon}\ DSources{\isacharunderscore}def\ AbstrLevel{\isadigit{0}}{\isacharcomma}\ auto{\isacharparenright}%
\endisatagproof
{\isafoldproof}%
\isadelimproof
\ \isanewline
\endisadelimproof
\isanewline
\isacommand{lemma}\isamarkupfalse%
\ DSourcesA{\isadigit{3}}{\isacharunderscore}L{\isadigit{0}}{\isacharcolon}\ {\isachardoublequoteopen}DSources\ level{\isadigit{0}}\ sA{\isadigit{3}}\ {\isacharequal}\ {\isacharbraceleft}\ sA{\isadigit{2}}\ {\isacharbraceright}{\isachardoublequoteclose}\isanewline
\isadelimproof
\endisadelimproof
\isatagproof
\isacommand{by}\isamarkupfalse%
\ {\isacharparenleft}simp\ add{\isacharcolon}\ DSources{\isacharunderscore}def\ AbstrLevel{\isadigit{0}}{\isacharcomma}\ auto{\isacharparenright}%
\endisatagproof
{\isafoldproof}%
\isadelimproof
\ \isanewline
\endisadelimproof
\isanewline
\isacommand{lemma}\isamarkupfalse%
\ DSourcesA{\isadigit{4}}{\isacharunderscore}L{\isadigit{0}}{\isacharcolon}\ {\isachardoublequoteopen}DSources\ level{\isadigit{0}}\ sA{\isadigit{4}}\ {\isacharequal}\ {\isacharbraceleft}\ sA{\isadigit{3}}\ {\isacharbraceright}{\isachardoublequoteclose}\isanewline
\isadelimproof
\endisadelimproof
\isatagproof
\isacommand{by}\isamarkupfalse%
\ {\isacharparenleft}simp\ add{\isacharcolon}\ DSources{\isacharunderscore}def\ AbstrLevel{\isadigit{0}}{\isacharcomma}\ auto{\isacharparenright}%
\endisatagproof
{\isafoldproof}%
\isadelimproof
\ \isanewline
\endisadelimproof
\isanewline
\isacommand{lemma}\isamarkupfalse%
\ DSourcesA{\isadigit{5}}{\isacharunderscore}L{\isadigit{0}}{\isacharcolon}\ {\isachardoublequoteopen}DSources\ level{\isadigit{0}}\ sA{\isadigit{5}}\ {\isacharequal}\ {\isacharbraceleft}\ sA{\isadigit{4}}\ {\isacharbraceright}{\isachardoublequoteclose}\isanewline
\isadelimproof
\endisadelimproof
\isatagproof
\isacommand{by}\isamarkupfalse%
\ {\isacharparenleft}simp\ add{\isacharcolon}\ DSources{\isacharunderscore}def\ AbstrLevel{\isadigit{0}}{\isacharcomma}\ auto{\isacharparenright}%
\endisatagproof
{\isafoldproof}%
\isadelimproof
\ \ \isanewline
\endisadelimproof
\isanewline
\isacommand{lemma}\isamarkupfalse%
\ DSourcesA{\isadigit{6}}{\isacharunderscore}L{\isadigit{0}}{\isacharcolon}\ {\isachardoublequoteopen}DSources\ level{\isadigit{0}}\ sA{\isadigit{6}}\ {\isacharequal}\ {\isacharbraceleft}{\isacharbraceright}{\isachardoublequoteclose}\isanewline
\isadelimproof
\endisadelimproof
\isatagproof
\isacommand{by}\isamarkupfalse%
\ {\isacharparenleft}simp\ add{\isacharcolon}\ DSources{\isacharunderscore}def{\isacharcomma}\ auto{\isacharcomma}\ case{\isacharunderscore}tac\ {\isachardoublequoteopen}x{\isachardoublequoteclose}{\isacharcomma}\ auto{\isacharparenright}%
\endisatagproof
{\isafoldproof}%
\isadelimproof
\ \isanewline
\endisadelimproof
\isanewline
\isacommand{lemma}\isamarkupfalse%
\ DSourcesA{\isadigit{7}}{\isacharunderscore}L{\isadigit{0}}{\isacharcolon}\ {\isachardoublequoteopen}DSources\ level{\isadigit{0}}\ sA{\isadigit{7}}\ {\isacharequal}\ {\isacharbraceleft}sA{\isadigit{6}}{\isacharbraceright}{\isachardoublequoteclose}\isanewline
\isadelimproof
\endisadelimproof
\isatagproof
\isacommand{by}\isamarkupfalse%
\ {\isacharparenleft}simp\ add{\isacharcolon}\ DSources{\isacharunderscore}def\ AbstrLevel{\isadigit{0}}{\isacharcomma}\ auto{\isacharparenright}%
\endisatagproof
{\isafoldproof}%
\isadelimproof
\ \isanewline
\endisadelimproof
\isanewline
\isacommand{lemma}\isamarkupfalse%
\ DSourcesA{\isadigit{8}}{\isacharunderscore}L{\isadigit{0}}{\isacharcolon}\ {\isachardoublequoteopen}DSources\ level{\isadigit{0}}\ sA{\isadigit{8}}\ {\isacharequal}\ {\isacharbraceleft}sA{\isadigit{7}}{\isacharcomma}\ sA{\isadigit{9}}{\isacharbraceright}{\isachardoublequoteclose}\isanewline
\isadelimproof
\endisadelimproof
\isatagproof
\isacommand{by}\isamarkupfalse%
\ {\isacharparenleft}simp\ add{\isacharcolon}\ DSources{\isacharunderscore}def\ AbstrLevel{\isadigit{0}}{\isacharcomma}\ force{\isacharparenright}%
\endisatagproof
{\isafoldproof}%
\isadelimproof
\isanewline
\endisadelimproof
\isanewline
\isacommand{lemma}\isamarkupfalse%
\ DSourcesA{\isadigit{9}}{\isacharunderscore}L{\isadigit{0}}{\isacharcolon}\ {\isachardoublequoteopen}DSources\ level{\isadigit{0}}\ sA{\isadigit{9}}\ {\isacharequal}\ {\isacharbraceleft}sA{\isadigit{8}}{\isacharbraceright}{\isachardoublequoteclose}\isanewline
\isadelimproof
\endisadelimproof
\isatagproof
\isacommand{by}\isamarkupfalse%
\ {\isacharparenleft}simp\ add{\isacharcolon}\ DSources{\isacharunderscore}def\ AbstrLevel{\isadigit{0}}{\isacharcomma}\ auto{\isacharparenright}%
\endisatagproof
{\isafoldproof}%
\isadelimproof
\ \isanewline
\endisadelimproof
\isanewline
\isacommand{lemma}\isamarkupfalse%
\ A{\isadigit{1}}{\isacharunderscore}DAcc{\isacharunderscore}level{\isadigit{0}}{\isacharcolon}\ {\isachardoublequoteopen}DAcc\ level{\isadigit{0}}\ sA{\isadigit{1}}\ {\isacharequal}\ {\isacharbraceleft}\ sA{\isadigit{2}}\ {\isacharbraceright}{\isachardoublequoteclose}\ \isanewline
\isadelimproof
\endisadelimproof
\isatagproof
\isacommand{by}\isamarkupfalse%
\ {\isacharparenleft}simp\ add{\isacharcolon}\ DAcc{\isacharunderscore}def\ \ AbstrLevel{\isadigit{0}}{\isacharcomma}\ auto{\isacharparenright}%
\endisatagproof
{\isafoldproof}%
\isadelimproof
\isanewline
\endisadelimproof
\isanewline
\isacommand{lemma}\isamarkupfalse%
\ A{\isadigit{2}}{\isacharunderscore}DAcc{\isacharunderscore}level{\isadigit{0}}{\isacharcolon}\ {\isachardoublequoteopen}DAcc\ level{\isadigit{0}}\ sA{\isadigit{2}}\ {\isacharequal}\ {\isacharbraceleft}\ sA{\isadigit{3}}\ {\isacharbraceright}{\isachardoublequoteclose}\ \isanewline
\isadelimproof
\endisadelimproof
\isatagproof
\isacommand{by}\isamarkupfalse%
\ {\isacharparenleft}simp\ add{\isacharcolon}\ DAcc{\isacharunderscore}def\ \ AbstrLevel{\isadigit{0}}{\isacharcomma}\ force{\isacharparenright}%
\endisatagproof
{\isafoldproof}%
\isadelimproof
\isanewline
\endisadelimproof
\isanewline
\isacommand{lemma}\isamarkupfalse%
\ A{\isadigit{3}}{\isacharunderscore}DAcc{\isacharunderscore}level{\isadigit{0}}{\isacharcolon}\ {\isachardoublequoteopen}DAcc\ level{\isadigit{0}}\ sA{\isadigit{3}}\ {\isacharequal}\ {\isacharbraceleft}\ sA{\isadigit{4}}\ {\isacharbraceright}{\isachardoublequoteclose}\ \isanewline
\isadelimproof
\endisadelimproof
\isatagproof
\isacommand{by}\isamarkupfalse%
\ {\isacharparenleft}simp\ add{\isacharcolon}\ DAcc{\isacharunderscore}def\ \ AbstrLevel{\isadigit{0}}{\isacharcomma}\ auto{\isacharparenright}%
\endisatagproof
{\isafoldproof}%
\isadelimproof
\isanewline
\endisadelimproof
\isanewline
\isacommand{lemma}\isamarkupfalse%
\ A{\isadigit{4}}{\isacharunderscore}DAcc{\isacharunderscore}level{\isadigit{0}}{\isacharcolon}\ {\isachardoublequoteopen}DAcc\ level{\isadigit{0}}\ sA{\isadigit{4}}\ {\isacharequal}\ {\isacharbraceleft}\ sA{\isadigit{2}}{\isacharcomma}\ sA{\isadigit{5}}\ {\isacharbraceright}{\isachardoublequoteclose}\ \isanewline
\isadelimproof
\endisadelimproof
\isatagproof
\isacommand{by}\isamarkupfalse%
\ {\isacharparenleft}simp\ add{\isacharcolon}\ DAcc{\isacharunderscore}def\ \ AbstrLevel{\isadigit{0}}{\isacharcomma}\ auto{\isacharparenright}%
\endisatagproof
{\isafoldproof}%
\isadelimproof
\isanewline
\endisadelimproof
\isanewline
\isacommand{lemma}\isamarkupfalse%
\ A{\isadigit{5}}{\isacharunderscore}DAcc{\isacharunderscore}level{\isadigit{0}}{\isacharcolon}\ {\isachardoublequoteopen}DAcc\ level{\isadigit{0}}\ sA{\isadigit{5}}\ {\isacharequal}\ {\isacharbraceleft}{\isacharbraceright}{\isachardoublequoteclose}\ \isanewline
\isadelimproof
\endisadelimproof
\isatagproof
\isacommand{by}\isamarkupfalse%
\ {\isacharparenleft}simp\ add{\isacharcolon}\ DAcc{\isacharunderscore}def\ \ AbstrLevel{\isadigit{0}}{\isacharcomma}\ auto{\isacharparenright}%
\endisatagproof
{\isafoldproof}%
\isadelimproof
\isanewline
\endisadelimproof
\isanewline
\isacommand{lemma}\isamarkupfalse%
\ A{\isadigit{6}}{\isacharunderscore}DAcc{\isacharunderscore}level{\isadigit{0}}{\isacharcolon}\ {\isachardoublequoteopen}DAcc\ level{\isadigit{0}}\ sA{\isadigit{6}}\ {\isacharequal}\ {\isacharbraceleft}\ sA{\isadigit{7}}\ {\isacharbraceright}{\isachardoublequoteclose}\ \isanewline
\isadelimproof
\endisadelimproof
\isatagproof
\isacommand{by}\isamarkupfalse%
\ {\isacharparenleft}simp\ add{\isacharcolon}\ DAcc{\isacharunderscore}def\ \ AbstrLevel{\isadigit{0}}{\isacharcomma}\ auto{\isacharparenright}%
\endisatagproof
{\isafoldproof}%
\isadelimproof
\isanewline
\endisadelimproof
\isanewline
\isacommand{lemma}\isamarkupfalse%
\ A{\isadigit{7}}{\isacharunderscore}DAcc{\isacharunderscore}level{\isadigit{0}}{\isacharcolon}\ {\isachardoublequoteopen}DAcc\ level{\isadigit{0}}\ sA{\isadigit{7}}\ {\isacharequal}\ {\isacharbraceleft}\ sA{\isadigit{8}}\ {\isacharbraceright}{\isachardoublequoteclose}\ \isanewline
\isadelimproof
\endisadelimproof
\isatagproof
\isacommand{by}\isamarkupfalse%
\ {\isacharparenleft}simp\ add{\isacharcolon}\ DAcc{\isacharunderscore}def\ \ AbstrLevel{\isadigit{0}}{\isacharcomma}\ auto{\isacharparenright}%
\endisatagproof
{\isafoldproof}%
\isadelimproof
\isanewline
\endisadelimproof
\isanewline
\isacommand{lemma}\isamarkupfalse%
\ A{\isadigit{8}}{\isacharunderscore}DAcc{\isacharunderscore}level{\isadigit{0}}{\isacharcolon}\ {\isachardoublequoteopen}DAcc\ level{\isadigit{0}}\ sA{\isadigit{8}}\ {\isacharequal}\ {\isacharbraceleft}\ sA{\isadigit{9}}\ {\isacharbraceright}{\isachardoublequoteclose}\ \isanewline
\isadelimproof
\endisadelimproof
\isatagproof
\isacommand{by}\isamarkupfalse%
\ {\isacharparenleft}simp\ add{\isacharcolon}\ DAcc{\isacharunderscore}def\ \ AbstrLevel{\isadigit{0}}{\isacharcomma}\ auto{\isacharparenright}%
\endisatagproof
{\isafoldproof}%
\isadelimproof
\isanewline
\endisadelimproof
\isanewline
\isacommand{lemma}\isamarkupfalse%
\ A{\isadigit{9}}{\isacharunderscore}DAcc{\isacharunderscore}level{\isadigit{0}}{\isacharcolon}\ {\isachardoublequoteopen}DAcc\ level{\isadigit{0}}\ sA{\isadigit{9}}\ {\isacharequal}\ {\isacharbraceleft}\ sA{\isadigit{8}}\ {\isacharbraceright}{\isachardoublequoteclose}\ \isanewline
\isadelimproof
\endisadelimproof
\isatagproof
\isacommand{by}\isamarkupfalse%
\ {\isacharparenleft}simp\ add{\isacharcolon}\ DAcc{\isacharunderscore}def\ \ AbstrLevel{\isadigit{0}}{\isacharcomma}\ force{\isacharparenright}%
\endisatagproof
{\isafoldproof}%
\isadelimproof
\isanewline
\endisadelimproof
\isanewline
\isacommand{lemma}\isamarkupfalse%
\ A{\isadigit{8}}{\isacharunderscore}NSources{\isacharcolon}\isanewline
{\isachardoublequoteopen}{\isasymforall}\ C\ {\isasymin}\ {\isacharparenleft}AbstrLevel\ level{\isadigit{0}}{\isacharparenright}{\isachardot}\ {\isacharparenleft}C\ {\isasymnoteq}\ sA{\isadigit{9}}\ {\isasymand}\ C\ {\isasymnoteq}\ sA{\isadigit{8}}\ {\isasymlongrightarrow}\ sA{\isadigit{8}}\ {\isasymnotin}\ {\isacharparenleft}Sources\ level{\isadigit{0}}\ C{\isacharparenright}{\isacharparenright}{\isachardoublequoteclose}\isanewline
\isadelimproof
\endisadelimproof
\isatagproof
\isacommand{by}\isamarkupfalse%
\ {\isacharparenleft}metis\ A{\isadigit{8}}{\isacharunderscore}DAcc{\isacharunderscore}level{\isadigit{0}}\ A{\isadigit{9}}{\isacharunderscore}DAcc{\isacharunderscore}level{\isadigit{0}}\ singleDSourceLoop{\isacharparenright}%
\endisatagproof
{\isafoldproof}%
\isadelimproof
\isanewline
\endisadelimproof
\isanewline
\isacommand{lemma}\isamarkupfalse%
\ A{\isadigit{9}}{\isacharunderscore}NSources{\isacharcolon}\isanewline
{\isachardoublequoteopen}{\isasymforall}\ C\ {\isasymin}\ {\isacharparenleft}AbstrLevel\ level{\isadigit{0}}{\isacharparenright}{\isachardot}\ {\isacharparenleft}C\ {\isasymnoteq}\ sA{\isadigit{9}}\ {\isasymand}\ C\ {\isasymnoteq}\ sA{\isadigit{8}}\ {\isasymlongrightarrow}\ sA{\isadigit{9}}\ {\isasymnotin}\ {\isacharparenleft}Sources\ level{\isadigit{0}}\ C{\isacharparenright}{\isacharparenright}{\isachardoublequoteclose}\isanewline
\isadelimproof
\endisadelimproof
\isatagproof
\isacommand{by}\isamarkupfalse%
\ {\isacharparenleft}metis\ A{\isadigit{8}}{\isacharunderscore}DAcc{\isacharunderscore}level{\isadigit{0}}\ A{\isadigit{9}}{\isacharunderscore}DAcc{\isacharunderscore}level{\isadigit{0}}\ singleDSourceLoop{\isacharparenright}%
\endisatagproof
{\isafoldproof}%
\isadelimproof
\isanewline
\endisadelimproof
\isanewline
\isacommand{lemma}\isamarkupfalse%
\ A{\isadigit{7}}{\isacharunderscore}Acc{\isacharcolon}\isanewline
{\isachardoublequoteopen}{\isacharparenleft}Acc\ level{\isadigit{0}}\ sA{\isadigit{7}}{\isacharparenright}\ {\isacharequal}\ {\isacharbraceleft}sA{\isadigit{8}}{\isacharcomma}\ sA{\isadigit{9}}{\isacharbraceright}{\isachardoublequoteclose}\isanewline
\isadelimproof
\ \ %
\endisadelimproof
\isatagproof
\isacommand{by}\isamarkupfalse%
\ {\isacharparenleft}metis\ A{\isadigit{7}}{\isacharunderscore}DAcc{\isacharunderscore}level{\isadigit{0}}\ A{\isadigit{8}}{\isacharunderscore}DAcc{\isacharunderscore}level{\isadigit{0}}\ A{\isadigit{9}}{\isacharunderscore}DAcc{\isacharunderscore}level{\isadigit{0}}\ AccDef\ AccSigleLoop\ insert{\isacharunderscore}commute{\isacharparenright}%
\endisatagproof
{\isafoldproof}%
\isadelimproof
\ \isanewline
\endisadelimproof
\isanewline
\isacommand{lemma}\isamarkupfalse%
\ A{\isadigit{7}}{\isacharunderscore}NSources{\isacharcolon}\isanewline
{\isachardoublequoteopen}{\isasymforall}\ C\ {\isasymin}\ {\isacharparenleft}AbstrLevel\ level{\isadigit{0}}{\isacharparenright}{\isachardot}\ {\isacharparenleft}C\ {\isasymnoteq}\ sA{\isadigit{9}}\ {\isasymand}\ C\ {\isasymnoteq}\ sA{\isadigit{8}}\ {\isasymlongrightarrow}\ sA{\isadigit{7}}\ {\isasymnotin}\ {\isacharparenleft}Sources\ level{\isadigit{0}}\ C{\isacharparenright}{\isacharparenright}{\isachardoublequoteclose}\isanewline
\isadelimproof
\endisadelimproof
\isatagproof
\isacommand{by}\isamarkupfalse%
\ {\isacharparenleft}metis\ A{\isadigit{7}}{\isacharunderscore}Acc\ Acc{\isacharunderscore}Sources\ insert{\isacharunderscore}iff\ singleton{\isacharunderscore}iff{\isacharparenright}%
\endisatagproof
{\isafoldproof}%
\isadelimproof
\isanewline
\endisadelimproof
\isanewline
\isacommand{lemma}\isamarkupfalse%
\ A{\isadigit{5}}{\isacharunderscore}Acc{\isacharcolon}\ {\isachardoublequoteopen}{\isacharparenleft}Acc\ level{\isadigit{0}}\ sA{\isadigit{5}}{\isacharparenright}\ {\isacharequal}\ {\isacharbraceleft}{\isacharbraceright}{\isachardoublequoteclose}\isanewline
\isadelimproof
\endisadelimproof
\isatagproof
\isacommand{by}\isamarkupfalse%
\ {\isacharparenleft}metis\ A{\isadigit{5}}{\isacharunderscore}NotDSource{\isacharunderscore}level{\isadigit{0}}\ isNotDSource{\isacharunderscore}EmptyAcc{\isacharparenright}%
\endisatagproof
{\isafoldproof}%
\isadelimproof
\isanewline
\endisadelimproof
\isanewline
\isacommand{lemma}\isamarkupfalse%
\ A{\isadigit{6}}{\isacharunderscore}Acc{\isacharcolon}\isanewline
{\isachardoublequoteopen}{\isacharparenleft}Acc\ level{\isadigit{0}}\ sA{\isadigit{6}}{\isacharparenright}\ {\isacharequal}\ {\isacharbraceleft}sA{\isadigit{7}}{\isacharcomma}\ sA{\isadigit{8}}{\isacharcomma}\ sA{\isadigit{9}}{\isacharbraceright}{\isachardoublequoteclose}\isanewline
\isadelimproof
\endisadelimproof
\isatagproof
\isacommand{proof}\isamarkupfalse%
\ {\isacharminus}\isanewline
\ \ \isacommand{have}\isamarkupfalse%
\ daA{\isadigit{6}}{\isacharcolon}\ \ {\isachardoublequoteopen}DAcc\ level{\isadigit{0}}\ sA{\isadigit{6}}\ {\isacharequal}\ {\isacharbraceleft}\ sA{\isadigit{7}}\ {\isacharbraceright}{\isachardoublequoteclose}\ \ \isacommand{by}\isamarkupfalse%
\ {\isacharparenleft}rule\ A{\isadigit{6}}{\isacharunderscore}DAcc{\isacharunderscore}level{\isadigit{0}}{\isacharparenright}\isanewline
\ \ \isacommand{hence}\isamarkupfalse%
\ {\isachardoublequoteopen}{\isacharparenleft}{\isasymUnion}\ S\ {\isasymin}\ {\isacharparenleft}DAcc\ level{\isadigit{0}}\ sA{\isadigit{6}}{\isacharparenright}{\isachardot}\ {\isacharparenleft}Acc\ level{\isadigit{0}}\ S{\isacharparenright}{\isacharparenright}\ {\isacharequal}\ {\isacharparenleft}Acc\ level{\isadigit{0}}\ sA{\isadigit{7}}{\isacharparenright}{\isachardoublequoteclose}\ \ \isacommand{by}\isamarkupfalse%
\ simp\isanewline
\ \ \isacommand{hence}\isamarkupfalse%
\ aA{\isadigit{6}}{\isacharcolon}{\isachardoublequoteopen}{\isacharparenleft}{\isasymUnion}\ S\ {\isasymin}\ {\isacharparenleft}DAcc\ level{\isadigit{0}}\ sA{\isadigit{6}}{\isacharparenright}{\isachardot}\ {\isacharparenleft}Acc\ level{\isadigit{0}}\ S{\isacharparenright}{\isacharparenright}\ {\isacharequal}\ {\isacharbraceleft}\ sA{\isadigit{8}}{\isacharcomma}\ sA{\isadigit{9}}\ {\isacharbraceright}{\isachardoublequoteclose}\ \isacommand{by}\isamarkupfalse%
\ {\isacharparenleft}simp\ add{\isacharcolon}\ A{\isadigit{7}}{\isacharunderscore}Acc{\isacharparenright}\isanewline
\ \ \isacommand{have}\isamarkupfalse%
\ {\isachardoublequoteopen}{\isacharparenleft}Acc\ level{\isadigit{0}}\ sA{\isadigit{6}}{\isacharparenright}\ {\isacharequal}\ {\isacharparenleft}DAcc\ level{\isadigit{0}}\ sA{\isadigit{6}}{\isacharparenright}\ {\isasymunion}\ {\isacharparenleft}{\isasymUnion}\ S\ {\isasymin}\ {\isacharparenleft}DAcc\ level{\isadigit{0}}\ sA{\isadigit{6}}{\isacharparenright}{\isachardot}\ {\isacharparenleft}Acc\ level{\isadigit{0}}\ S{\isacharparenright}{\isacharparenright}{\isachardoublequoteclose}\ \ \isanewline
\ \ \ \ \isacommand{by}\isamarkupfalse%
\ {\isacharparenleft}rule\ AccDef{\isacharparenright}\isanewline
\ \ \isacommand{with}\isamarkupfalse%
\ daA{\isadigit{6}}\ aA{\isadigit{6}}\ \isacommand{show}\isamarkupfalse%
\ {\isacharquery}thesis\ \isacommand{by}\isamarkupfalse%
\ auto\isanewline
\isacommand{qed}\isamarkupfalse%
\endisatagproof
{\isafoldproof}%
\isadelimproof
\isanewline
\endisadelimproof
\isanewline
\isacommand{lemma}\isamarkupfalse%
\ A{\isadigit{6}}{\isacharunderscore}NSources{\isacharcolon}\isanewline
{\isachardoublequoteopen}{\isasymforall}\ C\ {\isasymin}\ {\isacharparenleft}AbstrLevel\ level{\isadigit{0}}{\isacharparenright}{\isachardot}\ {\isacharparenleft}C\ {\isasymnoteq}\ sA{\isadigit{9}}\ {\isasymand}\ C\ {\isasymnoteq}\ sA{\isadigit{8}}\ {\isasymand}\ C\ {\isasymnoteq}\ sA{\isadigit{7}}\ {\isasymlongrightarrow}\ sA{\isadigit{6}}\ {\isasymnotin}\ {\isacharparenleft}Sources\ level{\isadigit{0}}\ C{\isacharparenright}{\isacharparenright}{\isachardoublequoteclose}\isanewline
\isadelimproof
\endisadelimproof
\isatagproof
\isacommand{by}\isamarkupfalse%
\ {\isacharparenleft}metis\ {\isacharparenleft}full{\isacharunderscore}types{\isacharparenright}\ A{\isadigit{6}}{\isacharunderscore}Acc\ A{\isadigit{7}}{\isacharunderscore}Acc\ Acc{\isacharunderscore}SourcesNOT\ insert{\isacharunderscore}iff\ singleton{\isacharunderscore}iff{\isacharparenright}%
\endisatagproof
{\isafoldproof}%
\isadelimproof
\isanewline
\endisadelimproof
\isanewline
\isacommand{lemma}\isamarkupfalse%
\ SourcesA{\isadigit{1}}{\isacharunderscore}L{\isadigit{0}}{\isacharcolon}\ {\isachardoublequoteopen}Sources\ level{\isadigit{0}}\ sA{\isadigit{1}}\ {\isacharequal}\ {\isacharbraceleft}{\isacharbraceright}{\isachardoublequoteclose}\ \ \isanewline
\isadelimproof
\endisadelimproof
\isatagproof
\isacommand{by}\isamarkupfalse%
\ {\isacharparenleft}simp\ add{\isacharcolon}\ DSourcesA{\isadigit{1}}{\isacharunderscore}L{\isadigit{0}}\ DSourcesEmptySources{\isacharparenright}%
\endisatagproof
{\isafoldproof}%
\isadelimproof
\ \isanewline
\endisadelimproof
\ \ \ \ \ \isanewline
\isacommand{lemma}\isamarkupfalse%
\ SourcesA{\isadigit{2}}{\isacharunderscore}L{\isadigit{0}}{\isacharcolon}\ {\isachardoublequoteopen}Sources\ level{\isadigit{0}}\ sA{\isadigit{2}}\ {\isacharequal}\ {\isacharbraceleft}\ sA{\isadigit{1}}{\isacharcomma}\ sA{\isadigit{2}}{\isacharcomma}\ sA{\isadigit{3}}{\isacharcomma}\ sA{\isadigit{4}}\ {\isacharbraceright}{\isachardoublequoteclose}\isanewline
\isadelimproof
\endisadelimproof
\isatagproof
\isacommand{proof}\isamarkupfalse%
\ \isanewline
\ \ \isacommand{show}\isamarkupfalse%
\ {\isachardoublequoteopen}Sources\ level{\isadigit{0}}\ sA{\isadigit{2}}\ {\isasymsubseteq}\ {\isacharbraceleft}sA{\isadigit{1}}{\isacharcomma}\ sA{\isadigit{2}}{\isacharcomma}\ sA{\isadigit{3}}{\isacharcomma}\ sA{\isadigit{4}}{\isacharbraceright}{\isachardoublequoteclose}\isanewline
\ \ \isacommand{proof}\isamarkupfalse%
\ {\isacharminus}\isanewline
\ \ \ \ \isacommand{have}\isamarkupfalse%
\ A{\isadigit{2}}level{\isadigit{0}}{\isacharcolon}{\isachardoublequoteopen}sA{\isadigit{2}}\ {\isasymin}\ {\isacharparenleft}AbstrLevel\ level{\isadigit{0}}{\isacharparenright}{\isachardoublequoteclose}\ \isacommand{by}\isamarkupfalse%
\ {\isacharparenleft}simp\ add{\isacharcolon}\ AbstrLevel{\isadigit{0}}{\isacharparenright}\isanewline
\ \ \ \ \isacommand{have}\isamarkupfalse%
\ sgA{\isadigit{5}}{\isacharcolon}{\isachardoublequoteopen}sA{\isadigit{5}}\ {\isasymnotin}\ Sources\ level{\isadigit{0}}\ sA{\isadigit{2}}{\isachardoublequoteclose}\ \isanewline
\ \ \ \ \ \ \isacommand{by}\isamarkupfalse%
\ {\isacharparenleft}metis\ A{\isadigit{5}}{\isacharunderscore}NotDSource{\isacharunderscore}level{\isadigit{0}}\ DSource{\isacharunderscore}level\ NoDSourceNoSource\ \isanewline
\ \ \ \ \ \ \ \ \ \ \ \ allNotDSource{\isacharunderscore}NotSource\ isNotSource{\isacharunderscore}Sources{\isacharparenright}\isanewline
\ \ \ \ \ \isacommand{from}\isamarkupfalse%
\ A{\isadigit{2}}level{\isadigit{0}}\ \isacommand{have}\isamarkupfalse%
\ sgA{\isadigit{6}}{\isacharcolon}{\isachardoublequoteopen}sA{\isadigit{6}}\ {\isasymnotin}\ Sources\ level{\isadigit{0}}\ sA{\isadigit{2}}{\isachardoublequoteclose}\ \isacommand{by}\isamarkupfalse%
\ {\isacharparenleft}simp\ add{\isacharcolon}\ A{\isadigit{6}}{\isacharunderscore}NSources{\isacharparenright}\isanewline
\ \ \ \ \ \isacommand{from}\isamarkupfalse%
\ A{\isadigit{2}}level{\isadigit{0}}\ \isacommand{have}\isamarkupfalse%
\ sgA{\isadigit{7}}{\isacharcolon}{\isachardoublequoteopen}sA{\isadigit{7}}\ {\isasymnotin}\ Sources\ level{\isadigit{0}}\ sA{\isadigit{2}}{\isachardoublequoteclose}\ \isacommand{by}\isamarkupfalse%
\ {\isacharparenleft}simp\ add{\isacharcolon}\ A{\isadigit{7}}{\isacharunderscore}NSources{\isacharparenright}\isanewline
\ \ \ \ \ \isacommand{from}\isamarkupfalse%
\ A{\isadigit{2}}level{\isadigit{0}}\ \isacommand{have}\isamarkupfalse%
\ sgA{\isadigit{8}}{\isacharcolon}{\isachardoublequoteopen}sA{\isadigit{8}}\ {\isasymnotin}\ Sources\ level{\isadigit{0}}\ sA{\isadigit{2}}{\isachardoublequoteclose}\ \isacommand{by}\isamarkupfalse%
\ {\isacharparenleft}simp\ add{\isacharcolon}\ A{\isadigit{8}}{\isacharunderscore}NSources{\isacharparenright}\isanewline
\ \ \ \ \ \isacommand{from}\isamarkupfalse%
\ A{\isadigit{2}}level{\isadigit{0}}\ \isacommand{have}\isamarkupfalse%
\ sgA{\isadigit{9}}{\isacharcolon}{\isachardoublequoteopen}sA{\isadigit{9}}\ {\isasymnotin}\ Sources\ level{\isadigit{0}}\ sA{\isadigit{2}}{\isachardoublequoteclose}\ \isacommand{by}\isamarkupfalse%
\ {\isacharparenleft}simp\ add{\isacharcolon}\ A{\isadigit{9}}{\isacharunderscore}NSources{\isacharparenright}\isanewline
\ \ \ \ \ \isacommand{have}\isamarkupfalse%
\ {\isachardoublequoteopen}Sources\ level{\isadigit{0}}\ sA{\isadigit{2}}\ {\isasymsubseteq}\ {\isacharbraceleft}sA{\isadigit{1}}{\isacharcomma}\ sA{\isadigit{2}}{\isacharcomma}\ sA{\isadigit{3}}{\isacharcomma}\ sA{\isadigit{4}}{\isacharcomma}\ sA{\isadigit{5}}{\isacharcomma}\ sA{\isadigit{6}}{\isacharcomma}\ sA{\isadigit{7}}{\isacharcomma}\ sA{\isadigit{8}}{\isacharcomma}\ sA{\isadigit{9}}{\isacharbraceright}{\isachardoublequoteclose}\isanewline
\ \ \ \ \ \ \ \ \isacommand{by}\isamarkupfalse%
\ {\isacharparenleft}metis\ AbstrLevel{\isadigit{0}}\ SourcesLevelX{\isacharparenright}\ \isanewline
\ \ \ \ \ \isacommand{with}\isamarkupfalse%
\ sgA{\isadigit{5}}\ sgA{\isadigit{6}}\ sgA{\isadigit{7}}\ sgA{\isadigit{8}}\ sgA{\isadigit{9}}\ \isacommand{show}\isamarkupfalse%
\ {\isachardoublequoteopen}Sources\ level{\isadigit{0}}\ sA{\isadigit{2}}\ {\isasymsubseteq}\ {\isacharbraceleft}sA{\isadigit{1}}{\isacharcomma}\ sA{\isadigit{2}}{\isacharcomma}\ sA{\isadigit{3}}{\isacharcomma}\ sA{\isadigit{4}}{\isacharbraceright}{\isachardoublequoteclose}\isanewline
\ \ \ \ \ \ \ \ \isacommand{by}\isamarkupfalse%
\ blast\ \ \ \isanewline
\ \ \isacommand{qed}\isamarkupfalse%
\isanewline
\isacommand{next}\isamarkupfalse%
\isanewline
\ \ \isacommand{show}\isamarkupfalse%
\ {\isachardoublequoteopen}{\isacharbraceleft}sA{\isadigit{1}}{\isacharcomma}\ sA{\isadigit{2}}{\isacharcomma}\ sA{\isadigit{3}}{\isacharcomma}\ sA{\isadigit{4}}{\isacharbraceright}\ {\isasymsubseteq}\ Sources\ level{\isadigit{0}}\ sA{\isadigit{2}}{\isachardoublequoteclose}\isanewline
\ \ \isacommand{proof}\isamarkupfalse%
\ {\isacharminus}\isanewline
\ \ \ \ \isacommand{have}\isamarkupfalse%
\ dsA{\isadigit{4}}{\isacharcolon}{\isachardoublequoteopen}{\isacharbraceleft}\ sA{\isadigit{3}}\ {\isacharbraceright}\ {\isasymsubseteq}\ Sources\ level{\isadigit{0}}\ sA{\isadigit{2}}{\isachardoublequoteclose}\isanewline
\ \ \ \ \ \ \ \isacommand{by}\isamarkupfalse%
\ {\isacharparenleft}metis\ DSource{\isacharunderscore}Sources\ DSourcesA{\isadigit{2}}{\isacharunderscore}L{\isadigit{0}}\ DSourcesA{\isadigit{4}}{\isacharunderscore}L{\isadigit{0}}\ \isanewline
\ \ \ \ \ \ \ \ \ \ \ \ \ Sources{\isacharunderscore}DSources\ insertI{\isadigit{1}}\ insert{\isacharunderscore}commute\ subset{\isacharunderscore}trans{\isacharparenright}\isanewline
\ \ \ \ \isacommand{have}\isamarkupfalse%
\ {\isachardoublequoteopen}{\isacharbraceleft}\ sA{\isadigit{2}}\ {\isacharbraceright}\ {\isasymsubseteq}\ Sources\ level{\isadigit{0}}\ sA{\isadigit{2}}{\isachardoublequoteclose}\isanewline
\ \ \ \ \ \ \isacommand{by}\isamarkupfalse%
\ {\isacharparenleft}metis\ DSource{\isacharunderscore}Sources\ DSourcesA{\isadigit{2}}{\isacharunderscore}L{\isadigit{0}}\ DSourcesA{\isadigit{3}}{\isacharunderscore}L{\isadigit{0}}\ \isanewline
\ \ \ \ \ \ \ \ \ \ \ \ \ DSourcesA{\isadigit{4}}{\isacharunderscore}L{\isadigit{0}}\ Sources{\isacharunderscore}DSources\ insertI{\isadigit{1}}\ \isanewline
\ \ \ \ \ \ \ \ \ \ \ \ \ insert{\isacharunderscore}commute\ subset{\isacharunderscore}trans{\isacharparenright}\isanewline
\ \ \ \ \isacommand{with}\isamarkupfalse%
\ dsA{\isadigit{4}}\ \isacommand{show}\isamarkupfalse%
\ {\isachardoublequoteopen}{\isacharbraceleft}sA{\isadigit{1}}{\isacharcomma}\ sA{\isadigit{2}}{\isacharcomma}\ sA{\isadigit{3}}{\isacharcomma}\ sA{\isadigit{4}}{\isacharbraceright}\ {\isasymsubseteq}\ Sources\ level{\isadigit{0}}\ sA{\isadigit{2}}{\isachardoublequoteclose}\isanewline
\ \ \ \ \ \ \ \isacommand{by}\isamarkupfalse%
\ {\isacharparenleft}metis\ DSourcesA{\isadigit{2}}{\isacharunderscore}L{\isadigit{0}}\ Sources{\isacharunderscore}DSources\ insert{\isacharunderscore}subset{\isacharparenright}\isanewline
\ \ \ \isacommand{qed}\isamarkupfalse%
\isanewline
\isacommand{qed}\isamarkupfalse%
\endisatagproof
{\isafoldproof}%
\isadelimproof
\isanewline
\endisadelimproof
\ \ \ \ \ \ \ \ \isanewline
\isacommand{lemma}\isamarkupfalse%
\ SourcesA{\isadigit{3}}{\isacharunderscore}L{\isadigit{0}}{\isacharcolon}\ {\isachardoublequoteopen}Sources\ level{\isadigit{0}}\ sA{\isadigit{3}}\ {\isacharequal}\ {\isacharbraceleft}\ sA{\isadigit{1}}{\isacharcomma}\ sA{\isadigit{2}}{\isacharcomma}\ sA{\isadigit{3}}{\isacharcomma}\ sA{\isadigit{4}}\ {\isacharbraceright}{\isachardoublequoteclose}\isanewline
\isadelimproof
\endisadelimproof
\isatagproof
\isacommand{proof}\isamarkupfalse%
\ \isanewline
\ \ \isacommand{show}\isamarkupfalse%
\ {\isachardoublequoteopen}Sources\ level{\isadigit{0}}\ sA{\isadigit{3}}\ {\isasymsubseteq}\ {\isacharbraceleft}sA{\isadigit{1}}{\isacharcomma}\ sA{\isadigit{2}}{\isacharcomma}\ sA{\isadigit{3}}{\isacharcomma}\ sA{\isadigit{4}}{\isacharbraceright}{\isachardoublequoteclose}\isanewline
\ \ \isacommand{proof}\isamarkupfalse%
\ {\isacharminus}\isanewline
\ \ \ \ \isacommand{have}\isamarkupfalse%
\ a{\isadigit{2}}{\isacharcolon}{\isachardoublequoteopen}Sources\ level{\isadigit{0}}\ sA{\isadigit{2}}\ {\isacharequal}\ {\isacharbraceleft}\ sA{\isadigit{1}}{\isacharcomma}\ sA{\isadigit{2}}{\isacharcomma}\ sA{\isadigit{3}}{\isacharcomma}\ sA{\isadigit{4}}{\isacharbraceright}{\isachardoublequoteclose}\ \isacommand{by}\isamarkupfalse%
\ {\isacharparenleft}simp\ add{\isacharcolon}\ SourcesA{\isadigit{2}}{\isacharunderscore}L{\isadigit{0}}{\isacharparenright}\isanewline
\ \ \ \ \isacommand{have}\isamarkupfalse%
\ {\isachardoublequoteopen}{\isacharbraceleft}\ sA{\isadigit{2}}\ {\isacharbraceright}\ {\isasymsubseteq}\ DSources\ level{\isadigit{0}}\ sA{\isadigit{3}}{\isachardoublequoteclose}\ \isacommand{by}\isamarkupfalse%
\ {\isacharparenleft}simp\ add{\isacharcolon}\ DSourcesA{\isadigit{3}}{\isacharunderscore}L{\isadigit{0}}{\isacharparenright}\isanewline
\ \ \ \ \isacommand{with}\isamarkupfalse%
\ a{\isadigit{2}}\ \isacommand{show}\isamarkupfalse%
\ {\isachardoublequoteopen}Sources\ level{\isadigit{0}}\ sA{\isadigit{3}}\ {\isasymsubseteq}\ {\isacharbraceleft}sA{\isadigit{1}}{\isacharcomma}\ sA{\isadigit{2}}{\isacharcomma}\ sA{\isadigit{3}}{\isacharcomma}\ sA{\isadigit{4}}{\isacharbraceright}{\isachardoublequoteclose}\isanewline
\ \ \ \ \ \ \ \isacommand{by}\isamarkupfalse%
\ {\isacharparenleft}metis\ DSource{\isacharunderscore}Sources\ DSourcesA{\isadigit{2}}{\isacharunderscore}L{\isadigit{0}}\ DSourcesA{\isadigit{4}}{\isacharunderscore}L{\isadigit{0}}\ insertI{\isadigit{1}}\ insert{\isacharunderscore}commute\ subset{\isacharunderscore}trans{\isacharparenright}\isanewline
\ \ \isacommand{qed}\isamarkupfalse%
\isanewline
\isacommand{next}\isamarkupfalse%
\isanewline
\ \ \ \isacommand{show}\isamarkupfalse%
\ {\isachardoublequoteopen}{\isacharbraceleft}sA{\isadigit{1}}{\isacharcomma}\ sA{\isadigit{2}}{\isacharcomma}\ sA{\isadigit{3}}{\isacharcomma}\ sA{\isadigit{4}}{\isacharbraceright}\ {\isasymsubseteq}\ Sources\ level{\isadigit{0}}\ sA{\isadigit{3}}{\isachardoublequoteclose}\isanewline
\ \ \ \isacommand{by}\isamarkupfalse%
\ {\isacharparenleft}metis\ {\isacharparenleft}full{\isacharunderscore}types{\isacharparenright}\ DSource{\isacharunderscore}Sources\ DSourcesA{\isadigit{3}}{\isacharunderscore}L{\isadigit{0}}\ \ SourcesA{\isadigit{2}}{\isacharunderscore}L{\isadigit{0}}\ insertI{\isadigit{1}}{\isacharparenright}\isanewline
\isacommand{qed}\isamarkupfalse%
\endisatagproof
{\isafoldproof}%
\isadelimproof
\ \ \ \ \isanewline
\endisadelimproof
\ \ \ \ \ \ \ \ \ \ \ \ \ \isanewline
\isacommand{lemma}\isamarkupfalse%
\ SourcesA{\isadigit{4}}{\isacharunderscore}L{\isadigit{0}}{\isacharcolon}\ {\isachardoublequoteopen}Sources\ level{\isadigit{0}}\ sA{\isadigit{4}}\ {\isacharequal}\ {\isacharbraceleft}\ sA{\isadigit{1}}{\isacharcomma}\ sA{\isadigit{2}}{\isacharcomma}\ sA{\isadigit{3}}{\isacharcomma}\ sA{\isadigit{4}}\ {\isacharbraceright}{\isachardoublequoteclose}\isanewline
\isadelimproof
\endisadelimproof
\isatagproof
\isacommand{proof}\isamarkupfalse%
\ {\isacharminus}\isanewline
\ \ \isacommand{have}\isamarkupfalse%
\ \ A{\isadigit{3}}s{\isacharcolon}{\isachardoublequoteopen}Sources\ level{\isadigit{0}}\ sA{\isadigit{3}}\ {\isacharequal}\ {\isacharbraceleft}\ sA{\isadigit{1}}{\isacharcomma}\ sA{\isadigit{2}}{\isacharcomma}\ sA{\isadigit{3}}{\isacharcomma}\ sA{\isadigit{4}}\ {\isacharbraceright}{\isachardoublequoteclose}\ \isacommand{by}\isamarkupfalse%
\ {\isacharparenleft}rule\ \ SourcesA{\isadigit{3}}{\isacharunderscore}L{\isadigit{0}}{\isacharparenright}\isanewline
\ \ \isacommand{have}\isamarkupfalse%
\ \ {\isachardoublequoteopen}Sources\ level{\isadigit{0}}\ sA{\isadigit{4}}\ {\isacharequal}\ {\isacharbraceleft}sA{\isadigit{3}}{\isacharbraceright}\ {\isasymunion}\ Sources\ level{\isadigit{0}}\ sA{\isadigit{3}}{\isachardoublequoteclose}\isanewline
\ \ \ \ \isacommand{by}\isamarkupfalse%
\ {\isacharparenleft}metis\ DSourcesA{\isadigit{4}}{\isacharunderscore}L{\isadigit{0}}\ Sources{\isacharunderscore}singleDSource{\isacharparenright}\ \isanewline
\ \ \isacommand{with}\isamarkupfalse%
\ A{\isadigit{3}}s\ \isacommand{show}\isamarkupfalse%
\ {\isacharquery}thesis\ \isacommand{by}\isamarkupfalse%
\ auto\isanewline
\isacommand{qed}\isamarkupfalse%
\endisatagproof
{\isafoldproof}%
\isadelimproof
\ \ \isanewline
\endisadelimproof
\isanewline
\isacommand{lemma}\isamarkupfalse%
\ SourcesA{\isadigit{5}}{\isacharunderscore}L{\isadigit{0}}{\isacharcolon}\ {\isachardoublequoteopen}Sources\ level{\isadigit{0}}\ sA{\isadigit{5}}\ {\isacharequal}\ {\isacharbraceleft}\ sA{\isadigit{1}}{\isacharcomma}\ sA{\isadigit{2}}{\isacharcomma}\ sA{\isadigit{3}}{\isacharcomma}\ sA{\isadigit{4}}\ {\isacharbraceright}{\isachardoublequoteclose}\isanewline
\isadelimproof
\endisadelimproof
\isatagproof
\isacommand{proof}\isamarkupfalse%
\ {\isacharminus}\ \ \isanewline
\ \ \isacommand{have}\isamarkupfalse%
\ \ A{\isadigit{4}}s{\isacharcolon}{\isachardoublequoteopen}Sources\ level{\isadigit{0}}\ sA{\isadigit{4}}\ {\isacharequal}\ {\isacharbraceleft}\ sA{\isadigit{1}}{\isacharcomma}\ sA{\isadigit{2}}{\isacharcomma}\ sA{\isadigit{3}}{\isacharcomma}\ sA{\isadigit{4}}\ {\isacharbraceright}{\isachardoublequoteclose}\ \isacommand{by}\isamarkupfalse%
\ {\isacharparenleft}rule\ \ SourcesA{\isadigit{4}}{\isacharunderscore}L{\isadigit{0}}{\isacharparenright}\isanewline
\ \ \isacommand{have}\isamarkupfalse%
\ \ {\isachardoublequoteopen}Sources\ level{\isadigit{0}}\ sA{\isadigit{5}}\ {\isacharequal}\ {\isacharbraceleft}sA{\isadigit{4}}{\isacharbraceright}\ {\isasymunion}\ Sources\ level{\isadigit{0}}\ sA{\isadigit{4}}{\isachardoublequoteclose}\isanewline
\ \ \ \ \isacommand{by}\isamarkupfalse%
\ {\isacharparenleft}metis\ DSourcesA{\isadigit{5}}{\isacharunderscore}L{\isadigit{0}}\ Sources{\isacharunderscore}singleDSource{\isacharparenright}\ \isanewline
\ \ \isacommand{with}\isamarkupfalse%
\ A{\isadigit{4}}s\ \isacommand{show}\isamarkupfalse%
\ {\isacharquery}thesis\ \isacommand{by}\isamarkupfalse%
\ auto\isanewline
\isacommand{qed}\isamarkupfalse%
\endisatagproof
{\isafoldproof}%
\isadelimproof
\ \ \isanewline
\endisadelimproof
\isanewline
\isacommand{lemma}\isamarkupfalse%
\ SourcesA{\isadigit{6}}{\isacharunderscore}L{\isadigit{0}}{\isacharcolon}\ {\isachardoublequoteopen}Sources\ level{\isadigit{0}}\ sA{\isadigit{6}}\ {\isacharequal}\ {\isacharbraceleft}{\isacharbraceright}{\isachardoublequoteclose}\ \ \isanewline
\isadelimproof
\endisadelimproof
\isatagproof
\isacommand{by}\isamarkupfalse%
\ {\isacharparenleft}simp\ add{\isacharcolon}\ DSourcesA{\isadigit{6}}{\isacharunderscore}L{\isadigit{0}}\ DSourcesEmptySources{\isacharparenright}%
\endisatagproof
{\isafoldproof}%
\isadelimproof
\ \isanewline
\endisadelimproof
\isanewline
\isacommand{lemma}\isamarkupfalse%
\ SourcesA{\isadigit{7}}{\isacharunderscore}L{\isadigit{0}}{\isacharcolon}\ {\isachardoublequoteopen}Sources\ level{\isadigit{0}}\ sA{\isadigit{7}}\ {\isacharequal}\ {\isacharbraceleft}\ sA{\isadigit{6}}\ {\isacharbraceright}{\isachardoublequoteclose}\ \ \isanewline
\isadelimproof
\endisadelimproof
\isatagproof
\isacommand{by}\isamarkupfalse%
\ {\isacharparenleft}metis\ DSourcesA{\isadigit{7}}{\isacharunderscore}L{\isadigit{0}}\ SourcesA{\isadigit{6}}{\isacharunderscore}L{\isadigit{0}}\ SourcesEmptyDSources\ SourcesOnlyDSources\ singleton{\isacharunderscore}iff{\isacharparenright}%
\endisatagproof
{\isafoldproof}%
\isadelimproof
\isanewline
\endisadelimproof
\isanewline
\ \isanewline
\isacommand{lemma}\isamarkupfalse%
\ SourcesA{\isadigit{8}}{\isacharunderscore}L{\isadigit{0}}{\isacharcolon}\ {\isachardoublequoteopen}Sources\ level{\isadigit{0}}\ sA{\isadigit{8}}\ {\isacharequal}\ {\isacharbraceleft}\ sA{\isadigit{6}}{\isacharcomma}\ sA{\isadigit{7}}{\isacharcomma}\ sA{\isadigit{8}}{\isacharcomma}\ sA{\isadigit{9}}\ {\isacharbraceright}{\isachardoublequoteclose}\ \ \isanewline
\isadelimproof
\endisadelimproof
\isatagproof
\isacommand{proof}\isamarkupfalse%
\ {\isacharminus}\ \isanewline
\ \ \isacommand{have}\isamarkupfalse%
\ \ dA{\isadigit{8}}{\isacharcolon}{\isachardoublequoteopen}DSources\ level{\isadigit{0}}\ sA{\isadigit{8}}\ {\isacharequal}\ {\isacharbraceleft}sA{\isadigit{7}}{\isacharcomma}\ sA{\isadigit{9}}{\isacharbraceright}{\isachardoublequoteclose}\ \isacommand{by}\isamarkupfalse%
\ {\isacharparenleft}rule\ DSourcesA{\isadigit{8}}{\isacharunderscore}L{\isadigit{0}}{\isacharparenright}\isanewline
\ \ \isacommand{have}\isamarkupfalse%
\ \ dA{\isadigit{9}}{\isacharcolon}{\isachardoublequoteopen}DSources\ level{\isadigit{0}}\ sA{\isadigit{9}}\ {\isacharequal}\ {\isacharbraceleft}sA{\isadigit{8}}{\isacharbraceright}{\isachardoublequoteclose}\ \isacommand{by}\isamarkupfalse%
\ {\isacharparenleft}rule\ DSourcesA{\isadigit{9}}{\isacharunderscore}L{\isadigit{0}}{\isacharparenright}\isanewline
\ \ \isacommand{have}\isamarkupfalse%
\ {\isachardoublequoteopen}{\isacharparenleft}Sources\ level{\isadigit{0}}\ sA{\isadigit{8}}{\isacharparenright}\ {\isacharequal}\ {\isacharparenleft}DSources\ level{\isadigit{0}}\ sA{\isadigit{8}}{\isacharparenright}\ {\isasymunion}\ {\isacharparenleft}{\isasymUnion}\ S\ {\isasymin}\ {\isacharparenleft}DSources\ level{\isadigit{0}}\ sA{\isadigit{8}}{\isacharparenright}{\isachardot}\ {\isacharparenleft}Sources\ level{\isadigit{0}}\ S{\isacharparenright}{\isacharparenright}{\isachardoublequoteclose}\ \isanewline
\ \ \ \ \isacommand{by}\isamarkupfalse%
\ {\isacharparenleft}rule\ SourcesDef{\isacharparenright}\isanewline
\ \ \isacommand{hence}\isamarkupfalse%
\ sourcesA{\isadigit{8}}{\isacharcolon}{\isachardoublequoteopen}{\isacharparenleft}Sources\ level{\isadigit{0}}\ sA{\isadigit{8}}{\isacharparenright}\ {\isacharequal}\ {\isacharparenleft}{\isacharbraceleft}sA{\isadigit{7}}{\isacharcomma}\ sA{\isadigit{9}}{\isacharcomma}\ sA{\isadigit{6}}{\isacharbraceright}\ {\isasymunion}\ {\isacharparenleft}Sources\ level{\isadigit{0}}\ sA{\isadigit{9}}{\isacharparenright}{\isacharparenright}{\isachardoublequoteclose}\ \isanewline
\ \ \ \ \isacommand{by}\isamarkupfalse%
\ {\isacharparenleft}simp\ add{\isacharcolon}\ \ DSourcesA{\isadigit{8}}{\isacharunderscore}L{\isadigit{0}}\ SourcesA{\isadigit{7}}{\isacharunderscore}L{\isadigit{0}}{\isacharcomma}\ auto{\isacharparenright}\isanewline
\ \ \isacommand{have}\isamarkupfalse%
\ {\isachardoublequoteopen}{\isacharparenleft}Sources\ level{\isadigit{0}}\ sA{\isadigit{9}}{\isacharparenright}\ {\isacharequal}\ {\isacharparenleft}DSources\ level{\isadigit{0}}\ sA{\isadigit{9}}{\isacharparenright}\ {\isasymunion}\ {\isacharparenleft}{\isasymUnion}\ S\ {\isasymin}\ {\isacharparenleft}DSources\ level{\isadigit{0}}\ sA{\isadigit{9}}{\isacharparenright}{\isachardot}\ {\isacharparenleft}Sources\ level{\isadigit{0}}\ S{\isacharparenright}{\isacharparenright}{\isachardoublequoteclose}\ \isanewline
\ \ \ \ \isacommand{by}\isamarkupfalse%
\ {\isacharparenleft}rule\ SourcesDef{\isacharparenright}\isanewline
\ \ \isacommand{hence}\isamarkupfalse%
\ {\isachardoublequoteopen}{\isacharparenleft}Sources\ level{\isadigit{0}}\ sA{\isadigit{9}}{\isacharparenright}\ {\isacharequal}\ {\isacharparenleft}{\isacharbraceleft}sA{\isadigit{8}}{\isacharbraceright}\ {\isasymunion}\ {\isacharparenleft}Sources\ level{\isadigit{0}}\ sA{\isadigit{8}}{\isacharparenright}{\isacharparenright}{\isachardoublequoteclose}\ \isanewline
\ \ \ \ \isacommand{by}\isamarkupfalse%
\ {\isacharparenleft}simp\ add{\isacharcolon}\ \ DSourcesA{\isadigit{9}}{\isacharunderscore}L{\isadigit{0}}{\isacharparenright}\isanewline
\ \ \isacommand{with}\isamarkupfalse%
\ sourcesA{\isadigit{8}}\ \isacommand{have}\isamarkupfalse%
\ {\isachardoublequoteopen}{\isacharparenleft}Sources\ level{\isadigit{0}}\ sA{\isadigit{8}}{\isacharparenright}\ {\isacharequal}\ {\isacharbraceleft}sA{\isadigit{7}}{\isacharcomma}\ sA{\isadigit{9}}{\isacharcomma}\ sA{\isadigit{6}}{\isacharbraceright}\ {\isasymunion}\ {\isacharbraceleft}sA{\isadigit{8}}{\isacharbraceright}\ {\isasymunion}\ {\isacharbraceleft}sA{\isadigit{8}}{\isacharcomma}\ sA{\isadigit{9}}{\isacharbraceright}{\isachardoublequoteclose}\isanewline
\ \ \ \ \isacommand{by}\isamarkupfalse%
\ {\isacharparenleft}metis\ SourcesLoop{\isacharparenright}\ \ \isanewline
\ \ \isacommand{thus}\isamarkupfalse%
\ \ {\isacharquery}thesis\ \isacommand{by}\isamarkupfalse%
\ auto\isanewline
\isacommand{qed}\isamarkupfalse%
\endisatagproof
{\isafoldproof}%
\isadelimproof
\isanewline
\endisadelimproof
\isanewline
\isacommand{lemma}\isamarkupfalse%
\ SourcesA{\isadigit{9}}{\isacharunderscore}L{\isadigit{0}}{\isacharcolon}\ {\isachardoublequoteopen}Sources\ level{\isadigit{0}}\ sA{\isadigit{9}}\ {\isacharequal}\ {\isacharbraceleft}\ sA{\isadigit{6}}{\isacharcomma}\ sA{\isadigit{7}}{\isacharcomma}\ sA{\isadigit{8}}{\isacharcomma}\ sA{\isadigit{9}}\ {\isacharbraceright}{\isachardoublequoteclose}\ \ \isanewline
\isadelimproof
\endisadelimproof
\isatagproof
\isacommand{proof}\isamarkupfalse%
\ {\isacharminus}\ \isanewline
\ \ \isacommand{have}\isamarkupfalse%
\ {\isachardoublequoteopen}{\isacharparenleft}Sources\ level{\isadigit{0}}\ sA{\isadigit{9}}{\isacharparenright}\ {\isacharequal}\ {\isacharparenleft}DSources\ level{\isadigit{0}}\ sA{\isadigit{9}}{\isacharparenright}\ {\isasymunion}\ {\isacharparenleft}{\isasymUnion}\ S\ {\isasymin}\ {\isacharparenleft}DSources\ level{\isadigit{0}}\ sA{\isadigit{9}}{\isacharparenright}{\isachardot}\ {\isacharparenleft}Sources\ level{\isadigit{0}}\ S{\isacharparenright}{\isacharparenright}{\isachardoublequoteclose}\ \isanewline
\ \ \ \ \isacommand{by}\isamarkupfalse%
\ {\isacharparenleft}rule\ SourcesDef{\isacharparenright}\isanewline
\ \ \isacommand{hence}\isamarkupfalse%
\ sourcesA{\isadigit{9}}{\isacharcolon}{\isachardoublequoteopen}{\isacharparenleft}Sources\ level{\isadigit{0}}\ sA{\isadigit{9}}{\isacharparenright}\ {\isacharequal}\ {\isacharparenleft}{\isacharbraceleft}sA{\isadigit{8}}{\isacharbraceright}\ {\isasymunion}\ {\isacharparenleft}Sources\ level{\isadigit{0}}\ sA{\isadigit{8}}{\isacharparenright}{\isacharparenright}{\isachardoublequoteclose}\ \isanewline
\ \ \ \ \isacommand{by}\isamarkupfalse%
\ {\isacharparenleft}simp\ add{\isacharcolon}\ \ DSourcesA{\isadigit{9}}{\isacharunderscore}L{\isadigit{0}}{\isacharparenright}\isanewline
\ \ \ \isacommand{thus}\isamarkupfalse%
\ {\isacharquery}thesis\ \ \isacommand{by}\isamarkupfalse%
\ {\isacharparenleft}metis\ SourcesA{\isadigit{8}}{\isacharunderscore}L{\isadigit{0}}\ Un{\isacharunderscore}insert{\isacharunderscore}right\ insert{\isacharunderscore}absorb{\isadigit{2}}\ insert{\isacharunderscore}is{\isacharunderscore}Un{\isacharparenright}\ \isanewline
\isacommand{qed}\isamarkupfalse%
\isanewline
\ \isanewline
\isanewline
\isamarkupcmt{Abstraction level 1%
}
\endisatagproof
{\isafoldproof}%
\isadelimproof
\isanewline
\endisadelimproof
\isanewline
\isacommand{lemma}\isamarkupfalse%
\ A{\isadigit{1}}{\isadigit{2}}{\isacharunderscore}NotSource{\isacharunderscore}level{\isadigit{1}}{\isacharcolon}\ {\isachardoublequoteopen}isNotDSource\ level{\isadigit{1}}\ sA{\isadigit{1}}{\isadigit{2}}{\isachardoublequoteclose}\isanewline
\isadelimproof
\endisadelimproof
\isatagproof
\isacommand{by}\isamarkupfalse%
\ {\isacharparenleft}simp\ add{\isacharcolon}\ isNotDSource{\isacharunderscore}def{\isacharcomma}\ auto{\isacharcomma}\ \ case{\isacharunderscore}tac\ {\isachardoublequoteopen}Z{\isachardoublequoteclose}{\isacharcomma}\ auto{\isacharparenright}%
\endisatagproof
{\isafoldproof}%
\isadelimproof
\isanewline
\endisadelimproof
\isanewline
\isacommand{lemma}\isamarkupfalse%
\ A{\isadigit{2}}{\isadigit{1}}{\isacharunderscore}NotSource{\isacharunderscore}level{\isadigit{1}}{\isacharcolon}\ {\isachardoublequoteopen}isNotDSource\ level{\isadigit{1}}\ sA{\isadigit{2}}{\isadigit{1}}{\isachardoublequoteclose}\isanewline
\isadelimproof
\endisadelimproof
\isatagproof
\isacommand{by}\isamarkupfalse%
\ {\isacharparenleft}simp\ add{\isacharcolon}\ isNotDSource{\isacharunderscore}def{\isacharcomma}\ auto{\isacharcomma}\ \ case{\isacharunderscore}tac\ {\isachardoublequoteopen}Z{\isachardoublequoteclose}{\isacharcomma}\ auto{\isacharparenright}%
\endisatagproof
{\isafoldproof}%
\isadelimproof
\isanewline
\endisadelimproof
\isanewline
\isacommand{lemma}\isamarkupfalse%
\ A{\isadigit{5}}{\isacharunderscore}NotSource{\isacharunderscore}level{\isadigit{1}}{\isacharcolon}\ {\isachardoublequoteopen}isNotDSource\ level{\isadigit{1}}\ sA{\isadigit{5}}{\isachardoublequoteclose}\isanewline
\isadelimproof
\endisadelimproof
\isatagproof
\isacommand{by}\isamarkupfalse%
\ {\isacharparenleft}simp\ add{\isacharcolon}\ isNotDSource{\isacharunderscore}def{\isacharcomma}\ auto{\isacharcomma}\ \ case{\isacharunderscore}tac\ {\isachardoublequoteopen}Z{\isachardoublequoteclose}{\isacharcomma}\ auto{\isacharparenright}%
\endisatagproof
{\isafoldproof}%
\isadelimproof
\isanewline
\endisadelimproof
\isanewline
\isacommand{lemma}\isamarkupfalse%
\ A{\isadigit{9}}{\isadigit{2}}{\isacharunderscore}NotSource{\isacharunderscore}level{\isadigit{1}}{\isacharcolon}\ {\isachardoublequoteopen}isNotDSource\ level{\isadigit{1}}\ sA{\isadigit{9}}{\isadigit{2}}{\isachardoublequoteclose}\isanewline
\isadelimproof
\endisadelimproof
\isatagproof
\isacommand{by}\isamarkupfalse%
\ {\isacharparenleft}simp\ add{\isacharcolon}\ isNotDSource{\isacharunderscore}def{\isacharcomma}\ auto{\isacharcomma}\ \ case{\isacharunderscore}tac\ {\isachardoublequoteopen}Z{\isachardoublequoteclose}{\isacharcomma}\ auto{\isacharparenright}%
\endisatagproof
{\isafoldproof}%
\isadelimproof
\isanewline
\endisadelimproof
\isanewline
\isacommand{lemma}\isamarkupfalse%
\ A{\isadigit{9}}{\isadigit{3}}{\isacharunderscore}NotSource{\isacharunderscore}level{\isadigit{1}}{\isacharcolon}\ {\isachardoublequoteopen}isNotDSource\ level{\isadigit{1}}\ sA{\isadigit{9}}{\isadigit{3}}{\isachardoublequoteclose}\isanewline
\isadelimproof
\endisadelimproof
\isatagproof
\isacommand{by}\isamarkupfalse%
\ {\isacharparenleft}simp\ add{\isacharcolon}\ isNotDSource{\isacharunderscore}def{\isacharcomma}\ auto{\isacharcomma}\ \ case{\isacharunderscore}tac\ {\isachardoublequoteopen}Z{\isachardoublequoteclose}{\isacharcomma}\ auto{\isacharparenright}%
\endisatagproof
{\isafoldproof}%
\isadelimproof
\isanewline
\endisadelimproof
\isanewline
\isacommand{lemma}\isamarkupfalse%
\ A{\isadigit{1}}{\isadigit{1}}{\isacharunderscore}DAcc{\isacharunderscore}level{\isadigit{1}}{\isacharcolon}\ {\isachardoublequoteopen}DAcc\ level{\isadigit{1}}\ sA{\isadigit{1}}{\isadigit{1}}\ {\isacharequal}\ {\isacharbraceleft}\ sA{\isadigit{2}}{\isadigit{1}}{\isacharcomma}\ sA{\isadigit{2}}{\isadigit{2}}{\isacharcomma}\ sA{\isadigit{2}}{\isadigit{3}}\ {\isacharbraceright}{\isachardoublequoteclose}\ \isanewline
\isadelimproof
\endisadelimproof
\isatagproof
\isacommand{by}\isamarkupfalse%
\ {\isacharparenleft}simp\ add{\isacharcolon}\ DAcc{\isacharunderscore}def\ \ AbstrLevel{\isadigit{1}}{\isacharcomma}\ auto{\isacharparenright}%
\endisatagproof
{\isafoldproof}%
\isadelimproof
\isanewline
\endisadelimproof
\isanewline
\isacommand{lemma}\isamarkupfalse%
\ A{\isadigit{1}}{\isadigit{2}}{\isacharunderscore}DAcc{\isacharunderscore}level{\isadigit{1}}{\isacharcolon}\ {\isachardoublequoteopen}DAcc\ level{\isadigit{1}}\ sA{\isadigit{1}}{\isadigit{2}}\ {\isacharequal}\ {\isacharbraceleft}{\isacharbraceright}{\isachardoublequoteclose}\ \isanewline
\isadelimproof
\endisadelimproof
\isatagproof
\isacommand{by}\isamarkupfalse%
\ {\isacharparenleft}simp\ add{\isacharcolon}\ DAcc{\isacharunderscore}def\ \ AbstrLevel{\isadigit{1}}{\isacharcomma}\ auto{\isacharparenright}%
\endisatagproof
{\isafoldproof}%
\isadelimproof
\isanewline
\endisadelimproof
\isanewline
\isacommand{lemma}\isamarkupfalse%
\ A{\isadigit{2}}{\isadigit{1}}{\isacharunderscore}DAcc{\isacharunderscore}level{\isadigit{1}}{\isacharcolon}\ {\isachardoublequoteopen}DAcc\ level{\isadigit{1}}\ sA{\isadigit{2}}{\isadigit{1}}\ {\isacharequal}\ {\isacharbraceleft}{\isacharbraceright}{\isachardoublequoteclose}\ \isanewline
\isadelimproof
\endisadelimproof
\isatagproof
\isacommand{by}\isamarkupfalse%
\ {\isacharparenleft}simp\ add{\isacharcolon}\ DAcc{\isacharunderscore}def\ \ AbstrLevel{\isadigit{1}}{\isacharcomma}\ auto{\isacharparenright}%
\endisatagproof
{\isafoldproof}%
\isadelimproof
\isanewline
\endisadelimproof
\isanewline
\isacommand{lemma}\isamarkupfalse%
\ A{\isadigit{2}}{\isadigit{2}}{\isacharunderscore}DAcc{\isacharunderscore}level{\isadigit{1}}{\isacharcolon}\ {\isachardoublequoteopen}DAcc\ level{\isadigit{1}}\ sA{\isadigit{2}}{\isadigit{2}}\ {\isacharequal}\ {\isacharbraceleft}sA{\isadigit{3}}{\isadigit{1}}{\isacharbraceright}{\isachardoublequoteclose}\ \isanewline
\isadelimproof
\endisadelimproof
\isatagproof
\isacommand{by}\isamarkupfalse%
\ {\isacharparenleft}simp\ add{\isacharcolon}\ DAcc{\isacharunderscore}def\ \ AbstrLevel{\isadigit{1}}{\isacharcomma}\ auto{\isacharparenright}%
\endisatagproof
{\isafoldproof}%
\isadelimproof
\isanewline
\endisadelimproof
\isanewline
\isacommand{lemma}\isamarkupfalse%
\ A{\isadigit{2}}{\isadigit{3}}{\isacharunderscore}DAcc{\isacharunderscore}level{\isadigit{1}}{\isacharcolon}\ {\isachardoublequoteopen}DAcc\ level{\isadigit{1}}\ sA{\isadigit{2}}{\isadigit{3}}\ {\isacharequal}\ {\isacharbraceleft}sA{\isadigit{3}}{\isadigit{2}}{\isacharbraceright}{\isachardoublequoteclose}\ \isanewline
\isadelimproof
\endisadelimproof
\isatagproof
\isacommand{by}\isamarkupfalse%
\ {\isacharparenleft}simp\ add{\isacharcolon}\ DAcc{\isacharunderscore}def\ \ AbstrLevel{\isadigit{1}}{\isacharcomma}\ auto{\isacharparenright}%
\endisatagproof
{\isafoldproof}%
\isadelimproof
\isanewline
\endisadelimproof
\isanewline
\isacommand{lemma}\isamarkupfalse%
\ A{\isadigit{3}}{\isadigit{1}}{\isacharunderscore}DAcc{\isacharunderscore}level{\isadigit{1}}{\isacharcolon}\ {\isachardoublequoteopen}DAcc\ level{\isadigit{1}}\ sA{\isadigit{3}}{\isadigit{1}}\ {\isacharequal}\ {\isacharbraceleft}sA{\isadigit{4}}{\isadigit{1}}{\isacharbraceright}{\isachardoublequoteclose}\ \isanewline
\isadelimproof
\endisadelimproof
\isatagproof
\isacommand{by}\isamarkupfalse%
\ {\isacharparenleft}simp\ add{\isacharcolon}\ DAcc{\isacharunderscore}def\ \ AbstrLevel{\isadigit{1}}{\isacharcomma}\ auto{\isacharparenright}%
\endisatagproof
{\isafoldproof}%
\isadelimproof
\isanewline
\endisadelimproof
\isanewline
\isacommand{lemma}\isamarkupfalse%
\ A{\isadigit{3}}{\isadigit{2}}{\isacharunderscore}DAcc{\isacharunderscore}level{\isadigit{1}}{\isacharcolon}\ {\isachardoublequoteopen}DAcc\ level{\isadigit{1}}\ sA{\isadigit{3}}{\isadigit{2}}\ {\isacharequal}\ {\isacharbraceleft}sA{\isadigit{4}}{\isadigit{1}}{\isacharbraceright}{\isachardoublequoteclose}\ \isanewline
\isadelimproof
\endisadelimproof
\isatagproof
\isacommand{by}\isamarkupfalse%
\ {\isacharparenleft}simp\ add{\isacharcolon}\ DAcc{\isacharunderscore}def\ \ AbstrLevel{\isadigit{1}}{\isacharcomma}\ auto{\isacharparenright}%
\endisatagproof
{\isafoldproof}%
\isadelimproof
\isanewline
\endisadelimproof
\isanewline
\isacommand{lemma}\isamarkupfalse%
\ A{\isadigit{4}}{\isadigit{1}}{\isacharunderscore}DAcc{\isacharunderscore}level{\isadigit{1}}{\isacharcolon}\ {\isachardoublequoteopen}DAcc\ level{\isadigit{1}}\ sA{\isadigit{4}}{\isadigit{1}}\ {\isacharequal}\ {\isacharbraceleft}sA{\isadigit{2}}{\isadigit{2}}{\isacharbraceright}{\isachardoublequoteclose}\ \isanewline
\isadelimproof
\endisadelimproof
\isatagproof
\isacommand{by}\isamarkupfalse%
\ {\isacharparenleft}simp\ add{\isacharcolon}\ DAcc{\isacharunderscore}def\ \ AbstrLevel{\isadigit{1}}{\isacharcomma}\ auto{\isacharparenright}%
\endisatagproof
{\isafoldproof}%
\isadelimproof
\isanewline
\endisadelimproof
\isanewline
\isacommand{lemma}\isamarkupfalse%
\ A{\isadigit{4}}{\isadigit{2}}{\isacharunderscore}DAcc{\isacharunderscore}level{\isadigit{1}}{\isacharcolon}\ {\isachardoublequoteopen}DAcc\ level{\isadigit{1}}\ sA{\isadigit{4}}{\isadigit{2}}\ {\isacharequal}\ {\isacharbraceleft}sA{\isadigit{5}}{\isacharbraceright}{\isachardoublequoteclose}\ \isanewline
\isadelimproof
\endisadelimproof
\isatagproof
\isacommand{by}\isamarkupfalse%
\ {\isacharparenleft}simp\ add{\isacharcolon}\ DAcc{\isacharunderscore}def\ \ AbstrLevel{\isadigit{1}}{\isacharcomma}\ auto{\isacharparenright}%
\endisatagproof
{\isafoldproof}%
\isadelimproof
\isanewline
\endisadelimproof
\isanewline
\isacommand{lemma}\isamarkupfalse%
\ A{\isadigit{5}}{\isacharunderscore}DAcc{\isacharunderscore}level{\isadigit{1}}{\isacharcolon}\ {\isachardoublequoteopen}DAcc\ level{\isadigit{1}}\ sA{\isadigit{5}}\ {\isacharequal}\ {\isacharbraceleft}{\isacharbraceright}{\isachardoublequoteclose}\ \isanewline
\isadelimproof
\endisadelimproof
\isatagproof
\isacommand{by}\isamarkupfalse%
\ {\isacharparenleft}simp\ add{\isacharcolon}\ DAcc{\isacharunderscore}def\ \ AbstrLevel{\isadigit{1}}{\isacharcomma}\ auto{\isacharparenright}%
\endisatagproof
{\isafoldproof}%
\isadelimproof
\isanewline
\endisadelimproof
\isanewline
\isacommand{lemma}\isamarkupfalse%
\ A{\isadigit{6}}{\isacharunderscore}DAcc{\isacharunderscore}level{\isadigit{1}}{\isacharcolon}\ {\isachardoublequoteopen}DAcc\ level{\isadigit{1}}\ sA{\isadigit{6}}\ {\isacharequal}\ {\isacharbraceleft}sA{\isadigit{7}}{\isadigit{1}}{\isacharcomma}\ sA{\isadigit{7}}{\isadigit{2}}{\isacharbraceright}{\isachardoublequoteclose}\ \isanewline
\isadelimproof
\endisadelimproof
\isatagproof
\isacommand{by}\isamarkupfalse%
\ {\isacharparenleft}simp\ add{\isacharcolon}\ DAcc{\isacharunderscore}def\ \ AbstrLevel{\isadigit{1}}{\isacharcomma}\ auto{\isacharparenright}%
\endisatagproof
{\isafoldproof}%
\isadelimproof
\isanewline
\endisadelimproof
\isanewline
\isacommand{lemma}\isamarkupfalse%
\ A{\isadigit{7}}{\isadigit{1}}{\isacharunderscore}DAcc{\isacharunderscore}level{\isadigit{1}}{\isacharcolon}\ {\isachardoublequoteopen}DAcc\ level{\isadigit{1}}\ sA{\isadigit{7}}{\isadigit{1}}\ {\isacharequal}\ {\isacharbraceleft}sA{\isadigit{8}}{\isadigit{1}}{\isacharbraceright}{\isachardoublequoteclose}\ \isanewline
\isadelimproof
\endisadelimproof
\isatagproof
\isacommand{by}\isamarkupfalse%
\ {\isacharparenleft}simp\ add{\isacharcolon}\ DAcc{\isacharunderscore}def\ \ AbstrLevel{\isadigit{1}}{\isacharcomma}\ auto{\isacharparenright}%
\endisatagproof
{\isafoldproof}%
\isadelimproof
\isanewline
\endisadelimproof
\isanewline
\isacommand{lemma}\isamarkupfalse%
\ A{\isadigit{7}}{\isadigit{2}}{\isacharunderscore}DAcc{\isacharunderscore}level{\isadigit{1}}{\isacharcolon}\ {\isachardoublequoteopen}DAcc\ level{\isadigit{1}}\ sA{\isadigit{7}}{\isadigit{2}}\ {\isacharequal}\ {\isacharbraceleft}sA{\isadigit{8}}{\isadigit{2}}{\isacharbraceright}{\isachardoublequoteclose}\ \isanewline
\isadelimproof
\endisadelimproof
\isatagproof
\isacommand{by}\isamarkupfalse%
\ {\isacharparenleft}simp\ add{\isacharcolon}\ DAcc{\isacharunderscore}def\ \ AbstrLevel{\isadigit{1}}{\isacharcomma}\ auto{\isacharparenright}%
\endisatagproof
{\isafoldproof}%
\isadelimproof
\isanewline
\endisadelimproof
\isanewline
\isacommand{lemma}\isamarkupfalse%
\ A{\isadigit{8}}{\isadigit{1}}{\isacharunderscore}DAcc{\isacharunderscore}level{\isadigit{1}}{\isacharcolon}\ {\isachardoublequoteopen}DAcc\ level{\isadigit{1}}\ sA{\isadigit{8}}{\isadigit{1}}\ {\isacharequal}\ {\isacharbraceleft}sA{\isadigit{9}}{\isadigit{1}}{\isacharcomma}\ sA{\isadigit{9}}{\isadigit{2}}{\isacharbraceright}{\isachardoublequoteclose}\ \isanewline
\isadelimproof
\endisadelimproof
\isatagproof
\isacommand{by}\isamarkupfalse%
\ {\isacharparenleft}simp\ add{\isacharcolon}\ DAcc{\isacharunderscore}def\ \ AbstrLevel{\isadigit{1}}{\isacharcomma}\ auto{\isacharparenright}%
\endisatagproof
{\isafoldproof}%
\isadelimproof
\isanewline
\endisadelimproof
\isanewline
\isacommand{lemma}\isamarkupfalse%
\ A{\isadigit{8}}{\isadigit{2}}{\isacharunderscore}DAcc{\isacharunderscore}level{\isadigit{1}}{\isacharcolon}\ {\isachardoublequoteopen}DAcc\ level{\isadigit{1}}\ sA{\isadigit{8}}{\isadigit{2}}\ {\isacharequal}\ {\isacharbraceleft}sA{\isadigit{9}}{\isadigit{3}}{\isacharbraceright}{\isachardoublequoteclose}\ \isanewline
\isadelimproof
\endisadelimproof
\isatagproof
\isacommand{by}\isamarkupfalse%
\ {\isacharparenleft}simp\ add{\isacharcolon}\ DAcc{\isacharunderscore}def\ \ AbstrLevel{\isadigit{1}}{\isacharcomma}\ auto{\isacharparenright}%
\endisatagproof
{\isafoldproof}%
\isadelimproof
\isanewline
\endisadelimproof
\isanewline
\isacommand{lemma}\isamarkupfalse%
\ A{\isadigit{9}}{\isadigit{1}}{\isacharunderscore}DAcc{\isacharunderscore}level{\isadigit{1}}{\isacharcolon}\ {\isachardoublequoteopen}DAcc\ level{\isadigit{1}}\ sA{\isadigit{9}}{\isadigit{1}}\ {\isacharequal}\ {\isacharbraceleft}sA{\isadigit{8}}{\isadigit{1}}{\isacharbraceright}{\isachardoublequoteclose}\ \isanewline
\isadelimproof
\endisadelimproof
\isatagproof
\isacommand{by}\isamarkupfalse%
\ {\isacharparenleft}simp\ add{\isacharcolon}\ DAcc{\isacharunderscore}def\ \ AbstrLevel{\isadigit{1}}{\isacharcomma}\ auto{\isacharparenright}%
\endisatagproof
{\isafoldproof}%
\isadelimproof
\isanewline
\endisadelimproof
\isanewline
\isacommand{lemma}\isamarkupfalse%
\ A{\isadigit{9}}{\isadigit{2}}{\isacharunderscore}DAcc{\isacharunderscore}level{\isadigit{1}}{\isacharcolon}\ {\isachardoublequoteopen}DAcc\ level{\isadigit{1}}\ sA{\isadigit{9}}{\isadigit{2}}\ {\isacharequal}\ {\isacharbraceleft}{\isacharbraceright}{\isachardoublequoteclose}\ \isanewline
\isadelimproof
\endisadelimproof
\isatagproof
\isacommand{by}\isamarkupfalse%
\ {\isacharparenleft}simp\ add{\isacharcolon}\ DAcc{\isacharunderscore}def\ \ AbstrLevel{\isadigit{1}}{\isacharcomma}\ auto{\isacharparenright}%
\endisatagproof
{\isafoldproof}%
\isadelimproof
\isanewline
\endisadelimproof
\isanewline
\isacommand{lemma}\isamarkupfalse%
\ A{\isadigit{9}}{\isadigit{3}}{\isacharunderscore}DAcc{\isacharunderscore}level{\isadigit{1}}{\isacharcolon}\ {\isachardoublequoteopen}DAcc\ level{\isadigit{1}}\ sA{\isadigit{9}}{\isadigit{3}}\ {\isacharequal}\ {\isacharbraceleft}{\isacharbraceright}{\isachardoublequoteclose}\ \isanewline
\isadelimproof
\endisadelimproof
\isatagproof
\isacommand{by}\isamarkupfalse%
\ {\isacharparenleft}simp\ add{\isacharcolon}\ DAcc{\isacharunderscore}def\ \ AbstrLevel{\isadigit{1}}{\isacharcomma}\ auto{\isacharparenright}%
\endisatagproof
{\isafoldproof}%
\isadelimproof
\isanewline
\endisadelimproof
\isanewline
\isacommand{lemma}\isamarkupfalse%
\ A{\isadigit{4}}{\isadigit{2}}{\isacharunderscore}NSources{\isacharunderscore}L{\isadigit{1}}{\isacharcolon}\isanewline
{\isachardoublequoteopen}{\isasymforall}\ C\ {\isasymin}\ {\isacharparenleft}AbstrLevel\ level{\isadigit{1}}{\isacharparenright}{\isachardot}\ C\ {\isasymnoteq}\ sA{\isadigit{5}}\ {\isasymlongrightarrow}\ sA{\isadigit{4}}{\isadigit{2}}\ {\isasymnotin}\ {\isacharparenleft}Sources\ level{\isadigit{1}}\ C{\isacharparenright}{\isachardoublequoteclose}\isanewline
\isadelimproof
\endisadelimproof
\isatagproof
\isacommand{by}\isamarkupfalse%
\ {\isacharparenleft}metis\ A{\isadigit{4}}{\isadigit{2}}{\isacharunderscore}DAcc{\isacharunderscore}level{\isadigit{1}}\ A{\isadigit{5}}{\isacharunderscore}NotSource{\isacharunderscore}level{\isadigit{1}}\ singleDSourceEmpty{\isadigit{4}}isNotSource{\isacharparenright}%
\endisatagproof
{\isafoldproof}%
\isadelimproof
\isanewline
\endisadelimproof
\isanewline
\isacommand{lemma}\isamarkupfalse%
\ A{\isadigit{5}}{\isacharunderscore}NotSourceSet{\isacharunderscore}level{\isadigit{1}}\ {\isacharcolon}\isanewline
{\isachardoublequoteopen}{\isasymforall}\ C\ \ {\isasymin}\ {\isacharparenleft}AbstrLevel\ level{\isadigit{1}}{\isacharparenright}{\isachardot}\ sA{\isadigit{5}}\ {\isasymnotin}\ {\isacharparenleft}Sources\ level{\isadigit{1}}\ C{\isacharparenright}{\isachardoublequoteclose}\isanewline
\isadelimproof
\endisadelimproof
\isatagproof
\isacommand{by}\isamarkupfalse%
\ {\isacharparenleft}metis\ A{\isadigit{5}}{\isacharunderscore}NotSource{\isacharunderscore}level{\isadigit{1}}\ isNotSource{\isacharunderscore}Sources{\isacharparenright}%
\endisatagproof
{\isafoldproof}%
\isadelimproof
\isanewline
\endisadelimproof
\isanewline
\isacommand{lemma}\isamarkupfalse%
\ A{\isadigit{9}}{\isadigit{2}}{\isacharunderscore}NotSourceSet{\isacharunderscore}level{\isadigit{1}}\ {\isacharcolon}\isanewline
{\isachardoublequoteopen}{\isasymforall}\ C\ \ {\isasymin}\ {\isacharparenleft}AbstrLevel\ level{\isadigit{1}}{\isacharparenright}{\isachardot}\ sA{\isadigit{9}}{\isadigit{2}}\ {\isasymnotin}\ {\isacharparenleft}Sources\ level{\isadigit{1}}\ C{\isacharparenright}{\isachardoublequoteclose}\ \isanewline
\isadelimproof
\endisadelimproof
\isatagproof
\isacommand{by}\isamarkupfalse%
\ {\isacharparenleft}metis\ A{\isadigit{9}}{\isadigit{2}}{\isacharunderscore}NotSource{\isacharunderscore}level{\isadigit{1}}\ isNotSource{\isacharunderscore}Sources{\isacharparenright}%
\endisatagproof
{\isafoldproof}%
\isadelimproof
\isanewline
\endisadelimproof
\isanewline
\isacommand{lemma}\isamarkupfalse%
\ A{\isadigit{9}}{\isadigit{3}}{\isacharunderscore}NotSourceSet{\isacharunderscore}level{\isadigit{1}}\ {\isacharcolon}\isanewline
{\isachardoublequoteopen}{\isasymforall}\ C\ \ {\isasymin}\ {\isacharparenleft}AbstrLevel\ level{\isadigit{1}}{\isacharparenright}{\isachardot}\ sA{\isadigit{9}}{\isadigit{3}}\ {\isasymnotin}\ {\isacharparenleft}Sources\ level{\isadigit{1}}\ C{\isacharparenright}{\isachardoublequoteclose}\ \isanewline
\isadelimproof
\endisadelimproof
\isatagproof
\isacommand{by}\isamarkupfalse%
\ {\isacharparenleft}metis\ A{\isadigit{9}}{\isadigit{3}}{\isacharunderscore}NotSource{\isacharunderscore}level{\isadigit{1}}\ isNotSource{\isacharunderscore}Sources{\isacharparenright}%
\endisatagproof
{\isafoldproof}%
\isadelimproof
\isanewline
\endisadelimproof
\isanewline
\isacommand{lemma}\isamarkupfalse%
\ DSourcesA{\isadigit{1}}{\isadigit{1}}{\isacharunderscore}L{\isadigit{1}}{\isacharcolon}\ {\isachardoublequoteopen}DSources\ level{\isadigit{1}}\ sA{\isadigit{1}}{\isadigit{1}}\ {\isacharequal}\ {\isacharbraceleft}{\isacharbraceright}{\isachardoublequoteclose}\isanewline
\isadelimproof
\endisadelimproof
\isatagproof
\isacommand{by}\isamarkupfalse%
\ {\isacharparenleft}simp\ add{\isacharcolon}\ DSources{\isacharunderscore}def{\isacharcomma}\ auto{\isacharcomma}\ case{\isacharunderscore}tac\ {\isachardoublequoteopen}x{\isachardoublequoteclose}{\isacharcomma}\ auto{\isacharparenright}%
\endisatagproof
{\isafoldproof}%
\isadelimproof
\ \isanewline
\endisadelimproof
\isanewline
\isacommand{lemma}\isamarkupfalse%
\ DSourcesA{\isadigit{1}}{\isadigit{2}}{\isacharunderscore}L{\isadigit{1}}{\isacharcolon}\ {\isachardoublequoteopen}DSources\ level{\isadigit{1}}\ sA{\isadigit{1}}{\isadigit{2}}\ {\isacharequal}\ {\isacharbraceleft}{\isacharbraceright}{\isachardoublequoteclose}\isanewline
\isadelimproof
\endisadelimproof
\isatagproof
\isacommand{by}\isamarkupfalse%
\ {\isacharparenleft}simp\ add{\isacharcolon}\ DSources{\isacharunderscore}def\ AbstrLevel{\isadigit{1}}{\isacharcomma}\ auto{\isacharparenright}%
\endisatagproof
{\isafoldproof}%
\isadelimproof
\ \isanewline
\endisadelimproof
\isanewline
\isacommand{lemma}\isamarkupfalse%
\ DSourcesA{\isadigit{2}}{\isadigit{1}}{\isacharunderscore}L{\isadigit{1}}{\isacharcolon}\ {\isachardoublequoteopen}DSources\ level{\isadigit{1}}\ sA{\isadigit{2}}{\isadigit{1}}\ {\isacharequal}\ {\isacharbraceleft}sA{\isadigit{1}}{\isadigit{1}}{\isacharbraceright}{\isachardoublequoteclose}\isanewline
\isadelimproof
\endisadelimproof
\isatagproof
\isacommand{by}\isamarkupfalse%
\ {\isacharparenleft}simp\ add{\isacharcolon}\ DSources{\isacharunderscore}def\ AbstrLevel{\isadigit{1}}{\isacharcomma}\ auto{\isacharparenright}%
\endisatagproof
{\isafoldproof}%
\isadelimproof
\ \isanewline
\endisadelimproof
\isanewline
\isacommand{lemma}\isamarkupfalse%
\ DSourcesA{\isadigit{2}}{\isadigit{2}}{\isacharunderscore}L{\isadigit{1}}{\isacharcolon}\ {\isachardoublequoteopen}DSources\ level{\isadigit{1}}\ sA{\isadigit{2}}{\isadigit{2}}\ {\isacharequal}\ {\isacharbraceleft}sA{\isadigit{1}}{\isadigit{1}}{\isacharcomma}\ sA{\isadigit{4}}{\isadigit{1}}{\isacharbraceright}{\isachardoublequoteclose}\isanewline
\isadelimproof
\endisadelimproof
\isatagproof
\isacommand{by}\isamarkupfalse%
\ {\isacharparenleft}simp\ add{\isacharcolon}\ DSources{\isacharunderscore}def\ AbstrLevel{\isadigit{1}}{\isacharcomma}\ auto{\isacharparenright}%
\endisatagproof
{\isafoldproof}%
\isadelimproof
\ \isanewline
\endisadelimproof
\isanewline
\isacommand{lemma}\isamarkupfalse%
\ DSourcesA{\isadigit{2}}{\isadigit{3}}{\isacharunderscore}L{\isadigit{1}}{\isacharcolon}\ {\isachardoublequoteopen}DSources\ level{\isadigit{1}}\ sA{\isadigit{2}}{\isadigit{3}}\ {\isacharequal}\ {\isacharbraceleft}sA{\isadigit{1}}{\isadigit{1}}{\isacharbraceright}{\isachardoublequoteclose}\isanewline
\isadelimproof
\endisadelimproof
\isatagproof
\isacommand{by}\isamarkupfalse%
\ {\isacharparenleft}simp\ add{\isacharcolon}\ DSources{\isacharunderscore}def\ AbstrLevel{\isadigit{1}}{\isacharcomma}\ auto{\isacharparenright}%
\endisatagproof
{\isafoldproof}%
\isadelimproof
\ \isanewline
\endisadelimproof
\isanewline
\isacommand{lemma}\isamarkupfalse%
\ DSourcesA{\isadigit{3}}{\isadigit{1}}{\isacharunderscore}L{\isadigit{1}}{\isacharcolon}\ {\isachardoublequoteopen}DSources\ level{\isadigit{1}}\ sA{\isadigit{3}}{\isadigit{1}}\ {\isacharequal}\ {\isacharbraceleft}\ sA{\isadigit{2}}{\isadigit{2}}\ {\isacharbraceright}{\isachardoublequoteclose}\isanewline
\isadelimproof
\endisadelimproof
\isatagproof
\isacommand{by}\isamarkupfalse%
\ {\isacharparenleft}simp\ add{\isacharcolon}\ DSources{\isacharunderscore}def\ AbstrLevel{\isadigit{1}}{\isacharcomma}\ auto{\isacharparenright}%
\endisatagproof
{\isafoldproof}%
\isadelimproof
\ \isanewline
\endisadelimproof
\isanewline
\isacommand{lemma}\isamarkupfalse%
\ DSourcesA{\isadigit{3}}{\isadigit{2}}{\isacharunderscore}L{\isadigit{1}}{\isacharcolon}\ {\isachardoublequoteopen}DSources\ level{\isadigit{1}}\ sA{\isadigit{3}}{\isadigit{2}}\ {\isacharequal}\ {\isacharbraceleft}\ sA{\isadigit{2}}{\isadigit{3}}\ {\isacharbraceright}{\isachardoublequoteclose}\isanewline
\isadelimproof
\endisadelimproof
\isatagproof
\isacommand{by}\isamarkupfalse%
\ {\isacharparenleft}simp\ add{\isacharcolon}\ DSources{\isacharunderscore}def\ AbstrLevel{\isadigit{1}}{\isacharcomma}\ auto{\isacharparenright}%
\endisatagproof
{\isafoldproof}%
\isadelimproof
\ \isanewline
\endisadelimproof
\isanewline
\isacommand{lemma}\isamarkupfalse%
\ DSourcesA{\isadigit{4}}{\isadigit{1}}{\isacharunderscore}L{\isadigit{1}}{\isacharcolon}\ {\isachardoublequoteopen}DSources\ level{\isadigit{1}}\ sA{\isadigit{4}}{\isadigit{1}}\ {\isacharequal}\ {\isacharbraceleft}\ sA{\isadigit{3}}{\isadigit{1}}{\isacharcomma}\ sA{\isadigit{3}}{\isadigit{2}}\ {\isacharbraceright}{\isachardoublequoteclose}\isanewline
\isadelimproof
\endisadelimproof
\isatagproof
\isacommand{by}\isamarkupfalse%
\ {\isacharparenleft}simp\ add{\isacharcolon}\ DSources{\isacharunderscore}def\ AbstrLevel{\isadigit{1}}{\isacharcomma}\ auto{\isacharparenright}%
\endisatagproof
{\isafoldproof}%
\isadelimproof
\ \isanewline
\endisadelimproof
\isanewline
\isacommand{lemma}\isamarkupfalse%
\ DSourcesA{\isadigit{4}}{\isadigit{2}}{\isacharunderscore}L{\isadigit{1}}{\isacharcolon}\ {\isachardoublequoteopen}DSources\ level{\isadigit{1}}\ sA{\isadigit{4}}{\isadigit{2}}\ {\isacharequal}\ {\isacharbraceleft}{\isacharbraceright}{\isachardoublequoteclose}\isanewline
\isadelimproof
\endisadelimproof
\isatagproof
\isacommand{by}\isamarkupfalse%
\ {\isacharparenleft}simp\ add{\isacharcolon}\ DSources{\isacharunderscore}def\ AbstrLevel{\isadigit{1}}{\isacharcomma}\ auto{\isacharparenright}%
\endisatagproof
{\isafoldproof}%
\isadelimproof
\ \isanewline
\endisadelimproof
\isanewline
\isacommand{lemma}\isamarkupfalse%
\ DSourcesA{\isadigit{5}}{\isacharunderscore}L{\isadigit{1}}{\isacharcolon}\ {\isachardoublequoteopen}DSources\ level{\isadigit{1}}\ sA{\isadigit{5}}\ {\isacharequal}\ {\isacharbraceleft}\ sA{\isadigit{4}}{\isadigit{2}}\ {\isacharbraceright}{\isachardoublequoteclose}\isanewline
\isadelimproof
\endisadelimproof
\isatagproof
\isacommand{by}\isamarkupfalse%
\ {\isacharparenleft}simp\ add{\isacharcolon}\ DSources{\isacharunderscore}def\ AbstrLevel{\isadigit{1}}{\isacharcomma}\ auto{\isacharparenright}%
\endisatagproof
{\isafoldproof}%
\isadelimproof
\ \ \isanewline
\endisadelimproof
\isanewline
\isacommand{lemma}\isamarkupfalse%
\ DSourcesA{\isadigit{6}}{\isacharunderscore}L{\isadigit{1}}{\isacharcolon}\ {\isachardoublequoteopen}DSources\ level{\isadigit{1}}\ sA{\isadigit{6}}\ {\isacharequal}\ {\isacharbraceleft}{\isacharbraceright}{\isachardoublequoteclose}\isanewline
\isadelimproof
\endisadelimproof
\isatagproof
\isacommand{by}\isamarkupfalse%
\ {\isacharparenleft}simp\ add{\isacharcolon}\ DSources{\isacharunderscore}def\ AbstrLevel{\isadigit{1}}{\isacharcomma}\ auto{\isacharparenright}%
\endisatagproof
{\isafoldproof}%
\isadelimproof
\ \ \isanewline
\endisadelimproof
\isanewline
\isacommand{lemma}\isamarkupfalse%
\ DSourcesA{\isadigit{7}}{\isadigit{1}}{\isacharunderscore}L{\isadigit{1}}{\isacharcolon}\ {\isachardoublequoteopen}DSources\ level{\isadigit{1}}\ sA{\isadigit{7}}{\isadigit{1}}\ {\isacharequal}\ {\isacharbraceleft}\ sA{\isadigit{6}}\ {\isacharbraceright}{\isachardoublequoteclose}\isanewline
\isadelimproof
\endisadelimproof
\isatagproof
\isacommand{by}\isamarkupfalse%
\ {\isacharparenleft}simp\ add{\isacharcolon}\ DSources{\isacharunderscore}def\ AbstrLevel{\isadigit{1}}{\isacharcomma}\ auto{\isacharparenright}%
\endisatagproof
{\isafoldproof}%
\isadelimproof
\ \ \isanewline
\endisadelimproof
\isanewline
\isacommand{lemma}\isamarkupfalse%
\ DSourcesA{\isadigit{7}}{\isadigit{2}}{\isacharunderscore}L{\isadigit{1}}{\isacharcolon}\ {\isachardoublequoteopen}DSources\ level{\isadigit{1}}\ sA{\isadigit{7}}{\isadigit{2}}\ {\isacharequal}\ {\isacharbraceleft}\ sA{\isadigit{6}}\ {\isacharbraceright}{\isachardoublequoteclose}\isanewline
\isadelimproof
\endisadelimproof
\isatagproof
\isacommand{by}\isamarkupfalse%
\ {\isacharparenleft}simp\ add{\isacharcolon}\ DSources{\isacharunderscore}def\ AbstrLevel{\isadigit{1}}{\isacharcomma}\ auto{\isacharparenright}%
\endisatagproof
{\isafoldproof}%
\isadelimproof
\ \ \isanewline
\endisadelimproof
\isanewline
\isacommand{lemma}\isamarkupfalse%
\ DSourcesA{\isadigit{8}}{\isadigit{1}}{\isacharunderscore}L{\isadigit{1}}{\isacharcolon}\ {\isachardoublequoteopen}DSources\ level{\isadigit{1}}\ sA{\isadigit{8}}{\isadigit{1}}\ {\isacharequal}\ {\isacharbraceleft}\ sA{\isadigit{7}}{\isadigit{1}}{\isacharcomma}\ sA{\isadigit{9}}{\isadigit{1}}\ {\isacharbraceright}{\isachardoublequoteclose}\isanewline
\isadelimproof
\endisadelimproof
\isatagproof
\isacommand{by}\isamarkupfalse%
\ {\isacharparenleft}simp\ add{\isacharcolon}\ DSources{\isacharunderscore}def\ AbstrLevel{\isadigit{1}}{\isacharcomma}\ auto{\isacharparenright}%
\endisatagproof
{\isafoldproof}%
\isadelimproof
\ \ \isanewline
\endisadelimproof
\isanewline
\isacommand{lemma}\isamarkupfalse%
\ DSourcesA{\isadigit{8}}{\isadigit{2}}{\isacharunderscore}L{\isadigit{1}}{\isacharcolon}\ {\isachardoublequoteopen}DSources\ level{\isadigit{1}}\ sA{\isadigit{8}}{\isadigit{2}}\ {\isacharequal}\ {\isacharbraceleft}\ sA{\isadigit{7}}{\isadigit{2}}\ {\isacharbraceright}{\isachardoublequoteclose}\isanewline
\isadelimproof
\endisadelimproof
\isatagproof
\isacommand{by}\isamarkupfalse%
\ {\isacharparenleft}simp\ add{\isacharcolon}\ DSources{\isacharunderscore}def\ AbstrLevel{\isadigit{1}}{\isacharcomma}\ auto{\isacharparenright}%
\endisatagproof
{\isafoldproof}%
\isadelimproof
\ \ \isanewline
\endisadelimproof
\isanewline
\isacommand{lemma}\isamarkupfalse%
\ DSourcesA{\isadigit{9}}{\isadigit{1}}{\isacharunderscore}L{\isadigit{1}}{\isacharcolon}\ {\isachardoublequoteopen}DSources\ level{\isadigit{1}}\ sA{\isadigit{9}}{\isadigit{1}}\ {\isacharequal}\ {\isacharbraceleft}\ sA{\isadigit{8}}{\isadigit{1}}\ {\isacharbraceright}{\isachardoublequoteclose}\isanewline
\isadelimproof
\endisadelimproof
\isatagproof
\isacommand{by}\isamarkupfalse%
\ {\isacharparenleft}simp\ add{\isacharcolon}\ DSources{\isacharunderscore}def\ AbstrLevel{\isadigit{1}}{\isacharcomma}\ auto{\isacharparenright}%
\endisatagproof
{\isafoldproof}%
\isadelimproof
\ \ \isanewline
\endisadelimproof
\isanewline
\isacommand{lemma}\isamarkupfalse%
\ DSourcesA{\isadigit{9}}{\isadigit{2}}{\isacharunderscore}L{\isadigit{1}}{\isacharcolon}\ {\isachardoublequoteopen}DSources\ level{\isadigit{1}}\ sA{\isadigit{9}}{\isadigit{2}}\ {\isacharequal}\ {\isacharbraceleft}\ sA{\isadigit{8}}{\isadigit{1}}\ {\isacharbraceright}{\isachardoublequoteclose}\isanewline
\isadelimproof
\endisadelimproof
\isatagproof
\isacommand{by}\isamarkupfalse%
\ {\isacharparenleft}simp\ add{\isacharcolon}\ DSources{\isacharunderscore}def\ AbstrLevel{\isadigit{1}}{\isacharcomma}\ auto{\isacharparenright}%
\endisatagproof
{\isafoldproof}%
\isadelimproof
\ \ \isanewline
\endisadelimproof
\isanewline
\isacommand{lemma}\isamarkupfalse%
\ DSourcesA{\isadigit{9}}{\isadigit{3}}{\isacharunderscore}L{\isadigit{1}}{\isacharcolon}\ {\isachardoublequoteopen}DSources\ level{\isadigit{1}}\ sA{\isadigit{9}}{\isadigit{3}}\ {\isacharequal}\ {\isacharbraceleft}\ sA{\isadigit{8}}{\isadigit{2}}\ {\isacharbraceright}{\isachardoublequoteclose}\isanewline
\isadelimproof
\endisadelimproof
\isatagproof
\isacommand{by}\isamarkupfalse%
\ {\isacharparenleft}simp\ add{\isacharcolon}\ DSources{\isacharunderscore}def\ AbstrLevel{\isadigit{1}}{\isacharcomma}\ auto{\isacharparenright}%
\endisatagproof
{\isafoldproof}%
\isadelimproof
\ \ \isanewline
\endisadelimproof
\isanewline
\isacommand{lemma}\isamarkupfalse%
\ A{\isadigit{8}}{\isadigit{2}}{\isacharunderscore}Acc{\isacharcolon}\ {\isachardoublequoteopen}{\isacharparenleft}Acc\ level{\isadigit{1}}\ sA{\isadigit{8}}{\isadigit{2}}{\isacharparenright}\ {\isacharequal}\ {\isacharbraceleft}sA{\isadigit{9}}{\isadigit{3}}{\isacharbraceright}{\isachardoublequoteclose}\isanewline
\isadelimproof
\endisadelimproof
\isatagproof
\isacommand{by}\isamarkupfalse%
\ {\isacharparenleft}metis\ A{\isadigit{8}}{\isadigit{2}}{\isacharunderscore}DAcc{\isacharunderscore}level{\isadigit{1}}\ A{\isadigit{9}}{\isadigit{3}}{\isacharunderscore}NotSource{\isacharunderscore}level{\isadigit{1}}\ singleDSourceEmpty{\isacharunderscore}Acc{\isacharparenright}%
\endisatagproof
{\isafoldproof}%
\isadelimproof
\ \isanewline
\endisadelimproof
\isanewline
\isacommand{lemma}\isamarkupfalse%
\ A{\isadigit{8}}{\isadigit{2}}{\isacharunderscore}NSources{\isacharunderscore}L{\isadigit{1}}{\isacharcolon}\isanewline
{\isachardoublequoteopen}{\isasymforall}\ C\ {\isasymin}\ {\isacharparenleft}AbstrLevel\ level{\isadigit{1}}{\isacharparenright}{\isachardot}\ {\isacharparenleft}C\ {\isasymnoteq}\ sA{\isadigit{9}}{\isadigit{3}}\ {\isasymlongrightarrow}\ sA{\isadigit{8}}{\isadigit{2}}\ {\isasymnotin}\ {\isacharparenleft}Sources\ level{\isadigit{1}}\ C{\isacharparenright}{\isacharparenright}{\isachardoublequoteclose}\isanewline
\isadelimproof
\endisadelimproof
\isatagproof
\isacommand{by}\isamarkupfalse%
\ {\isacharparenleft}metis\ A{\isadigit{8}}{\isadigit{2}}{\isacharunderscore}Acc\ Acc{\isacharunderscore}Sources\ singleton{\isacharunderscore}iff{\isacharparenright}%
\endisatagproof
{\isafoldproof}%
\isadelimproof
\ \isanewline
\endisadelimproof
\isanewline
\isacommand{lemma}\isamarkupfalse%
\ A{\isadigit{7}}{\isadigit{2}}{\isacharunderscore}Acc{\isacharcolon}\ {\isachardoublequoteopen}{\isacharparenleft}Acc\ level{\isadigit{1}}\ sA{\isadigit{7}}{\isadigit{2}}{\isacharparenright}\ {\isacharequal}\ {\isacharbraceleft}sA{\isadigit{8}}{\isadigit{2}}{\isacharcomma}\ sA{\isadigit{9}}{\isadigit{3}}{\isacharbraceright}{\isachardoublequoteclose}\isanewline
\isadelimproof
\endisadelimproof
\isatagproof
\isacommand{proof}\isamarkupfalse%
\ {\isacharminus}\isanewline
\ \ \isacommand{have}\isamarkupfalse%
\ daA{\isadigit{7}}{\isadigit{2}}{\isacharcolon}\ \ {\isachardoublequoteopen}DAcc\ level{\isadigit{1}}\ sA{\isadigit{7}}{\isadigit{2}}\ {\isacharequal}\ {\isacharbraceleft}\ sA{\isadigit{8}}{\isadigit{2}}\ {\isacharbraceright}{\isachardoublequoteclose}\ \ \isacommand{by}\isamarkupfalse%
\ {\isacharparenleft}rule\ A{\isadigit{7}}{\isadigit{2}}{\isacharunderscore}DAcc{\isacharunderscore}level{\isadigit{1}}{\isacharparenright}\isanewline
\ \ \isacommand{hence}\isamarkupfalse%
\ {\isachardoublequoteopen}{\isacharparenleft}{\isasymUnion}\ S\ {\isasymin}\ {\isacharparenleft}DAcc\ level{\isadigit{1}}\ sA{\isadigit{7}}{\isadigit{2}}{\isacharparenright}{\isachardot}\ {\isacharparenleft}Acc\ level{\isadigit{1}}\ S{\isacharparenright}{\isacharparenright}\ {\isacharequal}\ {\isacharparenleft}Acc\ level{\isadigit{1}}\ sA{\isadigit{8}}{\isadigit{2}}{\isacharparenright}{\isachardoublequoteclose}\ \ \isacommand{by}\isamarkupfalse%
\ simp\isanewline
\ \ \isacommand{hence}\isamarkupfalse%
\ aA{\isadigit{7}}{\isadigit{2}}{\isacharcolon}{\isachardoublequoteopen}{\isacharparenleft}{\isasymUnion}\ S\ {\isasymin}\ {\isacharparenleft}DAcc\ level{\isadigit{1}}\ sA{\isadigit{7}}{\isadigit{2}}{\isacharparenright}{\isachardot}\ {\isacharparenleft}Acc\ level{\isadigit{1}}\ S{\isacharparenright}{\isacharparenright}\ {\isacharequal}\ {\isacharbraceleft}\ sA{\isadigit{9}}{\isadigit{3}}\ {\isacharbraceright}{\isachardoublequoteclose}\ \isacommand{by}\isamarkupfalse%
\ {\isacharparenleft}simp\ add{\isacharcolon}\ A{\isadigit{8}}{\isadigit{2}}{\isacharunderscore}Acc{\isacharparenright}\isanewline
\ \ \isacommand{have}\isamarkupfalse%
\ {\isachardoublequoteopen}{\isacharparenleft}Acc\ level{\isadigit{1}}\ sA{\isadigit{7}}{\isadigit{2}}{\isacharparenright}\ {\isacharequal}\ {\isacharparenleft}DAcc\ level{\isadigit{1}}\ sA{\isadigit{7}}{\isadigit{2}}{\isacharparenright}\ {\isasymunion}\ {\isacharparenleft}{\isasymUnion}\ S\ {\isasymin}\ {\isacharparenleft}DAcc\ level{\isadigit{1}}\ sA{\isadigit{7}}{\isadigit{2}}{\isacharparenright}{\isachardot}\ {\isacharparenleft}Acc\ level{\isadigit{1}}\ S{\isacharparenright}{\isacharparenright}{\isachardoublequoteclose}\ \ \isanewline
\ \ \ \ \isacommand{by}\isamarkupfalse%
\ {\isacharparenleft}rule\ AccDef{\isacharparenright}\isanewline
\ \ \isacommand{with}\isamarkupfalse%
\ daA{\isadigit{7}}{\isadigit{2}}\ aA{\isadigit{7}}{\isadigit{2}}\ \isacommand{show}\isamarkupfalse%
\ {\isacharquery}thesis\ \isacommand{by}\isamarkupfalse%
\ auto\isanewline
\isacommand{qed}\isamarkupfalse%
\endisatagproof
{\isafoldproof}%
\isadelimproof
\isanewline
\endisadelimproof
\isanewline
\isacommand{lemma}\isamarkupfalse%
\ A{\isadigit{7}}{\isadigit{2}}{\isacharunderscore}NSources{\isacharunderscore}L{\isadigit{1}}{\isacharcolon}\isanewline
{\isachardoublequoteopen}{\isasymforall}\ C\ {\isasymin}\ {\isacharparenleft}AbstrLevel\ level{\isadigit{1}}{\isacharparenright}{\isachardot}\ {\isacharparenleft}C\ {\isasymnoteq}\ sA{\isadigit{9}}{\isadigit{3}}\ {\isasymand}\ C\ {\isasymnoteq}\ sA{\isadigit{8}}{\isadigit{2}}\ {\isasymlongrightarrow}\ sA{\isadigit{7}}{\isadigit{2}}\ {\isasymnotin}\ {\isacharparenleft}Sources\ level{\isadigit{1}}\ C{\isacharparenright}{\isacharparenright}{\isachardoublequoteclose}
\isadelimproof
\endisadelimproof
\isatagproof
\isacommand{by}\isamarkupfalse%
\ {\isacharparenleft}metis\ A{\isadigit{7}}{\isadigit{2}}{\isacharunderscore}Acc\ Acc{\isacharunderscore}Sources\ insert{\isacharunderscore}iff\ singleton{\isacharunderscore}iff{\isacharparenright}%
\endisatagproof
{\isafoldproof}%
\isadelimproof
\ \isanewline
\endisadelimproof
\isanewline
\isacommand{lemma}\isamarkupfalse%
\ A{\isadigit{9}}{\isadigit{2}}{\isacharunderscore}Acc{\isacharcolon}\ {\isachardoublequoteopen}{\isacharparenleft}Acc\ level{\isadigit{1}}\ sA{\isadigit{9}}{\isadigit{2}}{\isacharparenright}\ {\isacharequal}\ {\isacharbraceleft}{\isacharbraceright}{\isachardoublequoteclose}\isanewline
\isadelimproof
\endisadelimproof
\isatagproof
\isacommand{by}\isamarkupfalse%
\ {\isacharparenleft}metis\ A{\isadigit{9}}{\isadigit{2}}{\isacharunderscore}NotSource{\isacharunderscore}level{\isadigit{1}}\ isNotDSource{\isacharunderscore}EmptyAcc{\isacharparenright}%
\endisatagproof
{\isafoldproof}%
\isadelimproof
\isanewline
\endisadelimproof
\isanewline
\isacommand{lemma}\isamarkupfalse%
\ A{\isadigit{9}}{\isadigit{2}}{\isacharunderscore}NSources{\isacharunderscore}L{\isadigit{1}}{\isacharcolon}\isanewline
{\isachardoublequoteopen}{\isasymforall}\ C\ {\isasymin}\ {\isacharparenleft}AbstrLevel\ level{\isadigit{1}}{\isacharparenright}{\isachardot}\ {\isacharparenleft}sA{\isadigit{9}}{\isadigit{2}}\ {\isasymnotin}\ {\isacharparenleft}Sources\ level{\isadigit{1}}\ C{\isacharparenright}{\isacharparenright}{\isachardoublequoteclose}\isanewline
\isadelimproof
\endisadelimproof
\isatagproof
\isacommand{by}\isamarkupfalse%
\ {\isacharparenleft}metis\ A{\isadigit{9}}{\isadigit{2}}{\isacharunderscore}NotSourceSet{\isacharunderscore}level{\isadigit{1}}{\isacharparenright}%
\endisatagproof
{\isafoldproof}%
\isadelimproof
\isanewline
\endisadelimproof
\ \isanewline
\isacommand{lemma}\isamarkupfalse%
\ A{\isadigit{9}}{\isadigit{1}}{\isacharunderscore}Acc{\isacharcolon}\ {\isachardoublequoteopen}{\isacharparenleft}Acc\ level{\isadigit{1}}\ sA{\isadigit{9}}{\isadigit{1}}{\isacharparenright}\ {\isacharequal}\ {\isacharbraceleft}sA{\isadigit{8}}{\isadigit{1}}{\isacharcomma}\ sA{\isadigit{9}}{\isadigit{1}}{\isacharcomma}\ sA{\isadigit{9}}{\isadigit{2}}{\isacharbraceright}{\isachardoublequoteclose}\isanewline
\isadelimproof
\endisadelimproof
\isatagproof
\isacommand{proof}\isamarkupfalse%
\ {\isacharminus}\isanewline
\ \ \isacommand{have}\isamarkupfalse%
\ da{\isadigit{9}}{\isadigit{1}}{\isacharcolon}\ \ {\isachardoublequoteopen}DAcc\ level{\isadigit{1}}\ sA{\isadigit{9}}{\isadigit{1}}\ {\isacharequal}\ {\isacharbraceleft}\ sA{\isadigit{8}}{\isadigit{1}}\ {\isacharbraceright}{\isachardoublequoteclose}\ \ \isacommand{by}\isamarkupfalse%
\ {\isacharparenleft}rule\ A{\isadigit{9}}{\isadigit{1}}{\isacharunderscore}DAcc{\isacharunderscore}level{\isadigit{1}}{\isacharparenright}\isanewline
\ \ \isacommand{hence}\isamarkupfalse%
\ a{\isadigit{9}}{\isadigit{1}}{\isacharcolon}{\isachardoublequoteopen}{\isacharparenleft}{\isasymUnion}\ S\ {\isasymin}\ {\isacharparenleft}DAcc\ level{\isadigit{1}}\ sA{\isadigit{9}}{\isadigit{1}}{\isacharparenright}{\isachardot}\ {\isacharparenleft}Acc\ level{\isadigit{1}}\ S{\isacharparenright}{\isacharparenright}\ {\isacharequal}\ {\isacharparenleft}Acc\ level{\isadigit{1}}\ sA{\isadigit{8}}{\isadigit{1}}{\isacharparenright}{\isachardoublequoteclose}\ \ \isacommand{by}\isamarkupfalse%
\ simp\isanewline
\ \ \isacommand{have}\isamarkupfalse%
\ {\isachardoublequoteopen}{\isacharparenleft}Acc\ level{\isadigit{1}}\ sA{\isadigit{9}}{\isadigit{1}}{\isacharparenright}\ {\isacharequal}\ {\isacharparenleft}DAcc\ level{\isadigit{1}}\ sA{\isadigit{9}}{\isadigit{1}}{\isacharparenright}\ {\isasymunion}\ {\isacharparenleft}{\isasymUnion}\ S\ {\isasymin}\ {\isacharparenleft}DAcc\ level{\isadigit{1}}\ sA{\isadigit{9}}{\isadigit{1}}{\isacharparenright}{\isachardot}\ {\isacharparenleft}Acc\ level{\isadigit{1}}\ S{\isacharparenright}{\isacharparenright}{\isachardoublequoteclose}\ \ \isacommand{by}\isamarkupfalse%
\ {\isacharparenleft}rule\ AccDef{\isacharparenright}\isanewline
\ \ \isacommand{with}\isamarkupfalse%
\ da{\isadigit{9}}{\isadigit{1}}\ a{\isadigit{9}}{\isadigit{1}}\ \isacommand{have}\isamarkupfalse%
\ acc{\isadigit{9}}{\isadigit{1}}{\isacharcolon}{\isachardoublequoteopen}{\isacharparenleft}Acc\ level{\isadigit{1}}\ sA{\isadigit{9}}{\isadigit{1}}{\isacharparenright}\ {\isacharequal}\ {\isacharbraceleft}\ sA{\isadigit{8}}{\isadigit{1}}\ {\isacharbraceright}\ {\isasymunion}\ {\isacharparenleft}Acc\ level{\isadigit{1}}\ sA{\isadigit{8}}{\isadigit{1}}{\isacharparenright}{\isachardoublequoteclose}\ \isacommand{by}\isamarkupfalse%
\ simp\isanewline
\ \ \isacommand{have}\isamarkupfalse%
\ da{\isadigit{8}}{\isadigit{1}}{\isacharcolon}\ \ {\isachardoublequoteopen}DAcc\ level{\isadigit{1}}\ sA{\isadigit{8}}{\isadigit{1}}\ {\isacharequal}\ {\isacharbraceleft}\ sA{\isadigit{9}}{\isadigit{1}}{\isacharcomma}\ sA{\isadigit{9}}{\isadigit{2}}\ {\isacharbraceright}{\isachardoublequoteclose}\ \ \isacommand{by}\isamarkupfalse%
\ {\isacharparenleft}rule\ A{\isadigit{8}}{\isadigit{1}}{\isacharunderscore}DAcc{\isacharunderscore}level{\isadigit{1}}{\isacharparenright}\isanewline
\ \ \isacommand{hence}\isamarkupfalse%
\ a{\isadigit{8}}{\isadigit{1}}{\isacharcolon}{\isachardoublequoteopen}{\isacharparenleft}{\isasymUnion}\ S\ {\isasymin}\ {\isacharparenleft}DAcc\ level{\isadigit{1}}\ sA{\isadigit{8}}{\isadigit{1}}{\isacharparenright}{\isachardot}\ {\isacharparenleft}Acc\ level{\isadigit{1}}\ S{\isacharparenright}{\isacharparenright}\ {\isacharequal}\ {\isacharparenleft}Acc\ level{\isadigit{1}}\ sA{\isadigit{9}}{\isadigit{2}}{\isacharparenright}\ {\isasymunion}\ {\isacharparenleft}Acc\ level{\isadigit{1}}\ sA{\isadigit{9}}{\isadigit{1}}{\isacharparenright}{\isachardoublequoteclose}\ \ \isacommand{by}\isamarkupfalse%
\ auto\isanewline
\ \ \isacommand{have}\isamarkupfalse%
\ {\isachardoublequoteopen}{\isacharparenleft}Acc\ level{\isadigit{1}}\ sA{\isadigit{8}}{\isadigit{1}}{\isacharparenright}\ {\isacharequal}\ {\isacharparenleft}DAcc\ level{\isadigit{1}}\ sA{\isadigit{8}}{\isadigit{1}}{\isacharparenright}\ {\isasymunion}\ {\isacharparenleft}{\isasymUnion}\ S\ {\isasymin}\ {\isacharparenleft}DAcc\ level{\isadigit{1}}\ sA{\isadigit{8}}{\isadigit{1}}{\isacharparenright}{\isachardot}\ {\isacharparenleft}Acc\ level{\isadigit{1}}\ S{\isacharparenright}{\isacharparenright}{\isachardoublequoteclose}\ \ \isacommand{by}\isamarkupfalse%
\ {\isacharparenleft}rule\ AccDef{\isacharparenright}\isanewline
\ \ \isacommand{with}\isamarkupfalse%
\ da{\isadigit{8}}{\isadigit{1}}\ a{\isadigit{8}}{\isadigit{1}}\ \isacommand{have}\isamarkupfalse%
\ acc{\isadigit{8}}{\isadigit{1}}{\isacharcolon}\ {\isachardoublequoteopen}{\isacharparenleft}Acc\ level{\isadigit{1}}\ sA{\isadigit{8}}{\isadigit{1}}{\isacharparenright}\ {\isacharequal}\ {\isacharbraceleft}\ sA{\isadigit{9}}{\isadigit{1}}{\isacharcomma}\ sA{\isadigit{9}}{\isadigit{2}}\ {\isacharbraceright}\ \ {\isasymunion}\ {\isacharparenleft}Acc\ level{\isadigit{1}}\ sA{\isadigit{9}}{\isadigit{1}}{\isacharparenright}{\isachardoublequoteclose}\isanewline
\ \ \ \ \isacommand{by}\isamarkupfalse%
\ {\isacharparenleft}metis\ A{\isadigit{9}}{\isadigit{2}}{\isacharunderscore}Acc\ sup{\isacharunderscore}bot{\isachardot}left{\isacharunderscore}neutral{\isacharparenright}\isanewline
\ \ \isacommand{from}\isamarkupfalse%
\ acc{\isadigit{9}}{\isadigit{1}}\ acc{\isadigit{8}}{\isadigit{1}}\ \isacommand{have}\isamarkupfalse%
\ {\isachardoublequoteopen}{\isacharparenleft}Acc\ level{\isadigit{1}}\ sA{\isadigit{9}}{\isadigit{1}}{\isacharparenright}\ {\isacharequal}\ {\isacharbraceleft}\ sA{\isadigit{8}}{\isadigit{1}}\ {\isacharbraceright}\ {\isasymunion}\ {\isacharbraceleft}\ sA{\isadigit{9}}{\isadigit{1}}{\isacharcomma}\ sA{\isadigit{9}}{\isadigit{2}}\ {\isacharbraceright}\ \ {\isasymunion}\ {\isacharbraceleft}sA{\isadigit{9}}{\isadigit{1}}{\isacharcomma}\ sA{\isadigit{8}}{\isadigit{1}}{\isacharbraceright}{\isachardoublequoteclose}\isanewline
\ \ \ \isacommand{by}\isamarkupfalse%
\ {\isacharparenleft}metis\ AccLoop{\isacharparenright}\isanewline
\ \ \isacommand{thus}\isamarkupfalse%
\ {\isacharquery}thesis\ \isacommand{by}\isamarkupfalse%
\ auto\isanewline
\isacommand{qed}\isamarkupfalse%
\endisatagproof
{\isafoldproof}%
\isadelimproof
\isanewline
\endisadelimproof
\isanewline
\isacommand{lemma}\isamarkupfalse%
\ A{\isadigit{9}}{\isadigit{1}}{\isacharunderscore}NSources{\isacharunderscore}L{\isadigit{1}}{\isacharcolon}\isanewline
{\isachardoublequoteopen}{\isasymforall}\ C\ {\isasymin}\ {\isacharparenleft}AbstrLevel\ level{\isadigit{1}}{\isacharparenright}{\isachardot}\ {\isacharparenleft}C\ {\isasymnoteq}\ sA{\isadigit{9}}{\isadigit{2}}\ {\isasymand}\ C\ {\isasymnoteq}\ sA{\isadigit{9}}{\isadigit{1}}\ {\isasymand}\ C\ {\isasymnoteq}\ sA{\isadigit{8}}{\isadigit{1}}\ {\isasymlongrightarrow}\ sA{\isadigit{9}}{\isadigit{1}}\ {\isasymnotin}\ {\isacharparenleft}Sources\ level{\isadigit{1}}\ C{\isacharparenright}{\isacharparenright}{\isachardoublequoteclose}\isanewline
\isadelimproof
\endisadelimproof
\isatagproof
\isacommand{proof}\isamarkupfalse%
\ {\isacharminus}\isanewline
\ \ \isacommand{have}\isamarkupfalse%
\ {\isachardoublequoteopen}{\isasymforall}\ C\ {\isasymin}\ {\isacharparenleft}AbstrLevel\ level{\isadigit{1}}{\isacharparenright}{\isachardot}\ {\isacharparenleft}C\ {\isasymnoteq}\ sA{\isadigit{9}}{\isadigit{2}}\ {\isasymand}\ C\ {\isasymnoteq}\ sA{\isadigit{9}}{\isadigit{1}}\ {\isasymand}\ C\ {\isasymnoteq}\ sA{\isadigit{8}}{\isadigit{1}}\ \ {\isasymlongrightarrow}\ {\isacharparenleft}C\ {\isasymnotin}\ {\isacharparenleft}Acc\ level{\isadigit{1}}\ sA{\isadigit{9}}{\isadigit{1}}{\isacharparenright}{\isacharparenright}{\isacharparenright}{\isachardoublequoteclose}\isanewline
\ \ \ \ \isacommand{by}\isamarkupfalse%
\ {\isacharparenleft}metis\ A{\isadigit{9}}{\isadigit{1}}{\isacharunderscore}Acc\ insert{\isacharunderscore}iff\ singleton{\isacharunderscore}iff{\isacharparenright}\isanewline
\ \ \isacommand{thus}\isamarkupfalse%
\ {\isacharquery}thesis\ \ \isacommand{by}\isamarkupfalse%
\ {\isacharparenleft}metis\ Acc{\isacharunderscore}SourcesNOT{\isacharparenright}\ \isanewline
\isacommand{qed}\isamarkupfalse%
\endisatagproof
{\isafoldproof}%
\isadelimproof
\isanewline
\endisadelimproof
\isanewline
\isacommand{lemma}\isamarkupfalse%
\ A{\isadigit{8}}{\isadigit{1}}{\isacharunderscore}Acc{\isacharcolon}\ {\isachardoublequoteopen}{\isacharparenleft}Acc\ level{\isadigit{1}}\ sA{\isadigit{8}}{\isadigit{1}}{\isacharparenright}\ {\isacharequal}\ {\isacharbraceleft}sA{\isadigit{8}}{\isadigit{1}}{\isacharcomma}\ sA{\isadigit{9}}{\isadigit{1}}{\isacharcomma}\ sA{\isadigit{9}}{\isadigit{2}}{\isacharbraceright}{\isachardoublequoteclose}\isanewline
\isadelimproof
\endisadelimproof
\isatagproof
\isacommand{proof}\isamarkupfalse%
\ {\isacharminus}\isanewline
\ \ \isacommand{have}\isamarkupfalse%
\ da{\isadigit{9}}{\isadigit{1}}{\isacharcolon}\ \ {\isachardoublequoteopen}DAcc\ level{\isadigit{1}}\ sA{\isadigit{9}}{\isadigit{1}}\ {\isacharequal}\ {\isacharbraceleft}\ sA{\isadigit{8}}{\isadigit{1}}\ {\isacharbraceright}{\isachardoublequoteclose}\ \ \isacommand{by}\isamarkupfalse%
\ {\isacharparenleft}rule\ A{\isadigit{9}}{\isadigit{1}}{\isacharunderscore}DAcc{\isacharunderscore}level{\isadigit{1}}{\isacharparenright}\isanewline
\ \ \isacommand{hence}\isamarkupfalse%
\ a{\isadigit{9}}{\isadigit{1}}{\isacharcolon}{\isachardoublequoteopen}{\isacharparenleft}{\isasymUnion}\ S\ {\isasymin}\ {\isacharparenleft}DAcc\ level{\isadigit{1}}\ sA{\isadigit{9}}{\isadigit{1}}{\isacharparenright}{\isachardot}\ {\isacharparenleft}Acc\ level{\isadigit{1}}\ S{\isacharparenright}{\isacharparenright}\ {\isacharequal}\ {\isacharparenleft}Acc\ level{\isadigit{1}}\ sA{\isadigit{8}}{\isadigit{1}}{\isacharparenright}{\isachardoublequoteclose}\ \ \isacommand{by}\isamarkupfalse%
\ simp\isanewline
\ \ \isacommand{have}\isamarkupfalse%
\ {\isachardoublequoteopen}{\isacharparenleft}Acc\ level{\isadigit{1}}\ sA{\isadigit{9}}{\isadigit{1}}{\isacharparenright}\ {\isacharequal}\ {\isacharparenleft}DAcc\ level{\isadigit{1}}\ sA{\isadigit{9}}{\isadigit{1}}{\isacharparenright}\ {\isasymunion}\ {\isacharparenleft}{\isasymUnion}\ S\ {\isasymin}\ {\isacharparenleft}DAcc\ level{\isadigit{1}}\ sA{\isadigit{9}}{\isadigit{1}}{\isacharparenright}{\isachardot}\ {\isacharparenleft}Acc\ level{\isadigit{1}}\ S{\isacharparenright}{\isacharparenright}{\isachardoublequoteclose}\ \ \isacommand{by}\isamarkupfalse%
\ {\isacharparenleft}rule\ AccDef{\isacharparenright}\isanewline
\ \ \isacommand{with}\isamarkupfalse%
\ da{\isadigit{9}}{\isadigit{1}}\ a{\isadigit{9}}{\isadigit{1}}\ \isacommand{have}\isamarkupfalse%
\ acc{\isadigit{9}}{\isadigit{1}}{\isacharcolon}{\isachardoublequoteopen}{\isacharparenleft}Acc\ level{\isadigit{1}}\ sA{\isadigit{9}}{\isadigit{1}}{\isacharparenright}\ {\isacharequal}\ {\isacharbraceleft}\ sA{\isadigit{8}}{\isadigit{1}}\ {\isacharbraceright}\ {\isasymunion}\ {\isacharparenleft}Acc\ level{\isadigit{1}}\ sA{\isadigit{8}}{\isadigit{1}}{\isacharparenright}{\isachardoublequoteclose}\ \isacommand{by}\isamarkupfalse%
\ simp\isanewline
\ \ \isacommand{have}\isamarkupfalse%
\ da{\isadigit{8}}{\isadigit{1}}{\isacharcolon}\ \ {\isachardoublequoteopen}DAcc\ level{\isadigit{1}}\ sA{\isadigit{8}}{\isadigit{1}}\ {\isacharequal}\ {\isacharbraceleft}\ sA{\isadigit{9}}{\isadigit{1}}{\isacharcomma}\ sA{\isadigit{9}}{\isadigit{2}}\ {\isacharbraceright}{\isachardoublequoteclose}\ \ \isacommand{by}\isamarkupfalse%
\ {\isacharparenleft}rule\ A{\isadigit{8}}{\isadigit{1}}{\isacharunderscore}DAcc{\isacharunderscore}level{\isadigit{1}}{\isacharparenright}\isanewline
\ \ \isacommand{hence}\isamarkupfalse%
\ a{\isadigit{8}}{\isadigit{1}}{\isacharcolon}{\isachardoublequoteopen}{\isacharparenleft}{\isasymUnion}\ S\ {\isasymin}\ {\isacharparenleft}DAcc\ level{\isadigit{1}}\ sA{\isadigit{8}}{\isadigit{1}}{\isacharparenright}{\isachardot}\ {\isacharparenleft}Acc\ level{\isadigit{1}}\ S{\isacharparenright}{\isacharparenright}\ {\isacharequal}\ {\isacharparenleft}Acc\ level{\isadigit{1}}\ sA{\isadigit{9}}{\isadigit{2}}{\isacharparenright}\ {\isasymunion}\ {\isacharparenleft}Acc\ level{\isadigit{1}}\ sA{\isadigit{9}}{\isadigit{1}}{\isacharparenright}{\isachardoublequoteclose}\ \ \isacommand{by}\isamarkupfalse%
\ auto\isanewline
\ \ \isacommand{have}\isamarkupfalse%
\ {\isachardoublequoteopen}{\isacharparenleft}Acc\ level{\isadigit{1}}\ sA{\isadigit{8}}{\isadigit{1}}{\isacharparenright}\ {\isacharequal}\ {\isacharparenleft}DAcc\ level{\isadigit{1}}\ sA{\isadigit{8}}{\isadigit{1}}{\isacharparenright}\ {\isasymunion}\ {\isacharparenleft}{\isasymUnion}\ S\ {\isasymin}\ {\isacharparenleft}DAcc\ level{\isadigit{1}}\ sA{\isadigit{8}}{\isadigit{1}}{\isacharparenright}{\isachardot}\ {\isacharparenleft}Acc\ level{\isadigit{1}}\ S{\isacharparenright}{\isacharparenright}{\isachardoublequoteclose}\ \ \isacommand{by}\isamarkupfalse%
\ {\isacharparenleft}rule\ AccDef{\isacharparenright}\isanewline
\ \ \isacommand{with}\isamarkupfalse%
\ da{\isadigit{8}}{\isadigit{1}}\ a{\isadigit{8}}{\isadigit{1}}\ \isacommand{have}\isamarkupfalse%
\ acc{\isadigit{8}}{\isadigit{1}}{\isacharcolon}\ {\isachardoublequoteopen}{\isacharparenleft}Acc\ level{\isadigit{1}}\ sA{\isadigit{8}}{\isadigit{1}}{\isacharparenright}\ {\isacharequal}\ {\isacharbraceleft}\ sA{\isadigit{9}}{\isadigit{1}}{\isacharcomma}\ sA{\isadigit{9}}{\isadigit{2}}\ {\isacharbraceright}\ \ {\isasymunion}\ {\isacharparenleft}Acc\ level{\isadigit{1}}\ sA{\isadigit{9}}{\isadigit{1}}{\isacharparenright}{\isachardoublequoteclose}\isanewline
\ \ \ \ \isacommand{by}\isamarkupfalse%
\ {\isacharparenleft}metis\ A{\isadigit{9}}{\isadigit{2}}{\isacharunderscore}Acc\ sup{\isacharunderscore}bot{\isachardot}left{\isacharunderscore}neutral{\isacharparenright}\isanewline
\ \ \isacommand{from}\isamarkupfalse%
\ acc{\isadigit{8}}{\isadigit{1}}\ acc{\isadigit{9}}{\isadigit{1}}\ \isacommand{have}\isamarkupfalse%
\ {\isachardoublequoteopen}{\isacharparenleft}Acc\ level{\isadigit{1}}\ sA{\isadigit{8}}{\isadigit{1}}{\isacharparenright}\ {\isacharequal}\ \ {\isacharbraceleft}\ sA{\isadigit{9}}{\isadigit{1}}{\isacharcomma}\ sA{\isadigit{9}}{\isadigit{2}}\ {\isacharbraceright}\ \ {\isasymunion}\ {\isacharbraceleft}\ sA{\isadigit{8}}{\isadigit{1}}\ {\isacharbraceright}\ {\isasymunion}\ {\isacharbraceleft}sA{\isadigit{8}}{\isadigit{1}}{\isacharcomma}\ sA{\isadigit{9}}{\isadigit{1}}{\isacharbraceright}{\isachardoublequoteclose}\isanewline
\ \ \ \isacommand{by}\isamarkupfalse%
\ {\isacharparenleft}metis\ AccLoop{\isacharparenright}\isanewline
\ \ \isacommand{thus}\isamarkupfalse%
\ {\isacharquery}thesis\ \isacommand{by}\isamarkupfalse%
\ auto\isanewline
\isacommand{qed}\isamarkupfalse%
\endisatagproof
{\isafoldproof}%
\isadelimproof
\isanewline
\endisadelimproof
\isanewline
\isacommand{lemma}\isamarkupfalse%
\ A{\isadigit{8}}{\isadigit{1}}{\isacharunderscore}NSources{\isacharunderscore}L{\isadigit{1}}{\isacharcolon}\isanewline
{\isachardoublequoteopen}{\isasymforall}\ C\ {\isasymin}\ {\isacharparenleft}AbstrLevel\ level{\isadigit{1}}{\isacharparenright}{\isachardot}\ {\isacharparenleft}C\ {\isasymnoteq}\ sA{\isadigit{9}}{\isadigit{2}}\ {\isasymand}\ C\ {\isasymnoteq}\ sA{\isadigit{9}}{\isadigit{1}}\ {\isasymand}\ C\ {\isasymnoteq}\ sA{\isadigit{8}}{\isadigit{1}}\ {\isasymlongrightarrow}\ sA{\isadigit{8}}{\isadigit{1}}\ {\isasymnotin}\ {\isacharparenleft}Sources\ level{\isadigit{1}}\ C{\isacharparenright}{\isacharparenright}{\isachardoublequoteclose}\isanewline
\isadelimproof
\endisadelimproof
\isatagproof
\isacommand{proof}\isamarkupfalse%
\ {\isacharminus}\isanewline
\ \ \isacommand{have}\isamarkupfalse%
\ {\isachardoublequoteopen}{\isasymforall}\ C\ {\isasymin}\ {\isacharparenleft}AbstrLevel\ level{\isadigit{1}}{\isacharparenright}{\isachardot}\ {\isacharparenleft}C\ {\isasymnoteq}\ sA{\isadigit{9}}{\isadigit{2}}\ {\isasymand}\ C\ {\isasymnoteq}\ sA{\isadigit{9}}{\isadigit{1}}\ {\isasymand}\ C\ {\isasymnoteq}\ sA{\isadigit{8}}{\isadigit{1}}\ \ {\isasymlongrightarrow}\ {\isacharparenleft}C\ {\isasymnotin}\ {\isacharparenleft}Acc\ level{\isadigit{1}}\ sA{\isadigit{8}}{\isadigit{1}}{\isacharparenright}{\isacharparenright}{\isacharparenright}{\isachardoublequoteclose}\isanewline
\ \ \ \ \isacommand{by}\isamarkupfalse%
\ {\isacharparenleft}metis\ A{\isadigit{8}}{\isadigit{1}}{\isacharunderscore}Acc\ insert{\isacharunderscore}iff\ singleton{\isacharunderscore}iff{\isacharparenright}\isanewline
\ \ \isacommand{thus}\isamarkupfalse%
\ {\isacharquery}thesis\ \ \isacommand{by}\isamarkupfalse%
\ {\isacharparenleft}metis\ Acc{\isacharunderscore}SourcesNOT{\isacharparenright}\ \isanewline
\isacommand{qed}\isamarkupfalse%
\endisatagproof
{\isafoldproof}%
\isadelimproof
\isanewline
\endisadelimproof
\isanewline
\isacommand{lemma}\isamarkupfalse%
\ A{\isadigit{7}}{\isadigit{1}}{\isacharunderscore}Acc{\isacharcolon}\ {\isachardoublequoteopen}{\isacharparenleft}Acc\ level{\isadigit{1}}\ sA{\isadigit{7}}{\isadigit{1}}{\isacharparenright}\ {\isacharequal}\ {\isacharbraceleft}sA{\isadigit{8}}{\isadigit{1}}{\isacharcomma}\ sA{\isadigit{9}}{\isadigit{1}}{\isacharcomma}\ sA{\isadigit{9}}{\isadigit{2}}{\isacharbraceright}{\isachardoublequoteclose}\isanewline
\isadelimproof
\endisadelimproof
\isatagproof
\isacommand{proof}\isamarkupfalse%
\ {\isacharminus}\isanewline
\ \ \isacommand{have}\isamarkupfalse%
\ da{\isadigit{7}}{\isadigit{1}}{\isacharcolon}\ \ {\isachardoublequoteopen}DAcc\ level{\isadigit{1}}\ sA{\isadigit{7}}{\isadigit{1}}\ {\isacharequal}\ {\isacharbraceleft}\ sA{\isadigit{8}}{\isadigit{1}}\ {\isacharbraceright}{\isachardoublequoteclose}\ \ \isacommand{by}\isamarkupfalse%
\ {\isacharparenleft}rule\ A{\isadigit{7}}{\isadigit{1}}{\isacharunderscore}DAcc{\isacharunderscore}level{\isadigit{1}}{\isacharparenright}\isanewline
\ \ \isacommand{hence}\isamarkupfalse%
\ a{\isadigit{7}}{\isadigit{1}}{\isacharcolon}{\isachardoublequoteopen}{\isacharparenleft}{\isasymUnion}\ S\ {\isasymin}\ {\isacharparenleft}DAcc\ level{\isadigit{1}}\ sA{\isadigit{7}}{\isadigit{1}}{\isacharparenright}{\isachardot}\ {\isacharparenleft}Acc\ level{\isadigit{1}}\ S{\isacharparenright}{\isacharparenright}\ {\isacharequal}\ {\isacharparenleft}Acc\ level{\isadigit{1}}\ sA{\isadigit{8}}{\isadigit{1}}{\isacharparenright}{\isachardoublequoteclose}\ \ \isacommand{by}\isamarkupfalse%
\ simp\isanewline
\ \ \isacommand{have}\isamarkupfalse%
\ {\isachardoublequoteopen}{\isacharparenleft}Acc\ level{\isadigit{1}}\ sA{\isadigit{7}}{\isadigit{1}}{\isacharparenright}\ {\isacharequal}\ {\isacharparenleft}DAcc\ level{\isadigit{1}}\ sA{\isadigit{7}}{\isadigit{1}}{\isacharparenright}\ {\isasymunion}\ {\isacharparenleft}{\isasymUnion}\ S\ {\isasymin}\ {\isacharparenleft}DAcc\ level{\isadigit{1}}\ sA{\isadigit{7}}{\isadigit{1}}{\isacharparenright}{\isachardot}\ {\isacharparenleft}Acc\ level{\isadigit{1}}\ S{\isacharparenright}{\isacharparenright}{\isachardoublequoteclose}\ \ \isacommand{by}\isamarkupfalse%
\ {\isacharparenleft}rule\ AccDef{\isacharparenright}\isanewline
\ \ \isacommand{with}\isamarkupfalse%
\ da{\isadigit{7}}{\isadigit{1}}\ a{\isadigit{7}}{\isadigit{1}}\ \isacommand{show}\isamarkupfalse%
\ {\isacharquery}thesis\ \isacommand{by}\isamarkupfalse%
\ {\isacharparenleft}metis\ A{\isadigit{9}}{\isadigit{1}}{\isacharunderscore}Acc\ A{\isadigit{9}}{\isadigit{1}}{\isacharunderscore}DAcc{\isacharunderscore}level{\isadigit{1}}\ AccDef{\isacharparenright}\ \isanewline
\isacommand{qed}\isamarkupfalse%
\endisatagproof
{\isafoldproof}%
\isadelimproof
\isanewline
\endisadelimproof
\isanewline
\isacommand{lemma}\isamarkupfalse%
\ A{\isadigit{7}}{\isadigit{1}}{\isacharunderscore}NSources{\isacharunderscore}L{\isadigit{1}}{\isacharcolon}\isanewline
{\isachardoublequoteopen}{\isasymforall}\ C\ {\isasymin}\ {\isacharparenleft}AbstrLevel\ level{\isadigit{1}}{\isacharparenright}{\isachardot}\ {\isacharparenleft}C\ {\isasymnoteq}\ sA{\isadigit{9}}{\isadigit{2}}\ {\isasymand}\ C\ {\isasymnoteq}\ sA{\isadigit{9}}{\isadigit{1}}\ {\isasymand}\ C\ {\isasymnoteq}\ sA{\isadigit{8}}{\isadigit{1}}\ {\isasymlongrightarrow}\ sA{\isadigit{7}}{\isadigit{1}}\ {\isasymnotin}\ {\isacharparenleft}Sources\ level{\isadigit{1}}\ C{\isacharparenright}{\isacharparenright}{\isachardoublequoteclose}\isanewline
\isadelimproof
\endisadelimproof
\isatagproof
\isacommand{proof}\isamarkupfalse%
\ {\isacharminus}\isanewline
\ \ \isacommand{have}\isamarkupfalse%
\ {\isachardoublequoteopen}{\isasymforall}\ C\ {\isasymin}\ {\isacharparenleft}AbstrLevel\ level{\isadigit{1}}{\isacharparenright}{\isachardot}\ {\isacharparenleft}C\ {\isasymnoteq}\ sA{\isadigit{9}}{\isadigit{2}}\ {\isasymand}\ C\ {\isasymnoteq}\ sA{\isadigit{9}}{\isadigit{1}}\ {\isasymand}\ C\ {\isasymnoteq}\ sA{\isadigit{8}}{\isadigit{1}}\ \ {\isasymlongrightarrow}\ {\isacharparenleft}C\ {\isasymnotin}\ {\isacharparenleft}Acc\ level{\isadigit{1}}\ sA{\isadigit{7}}{\isadigit{1}}{\isacharparenright}{\isacharparenright}{\isacharparenright}{\isachardoublequoteclose}\isanewline
\ \ \ \ \isacommand{by}\isamarkupfalse%
\ {\isacharparenleft}metis\ A{\isadigit{7}}{\isadigit{1}}{\isacharunderscore}Acc\ insert{\isacharunderscore}iff\ singleton{\isacharunderscore}iff{\isacharparenright}\isanewline
\ \ \isacommand{thus}\isamarkupfalse%
\ {\isacharquery}thesis\ \ \isacommand{by}\isamarkupfalse%
\ {\isacharparenleft}metis\ Acc{\isacharunderscore}SourcesNOT{\isacharparenright}\ \isanewline
\isacommand{qed}\isamarkupfalse%
\endisatagproof
{\isafoldproof}%
\isadelimproof
\isanewline
\endisadelimproof
\isanewline
\isacommand{lemma}\isamarkupfalse%
\ A{\isadigit{6}}{\isacharunderscore}Acc{\isacharunderscore}L{\isadigit{1}}{\isacharcolon}\isanewline
{\isachardoublequoteopen}{\isacharparenleft}Acc\ level{\isadigit{1}}\ sA{\isadigit{6}}{\isacharparenright}\ {\isacharequal}\ {\isacharbraceleft}sA{\isadigit{7}}{\isadigit{1}}{\isacharcomma}\ sA{\isadigit{7}}{\isadigit{2}}{\isacharcomma}\ sA{\isadigit{8}}{\isadigit{1}}{\isacharcomma}\ sA{\isadigit{8}}{\isadigit{2}}{\isacharcomma}\ sA{\isadigit{9}}{\isadigit{1}}{\isacharcomma}\ sA{\isadigit{9}}{\isadigit{2}}{\isacharcomma}\ sA{\isadigit{9}}{\isadigit{3}}{\isacharbraceright}{\isachardoublequoteclose}\isanewline
\isadelimproof
\endisadelimproof
\isatagproof
\isacommand{proof}\isamarkupfalse%
\ {\isacharminus}\isanewline
\ \ \isacommand{have}\isamarkupfalse%
\ daA{\isadigit{6}}{\isacharcolon}\ \ {\isachardoublequoteopen}DAcc\ level{\isadigit{1}}\ sA{\isadigit{6}}\ {\isacharequal}\ {\isacharbraceleft}\ sA{\isadigit{7}}{\isadigit{1}}{\isacharcomma}\ sA{\isadigit{7}}{\isadigit{2}}\ {\isacharbraceright}{\isachardoublequoteclose}\ \ \isacommand{by}\isamarkupfalse%
\ {\isacharparenleft}rule\ A{\isadigit{6}}{\isacharunderscore}DAcc{\isacharunderscore}level{\isadigit{1}}{\isacharparenright}\isanewline
\ \ \isacommand{hence}\isamarkupfalse%
\ {\isachardoublequoteopen}{\isacharparenleft}{\isasymUnion}\ S\ {\isasymin}\ {\isacharparenleft}DAcc\ level{\isadigit{1}}\ sA{\isadigit{6}}{\isacharparenright}{\isachardot}\ {\isacharparenleft}Acc\ level{\isadigit{1}}\ S{\isacharparenright}{\isacharparenright}\ {\isacharequal}\ {\isacharparenleft}Acc\ level{\isadigit{1}}\ sA{\isadigit{7}}{\isadigit{1}}{\isacharparenright}\ {\isasymunion}\ {\isacharparenleft}Acc\ level{\isadigit{1}}\ sA{\isadigit{7}}{\isadigit{2}}{\isacharparenright}{\isachardoublequoteclose}\ \ \isacommand{by}\isamarkupfalse%
\ simp\isanewline
\ \ \isacommand{hence}\isamarkupfalse%
\ aA{\isadigit{6}}{\isacharcolon}{\isachardoublequoteopen}{\isacharparenleft}{\isasymUnion}\ S\ {\isasymin}\ {\isacharparenleft}DAcc\ level{\isadigit{1}}\ sA{\isadigit{6}}{\isacharparenright}{\isachardot}\ {\isacharparenleft}Acc\ level{\isadigit{1}}\ S{\isacharparenright}{\isacharparenright}\ {\isacharequal}\ {\isacharbraceleft}sA{\isadigit{8}}{\isadigit{1}}{\isacharcomma}\ sA{\isadigit{9}}{\isadigit{1}}{\isacharcomma}\ sA{\isadigit{9}}{\isadigit{2}}{\isacharbraceright}\ {\isasymunion}\ {\isacharbraceleft}sA{\isadigit{8}}{\isadigit{2}}{\isacharcomma}\ sA{\isadigit{9}}{\isadigit{3}}{\isacharbraceright}{\isachardoublequoteclose}\ \isanewline
\ \ \ \ \isacommand{by}\isamarkupfalse%
\ {\isacharparenleft}simp\ add{\isacharcolon}\ A{\isadigit{7}}{\isadigit{1}}{\isacharunderscore}Acc\ A{\isadigit{7}}{\isadigit{2}}{\isacharunderscore}Acc{\isacharparenright}\isanewline
\ \ \isacommand{have}\isamarkupfalse%
\ {\isachardoublequoteopen}{\isacharparenleft}Acc\ level{\isadigit{1}}\ sA{\isadigit{6}}{\isacharparenright}\ {\isacharequal}\ {\isacharparenleft}DAcc\ level{\isadigit{1}}\ sA{\isadigit{6}}{\isacharparenright}\ {\isasymunion}\ {\isacharparenleft}{\isasymUnion}\ S\ {\isasymin}\ {\isacharparenleft}DAcc\ level{\isadigit{1}}\ sA{\isadigit{6}}{\isacharparenright}{\isachardot}\ {\isacharparenleft}Acc\ level{\isadigit{1}}\ S{\isacharparenright}{\isacharparenright}{\isachardoublequoteclose}\ \ \isanewline
\ \ \ \ \isacommand{by}\isamarkupfalse%
\ {\isacharparenleft}rule\ AccDef{\isacharparenright}\isanewline
\ \ \isacommand{with}\isamarkupfalse%
\ daA{\isadigit{6}}\ aA{\isadigit{6}}\ \isacommand{show}\isamarkupfalse%
\ {\isacharquery}thesis\ \isacommand{by}\isamarkupfalse%
\ auto\isanewline
\isacommand{qed}\isamarkupfalse%
\endisatagproof
{\isafoldproof}%
\isadelimproof
\isanewline
\endisadelimproof
\isanewline
\isacommand{lemma}\isamarkupfalse%
\ A{\isadigit{6}}{\isacharunderscore}NSources{\isacharunderscore}L{\isadigit{1}}Acc{\isacharcolon}\isanewline
{\isachardoublequoteopen}{\isasymforall}\ C\ {\isasymin}\ {\isacharparenleft}AbstrLevel\ level{\isadigit{1}}{\isacharparenright}{\isachardot}\ {\isacharparenleft}C\ {\isasymnotin}\ {\isacharparenleft}Acc\ level{\isadigit{1}}\ sA{\isadigit{6}}{\isacharparenright}\ {\isasymlongrightarrow}\ sA{\isadigit{6}}\ {\isasymnotin}\ {\isacharparenleft}Sources\ level{\isadigit{1}}\ C{\isacharparenright}{\isacharparenright}{\isachardoublequoteclose}\isanewline
\isadelimproof
\endisadelimproof
\isatagproof
\isacommand{by}\isamarkupfalse%
\ {\isacharparenleft}metis\ Acc{\isacharunderscore}SourcesNOT{\isacharparenright}%
\endisatagproof
{\isafoldproof}%
\isadelimproof
\isanewline
\endisadelimproof
\isanewline
\isacommand{lemma}\isamarkupfalse%
\ A{\isadigit{6}}{\isacharunderscore}NSources{\isacharunderscore}L{\isadigit{1}}{\isacharcolon}\isanewline
{\isachardoublequoteopen}{\isasymforall}\ C\ {\isasymin}\ {\isacharparenleft}AbstrLevel\ level{\isadigit{1}}{\isacharparenright}{\isachardot}\ {\isacharparenleft}C\ {\isasymnoteq}\ sA{\isadigit{9}}{\isadigit{3}}\ {\isasymand}\ C\ {\isasymnoteq}\ sA{\isadigit{9}}{\isadigit{2}}\ {\isasymand}\ C\ {\isasymnoteq}\ sA{\isadigit{9}}{\isadigit{1}}\ {\isasymand}\ C\ {\isasymnoteq}\ sA{\isadigit{8}}{\isadigit{2}}\ \ {\isasymand}\ C\ {\isasymnoteq}\ sA{\isadigit{8}}{\isadigit{1}}\ {\isasymand}\ C\ {\isasymnoteq}\ sA{\isadigit{7}}{\isadigit{2}}\ {\isasymand}\ C\ {\isasymnoteq}\ sA{\isadigit{7}}{\isadigit{1}}\ \isanewline
{\isasymlongrightarrow}\ sA{\isadigit{6}}\ {\isasymnotin}\ {\isacharparenleft}Sources\ level{\isadigit{1}}\ C{\isacharparenright}{\isacharparenright}{\isachardoublequoteclose}\isanewline
\isadelimproof
\endisadelimproof
\isatagproof
\isacommand{proof}\isamarkupfalse%
\ {\isacharminus}\isanewline
\ \ \isacommand{have}\isamarkupfalse%
\ {\isachardoublequoteopen}{\isasymforall}\ C\ {\isasymin}\ {\isacharparenleft}AbstrLevel\ level{\isadigit{1}}{\isacharparenright}{\isachardot}\ \isanewline
\ \ {\isacharparenleft}C\ {\isasymnoteq}\ sA{\isadigit{9}}{\isadigit{3}}\ {\isasymand}\ C\ {\isasymnoteq}\ sA{\isadigit{9}}{\isadigit{2}}\ {\isasymand}\ C\ {\isasymnoteq}\ sA{\isadigit{9}}{\isadigit{1}}\ {\isasymand}\ C\ {\isasymnoteq}\ sA{\isadigit{8}}{\isadigit{2}}\ \ {\isasymand}\ C\ {\isasymnoteq}\ sA{\isadigit{8}}{\isadigit{1}}\ {\isasymand}\ C\ {\isasymnoteq}\ sA{\isadigit{7}}{\isadigit{2}}\ {\isasymand}\ C\ {\isasymnoteq}\ sA{\isadigit{7}}{\isadigit{1}}\ \isanewline
\ \ {\isasymlongrightarrow}\ {\isacharparenleft}C\ {\isasymnotin}\ {\isacharparenleft}Acc\ level{\isadigit{1}}\ sA{\isadigit{6}}{\isacharparenright}{\isacharparenright}{\isacharparenright}{\isachardoublequoteclose}\isanewline
\ \ \ \ \ \isacommand{by}\isamarkupfalse%
\ {\isacharparenleft}metis\ A{\isadigit{6}}{\isacharunderscore}Acc{\isacharunderscore}L{\isadigit{1}}\ empty{\isacharunderscore}iff\ insert{\isacharunderscore}iff{\isacharparenright}\isanewline
\ \ \isacommand{thus}\isamarkupfalse%
\ {\isacharquery}thesis\ \ \isacommand{by}\isamarkupfalse%
\ {\isacharparenleft}metis\ Acc{\isacharunderscore}SourcesNOT{\isacharparenright}\ \isanewline
\isacommand{qed}\isamarkupfalse%
\endisatagproof
{\isafoldproof}%
\isadelimproof
\isanewline
\endisadelimproof
\isanewline
\isacommand{lemma}\isamarkupfalse%
\ A{\isadigit{5}}{\isacharunderscore}Acc{\isacharunderscore}L{\isadigit{1}}{\isacharcolon}\ {\isachardoublequoteopen}{\isacharparenleft}Acc\ level{\isadigit{1}}\ sA{\isadigit{5}}{\isacharparenright}\ {\isacharequal}\ {\isacharbraceleft}{\isacharbraceright}{\isachardoublequoteclose}\isanewline
\isadelimproof
\endisadelimproof
\isatagproof
\isacommand{by}\isamarkupfalse%
\ {\isacharparenleft}metis\ A{\isadigit{5}}{\isacharunderscore}NotSource{\isacharunderscore}level{\isadigit{1}}\ isNotDSource{\isacharunderscore}EmptyAcc{\isacharparenright}%
\endisatagproof
{\isafoldproof}%
\isadelimproof
\isanewline
\endisadelimproof
\isanewline
\isacommand{lemma}\isamarkupfalse%
\ SourcesA{\isadigit{1}}{\isadigit{1}}{\isacharunderscore}L{\isadigit{1}}{\isacharcolon}\ {\isachardoublequoteopen}Sources\ level{\isadigit{1}}\ sA{\isadigit{1}}{\isadigit{1}}\ {\isacharequal}\ {\isacharbraceleft}{\isacharbraceright}{\isachardoublequoteclose}\ \ \isanewline
\isadelimproof
\endisadelimproof
\isatagproof
\isacommand{by}\isamarkupfalse%
\ {\isacharparenleft}simp\ add{\isacharcolon}\ DSourcesA{\isadigit{1}}{\isadigit{1}}{\isacharunderscore}L{\isadigit{1}}\ DSourcesEmptySources{\isacharparenright}%
\endisatagproof
{\isafoldproof}%
\isadelimproof
\ \isanewline
\endisadelimproof
\isanewline
\isacommand{lemma}\isamarkupfalse%
\ SourcesA{\isadigit{1}}{\isadigit{2}}{\isacharunderscore}L{\isadigit{1}}{\isacharcolon}\ {\isachardoublequoteopen}Sources\ level{\isadigit{1}}\ sA{\isadigit{1}}{\isadigit{2}}\ {\isacharequal}\ {\isacharbraceleft}{\isacharbraceright}{\isachardoublequoteclose}\ \ \isanewline
\isadelimproof
\endisadelimproof
\isatagproof
\isacommand{by}\isamarkupfalse%
\ {\isacharparenleft}simp\ add{\isacharcolon}\ DSourcesA{\isadigit{1}}{\isadigit{2}}{\isacharunderscore}L{\isadigit{1}}\ \ DSourcesEmptySources{\isacharparenright}%
\endisatagproof
{\isafoldproof}%
\isadelimproof
\ \isanewline
\endisadelimproof
\isanewline
\isacommand{lemma}\isamarkupfalse%
\ SourcesA{\isadigit{2}}{\isadigit{1}}{\isacharunderscore}L{\isadigit{1}}{\isacharcolon}\ {\isachardoublequoteopen}Sources\ level{\isadigit{1}}\ sA{\isadigit{2}}{\isadigit{1}}\ {\isacharequal}\ {\isacharbraceleft}sA{\isadigit{1}}{\isadigit{1}}{\isacharbraceright}{\isachardoublequoteclose}\isanewline
\isadelimproof
\endisadelimproof
\isatagproof
\isacommand{by}\isamarkupfalse%
\ {\isacharparenleft}simp\ add{\isacharcolon}\ DSourcesA{\isadigit{2}}{\isadigit{1}}{\isacharunderscore}L{\isadigit{1}}\ SourcesA{\isadigit{1}}{\isadigit{1}}{\isacharunderscore}L{\isadigit{1}}\ \ Sources{\isacharunderscore}singleDSource{\isacharparenright}%
\endisatagproof
{\isafoldproof}%
\isadelimproof
\ \isanewline
\endisadelimproof
\isanewline
\isacommand{lemma}\isamarkupfalse%
\ SourcesA{\isadigit{2}}{\isadigit{2}}{\isacharunderscore}L{\isadigit{1}}{\isacharcolon}\ {\isachardoublequoteopen}Sources\ level{\isadigit{1}}\ sA{\isadigit{2}}{\isadigit{2}}\ {\isacharequal}\ {\isacharbraceleft}sA{\isadigit{1}}{\isadigit{1}}{\isacharcomma}\ sA{\isadigit{2}}{\isadigit{2}}{\isacharcomma}\ \ sA{\isadigit{2}}{\isadigit{3}}{\isacharcomma}\ sA{\isadigit{3}}{\isadigit{1}}{\isacharcomma}\ sA{\isadigit{3}}{\isadigit{2}}{\isacharcomma}\ sA{\isadigit{4}}{\isadigit{1}}{\isacharbraceright}{\isachardoublequoteclose}\isanewline
\isadelimproof
\endisadelimproof
\isatagproof
\isacommand{proof}\isamarkupfalse%
\isanewline
\ \ \isacommand{show}\isamarkupfalse%
\ {\isachardoublequoteopen}Sources\ level{\isadigit{1}}\ sA{\isadigit{2}}{\isadigit{2}}\ {\isasymsubseteq}\ {\isacharbraceleft}sA{\isadigit{1}}{\isadigit{1}}{\isacharcomma}\ sA{\isadigit{2}}{\isadigit{2}}{\isacharcomma}\ sA{\isadigit{2}}{\isadigit{3}}{\isacharcomma}\ sA{\isadigit{3}}{\isadigit{1}}{\isacharcomma}\ sA{\isadigit{3}}{\isadigit{2}}{\isacharcomma}\ sA{\isadigit{4}}{\isadigit{1}}{\isacharbraceright}{\isachardoublequoteclose}\isanewline
\ \ \isacommand{proof}\isamarkupfalse%
\ {\isacharminus}\isanewline
\ \ \ \ \ \isacommand{have}\isamarkupfalse%
\ A{\isadigit{2}}level{\isadigit{1}}{\isacharcolon}{\isachardoublequoteopen}sA{\isadigit{2}}{\isadigit{2}}\ {\isasymin}\ {\isacharparenleft}AbstrLevel\ level{\isadigit{1}}{\isacharparenright}{\isachardoublequoteclose}\ \isacommand{by}\isamarkupfalse%
\ {\isacharparenleft}simp\ add{\isacharcolon}\ AbstrLevel{\isadigit{1}}{\isacharparenright}\isanewline
\ \ \ \ \ \isacommand{from}\isamarkupfalse%
\ A{\isadigit{2}}level{\isadigit{1}}\ \isacommand{have}\isamarkupfalse%
\ sgA{\isadigit{4}}{\isadigit{2}}{\isacharcolon}{\isachardoublequoteopen}sA{\isadigit{4}}{\isadigit{2}}\ {\isasymnotin}\ Sources\ level{\isadigit{1}}\ sA{\isadigit{2}}{\isadigit{2}}{\isachardoublequoteclose}\ \isacommand{by}\isamarkupfalse%
\ {\isacharparenleft}metis\ A{\isadigit{4}}{\isadigit{2}}{\isacharunderscore}NSources{\isacharunderscore}L{\isadigit{1}}\ CSet{\isachardot}distinct{\isacharparenleft}{\isadigit{3}}{\isadigit{4}}{\isadigit{7}}{\isacharparenright}{\isacharparenright}\ \isanewline
\ \ \ \ \ \isacommand{have}\isamarkupfalse%
\ sgA{\isadigit{5}}{\isacharcolon}{\isachardoublequoteopen}sA{\isadigit{5}}\ {\isasymnotin}\ Sources\ level{\isadigit{1}}\ sA{\isadigit{2}}{\isadigit{2}}{\isachardoublequoteclose}\isanewline
\ \ \ \ \ \isacommand{by}\isamarkupfalse%
\ {\isacharparenleft}metis\ A{\isadigit{5}}{\isacharunderscore}NotSource{\isacharunderscore}level{\isadigit{1}}\ Acc{\isacharunderscore}Sources\ all{\isacharunderscore}not{\isacharunderscore}in{\isacharunderscore}conv\ isNotDSource{\isacharunderscore}EmptyAcc{\isacharparenright}\ \ \isanewline
\ \ \ \ \ \isacommand{have}\isamarkupfalse%
\ sgA{\isadigit{1}}{\isadigit{2}}{\isacharcolon}{\isachardoublequoteopen}sA{\isadigit{1}}{\isadigit{2}}\ {\isasymnotin}\ Sources\ level{\isadigit{1}}\ sA{\isadigit{2}}{\isadigit{2}}{\isachardoublequoteclose}\ \isacommand{by}\isamarkupfalse%
\ {\isacharparenleft}metis\ A{\isadigit{1}}{\isadigit{2}}{\isacharunderscore}NotSource{\isacharunderscore}level{\isadigit{1}}\ A{\isadigit{2}}level{\isadigit{1}}\ isNotSource{\isacharunderscore}Sources{\isacharparenright}\ \ \ \isanewline
\ \ \ \ \ \isacommand{have}\isamarkupfalse%
\ sgA{\isadigit{2}}{\isadigit{1}}{\isacharcolon}{\isachardoublequoteopen}sA{\isadigit{2}}{\isadigit{1}}\ {\isasymnotin}\ Sources\ level{\isadigit{1}}\ sA{\isadigit{2}}{\isadigit{2}}{\isachardoublequoteclose}\isanewline
\ \ \ \ \ \isacommand{by}\isamarkupfalse%
\ {\isacharparenleft}metis\ A{\isadigit{2}}{\isadigit{1}}{\isacharunderscore}NotSource{\isacharunderscore}level{\isadigit{1}}\ DAcc{\isacharunderscore}DSourcesNOT\ NDSourceExistsDSource\ empty{\isacharunderscore}iff\ isNotDSource{\isacharunderscore}EmptyDAcc{\isacharparenright}\isanewline
\ \ \ \ \ \isacommand{from}\isamarkupfalse%
\ A{\isadigit{2}}level{\isadigit{1}}\ \isacommand{have}\isamarkupfalse%
\ sgA{\isadigit{6}}{\isacharcolon}{\isachardoublequoteopen}sA{\isadigit{6}}\ {\isasymnotin}\ Sources\ level{\isadigit{1}}\ sA{\isadigit{2}}{\isadigit{2}}{\isachardoublequoteclose}\ \isacommand{by}\isamarkupfalse%
\ {\isacharparenleft}simp\ add{\isacharcolon}\ A{\isadigit{6}}{\isacharunderscore}NSources{\isacharunderscore}L{\isadigit{1}}{\isacharparenright}\isanewline
\ \ \ \ \ \isacommand{from}\isamarkupfalse%
\ A{\isadigit{2}}level{\isadigit{1}}\ \isacommand{have}\isamarkupfalse%
\ sgA{\isadigit{7}}{\isadigit{1}}{\isacharcolon}{\isachardoublequoteopen}sA{\isadigit{7}}{\isadigit{1}}\ {\isasymnotin}\ Sources\ level{\isadigit{1}}\ sA{\isadigit{2}}{\isadigit{2}}{\isachardoublequoteclose}\ \isacommand{by}\isamarkupfalse%
\ {\isacharparenleft}simp\ add{\isacharcolon}\ A{\isadigit{7}}{\isadigit{1}}{\isacharunderscore}NSources{\isacharunderscore}L{\isadigit{1}}{\isacharparenright}\isanewline
\ \ \ \ \ \isacommand{from}\isamarkupfalse%
\ A{\isadigit{2}}level{\isadigit{1}}\ \isacommand{have}\isamarkupfalse%
\ sgA{\isadigit{7}}{\isadigit{2}}{\isacharcolon}{\isachardoublequoteopen}sA{\isadigit{7}}{\isadigit{2}}\ {\isasymnotin}\ Sources\ level{\isadigit{1}}\ sA{\isadigit{2}}{\isadigit{2}}{\isachardoublequoteclose}\ \isacommand{by}\isamarkupfalse%
\ {\isacharparenleft}simp\ add{\isacharcolon}\ A{\isadigit{7}}{\isadigit{2}}{\isacharunderscore}NSources{\isacharunderscore}L{\isadigit{1}}{\isacharparenright}\isanewline
\ \ \ \ \ \isacommand{from}\isamarkupfalse%
\ A{\isadigit{2}}level{\isadigit{1}}\ \isacommand{have}\isamarkupfalse%
\ sgA{\isadigit{8}}{\isadigit{1}}{\isacharcolon}{\isachardoublequoteopen}sA{\isadigit{8}}{\isadigit{1}}\ {\isasymnotin}\ Sources\ level{\isadigit{1}}\ sA{\isadigit{2}}{\isadigit{2}}{\isachardoublequoteclose}\ \isacommand{by}\isamarkupfalse%
\ {\isacharparenleft}simp\ add{\isacharcolon}\ A{\isadigit{8}}{\isadigit{1}}{\isacharunderscore}NSources{\isacharunderscore}L{\isadigit{1}}{\isacharparenright}\isanewline
\ \ \ \ \ \isacommand{from}\isamarkupfalse%
\ A{\isadigit{2}}level{\isadigit{1}}\ \isacommand{have}\isamarkupfalse%
\ sgA{\isadigit{8}}{\isadigit{2}}{\isacharcolon}{\isachardoublequoteopen}sA{\isadigit{8}}{\isadigit{2}}\ {\isasymnotin}\ Sources\ level{\isadigit{1}}\ sA{\isadigit{2}}{\isadigit{2}}{\isachardoublequoteclose}\ \isacommand{by}\isamarkupfalse%
\ {\isacharparenleft}simp\ add{\isacharcolon}\ A{\isadigit{8}}{\isadigit{2}}{\isacharunderscore}NSources{\isacharunderscore}L{\isadigit{1}}{\isacharparenright}\isanewline
\ \ \ \ \ \isacommand{from}\isamarkupfalse%
\ A{\isadigit{2}}level{\isadigit{1}}\ \isacommand{have}\isamarkupfalse%
\ sgA{\isadigit{9}}{\isadigit{1}}{\isacharcolon}{\isachardoublequoteopen}sA{\isadigit{9}}{\isadigit{1}}\ {\isasymnotin}\ Sources\ level{\isadigit{1}}\ sA{\isadigit{2}}{\isadigit{2}}{\isachardoublequoteclose}\ \isacommand{by}\isamarkupfalse%
\ {\isacharparenleft}simp\ add{\isacharcolon}\ A{\isadigit{9}}{\isadigit{1}}{\isacharunderscore}NSources{\isacharunderscore}L{\isadigit{1}}{\isacharparenright}\isanewline
\ \ \ \ \ \isacommand{from}\isamarkupfalse%
\ A{\isadigit{2}}level{\isadigit{1}}\ \isacommand{have}\isamarkupfalse%
\ sgA{\isadigit{9}}{\isadigit{2}}{\isacharcolon}{\isachardoublequoteopen}sA{\isadigit{9}}{\isadigit{2}}\ {\isasymnotin}\ Sources\ level{\isadigit{1}}\ sA{\isadigit{2}}{\isadigit{2}}{\isachardoublequoteclose}\ \isacommand{by}\isamarkupfalse%
\ {\isacharparenleft}simp\ add{\isacharcolon}\ A{\isadigit{9}}{\isadigit{2}}{\isacharunderscore}NSources{\isacharunderscore}L{\isadigit{1}}{\isacharparenright}\ \isanewline
\ \ \ \ \ \isacommand{from}\isamarkupfalse%
\ A{\isadigit{2}}level{\isadigit{1}}\ \isacommand{have}\isamarkupfalse%
\ sgA{\isadigit{9}}{\isadigit{3}}{\isacharcolon}{\isachardoublequoteopen}sA{\isadigit{9}}{\isadigit{3}}\ {\isasymnotin}\ Sources\ level{\isadigit{1}}\ sA{\isadigit{2}}{\isadigit{2}}{\isachardoublequoteclose}\ \isacommand{by}\isamarkupfalse%
\ {\isacharparenleft}metis\ A{\isadigit{9}}{\isadigit{3}}{\isacharunderscore}NotSourceSet{\isacharunderscore}level{\isadigit{1}}{\isacharparenright}\ \ \isanewline
\ \ \ \ \ \isacommand{have}\isamarkupfalse%
\ {\isachardoublequoteopen}Sources\ level{\isadigit{1}}\ sA{\isadigit{2}}{\isadigit{2}}\ {\isasymsubseteq}\ {\isacharbraceleft}sA{\isadigit{1}}{\isadigit{1}}{\isacharcomma}\ sA{\isadigit{1}}{\isadigit{2}}{\isacharcomma}\ sA{\isadigit{2}}{\isadigit{1}}{\isacharcomma}\ sA{\isadigit{2}}{\isadigit{2}}{\isacharcomma}\ sA{\isadigit{2}}{\isadigit{3}}{\isacharcomma}\ sA{\isadigit{3}}{\isadigit{1}}{\isacharcomma}\ sA{\isadigit{3}}{\isadigit{2}}{\isacharcomma}\ \isanewline
\ \ \ \ \ \ \ \ sA{\isadigit{4}}{\isadigit{1}}{\isacharcomma}\ sA{\isadigit{4}}{\isadigit{2}}{\isacharcomma}\ sA{\isadigit{5}}{\isacharcomma}\ sA{\isadigit{6}}{\isacharcomma}\ sA{\isadigit{7}}{\isadigit{1}}{\isacharcomma}\ sA{\isadigit{7}}{\isadigit{2}}{\isacharcomma}\ sA{\isadigit{8}}{\isadigit{1}}{\isacharcomma}\ sA{\isadigit{8}}{\isadigit{2}}{\isacharcomma}\ sA{\isadigit{9}}{\isadigit{1}}{\isacharcomma}\ sA{\isadigit{9}}{\isadigit{2}}{\isacharcomma}\ sA{\isadigit{9}}{\isadigit{3}}{\isacharbraceright}{\isachardoublequoteclose}\isanewline
\ \ \ \ \ \ \ \ \isacommand{by}\isamarkupfalse%
\ {\isacharparenleft}metis\ AbstrLevel{\isadigit{1}}\ SourcesLevelX{\isacharparenright}\ \isanewline
\ \ \ \ \ \isacommand{with}\isamarkupfalse%
\ sgA{\isadigit{5}}\ sgA{\isadigit{1}}{\isadigit{2}}\ sgA{\isadigit{2}}{\isadigit{1}}\ sgA{\isadigit{4}}{\isadigit{2}}\ sgA{\isadigit{6}}\ sgA{\isadigit{7}}{\isadigit{1}}\ sgA{\isadigit{7}}{\isadigit{2}}\ sgA{\isadigit{8}}{\isadigit{1}}\ sgA{\isadigit{8}}{\isadigit{2}}\ sgA{\isadigit{9}}{\isadigit{1}}\ sgA{\isadigit{9}}{\isadigit{2}}\ sgA{\isadigit{9}}{\isadigit{3}}\ \isacommand{show}\isamarkupfalse%
\ \isanewline
\ \ \ \ \ {\isachardoublequoteopen}Sources\ level{\isadigit{1}}\ sA{\isadigit{2}}{\isadigit{2}}\ {\isasymsubseteq}\ {\isacharbraceleft}sA{\isadigit{1}}{\isadigit{1}}{\isacharcomma}\ sA{\isadigit{2}}{\isadigit{2}}{\isacharcomma}\ sA{\isadigit{2}}{\isadigit{3}}{\isacharcomma}\ sA{\isadigit{3}}{\isadigit{1}}{\isacharcomma}\ sA{\isadigit{3}}{\isadigit{2}}{\isacharcomma}\ sA{\isadigit{4}}{\isadigit{1}}{\isacharbraceright}{\isachardoublequoteclose}\isanewline
\ \ \ \ \ \ \ \ \ \isacommand{by}\isamarkupfalse%
\ auto\ \isanewline
\ \ \ \ \isacommand{qed}\isamarkupfalse%
\isanewline
\isacommand{next}\isamarkupfalse%
\ \isanewline
\ \ \isacommand{show}\isamarkupfalse%
\ {\isachardoublequoteopen}{\isacharbraceleft}sA{\isadigit{1}}{\isadigit{1}}{\isacharcomma}\ sA{\isadigit{2}}{\isadigit{2}}{\isacharcomma}\ sA{\isadigit{2}}{\isadigit{3}}{\isacharcomma}\ sA{\isadigit{3}}{\isadigit{1}}{\isacharcomma}\ sA{\isadigit{3}}{\isadigit{2}}{\isacharcomma}\ sA{\isadigit{4}}{\isadigit{1}}{\isacharbraceright}\ {\isasymsubseteq}\ Sources\ level{\isadigit{1}}\ sA{\isadigit{2}}{\isadigit{2}}{\isachardoublequoteclose}\ \isanewline
\ \ \isacommand{proof}\isamarkupfalse%
\ {\isacharminus}\ \isanewline
\ \ \ \ \isacommand{have}\isamarkupfalse%
\ sDef{\isacharcolon}{\isachardoublequoteopen}{\isacharparenleft}Sources\ level{\isadigit{1}}\ sA{\isadigit{2}}{\isadigit{2}}{\isacharparenright}\ {\isacharequal}\ {\isacharparenleft}DSources\ level{\isadigit{1}}\ sA{\isadigit{2}}{\isadigit{2}}{\isacharparenright}\ {\isasymunion}\ {\isacharparenleft}{\isasymUnion}\ S\ {\isasymin}\ {\isacharparenleft}DSources\ level{\isadigit{1}}\ sA{\isadigit{2}}{\isadigit{2}}{\isacharparenright}{\isachardot}\ {\isacharparenleft}Sources\ level{\isadigit{1}}\ S{\isacharparenright}{\isacharparenright}{\isachardoublequoteclose}\ \isanewline
\ \ \ \ \ \ \ \isacommand{by}\isamarkupfalse%
\ {\isacharparenleft}rule\ SourcesDef{\isacharparenright}\ \isanewline
\ \ \ \ \isacommand{have}\isamarkupfalse%
\ A{\isadigit{1}}{\isadigit{1}}s{\isacharcolon}\ {\isachardoublequoteopen}sA{\isadigit{1}}{\isadigit{1}}\ {\isasymin}\ Sources\ level{\isadigit{1}}\ sA{\isadigit{2}}{\isadigit{2}}{\isachardoublequoteclose}\ \isacommand{by}\isamarkupfalse%
\ {\isacharparenleft}metis\ DSourceIsSource\ DSourcesA{\isadigit{2}}{\isadigit{2}}{\isacharunderscore}L{\isadigit{1}}\ insertI{\isadigit{1}}{\isacharparenright}\ \ \isanewline
\ \ \ \ \isacommand{have}\isamarkupfalse%
\ A{\isadigit{4}}{\isadigit{1}}s{\isacharcolon}\ {\isachardoublequoteopen}sA{\isadigit{4}}{\isadigit{1}}\ {\isasymin}\ Sources\ level{\isadigit{1}}\ sA{\isadigit{2}}{\isadigit{2}}{\isachardoublequoteclose}\ \isacommand{by}\isamarkupfalse%
\ {\isacharparenleft}metis\ {\isacharparenleft}full{\isacharunderscore}types{\isacharparenright}\ DSourceIsSource\ DSourcesA{\isadigit{2}}{\isadigit{2}}{\isacharunderscore}L{\isadigit{1}}\ insertCI{\isacharparenright}\isanewline
\ \ \ \ \isacommand{have}\isamarkupfalse%
\ A{\isadigit{3}}{\isadigit{1}}s{\isacharcolon}\ {\isachardoublequoteopen}sA{\isadigit{3}}{\isadigit{1}}\ {\isasymin}\ Sources\ level{\isadigit{1}}\ sA{\isadigit{2}}{\isadigit{2}}{\isachardoublequoteclose}\ \isanewline
\ \ \ \ \ \ \isacommand{by}\isamarkupfalse%
\ {\isacharparenleft}metis\ {\isacharparenleft}full{\isacharunderscore}types{\isacharparenright}\ A{\isadigit{4}}{\isadigit{1}}s\ DSourceIsSource\ DSourcesA{\isadigit{4}}{\isadigit{1}}{\isacharunderscore}L{\isadigit{1}}\ SourcesTrans\ insertCI{\isacharparenright}\isanewline
\ \ \ \ \isacommand{have}\isamarkupfalse%
\ A{\isadigit{3}}{\isadigit{2}}s{\isacharcolon}\ {\isachardoublequoteopen}sA{\isadigit{3}}{\isadigit{2}}\ {\isasymin}\ Sources\ level{\isadigit{1}}\ sA{\isadigit{2}}{\isadigit{2}}{\isachardoublequoteclose}\isanewline
\ \ \ \ \ \ \isacommand{by}\isamarkupfalse%
\ {\isacharparenleft}metis\ A{\isadigit{3}}{\isadigit{2}}{\isacharunderscore}DAcc{\isacharunderscore}level{\isadigit{1}}\ A{\isadigit{4}}{\isadigit{1}}s\ DAcc{\isacharunderscore}DSourcesNOT\ DSourceOfSource\ insertI{\isadigit{1}}{\isacharparenright}\isanewline
\ \ \ \ \isacommand{have}\isamarkupfalse%
\ A{\isadigit{2}}{\isadigit{3}}s{\isacharcolon}\ {\isachardoublequoteopen}sA{\isadigit{2}}{\isadigit{3}}\ {\isasymin}\ Sources\ level{\isadigit{1}}\ sA{\isadigit{2}}{\isadigit{2}}{\isachardoublequoteclose}\ \ \isacommand{by}\isamarkupfalse%
\ {\isacharparenleft}metis\ A{\isadigit{3}}{\isadigit{2}}s\ DSourceOfSource\ DSourcesA{\isadigit{3}}{\isadigit{2}}{\isacharunderscore}L{\isadigit{1}}\ insertI{\isadigit{1}}{\isacharparenright}\isanewline
\ \ \ \ \isacommand{have}\isamarkupfalse%
\ A{\isadigit{2}}{\isadigit{2}}s{\isacharcolon}\ {\isachardoublequoteopen}sA{\isadigit{2}}{\isadigit{2}}\ {\isasymin}\ Sources\ level{\isadigit{1}}\ sA{\isadigit{2}}{\isadigit{2}}{\isachardoublequoteclose}\ \ \isacommand{by}\isamarkupfalse%
\ {\isacharparenleft}metis\ A{\isadigit{3}}{\isadigit{1}}s\ DSourceOfSource\ DSourcesA{\isadigit{3}}{\isadigit{1}}{\isacharunderscore}L{\isadigit{1}}\ insertI{\isadigit{1}}{\isacharparenright}\isanewline
\ \ \ \ \isacommand{with}\isamarkupfalse%
\ A{\isadigit{1}}{\isadigit{1}}s\ A{\isadigit{2}}{\isadigit{2}}s\ A{\isadigit{2}}{\isadigit{3}}s\ A{\isadigit{3}}{\isadigit{1}}s\ A{\isadigit{3}}{\isadigit{2}}s\ A{\isadigit{4}}{\isadigit{1}}s\ \isacommand{show}\isamarkupfalse%
\ {\isacharquery}thesis\ \isacommand{by}\isamarkupfalse%
\ auto\isanewline
\ \ \ \ \isacommand{qed}\isamarkupfalse%
\isanewline
\isacommand{qed}\isamarkupfalse%
\endisatagproof
{\isafoldproof}%
\isadelimproof
\ \isanewline
\endisadelimproof
\isanewline
\isacommand{lemma}\isamarkupfalse%
\ SourcesA{\isadigit{2}}{\isadigit{3}}{\isacharunderscore}L{\isadigit{1}}{\isacharcolon}\ {\isachardoublequoteopen}Sources\ level{\isadigit{1}}\ sA{\isadigit{2}}{\isadigit{3}}\ {\isacharequal}\ {\isacharbraceleft}sA{\isadigit{1}}{\isadigit{1}}{\isacharbraceright}{\isachardoublequoteclose}\isanewline
\isadelimproof
\endisadelimproof
\isatagproof
\isacommand{by}\isamarkupfalse%
\ {\isacharparenleft}simp\ add{\isacharcolon}\ DSourcesA{\isadigit{2}}{\isadigit{3}}{\isacharunderscore}L{\isadigit{1}}\ SourcesA{\isadigit{1}}{\isadigit{1}}{\isacharunderscore}L{\isadigit{1}}\ \ Sources{\isacharunderscore}singleDSource{\isacharparenright}%
\endisatagproof
{\isafoldproof}%
\isadelimproof
\ \ \isanewline
\endisadelimproof
\isanewline
\isacommand{lemma}\isamarkupfalse%
\ SourcesA{\isadigit{3}}{\isadigit{1}}{\isacharunderscore}L{\isadigit{1}}{\isacharcolon}\ {\isachardoublequoteopen}Sources\ level{\isadigit{1}}\ sA{\isadigit{3}}{\isadigit{1}}\ {\isacharequal}\ {\isacharbraceleft}sA{\isadigit{1}}{\isadigit{1}}{\isacharcomma}\ sA{\isadigit{2}}{\isadigit{2}}{\isacharcomma}\ sA{\isadigit{2}}{\isadigit{3}}{\isacharcomma}\ sA{\isadigit{3}}{\isadigit{1}}{\isacharcomma}\ sA{\isadigit{3}}{\isadigit{2}}{\isacharcomma}\ sA{\isadigit{4}}{\isadigit{1}}{\isacharbraceright}{\isachardoublequoteclose}\isanewline
\isadelimproof
\endisadelimproof
\isatagproof
\isacommand{by}\isamarkupfalse%
\ {\isacharparenleft}metis\ DSourcesA{\isadigit{3}}{\isadigit{1}}{\isacharunderscore}L{\isadigit{1}}\ SourcesA{\isadigit{2}}{\isadigit{2}}{\isacharunderscore}L{\isadigit{1}}\ Sources{\isacharunderscore}singleDSource\ Un{\isacharunderscore}insert{\isacharunderscore}right\ insert{\isacharunderscore}absorb{\isadigit{2}}\ insert{\isacharunderscore}is{\isacharunderscore}Un{\isacharparenright}%
\endisatagproof
{\isafoldproof}%
\isadelimproof
\isanewline
\endisadelimproof
\isanewline
\isacommand{lemma}\isamarkupfalse%
\ SourcesA{\isadigit{3}}{\isadigit{2}}{\isacharunderscore}L{\isadigit{1}}{\isacharcolon}\ {\isachardoublequoteopen}Sources\ level{\isadigit{1}}\ sA{\isadigit{3}}{\isadigit{2}}\ {\isacharequal}\ {\isacharbraceleft}sA{\isadigit{1}}{\isadigit{1}}{\isacharcomma}\ sA{\isadigit{2}}{\isadigit{3}}{\isacharbraceright}{\isachardoublequoteclose}\isanewline
\isadelimproof
\endisadelimproof
\isatagproof
\isacommand{by}\isamarkupfalse%
\ {\isacharparenleft}metis\ DSourcesA{\isadigit{3}}{\isadigit{2}}{\isacharunderscore}L{\isadigit{1}}\ SourcesA{\isadigit{2}}{\isadigit{3}}{\isacharunderscore}L{\isadigit{1}}\ Sources{\isacharunderscore}singleDSource\ Un{\isacharunderscore}insert{\isacharunderscore}right\ insert{\isacharunderscore}is{\isacharunderscore}Un{\isacharparenright}%
\endisatagproof
{\isafoldproof}%
\isadelimproof
\isanewline
\endisadelimproof
\ \isanewline
\isacommand{lemma}\isamarkupfalse%
\ SourcesA{\isadigit{4}}{\isadigit{1}}{\isacharunderscore}L{\isadigit{1}}{\isacharcolon}\ {\isachardoublequoteopen}Sources\ level{\isadigit{1}}\ sA{\isadigit{4}}{\isadigit{1}}\ {\isacharequal}\ {\isacharbraceleft}sA{\isadigit{1}}{\isadigit{1}}{\isacharcomma}\ sA{\isadigit{2}}{\isadigit{2}}{\isacharcomma}\ sA{\isadigit{2}}{\isadigit{3}}{\isacharcomma}\ sA{\isadigit{3}}{\isadigit{1}}{\isacharcomma}\ sA{\isadigit{3}}{\isadigit{2}}{\isacharcomma}\ sA{\isadigit{4}}{\isadigit{1}}{\isacharbraceright}{\isachardoublequoteclose}\isanewline
\isadelimproof
\endisadelimproof
\isatagproof
\isacommand{by}\isamarkupfalse%
\ {\isacharparenleft}metis\ DSourcesA{\isadigit{4}}{\isadigit{1}}{\isacharunderscore}L{\isadigit{1}}\ SourcesA{\isadigit{3}}{\isadigit{1}}{\isacharunderscore}L{\isadigit{1}}\ SourcesA{\isadigit{3}}{\isadigit{2}}{\isacharunderscore}L{\isadigit{1}}\ Sources{\isacharunderscore}{\isadigit{2}}DSources\ Un{\isacharunderscore}absorb\ Un{\isacharunderscore}commute\ Un{\isacharunderscore}insert{\isacharunderscore}left{\isacharparenright}%
\endisatagproof
{\isafoldproof}%
\isadelimproof
\isanewline
\endisadelimproof
\isanewline
\isacommand{lemma}\isamarkupfalse%
\ SourcesA{\isadigit{4}}{\isadigit{2}}{\isacharunderscore}L{\isadigit{1}}{\isacharcolon}\ {\isachardoublequoteopen}Sources\ level{\isadigit{1}}\ sA{\isadigit{4}}{\isadigit{2}}\ {\isacharequal}\ {\isacharbraceleft}{\isacharbraceright}{\isachardoublequoteclose}\ \ \isanewline
\isadelimproof
\endisadelimproof
\isatagproof
\isacommand{by}\isamarkupfalse%
\ {\isacharparenleft}simp\ add{\isacharcolon}\ DSourcesA{\isadigit{4}}{\isadigit{2}}{\isacharunderscore}L{\isadigit{1}}\ \ DSourcesEmptySources{\isacharparenright}%
\endisatagproof
{\isafoldproof}%
\isadelimproof
\ \isanewline
\endisadelimproof
\isanewline
\isacommand{lemma}\isamarkupfalse%
\ SourcesA{\isadigit{5}}{\isacharunderscore}L{\isadigit{1}}{\isacharcolon}\ {\isachardoublequoteopen}Sources\ level{\isadigit{1}}\ sA{\isadigit{5}}\ {\isacharequal}\ {\isacharbraceleft}sA{\isadigit{4}}{\isadigit{2}}{\isacharbraceright}{\isachardoublequoteclose}\isanewline
\isadelimproof
\endisadelimproof
\isatagproof
\isacommand{by}\isamarkupfalse%
\ {\isacharparenleft}simp\ add{\isacharcolon}\ DSourcesA{\isadigit{5}}{\isacharunderscore}L{\isadigit{1}}\ SourcesA{\isadigit{4}}{\isadigit{2}}{\isacharunderscore}L{\isadigit{1}}\ \ Sources{\isacharunderscore}singleDSource{\isacharparenright}%
\endisatagproof
{\isafoldproof}%
\isadelimproof
\ \isanewline
\endisadelimproof
\isanewline
\isacommand{lemma}\isamarkupfalse%
\ SourcesA{\isadigit{6}}{\isacharunderscore}L{\isadigit{1}}{\isacharcolon}\ {\isachardoublequoteopen}Sources\ level{\isadigit{1}}\ sA{\isadigit{6}}\ {\isacharequal}\ {\isacharbraceleft}{\isacharbraceright}{\isachardoublequoteclose}\ \ \isanewline
\isadelimproof
\endisadelimproof
\isatagproof
\isacommand{by}\isamarkupfalse%
\ {\isacharparenleft}simp\ add{\isacharcolon}\ DSourcesA{\isadigit{6}}{\isacharunderscore}L{\isadigit{1}}\ DSourcesEmptySources{\isacharparenright}%
\endisatagproof
{\isafoldproof}%
\isadelimproof
\ \isanewline
\endisadelimproof
\isanewline
\isacommand{lemma}\isamarkupfalse%
\ SourcesA{\isadigit{7}}{\isadigit{1}}{\isacharunderscore}L{\isadigit{1}}{\isacharcolon}\ {\isachardoublequoteopen}Sources\ level{\isadigit{1}}\ sA{\isadigit{7}}{\isadigit{1}}\ {\isacharequal}\ {\isacharbraceleft}sA{\isadigit{6}}{\isacharbraceright}{\isachardoublequoteclose}\isanewline
\isadelimproof
\endisadelimproof
\isatagproof
\isacommand{by}\isamarkupfalse%
\ {\isacharparenleft}metis\ DSourcesA{\isadigit{7}}{\isadigit{1}}{\isacharunderscore}L{\isadigit{1}}\ SourcesA{\isadigit{6}}{\isacharunderscore}L{\isadigit{1}}\ SourcesEmptyDSources\ SourcesOnlyDSources\ singleton{\isacharunderscore}iff{\isacharparenright}%
\endisatagproof
{\isafoldproof}%
\isadelimproof
\ \ \ \isanewline
\endisadelimproof
\isanewline
\isacommand{lemma}\isamarkupfalse%
\ SourcesA{\isadigit{8}}{\isadigit{1}}{\isacharunderscore}L{\isadigit{1}}{\isacharcolon}\ {\isachardoublequoteopen}Sources\ level{\isadigit{1}}\ sA{\isadigit{8}}{\isadigit{1}}\ {\isacharequal}\ {\isacharbraceleft}sA{\isadigit{6}}{\isacharcomma}\ sA{\isadigit{7}}{\isadigit{1}}{\isacharcomma}\ sA{\isadigit{8}}{\isadigit{1}}{\isacharcomma}\ sA{\isadigit{9}}{\isadigit{1}}{\isacharbraceright}{\isachardoublequoteclose}\ \ \isanewline
\isadelimproof
\endisadelimproof
\isatagproof
\isacommand{proof}\isamarkupfalse%
\ {\isacharminus}\ \isanewline
\ \ \isacommand{have}\isamarkupfalse%
\ \ dA{\isadigit{8}}{\isadigit{1}}{\isacharcolon}{\isachardoublequoteopen}DSources\ level{\isadigit{1}}\ sA{\isadigit{8}}{\isadigit{1}}\ {\isacharequal}\ {\isacharbraceleft}sA{\isadigit{7}}{\isadigit{1}}{\isacharcomma}\ sA{\isadigit{9}}{\isadigit{1}}{\isacharbraceright}{\isachardoublequoteclose}\ \isacommand{by}\isamarkupfalse%
\ {\isacharparenleft}rule\ DSourcesA{\isadigit{8}}{\isadigit{1}}{\isacharunderscore}L{\isadigit{1}}{\isacharparenright}\isanewline
\ \ \isacommand{have}\isamarkupfalse%
\ \ dA{\isadigit{9}}{\isadigit{1}}{\isacharcolon}{\isachardoublequoteopen}DSources\ level{\isadigit{1}}\ sA{\isadigit{9}}{\isadigit{1}}\ {\isacharequal}\ {\isacharbraceleft}sA{\isadigit{8}}{\isadigit{1}}{\isacharbraceright}{\isachardoublequoteclose}\ \isacommand{by}\isamarkupfalse%
\ {\isacharparenleft}rule\ DSourcesA{\isadigit{9}}{\isadigit{1}}{\isacharunderscore}L{\isadigit{1}}{\isacharparenright}\isanewline
\ \ \isacommand{have}\isamarkupfalse%
\ {\isachardoublequoteopen}{\isacharparenleft}Sources\ level{\isadigit{1}}\ sA{\isadigit{8}}{\isadigit{1}}{\isacharparenright}\ {\isacharequal}\ {\isacharparenleft}DSources\ level{\isadigit{1}}\ sA{\isadigit{8}}{\isadigit{1}}{\isacharparenright}\ {\isasymunion}\ {\isacharparenleft}{\isasymUnion}\ S\ {\isasymin}\ {\isacharparenleft}DSources\ level{\isadigit{1}}\ sA{\isadigit{8}}{\isadigit{1}}{\isacharparenright}{\isachardot}\ {\isacharparenleft}Sources\ level{\isadigit{1}}\ S{\isacharparenright}{\isacharparenright}{\isachardoublequoteclose}\ \isanewline
\ \ \ \ \isacommand{by}\isamarkupfalse%
\ {\isacharparenleft}rule\ SourcesDef{\isacharparenright}\isanewline
\ \ \isacommand{with}\isamarkupfalse%
\ dA{\isadigit{8}}{\isadigit{1}}\ \isacommand{have}\isamarkupfalse%
\ \ {\isachardoublequoteopen}{\isacharparenleft}Sources\ level{\isadigit{1}}\ sA{\isadigit{8}}{\isadigit{1}}{\isacharparenright}\ {\isacharequal}\ {\isacharparenleft}{\isacharbraceleft}sA{\isadigit{7}}{\isadigit{1}}{\isacharcomma}\ sA{\isadigit{9}}{\isadigit{1}}{\isacharbraceright}\ {\isasymunion}\ {\isacharparenleft}Sources\ level{\isadigit{1}}\ sA{\isadigit{7}}{\isadigit{1}}{\isacharparenright}\ {\isasymunion}\ {\isacharparenleft}Sources\ level{\isadigit{1}}\ sA{\isadigit{9}}{\isadigit{1}}{\isacharparenright}{\isacharparenright}{\isachardoublequoteclose}\isanewline
\ \ \ \ \isacommand{by}\isamarkupfalse%
\ {\isacharparenleft}metis\ {\isacharparenleft}hide{\isacharunderscore}lams{\isacharcomma}\ no{\isacharunderscore}types{\isacharparenright}\ SUP{\isacharunderscore}empty\ UN{\isacharunderscore}insert\ Un{\isacharunderscore}insert{\isacharunderscore}left\ sup{\isacharunderscore}bot{\isachardot}left{\isacharunderscore}neutral\ sup{\isacharunderscore}commute{\isacharparenright}\isanewline
\ \ \isacommand{hence}\isamarkupfalse%
\ sourcesA{\isadigit{8}}{\isadigit{1}}{\isacharcolon}{\isachardoublequoteopen}{\isacharparenleft}Sources\ level{\isadigit{1}}\ sA{\isadigit{8}}{\isadigit{1}}{\isacharparenright}\ {\isacharequal}\ {\isacharparenleft}{\isacharbraceleft}sA{\isadigit{7}}{\isadigit{1}}{\isacharcomma}\ sA{\isadigit{9}}{\isadigit{1}}{\isacharcomma}\ sA{\isadigit{6}}{\isacharbraceright}\ {\isasymunion}\ {\isacharparenleft}Sources\ level{\isadigit{1}}\ sA{\isadigit{9}}{\isadigit{1}}{\isacharparenright}{\isacharparenright}{\isachardoublequoteclose}\isanewline
\ \ \ \ \isacommand{by}\isamarkupfalse%
\ {\isacharparenleft}metis\ SourcesA{\isadigit{7}}{\isadigit{1}}{\isacharunderscore}L{\isadigit{1}}\ insert{\isacharunderscore}is{\isacharunderscore}Un\ sup{\isacharunderscore}assoc{\isacharparenright}\isanewline
\ \ \isacommand{have}\isamarkupfalse%
\ {\isachardoublequoteopen}{\isacharparenleft}Sources\ level{\isadigit{1}}\ sA{\isadigit{9}}{\isadigit{1}}{\isacharparenright}\ {\isacharequal}\ {\isacharparenleft}DSources\ level{\isadigit{1}}\ sA{\isadigit{9}}{\isadigit{1}}{\isacharparenright}\ {\isasymunion}\ {\isacharparenleft}{\isasymUnion}\ S\ {\isasymin}\ {\isacharparenleft}DSources\ level{\isadigit{1}}\ sA{\isadigit{9}}{\isadigit{1}}{\isacharparenright}{\isachardot}\ {\isacharparenleft}Sources\ level{\isadigit{1}}\ S{\isacharparenright}{\isacharparenright}{\isachardoublequoteclose}\ \isanewline
\ \ \ \ \isacommand{by}\isamarkupfalse%
\ {\isacharparenleft}rule\ SourcesDef{\isacharparenright}\isanewline
\ \ \isacommand{with}\isamarkupfalse%
\ dA{\isadigit{9}}{\isadigit{1}}\ \isacommand{have}\isamarkupfalse%
\ {\isachardoublequoteopen}{\isacharparenleft}Sources\ level{\isadigit{1}}\ sA{\isadigit{9}}{\isadigit{1}}{\isacharparenright}\ {\isacharequal}\ {\isacharparenleft}{\isacharbraceleft}sA{\isadigit{8}}{\isadigit{1}}{\isacharbraceright}\ {\isasymunion}\ {\isacharparenleft}Sources\ level{\isadigit{1}}\ sA{\isadigit{8}}{\isadigit{1}}{\isacharparenright}{\isacharparenright}{\isachardoublequoteclose}\ \ \isacommand{by}\isamarkupfalse%
\ simp\isanewline
\ \ \isacommand{with}\isamarkupfalse%
\ sourcesA{\isadigit{8}}{\isadigit{1}}\ \isacommand{have}\isamarkupfalse%
\ {\isachardoublequoteopen}{\isacharparenleft}Sources\ level{\isadigit{1}}\ sA{\isadigit{8}}{\isadigit{1}}{\isacharparenright}\ {\isacharequal}\ {\isacharbraceleft}sA{\isadigit{7}}{\isadigit{1}}{\isacharcomma}\ sA{\isadigit{9}}{\isadigit{1}}{\isacharcomma}\ sA{\isadigit{6}}{\isacharbraceright}\ {\isasymunion}\ {\isacharbraceleft}sA{\isadigit{8}}{\isadigit{1}}{\isacharbraceright}\ {\isasymunion}\ {\isacharbraceleft}sA{\isadigit{8}}{\isadigit{1}}{\isacharcomma}\ sA{\isadigit{9}}{\isadigit{1}}{\isacharbraceright}{\isachardoublequoteclose}\isanewline
\ \ \ \ \isacommand{by}\isamarkupfalse%
\ {\isacharparenleft}metis\ SourcesLoop{\isacharparenright}\ \ \isanewline
\ \ \isacommand{thus}\isamarkupfalse%
\ \ {\isacharquery}thesis\ \isacommand{by}\isamarkupfalse%
\ auto\isanewline
\isacommand{qed}\isamarkupfalse%
\endisatagproof
{\isafoldproof}%
\isadelimproof
\isanewline
\endisadelimproof
\isanewline
\isanewline
\isacommand{lemma}\isamarkupfalse%
\ SourcesA{\isadigit{9}}{\isadigit{1}}{\isacharunderscore}L{\isadigit{1}}{\isacharcolon}\ {\isachardoublequoteopen}Sources\ level{\isadigit{1}}\ sA{\isadigit{9}}{\isadigit{1}}\ {\isacharequal}\ {\isacharbraceleft}sA{\isadigit{6}}{\isacharcomma}\ sA{\isadigit{7}}{\isadigit{1}}{\isacharcomma}\ sA{\isadigit{8}}{\isadigit{1}}{\isacharcomma}\ sA{\isadigit{9}}{\isadigit{1}}{\isacharbraceright}{\isachardoublequoteclose}\isanewline
\isadelimproof
\endisadelimproof
\isatagproof
\isacommand{proof}\isamarkupfalse%
\ {\isacharminus}\isanewline
\ \ \isacommand{have}\isamarkupfalse%
\ \ {\isachardoublequoteopen}DSources\ level{\isadigit{1}}\ sA{\isadigit{9}}{\isadigit{1}}\ {\isacharequal}\ {\isacharbraceleft}sA{\isadigit{8}}{\isadigit{1}}{\isacharbraceright}{\isachardoublequoteclose}\ \isacommand{by}\isamarkupfalse%
\ {\isacharparenleft}rule\ DSourcesA{\isadigit{9}}{\isadigit{1}}{\isacharunderscore}L{\isadigit{1}}{\isacharparenright}\isanewline
\ \ \isacommand{thus}\isamarkupfalse%
\ {\isacharquery}thesis\ \isacommand{by}\isamarkupfalse%
\ {\isacharparenleft}metis\ SourcesA{\isadigit{8}}{\isadigit{1}}{\isacharunderscore}L{\isadigit{1}}\ Sources{\isacharunderscore}singleDSource\ \isanewline
\ \ \ \ \ \ \ \ \ \ Un{\isacharunderscore}empty{\isacharunderscore}left\ Un{\isacharunderscore}insert{\isacharunderscore}left\ insert{\isacharunderscore}absorb{\isadigit{2}}\ insert{\isacharunderscore}commute{\isacharparenright}\ \isanewline
\isacommand{qed}\isamarkupfalse%
\endisatagproof
{\isafoldproof}%
\isadelimproof
\isanewline
\endisadelimproof
\isanewline
\isacommand{lemma}\isamarkupfalse%
\ SourcesA{\isadigit{9}}{\isadigit{2}}{\isacharunderscore}L{\isadigit{1}}{\isacharcolon}\ {\isachardoublequoteopen}Sources\ level{\isadigit{1}}\ sA{\isadigit{9}}{\isadigit{2}}\ {\isacharequal}\ {\isacharbraceleft}sA{\isadigit{6}}{\isacharcomma}\ sA{\isadigit{7}}{\isadigit{1}}{\isacharcomma}\ sA{\isadigit{8}}{\isadigit{1}}{\isacharcomma}\ sA{\isadigit{9}}{\isadigit{1}}{\isacharbraceright}{\isachardoublequoteclose}\isanewline
\isadelimproof
\endisadelimproof
\isatagproof
\isacommand{by}\isamarkupfalse%
\ {\isacharparenleft}metis\ DSourcesA{\isadigit{9}}{\isadigit{1}}{\isacharunderscore}L{\isadigit{1}}\ DSourcesA{\isadigit{9}}{\isadigit{2}}{\isacharunderscore}L{\isadigit{1}}\ SourcesA{\isadigit{9}}{\isadigit{1}}{\isacharunderscore}L{\isadigit{1}}\ Sources{\isacharunderscore}singleDSource{\isacharparenright}%
\endisatagproof
{\isafoldproof}%
\isadelimproof
\ \isanewline
\endisadelimproof
\isanewline
\isacommand{lemma}\isamarkupfalse%
\ SourcesA{\isadigit{7}}{\isadigit{2}}{\isacharunderscore}L{\isadigit{1}}{\isacharcolon}\ {\isachardoublequoteopen}Sources\ level{\isadigit{1}}\ sA{\isadigit{7}}{\isadigit{2}}\ {\isacharequal}\ {\isacharbraceleft}sA{\isadigit{6}}{\isacharbraceright}{\isachardoublequoteclose}\isanewline
\isadelimproof
\endisadelimproof
\isatagproof
\isacommand{by}\isamarkupfalse%
\ {\isacharparenleft}metis\ DSourcesA{\isadigit{6}}{\isacharunderscore}L{\isadigit{1}}\ DSourcesA{\isadigit{7}}{\isadigit{2}}{\isacharunderscore}L{\isadigit{1}}\ SourcesOnlyDSources\ singleton{\isacharunderscore}iff{\isacharparenright}%
\endisatagproof
{\isafoldproof}%
\isadelimproof
\ \ \ \isanewline
\endisadelimproof
\isanewline
\isacommand{lemma}\isamarkupfalse%
\ SourcesA{\isadigit{8}}{\isadigit{2}}{\isacharunderscore}L{\isadigit{1}}{\isacharcolon}\ {\isachardoublequoteopen}Sources\ level{\isadigit{1}}\ sA{\isadigit{8}}{\isadigit{2}}\ {\isacharequal}\ {\isacharbraceleft}sA{\isadigit{6}}{\isacharcomma}\ sA{\isadigit{7}}{\isadigit{2}}{\isacharbraceright}{\isachardoublequoteclose}\ \ \isanewline
\isadelimproof
\endisadelimproof
\isatagproof
\isacommand{proof}\isamarkupfalse%
\ {\isacharminus}\ \isanewline
\ \ \isacommand{have}\isamarkupfalse%
\ \ dA{\isadigit{8}}{\isadigit{2}}{\isacharcolon}{\isachardoublequoteopen}DSources\ level{\isadigit{1}}\ sA{\isadigit{8}}{\isadigit{2}}\ {\isacharequal}\ {\isacharbraceleft}sA{\isadigit{7}}{\isadigit{2}}{\isacharbraceright}{\isachardoublequoteclose}\ \isacommand{by}\isamarkupfalse%
\ {\isacharparenleft}rule\ DSourcesA{\isadigit{8}}{\isadigit{2}}{\isacharunderscore}L{\isadigit{1}}{\isacharparenright}\isanewline
\ \ \isacommand{have}\isamarkupfalse%
\ {\isachardoublequoteopen}{\isacharparenleft}Sources\ level{\isadigit{1}}\ sA{\isadigit{8}}{\isadigit{2}}{\isacharparenright}\ {\isacharequal}\ {\isacharparenleft}DSources\ level{\isadigit{1}}\ sA{\isadigit{8}}{\isadigit{2}}{\isacharparenright}\ {\isasymunion}\ {\isacharparenleft}{\isasymUnion}\ S\ {\isasymin}\ {\isacharparenleft}DSources\ level{\isadigit{1}}\ sA{\isadigit{8}}{\isadigit{2}}{\isacharparenright}{\isachardot}\ {\isacharparenleft}Sources\ level{\isadigit{1}}\ S{\isacharparenright}{\isacharparenright}{\isachardoublequoteclose}\ \isanewline
\ \ \ \ \isacommand{by}\isamarkupfalse%
\ {\isacharparenleft}rule\ SourcesDef{\isacharparenright}\isanewline
\ \ \isacommand{with}\isamarkupfalse%
\ dA{\isadigit{8}}{\isadigit{2}}\ \isacommand{have}\isamarkupfalse%
\ {\isachardoublequoteopen}{\isacharparenleft}Sources\ level{\isadigit{1}}\ sA{\isadigit{8}}{\isadigit{2}}{\isacharparenright}\ {\isacharequal}\ \ {\isacharbraceleft}sA{\isadigit{7}}{\isadigit{2}}{\isacharbraceright}\ {\isasymunion}\ {\isacharparenleft}Sources\ level{\isadigit{1}}\ sA{\isadigit{7}}{\isadigit{2}}{\isacharparenright}{\isachardoublequoteclose}\ \ \isacommand{by}\isamarkupfalse%
\ simp\isanewline
\ \ \isacommand{thus}\isamarkupfalse%
\ {\isacharquery}thesis\ \isacommand{by}\isamarkupfalse%
\ {\isacharparenleft}metis\ SourcesA{\isadigit{7}}{\isadigit{2}}{\isacharunderscore}L{\isadigit{1}}\ Un{\isacharunderscore}commute\ insert{\isacharunderscore}is{\isacharunderscore}Un{\isacharparenright}\ \isanewline
\isacommand{qed}\isamarkupfalse%
\endisatagproof
{\isafoldproof}%
\isadelimproof
\isanewline
\endisadelimproof
\isanewline
\isacommand{lemma}\isamarkupfalse%
\ SourcesA{\isadigit{9}}{\isadigit{3}}{\isacharunderscore}L{\isadigit{1}}{\isacharcolon}\ {\isachardoublequoteopen}Sources\ level{\isadigit{1}}\ sA{\isadigit{9}}{\isadigit{3}}\ {\isacharequal}\ {\isacharbraceleft}sA{\isadigit{6}}{\isacharcomma}\ sA{\isadigit{7}}{\isadigit{2}}{\isacharcomma}\ sA{\isadigit{8}}{\isadigit{2}}{\isacharbraceright}{\isachardoublequoteclose}\isanewline
\isadelimproof
\endisadelimproof
\isatagproof
\isacommand{by}\isamarkupfalse%
\ {\isacharparenleft}metis\ DSourcesA{\isadigit{9}}{\isadigit{3}}{\isacharunderscore}L{\isadigit{1}}\ SourcesA{\isadigit{8}}{\isadigit{2}}{\isacharunderscore}L{\isadigit{1}}\ Sources{\isacharunderscore}singleDSource\ Un{\isacharunderscore}insert{\isacharunderscore}right\ insert{\isacharunderscore}is{\isacharunderscore}Un{\isacharparenright}\ \ \ \isanewline
\isanewline
\isanewline
\isamarkupcmt{Abstraction level 2%
}
\endisatagproof
{\isafoldproof}%
\isadelimproof
\isanewline
\endisadelimproof
\isanewline
\isacommand{lemma}\isamarkupfalse%
\ SourcesS{\isadigit{1}}{\isacharunderscore}L{\isadigit{2}}{\isacharcolon}\ {\isachardoublequoteopen}Sources\ level{\isadigit{2}}\ sS{\isadigit{1}}\ {\isacharequal}\ {\isacharbraceleft}{\isacharbraceright}{\isachardoublequoteclose}\isanewline
\isadelimproof
\endisadelimproof
\isatagproof
\isacommand{proof}\isamarkupfalse%
\ {\isacharminus}\isanewline
\ \ \isacommand{have}\isamarkupfalse%
\ {\isachardoublequoteopen}DSources\ level{\isadigit{2}}\ sS{\isadigit{1}}\ {\isacharequal}\ {\isacharbraceleft}{\isacharbraceright}{\isachardoublequoteclose}\ \ \isacommand{by}\isamarkupfalse%
\ {\isacharparenleft}simp\ add{\isacharcolon}\ DSources{\isacharunderscore}def\ AbstrLevel{\isadigit{2}}{\isacharcomma}\ auto{\isacharparenright}\isanewline
\ \ \isacommand{thus}\isamarkupfalse%
\ {\isacharquery}thesis\ \ \isacommand{by}\isamarkupfalse%
\ {\isacharparenleft}simp\ add{\isacharcolon}\ DSourcesEmptySources{\isacharparenright}\isanewline
\isacommand{qed}\isamarkupfalse%
\endisatagproof
{\isafoldproof}%
\isadelimproof
\isanewline
\endisadelimproof
\isanewline
\isacommand{lemma}\isamarkupfalse%
\ SourcesS{\isadigit{2}}{\isacharunderscore}L{\isadigit{2}}{\isacharcolon}\ {\isachardoublequoteopen}Sources\ level{\isadigit{2}}\ sS{\isadigit{2}}\ {\isacharequal}\ {\isacharbraceleft}{\isacharbraceright}{\isachardoublequoteclose}\isanewline
\isadelimproof
\endisadelimproof
\isatagproof
\isacommand{proof}\isamarkupfalse%
\ {\isacharminus}\isanewline
\ \ \isacommand{have}\isamarkupfalse%
\ {\isachardoublequoteopen}DSources\ level{\isadigit{2}}\ sS{\isadigit{2}}\ {\isacharequal}\ {\isacharbraceleft}{\isacharbraceright}{\isachardoublequoteclose}\ \ \isacommand{by}\isamarkupfalse%
\ {\isacharparenleft}simp\ add{\isacharcolon}\ DSources{\isacharunderscore}def\ AbstrLevel{\isadigit{2}}{\isacharcomma}\ auto{\isacharparenright}\isanewline
\ \ \isacommand{thus}\isamarkupfalse%
\ {\isacharquery}thesis\ \ \isacommand{by}\isamarkupfalse%
\ {\isacharparenleft}simp\ add{\isacharcolon}\ DSourcesEmptySources{\isacharparenright}\isanewline
\isacommand{qed}\isamarkupfalse%
\endisatagproof
{\isafoldproof}%
\isadelimproof
\isanewline
\endisadelimproof
\isanewline
\isacommand{lemma}\isamarkupfalse%
\ SourcesS{\isadigit{3}}{\isacharunderscore}L{\isadigit{2}}{\isacharcolon}\ {\isachardoublequoteopen}Sources\ level{\isadigit{2}}\ sS{\isadigit{3}}\ {\isacharequal}\ {\isacharbraceleft}sS{\isadigit{2}}{\isacharbraceright}{\isachardoublequoteclose}\isanewline
\isadelimproof
\endisadelimproof
\isatagproof
\isacommand{proof}\isamarkupfalse%
\ {\isacharminus}\isanewline
\ \ \isacommand{have}\isamarkupfalse%
\ DSourcesS{\isadigit{3}}{\isacharcolon}{\isachardoublequoteopen}DSources\ level{\isadigit{2}}\ sS{\isadigit{3}}\ {\isacharequal}\ {\isacharbraceleft}sS{\isadigit{2}}{\isacharbraceright}{\isachardoublequoteclose}\ \ \isacommand{by}\isamarkupfalse%
\ {\isacharparenleft}simp\ add{\isacharcolon}\ DSources{\isacharunderscore}def\ AbstrLevel{\isadigit{2}}{\isacharcomma}\ auto{\isacharparenright}\isanewline
\ \ \isacommand{have}\isamarkupfalse%
\ {\isachardoublequoteopen}Sources\ level{\isadigit{2}}\ sS{\isadigit{2}}\ {\isacharequal}\ {\isacharbraceleft}{\isacharbraceright}{\isachardoublequoteclose}\ \ \isacommand{by}\isamarkupfalse%
\ {\isacharparenleft}rule\ SourcesS{\isadigit{2}}{\isacharunderscore}L{\isadigit{2}}{\isacharparenright}\isanewline
\ \ \isacommand{with}\isamarkupfalse%
\ DSourcesS{\isadigit{3}}\ \isacommand{show}\isamarkupfalse%
\ {\isacharquery}thesis\ \ \isacommand{by}\isamarkupfalse%
\ \ {\isacharparenleft}simp\ add{\isacharcolon}\ Sources{\isacharunderscore}singleDSource{\isacharparenright}\isanewline
\isacommand{qed}\isamarkupfalse%
\endisatagproof
{\isafoldproof}%
\isadelimproof
\isanewline
\endisadelimproof
\isanewline
\isacommand{lemma}\isamarkupfalse%
\ SourcesS{\isadigit{4}}{\isacharunderscore}L{\isadigit{2}}{\isacharcolon}\ \ {\isachardoublequoteopen}Sources\ level{\isadigit{2}}\ sS{\isadigit{4}}\ {\isacharequal}\ {\isacharbraceleft}sS{\isadigit{2}}{\isacharbraceright}{\isachardoublequoteclose}\isanewline
\isadelimproof
\endisadelimproof
\isatagproof
\isacommand{proof}\isamarkupfalse%
\ {\isacharminus}\isanewline
\ \ \isacommand{have}\isamarkupfalse%
\ DSourcesS{\isadigit{4}}{\isacharcolon}{\isachardoublequoteopen}DSources\ level{\isadigit{2}}\ sS{\isadigit{4}}\ {\isacharequal}\ {\isacharbraceleft}sS{\isadigit{2}}{\isacharbraceright}{\isachardoublequoteclose}\ \isacommand{by}\isamarkupfalse%
\ {\isacharparenleft}simp\ add{\isacharcolon}\ DSources{\isacharunderscore}def\ AbstrLevel{\isadigit{2}}{\isacharcomma}\ auto{\isacharparenright}\isanewline
\ \ \isacommand{have}\isamarkupfalse%
\ {\isachardoublequoteopen}Sources\ level{\isadigit{2}}\ sS{\isadigit{2}}\ {\isacharequal}\ {\isacharbraceleft}{\isacharbraceright}{\isachardoublequoteclose}\ \ \isacommand{by}\isamarkupfalse%
\ {\isacharparenleft}rule\ SourcesS{\isadigit{2}}{\isacharunderscore}L{\isadigit{2}}{\isacharparenright}\isanewline
\ \ \isacommand{with}\isamarkupfalse%
\ DSourcesS{\isadigit{4}}\ \isacommand{show}\isamarkupfalse%
\ {\isacharquery}thesis\ \ \isacommand{by}\isamarkupfalse%
\ \ {\isacharparenleft}simp\ add{\isacharcolon}\ Sources{\isacharunderscore}singleDSource{\isacharparenright}\isanewline
\isacommand{qed}\isamarkupfalse%
\endisatagproof
{\isafoldproof}%
\isadelimproof
\isanewline
\endisadelimproof
\isanewline
\isacommand{lemma}\isamarkupfalse%
\ SourcesS{\isadigit{5}}{\isacharunderscore}L{\isadigit{2}}{\isacharcolon}\ \ {\isachardoublequoteopen}Sources\ level{\isadigit{2}}\ sS{\isadigit{5}}\ {\isacharequal}\ {\isacharbraceleft}sS{\isadigit{2}}{\isacharcomma}\ sS{\isadigit{4}}{\isacharbraceright}{\isachardoublequoteclose}\isanewline
\isadelimproof
\endisadelimproof
\isatagproof
\isacommand{proof}\isamarkupfalse%
\ {\isacharminus}\isanewline
\ \ \isacommand{have}\isamarkupfalse%
\ DSourcesS{\isadigit{5}}{\isacharcolon}{\isachardoublequoteopen}DSources\ level{\isadigit{2}}\ sS{\isadigit{5}}\ {\isacharequal}\ {\isacharbraceleft}sS{\isadigit{4}}{\isacharbraceright}{\isachardoublequoteclose}\ \ \isacommand{by}\isamarkupfalse%
\ {\isacharparenleft}simp\ add{\isacharcolon}\ DSources{\isacharunderscore}def\ AbstrLevel{\isadigit{2}}{\isacharcomma}\ auto{\isacharparenright}\isanewline
\ \ \isacommand{have}\isamarkupfalse%
\ {\isachardoublequoteopen}Sources\ level{\isadigit{2}}\ sS{\isadigit{4}}\ {\isacharequal}\ {\isacharbraceleft}sS{\isadigit{2}}{\isacharbraceright}{\isachardoublequoteclose}\ \isacommand{by}\isamarkupfalse%
\ {\isacharparenleft}rule\ SourcesS{\isadigit{4}}{\isacharunderscore}L{\isadigit{2}}{\isacharparenright}\isanewline
\ \ \isacommand{with}\isamarkupfalse%
\ DSourcesS{\isadigit{5}}\ \isacommand{show}\isamarkupfalse%
\ {\isacharquery}thesis\ \ \isacommand{by}\isamarkupfalse%
\ \ {\isacharparenleft}simp\ add{\isacharcolon}\ Sources{\isacharunderscore}singleDSource{\isacharparenright}\isanewline
\isacommand{qed}\isamarkupfalse%
\endisatagproof
{\isafoldproof}%
\isadelimproof
\isanewline
\endisadelimproof
\isanewline
\isacommand{lemma}\isamarkupfalse%
\ SourcesS{\isadigit{6}}{\isacharunderscore}L{\isadigit{2}}{\isacharcolon}\ \ {\isachardoublequoteopen}Sources\ level{\isadigit{2}}\ sS{\isadigit{6}}\ {\isacharequal}\ {\isacharbraceleft}sS{\isadigit{2}}{\isacharcomma}\ sS{\isadigit{4}}{\isacharcomma}\ sS{\isadigit{5}}{\isacharbraceright}{\isachardoublequoteclose}\isanewline
\isadelimproof
\endisadelimproof
\isatagproof
\isacommand{proof}\isamarkupfalse%
\ {\isacharminus}\isanewline
\ \ \isacommand{have}\isamarkupfalse%
\ DSourcesS{\isadigit{6}}{\isacharcolon}{\isachardoublequoteopen}DSources\ level{\isadigit{2}}\ sS{\isadigit{6}}\ {\isacharequal}\ {\isacharbraceleft}sS{\isadigit{2}}{\isacharcomma}\ sS{\isadigit{5}}{\isacharbraceright}{\isachardoublequoteclose}\ \ \isacommand{by}\isamarkupfalse%
\ {\isacharparenleft}simp\ add{\isacharcolon}\ DSources{\isacharunderscore}def\ AbstrLevel{\isadigit{2}}{\isacharcomma}\ auto{\isacharparenright}\isanewline
\ \ \isacommand{have}\isamarkupfalse%
\ SourcesS{\isadigit{2}}{\isacharcolon}{\isachardoublequoteopen}Sources\ level{\isadigit{2}}\ sS{\isadigit{2}}\ {\isacharequal}\ {\isacharbraceleft}{\isacharbraceright}{\isachardoublequoteclose}\ \ \isacommand{by}\isamarkupfalse%
\ {\isacharparenleft}rule\ SourcesS{\isadigit{2}}{\isacharunderscore}L{\isadigit{2}}{\isacharparenright}\isanewline
\ \ \isacommand{have}\isamarkupfalse%
\ {\isachardoublequoteopen}Sources\ level{\isadigit{2}}\ sS{\isadigit{5}}\ {\isacharequal}\ {\isacharbraceleft}sS{\isadigit{2}}{\isacharcomma}\ sS{\isadigit{4}}{\isacharbraceright}{\isachardoublequoteclose}\ \ \isacommand{by}\isamarkupfalse%
\ {\isacharparenleft}rule\ SourcesS{\isadigit{5}}{\isacharunderscore}L{\isadigit{2}}{\isacharparenright}\isanewline
\ \ \isacommand{with}\isamarkupfalse%
\ \ SourcesS{\isadigit{2}}\ DSourcesS{\isadigit{6}}\ \isacommand{show}\isamarkupfalse%
\ {\isacharquery}thesis\ \ \isacommand{by}\isamarkupfalse%
\ {\isacharparenleft}simp\ add{\isacharcolon}\ Sources{\isacharunderscore}{\isadigit{2}}DSources{\isacharcomma}\ auto{\isacharparenright}\isanewline
\isacommand{qed}\isamarkupfalse%
\endisatagproof
{\isafoldproof}%
\isadelimproof
\isanewline
\endisadelimproof
\isanewline
\isacommand{lemma}\isamarkupfalse%
\ SourcesS{\isadigit{7}}{\isacharunderscore}L{\isadigit{2}}{\isacharcolon}\ \ {\isachardoublequoteopen}Sources\ level{\isadigit{2}}\ sS{\isadigit{7}}\ {\isacharequal}\ {\isacharbraceleft}{\isacharbraceright}{\isachardoublequoteclose}\isanewline
\isadelimproof
\endisadelimproof
\isatagproof
\isacommand{proof}\isamarkupfalse%
\ {\isacharminus}\isanewline
\ \ \isacommand{have}\isamarkupfalse%
\ {\isachardoublequoteopen}DSources\ level{\isadigit{2}}\ sS{\isadigit{7}}\ {\isacharequal}\ {\isacharbraceleft}{\isacharbraceright}{\isachardoublequoteclose}\ \ \isacommand{by}\isamarkupfalse%
\ {\isacharparenleft}simp\ add{\isacharcolon}\ DSources{\isacharunderscore}def\ AbstrLevel{\isadigit{2}}{\isacharcomma}\ auto{\isacharparenright}\isanewline
\ \ \isacommand{thus}\isamarkupfalse%
\ {\isacharquery}thesis\ \ \isacommand{by}\isamarkupfalse%
\ {\isacharparenleft}simp\ add{\isacharcolon}\ DSourcesEmptySources{\isacharparenright}\isanewline
\isacommand{qed}\isamarkupfalse%
\endisatagproof
{\isafoldproof}%
\isadelimproof
\isanewline
\endisadelimproof
\isanewline
\isacommand{lemma}\isamarkupfalse%
\ SourcesS{\isadigit{8}}{\isacharunderscore}L{\isadigit{2}}{\isacharcolon}\isanewline
\ {\isachardoublequoteopen}Sources\ level{\isadigit{2}}\ sS{\isadigit{8}}\ {\isacharequal}\ {\isacharbraceleft}sS{\isadigit{7}}{\isacharbraceright}{\isachardoublequoteclose}\isanewline
\isadelimproof
\endisadelimproof
\isatagproof
\isacommand{proof}\isamarkupfalse%
\ {\isacharminus}\isanewline
\ \ \isacommand{have}\isamarkupfalse%
\ DSourcesS{\isadigit{8}}{\isacharcolon}{\isachardoublequoteopen}DSources\ level{\isadigit{2}}\ sS{\isadigit{8}}\ {\isacharequal}\ {\isacharbraceleft}sS{\isadigit{7}}{\isacharbraceright}{\isachardoublequoteclose}\ \ \isacommand{by}\isamarkupfalse%
\ {\isacharparenleft}simp\ add{\isacharcolon}\ DSources{\isacharunderscore}def\ AbstrLevel{\isadigit{2}}{\isacharcomma}\ auto{\isacharparenright}\isanewline
\ \ \isacommand{have}\isamarkupfalse%
\ {\isachardoublequoteopen}Sources\ level{\isadigit{2}}\ sS{\isadigit{7}}\ {\isacharequal}\ {\isacharbraceleft}{\isacharbraceright}{\isachardoublequoteclose}\ \ \isacommand{by}\isamarkupfalse%
\ {\isacharparenleft}rule\ SourcesS{\isadigit{7}}{\isacharunderscore}L{\isadigit{2}}{\isacharparenright}\isanewline
\ \ \isacommand{with}\isamarkupfalse%
\ DSourcesS{\isadigit{8}}\ \isacommand{show}\isamarkupfalse%
\ {\isacharquery}thesis\ \ \isacommand{by}\isamarkupfalse%
\ \ {\isacharparenleft}simp\ add{\isacharcolon}\ Sources{\isacharunderscore}singleDSource{\isacharparenright}\isanewline
\isacommand{qed}\isamarkupfalse%
\endisatagproof
{\isafoldproof}%
\isadelimproof
\isanewline
\endisadelimproof
\isanewline
\isacommand{lemma}\isamarkupfalse%
\ SourcesS{\isadigit{9}}{\isacharunderscore}L{\isadigit{2}}{\isacharcolon}\isanewline
\ {\isachardoublequoteopen}Sources\ level{\isadigit{2}}\ sS{\isadigit{9}}\ {\isacharequal}\ {\isacharbraceleft}{\isacharbraceright}{\isachardoublequoteclose}\isanewline
\isadelimproof
\endisadelimproof
\isatagproof
\isacommand{proof}\isamarkupfalse%
\ {\isacharminus}\isanewline
\ \ \isacommand{have}\isamarkupfalse%
\ {\isachardoublequoteopen}DSources\ level{\isadigit{2}}\ sS{\isadigit{9}}\ {\isacharequal}\ {\isacharbraceleft}{\isacharbraceright}{\isachardoublequoteclose}\ \ \isacommand{by}\isamarkupfalse%
\ {\isacharparenleft}simp\ add{\isacharcolon}\ DSources{\isacharunderscore}def\ AbstrLevel{\isadigit{2}}{\isacharcomma}\ auto{\isacharparenright}\isanewline
\ \ \isacommand{thus}\isamarkupfalse%
\ {\isacharquery}thesis\ \ \isacommand{by}\isamarkupfalse%
\ {\isacharparenleft}simp\ add{\isacharcolon}\ DSourcesEmptySources{\isacharparenright}\isanewline
\isacommand{qed}\isamarkupfalse%
\endisatagproof
{\isafoldproof}%
\isadelimproof
\isanewline
\endisadelimproof
\isanewline
\isacommand{lemma}\isamarkupfalse%
\ SourcesS{\isadigit{1}}{\isadigit{0}}{\isacharunderscore}L{\isadigit{2}}{\isacharcolon}\ {\isachardoublequoteopen}Sources\ level{\isadigit{2}}\ sS{\isadigit{1}}{\isadigit{0}}\ {\isacharequal}\ {\isacharbraceleft}sS{\isadigit{9}}{\isacharbraceright}{\isachardoublequoteclose}\isanewline
\isadelimproof
\endisadelimproof
\isatagproof
\isacommand{proof}\isamarkupfalse%
\ {\isacharminus}\isanewline
\ \ \isacommand{have}\isamarkupfalse%
\ DSourcesS{\isadigit{1}}{\isadigit{0}}{\isacharcolon}{\isachardoublequoteopen}DSources\ level{\isadigit{2}}\ sS{\isadigit{1}}{\isadigit{0}}\ {\isacharequal}\ {\isacharbraceleft}sS{\isadigit{9}}{\isacharbraceright}{\isachardoublequoteclose}\ \isacommand{by}\isamarkupfalse%
\ {\isacharparenleft}simp\ add{\isacharcolon}\ DSources{\isacharunderscore}def\ AbstrLevel{\isadigit{2}}{\isacharcomma}\ auto{\isacharparenright}\isanewline
\ \ \isacommand{have}\isamarkupfalse%
\ {\isachardoublequoteopen}Sources\ level{\isadigit{2}}\ sS{\isadigit{9}}\ {\isacharequal}\ {\isacharbraceleft}{\isacharbraceright}{\isachardoublequoteclose}\ \isacommand{by}\isamarkupfalse%
\ {\isacharparenleft}rule\ SourcesS{\isadigit{9}}{\isacharunderscore}L{\isadigit{2}}{\isacharparenright}\isanewline
\ \ \isacommand{with}\isamarkupfalse%
\ DSourcesS{\isadigit{1}}{\isadigit{0}}\ \isacommand{show}\isamarkupfalse%
\ {\isacharquery}thesis\ \ \isacommand{by}\isamarkupfalse%
\ \ {\isacharparenleft}simp\ add{\isacharcolon}\ Sources{\isacharunderscore}singleDSource{\isacharparenright}\isanewline
\isacommand{qed}\isamarkupfalse%
\endisatagproof
{\isafoldproof}%
\isadelimproof
\isanewline
\endisadelimproof
\isanewline
\isacommand{lemma}\isamarkupfalse%
\ SourcesS{\isadigit{1}}{\isadigit{1}}{\isacharunderscore}L{\isadigit{2}}{\isacharcolon}\ {\isachardoublequoteopen}Sources\ level{\isadigit{2}}\ sS{\isadigit{1}}{\isadigit{1}}\ {\isacharequal}\ {\isacharbraceleft}sS{\isadigit{9}}{\isacharbraceright}{\isachardoublequoteclose}\isanewline
\isadelimproof
\endisadelimproof
\isatagproof
\isacommand{proof}\isamarkupfalse%
\ {\isacharminus}\isanewline
\ \ \isacommand{have}\isamarkupfalse%
\ DSourcesS{\isadigit{1}}{\isadigit{1}}{\isacharcolon}{\isachardoublequoteopen}DSources\ level{\isadigit{2}}\ sS{\isadigit{1}}{\isadigit{1}}\ {\isacharequal}\ {\isacharbraceleft}sS{\isadigit{9}}{\isacharbraceright}{\isachardoublequoteclose}\ \ \isacommand{by}\isamarkupfalse%
\ {\isacharparenleft}simp\ add{\isacharcolon}\ DSources{\isacharunderscore}def\ AbstrLevel{\isadigit{2}}{\isacharcomma}\ auto{\isacharparenright}\isanewline
\ \ \isacommand{have}\isamarkupfalse%
\ {\isachardoublequoteopen}Sources\ level{\isadigit{2}}\ sS{\isadigit{9}}\ {\isacharequal}\ {\isacharbraceleft}{\isacharbraceright}{\isachardoublequoteclose}\ \ \isacommand{by}\isamarkupfalse%
\ {\isacharparenleft}rule\ SourcesS{\isadigit{9}}{\isacharunderscore}L{\isadigit{2}}{\isacharparenright}\isanewline
\ \ \isacommand{with}\isamarkupfalse%
\ DSourcesS{\isadigit{1}}{\isadigit{1}}\ \isacommand{show}\isamarkupfalse%
\ {\isacharquery}thesis\ \ \isacommand{by}\isamarkupfalse%
\ \ {\isacharparenleft}simp\ add{\isacharcolon}\ Sources{\isacharunderscore}singleDSource{\isacharparenright}\isanewline
\isacommand{qed}\isamarkupfalse%
\endisatagproof
{\isafoldproof}%
\isadelimproof
\isanewline
\endisadelimproof
\isanewline
\isacommand{lemma}\isamarkupfalse%
\ SourcesS{\isadigit{1}}{\isadigit{2}}{\isacharunderscore}L{\isadigit{2}}{\isacharcolon}\ {\isachardoublequoteopen}Sources\ level{\isadigit{2}}\ sS{\isadigit{1}}{\isadigit{2}}\ {\isacharequal}\ {\isacharbraceleft}sS{\isadigit{9}}{\isacharcomma}\ sS{\isadigit{1}}{\isadigit{0}}{\isacharbraceright}{\isachardoublequoteclose}\isanewline
\isadelimproof
\endisadelimproof
\isatagproof
\isacommand{proof}\isamarkupfalse%
\ {\isacharminus}\isanewline
\ \ \isacommand{have}\isamarkupfalse%
\ DSourcesS{\isadigit{1}}{\isadigit{2}}{\isacharcolon}{\isachardoublequoteopen}DSources\ level{\isadigit{2}}\ sS{\isadigit{1}}{\isadigit{2}}\ {\isacharequal}\ {\isacharbraceleft}sS{\isadigit{1}}{\isadigit{0}}{\isacharbraceright}{\isachardoublequoteclose}\ \isacommand{by}\isamarkupfalse%
\ {\isacharparenleft}simp\ add{\isacharcolon}\ DSources{\isacharunderscore}def\ AbstrLevel{\isadigit{2}}{\isacharcomma}\ auto{\isacharparenright}\isanewline
\ \ \isacommand{have}\isamarkupfalse%
\ {\isachardoublequoteopen}Sources\ level{\isadigit{2}}\ sS{\isadigit{1}}{\isadigit{0}}\ {\isacharequal}\ {\isacharbraceleft}sS{\isadigit{9}}{\isacharbraceright}{\isachardoublequoteclose}\ \ \isacommand{by}\isamarkupfalse%
\ {\isacharparenleft}rule\ SourcesS{\isadigit{1}}{\isadigit{0}}{\isacharunderscore}L{\isadigit{2}}{\isacharparenright}\isanewline
\ \ \isacommand{with}\isamarkupfalse%
\ DSourcesS{\isadigit{1}}{\isadigit{2}}\ \isacommand{show}\isamarkupfalse%
\ {\isacharquery}thesis\ \ \isacommand{by}\isamarkupfalse%
\ \ {\isacharparenleft}simp\ add{\isacharcolon}\ Sources{\isacharunderscore}singleDSource{\isacharparenright}\isanewline
\isacommand{qed}\isamarkupfalse%
\endisatagproof
{\isafoldproof}%
\isadelimproof
\isanewline
\endisadelimproof
\isanewline
\isacommand{lemma}\isamarkupfalse%
\ SourcesS{\isadigit{1}}{\isadigit{3}}{\isacharunderscore}L{\isadigit{2}}{\isacharcolon}\ {\isachardoublequoteopen}Sources\ level{\isadigit{2}}\ sS{\isadigit{1}}{\isadigit{3}}\ {\isacharequal}\ {\isacharbraceleft}sS{\isadigit{9}}{\isacharcomma}\ sS{\isadigit{1}}{\isadigit{0}}{\isacharcomma}\ sS{\isadigit{1}}{\isadigit{2}}{\isacharbraceright}{\isachardoublequoteclose}\isanewline
\isadelimproof
\endisadelimproof
\isatagproof
\isacommand{proof}\isamarkupfalse%
\ {\isacharminus}\isanewline
\ \ \isacommand{have}\isamarkupfalse%
\ DSourcesS{\isadigit{1}}{\isadigit{3}}{\isacharcolon}{\isachardoublequoteopen}DSources\ level{\isadigit{2}}\ sS{\isadigit{1}}{\isadigit{3}}\ {\isacharequal}\ {\isacharbraceleft}sS{\isadigit{1}}{\isadigit{2}}{\isacharbraceright}{\isachardoublequoteclose}\ \ \isacommand{by}\isamarkupfalse%
\ {\isacharparenleft}simp\ add{\isacharcolon}\ DSources{\isacharunderscore}def\ AbstrLevel{\isadigit{2}}{\isacharcomma}\ auto{\isacharparenright}\isanewline
\ \ \isacommand{have}\isamarkupfalse%
\ {\isachardoublequoteopen}Sources\ level{\isadigit{2}}\ sS{\isadigit{1}}{\isadigit{2}}\ {\isacharequal}\ {\isacharbraceleft}sS{\isadigit{9}}{\isacharcomma}\ sS{\isadigit{1}}{\isadigit{0}}{\isacharbraceright}{\isachardoublequoteclose}\ \isacommand{by}\isamarkupfalse%
\ {\isacharparenleft}rule\ SourcesS{\isadigit{1}}{\isadigit{2}}{\isacharunderscore}L{\isadigit{2}}{\isacharparenright}\isanewline
\ \ \isacommand{with}\isamarkupfalse%
\ DSourcesS{\isadigit{1}}{\isadigit{3}}\ \isacommand{show}\isamarkupfalse%
\ {\isacharquery}thesis\ \isacommand{by}\isamarkupfalse%
\ \ {\isacharparenleft}simp\ add{\isacharcolon}\ Sources{\isacharunderscore}singleDSource{\isacharparenright}\isanewline
\isacommand{qed}\isamarkupfalse%
\endisatagproof
{\isafoldproof}%
\isadelimproof
\isanewline
\endisadelimproof
\isanewline
\isacommand{lemma}\isamarkupfalse%
\ SourcesS{\isadigit{1}}{\isadigit{4}}{\isacharunderscore}L{\isadigit{2}}{\isacharcolon}\ {\isachardoublequoteopen}Sources\ level{\isadigit{2}}\ sS{\isadigit{1}}{\isadigit{4}}\ {\isacharequal}\ {\isacharbraceleft}sS{\isadigit{9}}{\isacharcomma}\ sS{\isadigit{1}}{\isadigit{1}}{\isacharbraceright}{\isachardoublequoteclose}\isanewline
\isadelimproof
\endisadelimproof
\isatagproof
\isacommand{proof}\isamarkupfalse%
\ {\isacharminus}\isanewline
\ \ \isacommand{have}\isamarkupfalse%
\ DSourcesS{\isadigit{1}}{\isadigit{4}}{\isacharcolon}{\isachardoublequoteopen}DSources\ level{\isadigit{2}}\ sS{\isadigit{1}}{\isadigit{4}}\ {\isacharequal}\ {\isacharbraceleft}sS{\isadigit{1}}{\isadigit{1}}{\isacharbraceright}{\isachardoublequoteclose}\ \ \isacommand{by}\isamarkupfalse%
\ {\isacharparenleft}simp\ add{\isacharcolon}\ DSources{\isacharunderscore}def\ AbstrLevel{\isadigit{2}}{\isacharcomma}\ auto{\isacharparenright}\isanewline
\ \ \isacommand{have}\isamarkupfalse%
\ {\isachardoublequoteopen}Sources\ level{\isadigit{2}}\ sS{\isadigit{1}}{\isadigit{1}}\ {\isacharequal}\ {\isacharbraceleft}sS{\isadigit{9}}{\isacharbraceright}{\isachardoublequoteclose}\ \ \isacommand{by}\isamarkupfalse%
\ {\isacharparenleft}rule\ SourcesS{\isadigit{1}}{\isadigit{1}}{\isacharunderscore}L{\isadigit{2}}{\isacharparenright}\isanewline
\ \ \isacommand{with}\isamarkupfalse%
\ DSourcesS{\isadigit{1}}{\isadigit{4}}\ \isacommand{show}\isamarkupfalse%
\ {\isacharquery}thesis\ \ \isacommand{by}\isamarkupfalse%
\ \ {\isacharparenleft}simp\ add{\isacharcolon}\ Sources{\isacharunderscore}singleDSource{\isacharparenright}\isanewline
\isacommand{qed}\isamarkupfalse%
\endisatagproof
{\isafoldproof}%
\isadelimproof
\isanewline
\endisadelimproof
\isanewline
\isacommand{lemma}\isamarkupfalse%
\ SourcesS{\isadigit{1}}{\isadigit{5}}{\isacharunderscore}L{\isadigit{2}}{\isacharcolon}\ {\isachardoublequoteopen}Sources\ level{\isadigit{2}}\ sS{\isadigit{1}}{\isadigit{5}}\ {\isacharequal}\ {\isacharbraceleft}sS{\isadigit{9}}{\isacharcomma}\ sS{\isadigit{1}}{\isadigit{1}}{\isacharcomma}\ sS{\isadigit{1}}{\isadigit{4}}{\isacharbraceright}{\isachardoublequoteclose}\isanewline
\isadelimproof
\endisadelimproof
\isatagproof
\isacommand{proof}\isamarkupfalse%
\ {\isacharminus}\isanewline
\ \ \isacommand{have}\isamarkupfalse%
\ DSourcesS{\isadigit{1}}{\isadigit{5}}{\isacharcolon}{\isachardoublequoteopen}DSources\ level{\isadigit{2}}\ sS{\isadigit{1}}{\isadigit{5}}{\isacharequal}\ {\isacharbraceleft}sS{\isadigit{1}}{\isadigit{4}}{\isacharbraceright}{\isachardoublequoteclose}\ \ \isacommand{by}\isamarkupfalse%
\ {\isacharparenleft}simp\ add{\isacharcolon}\ DSources{\isacharunderscore}def\ AbstrLevel{\isadigit{2}}{\isacharcomma}\ auto{\isacharparenright}\isanewline
\ \ \isacommand{have}\isamarkupfalse%
\ {\isachardoublequoteopen}Sources\ level{\isadigit{2}}\ sS{\isadigit{1}}{\isadigit{4}}\ {\isacharequal}\ {\isacharbraceleft}sS{\isadigit{9}}{\isacharcomma}\ sS{\isadigit{1}}{\isadigit{1}}{\isacharbraceright}{\isachardoublequoteclose}\ \ \isacommand{by}\isamarkupfalse%
\ {\isacharparenleft}rule\ SourcesS{\isadigit{1}}{\isadigit{4}}{\isacharunderscore}L{\isadigit{2}}{\isacharparenright}\isanewline
\ \ \isacommand{with}\isamarkupfalse%
\ DSourcesS{\isadigit{1}}{\isadigit{5}}\ \isacommand{show}\isamarkupfalse%
\ {\isacharquery}thesis\ \isacommand{by}\isamarkupfalse%
\ \ {\isacharparenleft}simp\ add{\isacharcolon}\ Sources{\isacharunderscore}singleDSource{\isacharparenright}\isanewline
\isacommand{qed}\isamarkupfalse%
\endisatagproof
{\isafoldproof}%
\isadelimproof
\endisadelimproof
\isamarkupsubsection{Minimal sets of components to prove certain properties%
}
\isamarkuptrue%
\isacommand{lemma}\isamarkupfalse%
\ minSetOfComponentsTestL{\isadigit{2}}p{\isadigit{1}}{\isacharcolon}\isanewline
{\isachardoublequoteopen}minSetOfComponents\ level{\isadigit{2}}\ {\isacharbraceleft}data{\isadigit{1}}{\isadigit{0}}{\isacharcomma}\ data{\isadigit{1}}{\isadigit{3}}{\isacharbraceright}\ {\isacharequal}\ {\isacharbraceleft}sS{\isadigit{1}}{\isacharbraceright}{\isachardoublequoteclose}\isanewline
\isadelimproof
\endisadelimproof
\isatagproof
\isacommand{proof}\isamarkupfalse%
\ {\isacharminus}\ \isanewline
\ \ \isacommand{have}\isamarkupfalse%
\ outL{\isadigit{2}}{\isacharcolon}{\isachardoublequoteopen}outSetOfComponents\ level{\isadigit{2}}\ {\isacharbraceleft}data{\isadigit{1}}{\isadigit{0}}{\isacharcomma}\ data{\isadigit{1}}{\isadigit{3}}{\isacharbraceright}\ {\isacharequal}\ {\isacharbraceleft}sS{\isadigit{1}}{\isacharbraceright}{\isachardoublequoteclose}\isanewline
\ \ \ \ \isacommand{by}\isamarkupfalse%
\ {\isacharparenleft}simp\ add{\isacharcolon}\ outSetOfComponents{\isacharunderscore}def\ \ AbstrLevel{\isadigit{2}}{\isacharcomma}\ auto{\isacharparenright}\ \isanewline
\ \ \isacommand{have}\isamarkupfalse%
\ {\isachardoublequoteopen}Sources\ level{\isadigit{2}}\ sS{\isadigit{1}}\ {\isacharequal}\ {\isacharbraceleft}{\isacharbraceright}{\isachardoublequoteclose}\ \isacommand{by}\isamarkupfalse%
\ {\isacharparenleft}simp\ add{\isacharcolon}\ SourcesS{\isadigit{1}}{\isacharunderscore}L{\isadigit{2}}{\isacharparenright}\isanewline
\ \ \isacommand{with}\isamarkupfalse%
\ outL{\isadigit{2}}\ \isacommand{show}\isamarkupfalse%
\ {\isacharquery}thesis\ \isacommand{by}\isamarkupfalse%
\ {\isacharparenleft}simp\ add{\isacharcolon}\ \ minSetOfComponents{\isacharunderscore}def{\isacharparenright}\isanewline
\isacommand{qed}\isamarkupfalse%
\endisatagproof
{\isafoldproof}%
\isadelimproof
\isanewline
\endisadelimproof
\isanewline
\isacommand{lemma}\isamarkupfalse%
\ NOT{\isacharunderscore}noIrrelevantChannelsTestL{\isadigit{2}}p{\isadigit{1}}{\isacharcolon}\isanewline
{\isachardoublequoteopen}\ {\isasymnot}\ noIrrelevantChannels\ level{\isadigit{2}}\ {\isacharbraceleft}data{\isadigit{1}}{\isadigit{0}}{\isacharcomma}\ data{\isadigit{1}}{\isadigit{3}}{\isacharbraceright}{\isachardoublequoteclose}\isanewline
\isadelimproof
\endisadelimproof
\isatagproof
\isacommand{by}\isamarkupfalse%
\ {\isacharparenleft}simp\ add{\isacharcolon}\ noIrrelevantChannels{\isacharunderscore}def\ systemIN{\isacharunderscore}def\ minSetOfComponentsTestL{\isadigit{2}}p{\isadigit{1}}\ AbstrLevel{\isadigit{2}}{\isacharparenright}%
\endisatagproof
{\isafoldproof}%
\isadelimproof
\isanewline
\endisadelimproof
\isanewline
\isacommand{lemma}\isamarkupfalse%
\ NOT{\isacharunderscore}allNeededINChannelsTestL{\isadigit{2}}p{\isadigit{1}}{\isacharcolon}\isanewline
{\isachardoublequoteopen}{\isasymnot}\ allNeededINChannels\ \ level{\isadigit{2}}\ {\isacharbraceleft}data{\isadigit{1}}{\isadigit{0}}{\isacharcomma}\ data{\isadigit{1}}{\isadigit{3}}{\isacharbraceright}{\isachardoublequoteclose}\isanewline
\isadelimproof
\endisadelimproof
\isatagproof
\isacommand{by}\isamarkupfalse%
\ {\isacharparenleft}simp\ add{\isacharcolon}\ allNeededINChannels{\isacharunderscore}def\ minSetOfComponentsTestL{\isadigit{2}}p{\isadigit{1}}\ \ systemIN{\isacharunderscore}def\ AbstrLevel{\isadigit{2}}{\isacharparenright}%
\endisatagproof
{\isafoldproof}%
\isadelimproof
\isanewline
\endisadelimproof
\isanewline
\isacommand{lemma}\isamarkupfalse%
\ minSetOfComponentsTestL{\isadigit{2}}p{\isadigit{2}}{\isacharcolon}\isanewline
{\isachardoublequoteopen}minSetOfComponents\ level{\isadigit{2}}\ {\isacharbraceleft}data{\isadigit{1}}{\isacharcomma}\ data{\isadigit{1}}{\isadigit{2}}{\isacharbraceright}\ {\isacharequal}\ {\isacharbraceleft}sS{\isadigit{2}}{\isacharcomma}\ sS{\isadigit{4}}{\isacharcomma}\ sS{\isadigit{5}}{\isacharcomma}\ sS{\isadigit{6}}{\isacharbraceright}{\isachardoublequoteclose}\isanewline
\isadelimproof
\endisadelimproof
\isatagproof
\isacommand{proof}\isamarkupfalse%
\ {\isacharminus}\ \isanewline
\ \ \isacommand{have}\isamarkupfalse%
\ outL{\isadigit{2}}{\isacharcolon}{\isachardoublequoteopen}outSetOfComponents\ level{\isadigit{2}}\ {\isacharbraceleft}data{\isadigit{1}}{\isacharcomma}\ data{\isadigit{1}}{\isadigit{2}}{\isacharbraceright}\ {\isacharequal}\ {\isacharbraceleft}sS{\isadigit{6}}{\isacharbraceright}{\isachardoublequoteclose}\isanewline
\ \ \ \ \isacommand{by}\isamarkupfalse%
\ {\isacharparenleft}simp\ add{\isacharcolon}\ outSetOfComponents{\isacharunderscore}def\ \ AbstrLevel{\isadigit{2}}{\isacharcomma}\ auto{\isacharparenright}\ \isanewline
\ \ \isacommand{have}\isamarkupfalse%
\ {\isachardoublequoteopen}Sources\ level{\isadigit{2}}\ sS{\isadigit{6}}\ {\isacharequal}\ {\isacharbraceleft}sS{\isadigit{2}}{\isacharcomma}\ sS{\isadigit{4}}{\isacharcomma}\ sS{\isadigit{5}}{\isacharbraceright}{\isachardoublequoteclose}\isanewline
\ \ \ \ \isacommand{by}\isamarkupfalse%
\ \ {\isacharparenleft}simp\ add{\isacharcolon}\ SourcesS{\isadigit{6}}{\isacharunderscore}L{\isadigit{2}}{\isacharparenright}\ \isanewline
\ \ \isacommand{with}\isamarkupfalse%
\ outL{\isadigit{2}}\ \isacommand{show}\isamarkupfalse%
\ {\isacharquery}thesis\ \isanewline
\ \ \ \ \isacommand{by}\isamarkupfalse%
\ {\isacharparenleft}simp\ add{\isacharcolon}\ \ minSetOfComponents{\isacharunderscore}def{\isacharparenright}\ \isanewline
\isacommand{qed}\isamarkupfalse%
\endisatagproof
{\isafoldproof}%
\isadelimproof
\isanewline
\endisadelimproof
\ \isanewline
\isacommand{lemma}\isamarkupfalse%
\ noIrrelevantChannelsTestL{\isadigit{2}}p{\isadigit{2}}{\isacharcolon}\isanewline
{\isachardoublequoteopen}noIrrelevantChannels\ level{\isadigit{2}}\ \ {\isacharbraceleft}data{\isadigit{1}}{\isacharcomma}\ data{\isadigit{1}}{\isadigit{2}}{\isacharbraceright}{\isachardoublequoteclose}\isanewline
\isadelimproof
\endisadelimproof
\isatagproof
\isacommand{by}\isamarkupfalse%
\ {\isacharparenleft}simp\ add{\isacharcolon}\ noIrrelevantChannels{\isacharunderscore}def\ systemIN{\isacharunderscore}def\ minSetOfComponentsTestL{\isadigit{2}}p{\isadigit{2}}\ AbstrLevel{\isadigit{2}}{\isacharparenright}%
\endisatagproof
{\isafoldproof}%
\isadelimproof
\isanewline
\endisadelimproof
\isanewline
\isacommand{lemma}\isamarkupfalse%
\ allNeededINChannelsTestL{\isadigit{2}}p{\isadigit{2}}{\isacharcolon}\isanewline
{\isachardoublequoteopen}allNeededINChannels\ \ level{\isadigit{2}}\ {\isacharbraceleft}data{\isadigit{1}}{\isacharcomma}\ data{\isadigit{1}}{\isadigit{2}}{\isacharbraceright}{\isachardoublequoteclose}\isanewline
\isadelimproof
\endisadelimproof
\isatagproof
\isacommand{by}\isamarkupfalse%
\ {\isacharparenleft}simp\ add{\isacharcolon}\ allNeededINChannels{\isacharunderscore}def\ minSetOfComponentsTestL{\isadigit{2}}p{\isadigit{2}}\ \ systemIN{\isacharunderscore}def\ AbstrLevel{\isadigit{2}}{\isacharparenright}%
\endisatagproof
{\isafoldproof}%
\isadelimproof
\isanewline
\endisadelimproof
\isanewline
\isacommand{lemma}\isamarkupfalse%
\ minSetOfComponentsTestL{\isadigit{1}}p{\isadigit{3}}{\isacharcolon}\isanewline
{\isachardoublequoteopen}minSetOfComponents\ level{\isadigit{1}}\ {\isacharbraceleft}data{\isadigit{1}}{\isacharcomma}\ data{\isadigit{1}}{\isadigit{0}}{\isacharcomma}\ data{\isadigit{1}}{\isadigit{1}}{\isacharbraceright}\ {\isacharequal}\ {\isacharbraceleft}sA{\isadigit{1}}{\isadigit{2}}{\isacharcomma}\ sA{\isadigit{1}}{\isadigit{1}}{\isacharcomma}\ sA{\isadigit{2}}{\isadigit{1}}{\isacharbraceright}{\isachardoublequoteclose}\isanewline
\isadelimproof
\endisadelimproof
\isatagproof
\isacommand{proof}\isamarkupfalse%
\ {\isacharminus}\ \isanewline
\ \ \isacommand{have}\isamarkupfalse%
\ sg{\isadigit{1}}{\isacharcolon}{\isachardoublequoteopen}outSetOfComponents\ level{\isadigit{1}}\ {\isacharbraceleft}data{\isadigit{1}}{\isacharcomma}\ data{\isadigit{1}}{\isadigit{0}}{\isacharcomma}\ data{\isadigit{1}}{\isadigit{1}}{\isacharbraceright}\ {\isacharequal}\ {\isacharbraceleft}sA{\isadigit{1}}{\isadigit{2}}{\isacharcomma}\ sA{\isadigit{2}}{\isadigit{1}}{\isacharbraceright}{\isachardoublequoteclose}\isanewline
\ \ \ \ \isacommand{by}\isamarkupfalse%
\ {\isacharparenleft}simp\ add{\isacharcolon}\ outSetOfComponents{\isacharunderscore}def\ \ AbstrLevel{\isadigit{1}}{\isacharcomma}\ auto{\isacharparenright}\ \ \isanewline
\ \ \isacommand{have}\isamarkupfalse%
\ {\isachardoublequoteopen}DSources\ level{\isadigit{1}}\ sA{\isadigit{1}}{\isadigit{2}}\ {\isacharequal}\ {\isacharbraceleft}{\isacharbraceright}{\isachardoublequoteclose}\isanewline
\ \ \ \ \isacommand{by}\isamarkupfalse%
\ {\isacharparenleft}simp\ add{\isacharcolon}\ DSources{\isacharunderscore}def\ AbstrLevel{\isadigit{1}}{\isacharcomma}\ auto{\isacharparenright}\ \isanewline
\ \ \isacommand{hence}\isamarkupfalse%
\ sg{\isadigit{2}}{\isacharcolon}{\isachardoublequoteopen}Sources\ level{\isadigit{1}}\ sA{\isadigit{1}}{\isadigit{2}}\ {\isacharequal}\ {\isacharbraceleft}{\isacharbraceright}{\isachardoublequoteclose}\isanewline
\ \ \ \ \isacommand{by}\isamarkupfalse%
\ {\isacharparenleft}simp\ add{\isacharcolon}\ DSourcesEmptySources{\isacharparenright}\ \ \isanewline
\ \ \isacommand{have}\isamarkupfalse%
\ sg{\isadigit{3}}{\isacharcolon}{\isachardoublequoteopen}DSources\ level{\isadigit{1}}\ sA{\isadigit{2}}{\isadigit{1}}\ {\isacharequal}\ {\isacharbraceleft}sA{\isadigit{1}}{\isadigit{1}}{\isacharbraceright}{\isachardoublequoteclose}\isanewline
\ \ \ \ \isacommand{by}\isamarkupfalse%
\ {\isacharparenleft}simp\ add{\isacharcolon}\ DSources{\isacharunderscore}def\ AbstrLevel{\isadigit{1}}{\isacharcomma}\ auto{\isacharparenright}\ \isanewline
\ \ \isacommand{have}\isamarkupfalse%
\ sg{\isadigit{4}}{\isacharcolon}{\isachardoublequoteopen}DSources\ level{\isadigit{1}}\ sA{\isadigit{1}}{\isadigit{1}}\ {\isacharequal}\ {\isacharbraceleft}{\isacharbraceright}{\isachardoublequoteclose}\isanewline
\ \ \ \ \isacommand{by}\isamarkupfalse%
\ {\isacharparenleft}simp\ add{\isacharcolon}\ DSources{\isacharunderscore}def\ AbstrLevel{\isadigit{1}}{\isacharcomma}\ auto{\isacharparenright}\ \ \isanewline
\ \ \isacommand{hence}\isamarkupfalse%
\ {\isachardoublequoteopen}Sources\ level{\isadigit{1}}\ sA{\isadigit{2}}{\isadigit{1}}\ {\isacharequal}\ {\isacharbraceleft}sA{\isadigit{1}}{\isadigit{1}}{\isacharbraceright}{\isachardoublequoteclose}\isanewline
\ \ \ \ \isacommand{by}\isamarkupfalse%
\ {\isacharparenleft}metis\ SourcesOnlyDSources\ sg{\isadigit{3}}\ singleton{\isacharunderscore}iff{\isacharparenright}\isanewline
\ \ \isacommand{from}\isamarkupfalse%
\ this\ \isakeyword{and}\ sg{\isadigit{1}}\ \isakeyword{and}\ sg{\isadigit{2}}\ \isacommand{show}\isamarkupfalse%
\ {\isacharquery}thesis\isanewline
\ \ \ \ \ \isacommand{by}\isamarkupfalse%
\ {\isacharparenleft}simp\ add{\isacharcolon}\ \ minSetOfComponents{\isacharunderscore}def{\isacharcomma}\ blast{\isacharparenright}\ \isanewline
\isacommand{qed}\isamarkupfalse%
\endisatagproof
{\isafoldproof}%
\isadelimproof
\isanewline
\endisadelimproof
\isanewline
\isacommand{lemma}\isamarkupfalse%
\ noIrrelevantChannelsTestL{\isadigit{1}}p{\isadigit{3}}{\isacharcolon}\isanewline
{\isachardoublequoteopen}noIrrelevantChannels\ level{\isadigit{1}}\ \ {\isacharbraceleft}data{\isadigit{1}}{\isacharcomma}\ data{\isadigit{1}}{\isadigit{0}}{\isacharcomma}\ data{\isadigit{1}}{\isadigit{1}}{\isacharbraceright}{\isachardoublequoteclose}\isanewline
\isadelimproof
\endisadelimproof
\isatagproof
\isacommand{by}\isamarkupfalse%
\ {\isacharparenleft}simp\ add{\isacharcolon}\ noIrrelevantChannels{\isacharunderscore}def\ systemIN{\isacharunderscore}def\ minSetOfComponentsTestL{\isadigit{1}}p{\isadigit{3}}\ AbstrLevel{\isadigit{1}}{\isacharparenright}%
\endisatagproof
{\isafoldproof}%
\isadelimproof
\isanewline
\endisadelimproof
\isanewline
\isacommand{lemma}\isamarkupfalse%
\ allNeededINChannelsTestL{\isadigit{1}}p{\isadigit{3}}{\isacharcolon}\isanewline
{\isachardoublequoteopen}allNeededINChannels\ \ level{\isadigit{1}}\ {\isacharbraceleft}data{\isadigit{1}}{\isacharcomma}\ data{\isadigit{1}}{\isadigit{0}}{\isacharcomma}\ data{\isadigit{1}}{\isadigit{1}}{\isacharbraceright}{\isachardoublequoteclose}\isanewline
\isadelimproof
\endisadelimproof
\isatagproof
\isacommand{by}\isamarkupfalse%
\ {\isacharparenleft}simp\ add{\isacharcolon}\ allNeededINChannels{\isacharunderscore}def\ minSetOfComponentsTestL{\isadigit{1}}p{\isadigit{3}}\ \ systemIN{\isacharunderscore}def\ AbstrLevel{\isadigit{1}}{\isacharparenright}%
\endisatagproof
{\isafoldproof}%
\isadelimproof
\isanewline
\endisadelimproof
\isanewline
\isacommand{lemma}\isamarkupfalse%
\ minSetOfComponentsTestL{\isadigit{2}}p{\isadigit{3}}{\isacharcolon}\isanewline
{\isachardoublequoteopen}minSetOfComponents\ level{\isadigit{2}}\ {\isacharbraceleft}data{\isadigit{1}}{\isacharcomma}\ data{\isadigit{1}}{\isadigit{0}}{\isacharcomma}\ data{\isadigit{1}}{\isadigit{1}}{\isacharbraceright}\ {\isacharequal}\ {\isacharbraceleft}sS{\isadigit{1}}{\isacharcomma}\ sS{\isadigit{2}}{\isacharcomma}\ sS{\isadigit{3}}{\isacharbraceright}{\isachardoublequoteclose}\isanewline
\isadelimproof
\endisadelimproof
\isatagproof
\isacommand{proof}\isamarkupfalse%
\ {\isacharminus}\ \isanewline
\ \ \isacommand{have}\isamarkupfalse%
\ sg{\isadigit{1}}{\isacharcolon}{\isachardoublequoteopen}outSetOfComponents\ level{\isadigit{2}}\ {\isacharbraceleft}data{\isadigit{1}}{\isacharcomma}\ data{\isadigit{1}}{\isadigit{0}}{\isacharcomma}\ data{\isadigit{1}}{\isadigit{1}}{\isacharbraceright}\ {\isacharequal}\ {\isacharbraceleft}sS{\isadigit{1}}{\isacharcomma}\ sS{\isadigit{3}}{\isacharbraceright}{\isachardoublequoteclose}\isanewline
\ \ \ \ \isacommand{by}\isamarkupfalse%
\ {\isacharparenleft}simp\ add{\isacharcolon}\ outSetOfComponents{\isacharunderscore}def\ \ AbstrLevel{\isadigit{2}}{\isacharcomma}\ auto{\isacharparenright}\ \ \isanewline
\ \ \isacommand{have}\isamarkupfalse%
\ sS{\isadigit{1}}{\isacharcolon}{\isachardoublequoteopen}Sources\ level{\isadigit{2}}\ sS{\isadigit{1}}\ {\isacharequal}\ {\isacharbraceleft}{\isacharbraceright}{\isachardoublequoteclose}\ \isacommand{by}\isamarkupfalse%
\ {\isacharparenleft}simp\ add{\isacharcolon}\ SourcesS{\isadigit{1}}{\isacharunderscore}L{\isadigit{2}}{\isacharparenright}\isanewline
\ \ \isacommand{have}\isamarkupfalse%
\ {\isachardoublequoteopen}Sources\ level{\isadigit{2}}\ sS{\isadigit{3}}\ {\isacharequal}\ {\isacharbraceleft}sS{\isadigit{2}}{\isacharbraceright}{\isachardoublequoteclose}\ \isacommand{by}\isamarkupfalse%
\ {\isacharparenleft}simp\ add{\isacharcolon}\ SourcesS{\isadigit{3}}{\isacharunderscore}L{\isadigit{2}}{\isacharparenright}\isanewline
\ \ \ \isacommand{with}\isamarkupfalse%
\ sg{\isadigit{1}}\ sS{\isadigit{1}}\ \isacommand{show}\isamarkupfalse%
\ {\isacharquery}thesis\isanewline
\ \ \ \ \ \isacommand{by}\isamarkupfalse%
\ {\isacharparenleft}simp\ add{\isacharcolon}\ \ minSetOfComponents{\isacharunderscore}def{\isacharcomma}\ blast{\isacharparenright}\ \isanewline
\isacommand{qed}\isamarkupfalse%
\endisatagproof
{\isafoldproof}%
\isadelimproof
\isanewline
\endisadelimproof
\ \isanewline
\isacommand{lemma}\isamarkupfalse%
\ noIrrelevantChannelsTestL{\isadigit{2}}p{\isadigit{3}}{\isacharcolon}\isanewline
{\isachardoublequoteopen}noIrrelevantChannels\ level{\isadigit{2}}\ \ {\isacharbraceleft}data{\isadigit{1}}{\isacharcomma}\ data{\isadigit{1}}{\isadigit{0}}{\isacharcomma}\ data{\isadigit{1}}{\isadigit{1}}{\isacharbraceright}{\isachardoublequoteclose}\isanewline
\isadelimproof
\endisadelimproof
\isatagproof
\isacommand{by}\isamarkupfalse%
\ {\isacharparenleft}simp\ add{\isacharcolon}\ noIrrelevantChannels{\isacharunderscore}def\ systemIN{\isacharunderscore}def\ minSetOfComponentsTestL{\isadigit{2}}p{\isadigit{3}}\ AbstrLevel{\isadigit{2}}{\isacharparenright}%
\endisatagproof
{\isafoldproof}%
\isadelimproof
\isanewline
\endisadelimproof
\isanewline
\isacommand{lemma}\isamarkupfalse%
\ allNeededINChannelsTestL{\isadigit{2}}p{\isadigit{3}}{\isacharcolon}\isanewline
{\isachardoublequoteopen}allNeededINChannels\ \ level{\isadigit{2}}\ {\isacharbraceleft}data{\isadigit{1}}{\isacharcomma}\ data{\isadigit{1}}{\isadigit{0}}{\isacharcomma}\ data{\isadigit{1}}{\isadigit{1}}{\isacharbraceright}{\isachardoublequoteclose}\isanewline
\isadelimproof
\endisadelimproof
\isatagproof
\isacommand{by}\isamarkupfalse%
\ {\isacharparenleft}simp\ add{\isacharcolon}\ allNeededINChannels{\isacharunderscore}def\ minSetOfComponentsTestL{\isadigit{2}}p{\isadigit{3}}\ \ systemIN{\isacharunderscore}def\ AbstrLevel{\isadigit{2}}{\isacharparenright}%
\endisatagproof
{\isafoldproof}%
\isadelimproof
\isanewline
\endisadelimproof
\isadelimtheory
\isanewline
\endisadelimtheory
\isatagtheory
\isacommand{end}\isamarkupfalse%
\endisatagtheory
{\isafoldtheory}%
\isadelimtheory
\isanewline
\endisadelimtheory
\end{isabellebody}%

\newpage
\thispagestyle{empty}
\bibliographystyle{abbrv}

\end{document}